# The genetic code, 8-dimensional hypercomplex numbers and dyadic shifts

Sergey V. Petoukhov

Head of Laboratory of Biomechanical System, Mechanical Engineering Research Institute
of the Russian Academy of Sciences, Moscow
spetoukhov@gmail.com, http://petoukhov.com/

**Comment**: Some materials of this article were presented in the International Workshop on Genomic Signal Processing (Bucharest, Romania, June 27-28, 2011), the 2011 International Conference on Bioinformatics & Computational Biology (Las Vegas, USA, July 18-21, 2011) and the International Conference «The information society at the crossroads» (Vienna, Austria, 3-7 June 2015).

**Abstract**: Matrix forms of the representation of the multi-level system of molecular-genetic alphabets have revealed algebraic properties of this system. Families of genetic (4*4)- and (8*8)-matrices show unexpected connections of the genetic system with Walsh functions and Hadamard matrices, which are known in theory of noise-immunity coding, digital communication and digital holography. Dyadic-shift decompositions of such genetic matrices lead to sets of sparse matrices. Each of these sets is closed in relation to multiplication and defines relevant algebra of hypercomplex numbers. It is shown that genetic Hadamard matrices are identical to matrix representations of Hamilton quaternions and its complexification in the case of unit coordinates. The diversity of known dialects of the genetic code is analyzed from the viewpoint of the genetic algebras. An algebraic analogy with Punnett squares for inherited traits is shown. Our results are used in analyzing genetic phenomena. The statement about existence of the geno-logic code in DNA and epigenetics on the base of the spectral logic of systems of Boolean functions is put forward. Our results show promising ways to develop algebraic-logical biology, in particular, in connection with the logic holography on Walsh functions.

## Contents

## 1. Introduction

Science has led to a new understanding of life itself: «*Life is a partnership between genes and mathematics*» [Stewart, 1999]. But what kind of mathematics is a partner with the genetic code? Trying to find such mathematics, we have turned to study the multi-level system of interrelated molecular-genetic alphabets. On this way we were surprised to find connections of this genetic system with well-known formalisms of the engineering theory of noise-immunity coding: Kronecker products of matrices; orthogonal systems of Walsh functions; Hadamard matrices; a group of dyadic shifts; hypercomplex numbers, etc. This article is devoted to some of our results of such studing of the phenomenologic system of interrelated genetic alphabets.

Alphabets play a basic role in communication technologies. In any communication system of "transmitter-receiver" the receiver always knows the alphabet of signals, which are used by the transmitter. In linguistics, each alphabet has a complex multi-level structure because it contains sets of vowels and consonants where, in some languages, the set of vowels is divided into sub-sets of short and long sounds, and the set of consonants is divided into subsets of voiced and voiceless consonants, etc. Quantities of members in all of these parts of linguistic alphabets are not interrelated by means of known regularities of algebraic connections. We have discovered that the situation in the multi-level system of genetic alphabets is quite different: many parts of this system are closely interconnected by means of deep algebraic regularities and formalisms which are well-known in communication technologies as said before.

It is known that the molecular-genetic system of living matter includes the following alphabets, each of which can be considered as a part of a complex alphabetic system:

- 4-letter alphabet of nitrogenous bases;
- 64-letter alphabet of triplets;- 2-letter alphabet of "weak and strong roots" of triplets;
- 20-letter alphabet of amino acids;
- 2-letter alphabet "purines vs. pyrimidines";
- 2-letter alphabet "strong vs. weak hydrogen bonds''
- 2-letter alphabet "keto vs. amino", etc.

(See the wide list of genetic alphabets in [Karlin, Ost, Blaisdell, 1989]).

So, the molecular-genetic system is a multi-lingual system. Any sequence of nucleotides can be read from viewpoints of different genetic languages depending on the reader alphabet. It can be added that the typical expression "the genetic code" means an interrelation between elements of two of these genetic alphabets: the alphabet of triplets and the alphabet of amino acids and stop-codons.

Genetic information from the micro-world of genetic molecules dictates constructions in the macro-world of living organisms under strong noise and interference. The Mendel's law of independent inheritance of different traits (for example, colors of hair, skin and eyes are inherited independently of each other) testifies that this dictation is realized through different independent channels by means of unknown algorithms of multi-channel noise-immunity coding.

It means that each living organism is an algorithmic machine of multi-channel noise-immunity coding. To understand this machine we should use the theory of noise-immunity coding.

Genetic information is transferred by means of discrete elements. General theory of signal processing utilizes the encoding of discrete signals by means of special mathematical matrices and spectral representations of signals to increase reliability and efficiency of information transfer [Ahmed, Rao, 1975; Sklar, 2001; etc]. A typical example of such matrices is the family of Hadamard matrices. Rows of Hadamard matrices form an orthogonal system of Walsh functions which is used for the spectral representation and transfer of discrete signals [Ahmed, Rao, 1975; Geramita, 1979; Yarlagadda, Hershey, 1997]. An investigation of structural analogies between digital informatics and genetic informatics is one of the important tasks of modern science in connection with the development of DNA-computers and bioinformatics. The author investigates molecular structures of the system of genetic alphabets by means of matrix methods of discrete signal processing [Petoukhov, 2001, 2005a,b, 2008a-c; Petoukhov, He, 2010, etc.].

The article describes author's results about relations of matrix forms of representation of the system of genetic alphabets with special systems of 8-dimensional hypercomplex numbers (they differ from the Cayley's octonions). The discovery of these relationships is significant from some viewpoints. For example, it is interesting because systems of 8-dimensional hypercomplex numbers (first of all, Cayley's octonions and split-octonions)–are one of key objects of mathematical natural sciences today. They relate to a number of exceptional structures in mathematics, among them the exceptional Lie groups; they have applications in many fields such as string theory, special relativity, the supersymmetric quantum mechanics, quantum logic, etc. (see for example, [http://en.wikipedia.org/wiki/Octonion; http://en.wikipedia.org/wiki/Split-octonion; Dixon, 1994; Dray & Manoque, 2009; Lisi, 2007; Nurowski, 2009]). The term "octet" is also used frequently in phenomenologic laws of science: the Eightfold way by M.Gell-Mann and Y.Ne'eman (1964) in physics; the octet rule in chemistry [http://en.wikipedia.org/wiki/Octet_rule], etc. In view of these facts one can think that genetic systems of 8-dimensional numbers will become one of the interesting parts of mathematical natural sciences.

In addition, hypercomplex numbers are widely used in digital signal processing [Bulow, 1999, 2001; Chernov, 2002; Felberg, 2001; Furman et al., 2003; McCarthy, 2000; Sin'kov, 2010; Toyoshima, 1999, 2002; etc.]. Formalisms of multi-dimensional vector spaces are one of basic formalisms in digital communication technologies, systems of artificial intelligence, pattern recognition, training of robots, detection of errors in the transmission of information, etc. Revealed genetic types of hypercomplex numbers can be useful to answer many questions of bioinformatics and to develop new kinds of genetic algorithms.

Hadamard matrices and orthogonal systems of Walsh functions are among the most used tools for error-correcting coding information, and for many other applications in digital signal processing [Ahmed, Rao, 1975; Geramita, 1979; Yarlagadda, Hershey, 1997]. As noted in the article [Seberry, et al., 2005], many tens of thousands of works are devoted to diverse applications of Hadamard matrices for signal processing. Our discovery of relations of the system of genetic alphabets with the special systems of 4-dimensional and 8-dimensional hypercomplex numbers and with special Hadamard matrices helps to establish the kind of mathematics which is a partner of the molecular-genetic system.

Hypercomplex numbers, first of all, Hamilton quaternions and their complexification (biquaternions) are widely applied in theoretical physics. The author shows that matrix genetics reveals a special connection of the system of genetic alphabets with Hamilton quaternions and their complexification. These results give new promising ways to build a bridge between theoretical physics and mathematical biology. They can be considered as a new step to knowledge of a mathematical unity of the nature.

## 2. Matrices of genetic multiplets and matrices of diadic shifts

For special decompositions of genetic matrices we will use structures of matrices of dyadic shifts long known in theory of discrete signal processing, sequency theory by Harmuth, etc. [Ahmed, Rao, 1975; Harmuth, 1977, §1.2.6]. Some relations of matrices of dyadic shifts with the genetic code have been described by the author previously [Petoukhov, 2008a, § 2.7; Petoukhov, 2010; Petoukhov & He, 2010, Chapter 1, Figure 5]. These matrices are constructed on the basis of mathematical operation of modulo-2 addition for binary numbers.

Modulo-2 addition is utilized broadly in the theory of discrete signal processing as a fundamental operation for binary variables. By definition, the modulo-2 addition of two numbers written in binary notation is made in a bitwise manner in accordance with the following rules:

$$0 + 0 = 0,\ 0 + 1 = 1,\ 1+ 0 = 1,\ 1+ 1 = 0 \qquad (1)$$

For example, modulo-2 addition of two binary numbers 110 and 101, which are equal to 6 and 5 respectively in decimal notation, gives the result 110 $\oplus$ 101 = 011, which is equal to 3 in decimal notation ($\oplus$ is the symbol for modulo-2 addition). The set of binary numbers

$$000, 001, 010, 011, 100, 101, 110, 111 \qquad (2)$$

forms a diadic group, in which modulo-2 addition serves as the group operation [Harmuth, 1989]. The distance in this symmetry group is known as the Hamming distance. Since the Hamming distance satisfies the conditions of a metric group, the diadic group is a metric group. The modulo-2 addition of any two binary numbers from (2) always results in a new number from the same series. The number 000 serves as the unit element of this group: for example, 010 $\oplus$ 000 = 010. The reverse element for any number in this group is the number itself: for example, 010 $\oplus$ 010 = 000. The series (2) is transformed by modulo-2 addition with the binary number 001 into a new series of the same numbers:

$$001, 000, 011, 010, 101, 100, 111, 110 \qquad (3)$$

Such changes in the initial binary sequence, produced by modulo-2 addition of its members with any binary numbers (2), are termed diadic shifts [Ahmed and Rao, 1975; Harmuth, 1989]. If any system of elements demonstrates its connection with diadic shifts, it indicates that the structural organization of its system is related to the logic of modulo-2 addition. The article shows that the structural organization of genetic systems is related to logic of modulo-2 addition.

| | 000 (0) | 001 (1) | 010 (2) | 011 (3) | 100 (4) | 101 (5) | 110 (6) | 111 (7) |
|---|---|---|---|---|---|---|---|---|
| 000 (0) | 000 (0) | 001 (1) | 010 (2) | 011 (3) | 100 (4) | 101 (5) | 110 (6) | 111 (7) |
| 001 (1) | 001 (1) | 000 (0) | 011 (3) | 010 (2) | 101 (5) | 100 (4) | 111 (7) | 110 (6) |
| 010 (2) | 010 (2) | 011 (3) | 000 (0) | 001 (1) | 110 (6) | 111 (7) | 100 (4) | 101 (5) |
| 011 (3) | 011 (3) | 010 (2) | 001 (1) | 000 (0) | 111 (7) | 110 (6) | 101 (5) | 100 (4) |
| 100 (4) | 100 (4) | 101 (5) | 110 (6) | 111 (7) | 000 (0) | 001 (1) | 010 (2) | 011 (3) |
| 101 (5) | 101 (5) | 100 (4) | 111 (7) | 110 (6) | 001 (1) | 000 (0) | 011 (3) | 010 (2) |
| 110 (6) | 110 (6) | 111 (7) | 100 (4) | 101 (5) | 010 (2) | 011 (3) | 000 (0) | 001 (1) |
| 111 (7) | 111 (7) | 110 (6) | 101 (5) | 100 (4) | 011 (3) | 010 (2) | 001 (1) | 000 (0) |

Figure 1. An example of a matrix of dyadic shifts. Parentheses contain expressions of numbers in decimal notation.

Figure 1 shows an example of matrices of dyadic shifts. Each row and each column are numerated by means of binary numbers from the dyadic group (2). Each of matrix cells has its binary numeration from the same dyadic group (2). This numeration of any cell is the result of modulo-2 addition of binary numerations of the column and the row of this cell. For example, the cell from the column 110 and the row 101 obtains the binary numeration 011 by means of such addition. Such numerations of matrix cells are termed "dyadic-shift numerations" (or simply "dyadic numeration"). One can see that the disposition of even and odd numbers 0, 1, 2,…, 7 of the dyadic group in this matrix of dyadic shifts is identical to the disposition of white and black cells in a chessboard.

Now let us proceed to the system of genetic alphabets. All living organisms have the same main set of molecular-genetic alphabets in which genetic multiplets play a significant role. Four monoplets forms the alphabet of nitrogenous bases: A (adenine), C (cytosine), G (guanine), U/T (uracil in RNA or thymine in DNA); 64 triplets encode amino acids and termination signals; each protein is encoded by more or less long multiplets (n-plets). Each complete set of n-plets contains $4^n$ different n-plets and can be considered as a separate alphabet. On the basis of the idea about analogies between computer informatics and genetic informatics, each of the formal alphabets of genetic n-plets can be presented in a general form of the relevant square matrix (genomatrix) $[C\ A;\ U\ G]^{(n)}$ of the Kronecker family (Figure 4) [Petoukhov, 2001, 2005, 2208a,b]. (One should note that, by analogy with linguistic alphabets in different languages including dead languages, not all the formal genetic alphabets and not all the elements of each genetic alphabet are necessarily used in phenomenological constructions of molecular genetics). Here A, C, G, U are the letters of the alphabet of 1-plets of nitrogenous bases (Figure 2), (n) means the Kronecker exponentiation. One can remind that Kronecker product of matrices is an important tool in the theory of linear spaces and operators; it is associated with the tasks of a special union of two linear spaces with dimensions "n" and "m" into a linear space with the higher dimension "n*m", and also with the tasks of constructing the matrix operators in this (n*m)-space on the basis of operators of the initial spaces of smaller dimensions "n" and "m" [Halmos, 1974]. Each genomatrix $[C\ A;\ U\ G]^{(n)}$ contains a complete set of n-plets as its matrix elements (Figure 4). For example, the (8x8)-genomatrix $[C\ A;\ U\ G]^{(3)}$ contains all 64 triplets which encode 20 amino acids and stop-signals. We will show that these genetic matrices have phenomenological relations with the logic of dyadic shifts.

The 4-letter alphabet of nitrogenous bases contains the 4 specific polyatomic constructions with the special biochemical properties (Figure 2). The set of these 4 constructions is not absolutely heterogeneous, but it bears the substantial symmetric system of distinctive-uniting attributes (or, more precisely, pairs of "attribute-antiattribute"). This system of pairs of opposite attributes divides this 4-letter alphabet into various three pairs of letters by all three possible ways; letters of each such pair are equivalent to each other in accordance with one of these attributes or with its absence.

The system of such attributes divides this 4-letter alphabet into the following three pairs of letters, which are equivalent from a viewpoint of one of these attributes or its absence: 1) C = U & A = G (according to the binary-opposite attributes: "pyrimidine" or "non-pyrimidine", which is purine); 2) A = C & G = U (according to the attributes: amino-mutating or non-amino-mutating under action of nitrous acid $HNO_2$ [Wittmann, 1961; Ycas, 1969]; the same division is given by the attributes "keto" or "amino" [Karlin, Ost, Blaisdell, 1989]; 3) C = G & A = U (according to the attributes: three or two hydrogen bonds are materialized in these complementary pairs). The possibility of such division of the genetic alphabet into three binary sub-alphabets is known [Karlin, Ost, Blaisdell, 1989]. We utilize these known sub-alphabets in the field of matrix genetics which studies matrix forms of representation of the genetic alphabets. Each of the monoplets A, C, G, U/T can be symbolized by appropriate binary symbols "0" or "1" on the basis of these sub-alphabets. Then these binary symbols can be used for a binary numeration of columns and rows of the matrices in the Kronecker family of genomatrices of n-plets.

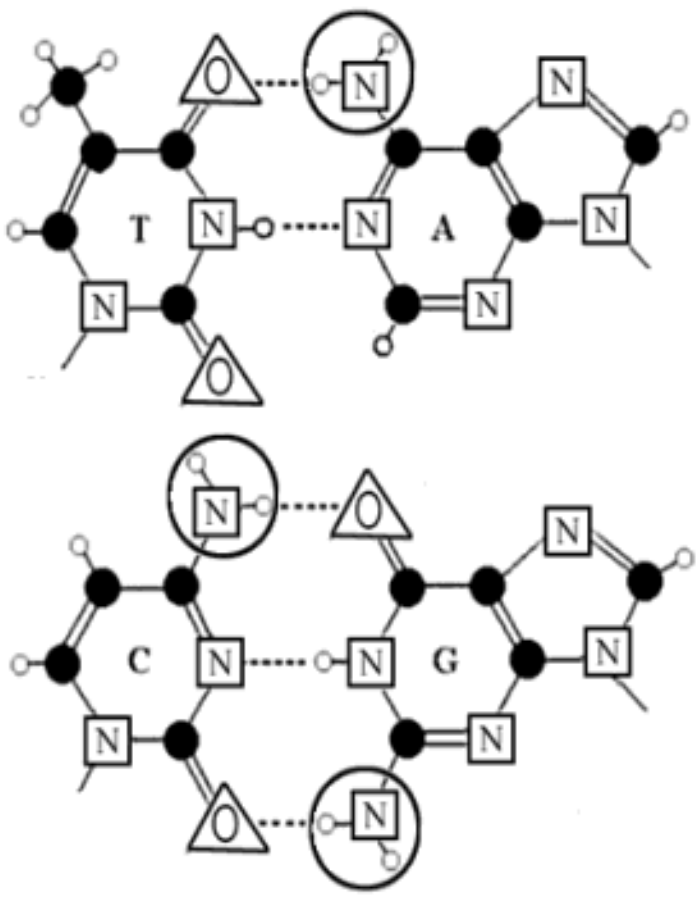

Figure 2. The complementary pairs of the 4 nitrogenous bases in DNA: A-T (adenine and thymine), C-G (cytosine and guanine). Hydrogen bonds in these pairs are shown by dotted lines. Black circles are atoms of carbon; small white circles are atoms of hydrogen; squares with the letter N are atoms of nitrogen; triangles with the letter O are atoms of oxygen. Amides (or amino-groups) $NH_2$ are marked by big circles.

Let us mark these three kinds of binary-opposite attributes by numbers N = 1, 2, 3 and ascribe to each of the four genetic letters the symbol "$0_N$" (the symbol "$1_N$") in a case of presence (or of absence correspondingly) of the attribute under number "N" in this letter. As a result we obtain the representation of the genetic 4-letter alphabet in the system of its three "binary sub-alphabets corresponding to attributes" (Figure 3).

| | **Symbols of a genetic letter from a viewpoint of a kind of the binary-opposite attributes** | **C** | **A** | **G** | **U/T** | |
|---|---|---|---|---|---|---|
| **№1** | **$0_1$** – pyrimidines (one ring in a molecule);<br>**$1_1$** – purines (two rings in a molecule) | **$0_1$** | **$1_1$** | **$1_1$** | **$0_1$** | 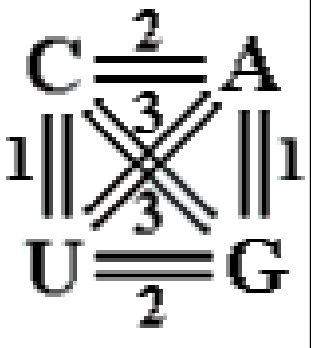 |
| **№2** | **$0_2$** – a letter with amino-mutating property (amino);<br>**$1_2$** – a letter without it (keto) | **$0_2$** | **$0_2$** | **$1_2$** | **$1_2$** | |
| **№3** | **$0_3$** – a letter with three hydrogen bonds;<br>**$1_3$** – a letter with two hydrogen bonds | **$0_3$** | **$1_3$** | **$0_3$** | **$1_3$** | |

Figure 3. Three binary sub-alphabets according to three kinds of binary-opposite attributes in the 4-letter alphabet of nitrogenous bases C, A, G, U. The scheme on the right side explains graphically the symmetric relations of equivalence between the pairs of letters from the viewpoint of the separate attributes 1, 2, 3.

The table on Figure 3 shows that, on the basis of each kind of the attributes, each of the letters A, C, G, U/T possesses three "faces" or meanings in the three binary sub-alphabets. On the basis of each kind of the attributes, the 4-letter alphabet is curtailed into the 2-letter alphabet. For example, on the basis of the first kind of binary-opposite attributes we have (instead of the 4-letter alphabet) the alphabet from two letters $0_1$ and $1_1$, which one can name "the binary sub-alphabet of the first kind of the binary attributes".

On the basis of the idea about a possible analogy between discrete signals processing in computers and in the genetic code system, one can present the genetic 4-letter alphabet in the following matrix form [C A; U G] (Figure 4). Then the Kronecker family of matrices with such alphabetical kernel can be considered: $[C\ A;\ U\ G]^{(n)}$, where (n) means the integer Kronecker (or tensor) power. This form possesses the analogy with the Kronecker family of Hadamard matrices $[\ 1\ 1;\ -1\ 1]^{(n)}$ from discrete signals processing [Ahmed, Rao, 1975]. Figure 4 shows the first genetic matrices of such a family. One can see on this Figure that each of the matrices contains all the genetic multiplets of equal length in a strict order: [C A; U G] contains all the 4 monoplets; $[C\ A;\ U\ G]^{(2)}$ contains all the 16 duplets; $[C\ A;\ U\ G]^{(3)}$ contains all the 64 triplets.

All the columns and rows of the matrices on Figure 4 are binary numerated and disposed in a monotonic order by the following algorithm which uses biochemical features of the genetic nitrogenous bases and which can be used in bio-computers of any organism. Numerations of columns and rows are formed automatically if one interprets multiplets of each column from the viewpoint of the first binary sub-alphabet (Figure 3) and if one interprets multiplets of each row from the viewpoint of the second binary sub-alphabet. For example, the column 010 contains all the triplets of the form "pyrimidine-purine-pyrimidine"; the row 010 contains all the triplets of the form “amino-keto-amino”. Each of the triplets in the matrix $[C\ A;\ U\ G]^{(3)}$ receives its dyadic-shift numeration by means of modulo-2 addition of numerations of its column and row. For example, the triplet CAG receives its dyadic-shift numeration 010 (or 2 in decimal notation) because it belongs to the column 011 and the row 001. Any codon and its anti-codon are disposed in inversion-symmetrical manner in relation to the centre of the genomatrix $[C\ A;\ U\ G]^{(3)}$ (Figure 4). In this genomatrix any codon and its anti-codon possess the same dyadic-shift numeration; each of the dyadic-shift numerations denotes a subset of 8 triplets with 4 pairs “codon-anticodon”.

**[C A; U G] =**

| | **0** | **1** |
|---|---|---|
| **0** | **C 0** | **A 1** |
| **1** | **U 1** | **G 0** |

**; $[C\ A;\ U\ G]^{(2)}$ =**

| | **00(0)** | **01(1)** | **10(2)** | **11(3)** |
|---|---|---|---|---|
| **00 (0)** | **CC 00 (0)** | **CA 01(1)** | **AC 10 (2)** | **AA 11 (3)** |
| **01 (1)** | **CU 01 (1)** | **CG 00 (0)** | **AU 11 (3)** | **AG 10 (2)** |
| **10 (2)** | **UC 10 (2)** | **UA 11(3)** | **GC 00 (0)** | **GA 01 (1)** |
| **11 (3)** | **UU 11 (3)** | **UG 10 (2)** | **GU 01 (1)** | **GG 00 (0)** |

**$[C\ A;\ U\ G]^{(3)}$ =**

| | **000 (0)** | **001 (1)** | **010 (2)** | **011 (3)** | **100 (4)** | **101 (5)** | **110 (6)** | **111 (7)** |
|---|---|---|---|---|---|---|---|---|
| **000 (0)** | **CCC 000 (0)** | **CCA 001 (1)** | **CAC 010 (2)** | **CAA 011 (3)** | **ACC 100 (4)** | **ACA 101 (5)** | **AAC 110 (6)** | **AAA 111 (7)** |
| **001 (1)** | **CCU 001 (1)** | **CCG 000 (0)** | **CAU 011 (3)** | **CAG 010 (2)** | **ACU 101 (5)** | **ACG 100 (4)** | **AAU 111 (7)** | **AAG 110 (6)** |
| **010 (2)** | **CUC 010 (2)** | **CUA 011 (3)** | **CGC 000 (0)** | **CGA 001 (1)** | **AUC 110 (6)** | **AUA 111 (7)** | **AGC 100 (4)** | **AGA 101 (5)** |
| **011 (3)** | **CUU 011 (3)** | **CUG 010 (2)** | **CGU 001 (1)** | **CGG 000 (0)** | **AUU 111 (7)** | **AUG 110 (6)** | **AGU 101 (5)** | **AGG 100 (4)** |
| **100 (4)** | **UCC 100 (4)** | **UCA 101 (5))** | **UAC 110 (6)** | **UAA 111 (7)** | **GCC 000 (0)** | **GCA 001 (1)** | **GAC 010 (2)** | **GAA 011 (3)** |
| **101 (5)** | **UCU 101 (5)** | **UCG 100 (4)** | **UAU 111 (7)** | **UAG 110 (6)** | **GCU 001 (1)** | **GCG 000 (0)** | **GAU 011 (3)** | **GAG 010 (2)** |
| **110 (6)** | **UUC 110 (6)** | **UUA 111 (7)** | **UGC 100 (4)** | **UGA 101 (5)** | **GUC 010 (2)** | **GUA 011 (3)** | **GGC 000 (0)** | **GGA 001 (1)** |
| **111 (7)** | **UUU 111 (7)** | **UUG 110 (6)** | **UGU 101 (5)** | **UGG 100 (4)** | **GUU 011 (3)** | **GUG 010 (2)** | **GGU 001 (1)** | **GGG 000 (0)** |

Figure 4. The first genetic matrices of the Kronecker family $[C\ A;\ U\ G]^{(n)}$ with binary numerations of their columns and rows on the basis of the binary sub-alphabets № 1 and № 2 from Figure 3. The lower matrix is the genomatrix $[C\ A;\ U\ G]^{(3)}$. Each of matrix cells contains a symbol of a multiplet, a dyadic-shift numeration of this multiplet and its expression in decimal notation. Decimal numerations of columns, rows and multiplets are written in brackets. Black

and white cells contain triplets and duplets with strong and weak roots correspondingly (see the text).

It should be noted additionally that each of 64 triplets in the genomatrix $[C\ A;\ U\ G]^{(3)}$ receives the same dyadic-shift numeration if each of its three nitrogenous bases is numerated from the viewpoint of the binary sub-alphabet № 3 from Figure 3 in which C=G=0, A=U/T=1. For example in this way the triplet CAG receives again the dyadic-shift numeration 010 if each of its letters is replaced by its binary symbol (C=G=0, A=U/T=1). This alternative way of dyadic-shift numeration of multiplets is simpler for using and it is termed "the direct dyadic-shift numeration".

### 3. The 2-letter alphabet of strong and weak roots of triplets and Rademacher functions

The genetic code is termed the degeneracy code because its 64 triplets encode 20 amino acids, and different amino acids are encoded by means of different quantities of triplets. Modern science knows many variants (or dialects) of the genetic code, which are variants of the correspondence between the alphabet of triplets and the alphabet of amino acids and stop-codons. Data about these variants are presented on NCBI's Web site, http://www.ncbi.nlm.nih.gov/Taxonomy/Utils/wprintgc.cgi. Seventeen variants (or dialects) of the genetic code exist, which differ one from another by some details of correspondences between triplets and objects encoded by them. Most of these dialects (including the Standard Code and the Vertebrate Mitochondrial Code which are presented on Figure 5) have a general scheme of their degeneracy, where 32 triplets with "strong roots" and 32 triplets with "weak roots" exist (see more details in [Petoukhov, 2001, 2005, 2008; Petoukhov, He, 2010]). Cells with these triplets are marked by black and white colors correspondingly inside the matrix $[C\ A;\ U\ G]^{(3)}$ on Figure 4. Here it should be mentioned that a combination of letters on the two first positions of each triplet is termed a "root" of this triplet; a letter on its third position is termed a "suffix".

The set of 64 triplets contains 16 possible variants of such roots. Taking into account properties of triplets, the set of 16 possible roots is divided into two subsets with 8 roots in each. The first of such octets contains roots CC, CU, CG, AC, UC, GC, GU and GG. These roots are termed "strong roots" [Konopel'chenko, Rumer, 1975] because each of them defines four triplets with this root, coding values of which are independent on their suffix. For example, four triplets CGC, CGA, CGU, CGG, which have the identical strong root CG, encode the same amino acid Arg, although they have different suffixes (Figure 5). The second octet contains roots CA, AA, AU, AG, UA, UU, UG and GA. These roots are termed "weak roots" because each of them defines four triplets with this root, coding values of which depend on their suffix. An example of such a subfamily in Figure 5 is represented by four triplets CAC, CAA, CAU and CAC, two of which (CAC, CAU) encode the amino acid His and the other two (CAA, CAG) encode the amino acid Gln.

How these two subsets of triplets with strong and weak roots are disposed in the genomatrix $[C\ A;\ U\ G]^{(3)}$ (Figure 4) which was constructed formally on the base of the 4-letter alphabet of nitrogenous bases and Kronecher multiplications without any mention about the degeneracy of the genetic code and about amino acids? Can one anticipate any symmetry in their disposition? It should be noted that the huge quantity $64! \approx 10^{89}$ of variants exists for dispositions of 64 triplets in the (8x8)-matrix. One can note for comparison, that the modern physics estimates the time of existence of the Universe in $10^{17}$ seconds. It is obvious that in such a situation an accidental disposition of the 20 amino acids and the corresponding triplets in a (8x8)-matrix will give almost never any symmetry in their disposition in matrix halves, quadrants and rows.

| THE STANDARD CODE | |
|---|---|
| 8 subfamilies of triplets with strong roots (“black triplets”) and the amino acids, which are encoded by them | 8 subfamilies of triplets with weal roots (“white triplets”) and the amino acids, which are encoded by them |
| CCC, CCU, CCA, CCG → Pro | CAC, CAU, CAA, CAG → His, His, Gln, Gln |
| CUC, CUU, CUA, CUG → Leu | AAC, AAU, AAA, AAG → Asn, Asn, Lys, Lys |
| CGC, CGU, CGA, CGG → Arg | AUC, AUU, AUA, AUG → Ile, Ile, Ile, Met |
| ACC, ACU, ACA, ACG → Thr | AGC, AGU, AGA, AGG → Ser, Ser, Arg, Arg |
| UCC, UCU, UCA, UCG → Ser | UAC, UAU, UAA, UAG → Tyr, Tyr, Stop, Stop |
| GCC, GCU, GCA, GCG → Ala | UUC, UUU, UUA, UUG → Phe, Phe, Leu, Leu |
| GUC, GUU, GUA, GUG → Val | UGC, UGU, UGA, UGG → Cys, Cys, Stop, Trp |
| GGC, GGU, GGA, GGG → Gly | GAC, GAU, GAA, GAG → Asp, Asp, Glu, Glu |
| THE VERTEBRATE MITOCHONDRIAL CODE | |
| CCC, CCU, CCA, CCG → Pro | CAC, CAU, CAA, CAG → His, His, Gln, Gln |
| CUC, CUU, CUA, CUG → Leu | AAC, AAU, AAA, AAG → Asn, Asn, Lys, Lys |
| CGC, CGU, CGA, CGG → Arg | AUC, AUU, AUA, AUG → Ile, Ile, Met, Met |
| ACC, ACU, ACA, ACG → Thr | AGC, AGU, AGA, AGG → Ser, Ser, Stop, Stop |
| UCC, UCU, UCA, UCG → Ser | UAC, UAU, UAA, UAG → Tyr, Tyr, Stop, Stop |
| GCC, GCU, GCA, GCG → Ala | UUC, UUU, UUA, UUG → Phe, Phe, Leu, Leu |
| GUC, GUU, GUA, GUG → Val | UGC, UGU, UGA, UGG → Cys, Cys, Trp, Trp |
| GGC, GGU, GGA, GGG → Gly | GAC, GAU, GAA, GAG → Asp, Asp, Glu, Glu |

Figure 5. The Standard Code and the Vertebrate Mitochondrial Code possess the basic scheme of the genetic code degeneracy with 32 triplets of strong roots and 32 triplets of weak roots (Initial data from http://www.ncbi.nlm.nih.gov/Taxonomy/Utils/wprintgc.cgi.)

| | | | | | | | |
|---|---|---|---|---|---|---|---|
| 1 | 1 | −1 | −1 | 1 | 1 | −1 | −1 |
| 1 | 1 | −1 | −1 | 1 | 1 | −1 | −1 |
| 1 | 1 | 1 | 1 | −1 | −1 | −1 | −1 |
| 1 | 1 | 1 | 1 | −1 | −1 | −1 | −1 |
| 1 | 1 | −1 | −1 | 1 | 1 | −1 | −1 |
| 1 | 1 | −1 | −1 | 1 | 1 | −1 | −1 |
| −1 | −1 | −1 | −1 | 1 | 1 | 1 | 1 |
| −1 | −1 | −1 | −1 | 1 | 1 | 1 | 1 |

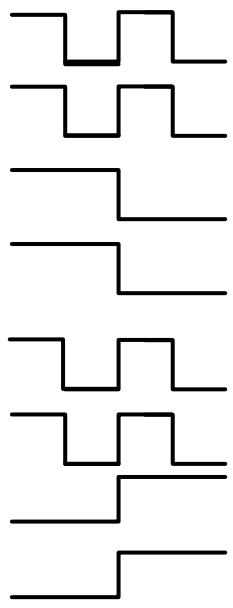

Figure 6. The Rademacher form R of representation of the genomatrix [C A; U G]$^{(3)}$ from Figure 4. A relevant system of Rademacher functions is shown at the right side.

But it is a phenomenological fact that the disposition of the 32 triplets with strong roots (“black triplets” in Figure 4) and the 32 triplets with weak roots (“white triplets”) has a unique and exceptional symmetric character (see Figure 4). For example the left and right halves of the matrix mosaic are mirror-anti-symmetric to each other in its colors: any pair of cells, disposed by mirror-symmetrical manner in these halves, possesses the opposite colors. One can say that each row of this mosaic matrix corresponds to an odd function. In addition each row of the mosaic matrix [C A; U G]$^{(3)}$ has a meander-line character (the term “meander-line” means here that lengths of black and white fragments are equal to each other along each row). But the theory of discrete signal processing uses such odd meander functions for a long time under the known name “Rademacher functions”. Rademacher functions contain elements “+1” and “-1” only.

Each of the matrix rows presents one of the Rademacher functions if each black (white) cell is interpreted such that it contains the number +1 (−1). Figure 6 shows a transformation of the mosaic matrix [C A; U G]$^{(3)}$ of (Figure 4) into a numeric matrix in the result of such replacements of black and white triplets by numbers "+1" and "-1" correspondingly. Rademacher functions are particular cases of Walsh functions considered below.

### 4. The dyadic-shift decomposition of the genomatrix [C A; U G]$^{(3)}$ and the first type of 8-dimensional hypercomplex numbers

The Rademacher form R of the genomatrix [C A; U G]$^{(3)}$ (Figure 6) can be decomposed into the sum of 8 sparse matrices $r_0$, $r_1$, $r_2$, $r_3$, $r_4$, $r_5$, $r_6$, $r_7$ (Figure 7) in accordance with the structural principle of the dyadic-shift matrix from Figure 1. More precisely any sparse matrix $r_k$ (k=0, 1, …, 7) contains entries "+1" or "-1" from the matrix R on Figure 6 in those cells which correspond to cells of the dyadic-shift matrix with the same dyadic-shift numeration "k" (Figure 1); all the other cells of the matrix $r_k$ contain zero. For example, the sparse matrix $r_2$ contains entries "+1" and "-1" from the matrix R (Figure 6) only in those cells which correspond to cells with the dyadic numeration "2" in the dyadic-shift matrix on Figure 1. Determinants of all the sparse matrices $r_k$ are equal to 1.

$$R = r_0+r_1+r_2+r_3+r_4+r_5+r_6+r_7 =$$

$$\begin{vmatrix} 1&0&0&0&0&0&0&0\\ 0&1&0&0&0&0&0&0\\ 0&0&1&0&0&0&0&0\\ 0&0&0&1&0&0&0&0\\ 0&0&0&0&1&0&0&0\\ 0&0&0&0&0&1&0&0\\ 0&0&0&0&0&0&1&0\\ 0&0&0&0&0&0&0&1 \end{vmatrix} + \begin{vmatrix} 0&1&0&0&0&0&0&0\\ 1&0&0&0&0&0&0&0\\ 0&0&0&1&0&0&0&0\\ 0&0&1&0&0&0&0&0\\ 0&0&0&0&0&1&0&0\\ 0&0&0&0&1&0&0&0\\ 0&0&0&0&0&0&0&1\\ 0&0&0&0&0&0&1&0 \end{vmatrix} + \begin{vmatrix} 0&0&-1&0&0&0&0&0\\ 0&0&0&-1&0&0&0&0\\ 1&0&0&0&0&0&0&0\\ 0&1&0&0&0&0&0&0\\ 0&0&0&0&0&0&-1&0\\ 0&0&0&0&0&0&0&-1\\ 0&0&0&0&1&0&0&0\\ 0&0&0&0&0&1&0&0 \end{vmatrix} + \begin{vmatrix} 0&0&0&-1&0&0&0&0\\ 0&0&-1&0&0&0&0&0\\ 0&1&0&0&0&0&0&0\\ 1&0&0&0&0&0&0&0\\ 0&0&0&0&0&0&0&-1\\ 0&0&0&0&0&0&-1&0\\ 0&0&0&0&0&1&0&0\\ 0&0&0&0&1&0&0&0 \end{vmatrix} +$$

$$\begin{vmatrix} 0&0&0&0&1&0&0&0\\ 0&0&0&0&0&1&0&0\\ 0&0&0&0&0&0&-1&0\\ 0&0&0&0&0&0&0&-1\\ 1&0&0&0&0&0&0&0\\ 0&1&0&0&0&0&0&0\\ 0&0&-1&0&0&0&0&0\\ 0&0&0&-1&0&0&0&0 \end{vmatrix} + \begin{vmatrix} 0&0&0&0&0&1&0&0\\ 0&0&0&0&1&0&0&0\\ 0&0&0&0&0&0&0&-1\\ 0&0&0&0&0&0&-1&0\\ 0&1&0&0&0&0&0&0\\ 1&0&0&0&0&0&0&0\\ 0&0&0&-1&0&0&0&0\\ 0&0&-1&0&0&0&0&0 \end{vmatrix} + \begin{vmatrix} 0&0&0&0&0&0&-1&0\\ 0&0&0&0&0&0&0&-1\\ 0&0&0&0&-1&0&0&0\\ 0&0&0&0&0&-1&0&0\\ 0&0&-1&0&0&0&0&0\\ 0&0&0&-1&0&0&0&0\\ -1&0&0&0&0&0&0&0\\ 0&-1&0&0&0&0&0&0 \end{vmatrix} + \begin{vmatrix} 0&0&0&0&0&0&0&-1\\ 0&0&0&0&0&0&-1&0\\ 0&0&0&0&0&-1&0&0\\ 0&0&0&0&-1&0&0&0\\ 0&0&0&-1&0&0&0&0\\ 0&0&-1&0&0&0&0&0\\ 0&-1&0&0&0&0&0&0\\ -1&0&0&0&0&0&0&0 \end{vmatrix}$$

Figure 7. The dyadic-shift decomposition of the Rademacher form R (Figure 6) of the genomatrix [C A; U G]$^{(3)}$ into the sum of 8 sparse matrices $r_0$, $r_1$,…, $r_7$.

The author has discovered that this set of 8 matrices $r_0$, $r_1$,… , $r_7$ (where $r_0$ is the identity matrix) is closed in relation to multiplication and it satisfies the table on Figure 8.

| | **1** | **$r_1$** | **$r_2$** | **$r_3$** | **$r_4$** | **$r_5$** | **$r_6$** | **$r_7$** |
|---|---|---|---|---|---|---|---|---|
| **1** | **1** | **$r_1$** | **$r_2$** | **$r_3$** | **$r_4$** | **$r_5$** | **$r_6$** | **$r_7$** |
| **$r_1$** | **$r_1$** | **1** | **$r_3$** | **$r_2$** | **$r_5$** | **$r_4$** | **$r_7$** | **$r_6$** |
| **$r_2$** | **$r_2$** | **$r_3$** | **-1** | **$-r_1$** | **$-r_6$** | **$-r_7$** | **$r_4$** | **$r_5$** |
| **$r_3$** | **$r_3$** | **$r_2$** | **$-r_1$** | **-1** | **$-r_7$** | **$-r_6$** | **$r_5$** | **$r_4$** |
| **$r_4$** | **$r_4$** | **$r_5$** | **$r_6$** | **$r_7$** | **1** | **$r_1$** | **$r_2$** | **$r_3$** |
| **$r_5$** | **$r_5$** | **$r_4$** | **$r_7$** | **$r_6$** | **$r_1$** | **1** | **$r_3$** | **$r_2$** |
| **$r_6$** | **$r_6$** | **$r_7$** | **$-r_4$** | **$-r_5$** | **$-r_2$** | **$-r_3$** | **1** | **$r_1$** |
| **$r_7$** | **$r_7$** | **$r_6$** | **$-r_5$** | **$-r_4$** | **$-r_3$** | **$-r_2$** | **$r_1$** | **1** |

Figure 8. The multiplication table of basic matrices $r_0, r_1, \ldots, r_7$ (where $r_0$ is the identity matrix) which corresponds to the 8-dimensional algebra over the field of real numbers. It defines the 8-dimensional numeric system of genetic $R_{123}$-octetons.

The multiplication table on Figure 8 is asymmetrical in relation to the main diagonal and it corresponds to the non-commutative associative algebra of 8-dimensional hypercomplex numbers, which are an extension of double numbers (or numbers by Lorentz). This matrix algebra is non-division algebra because it has zero divisors. It means that such non-zero hypercomplex numbers exist whose product is equal to zero. For example, $(r_3+r_4)$ and $(r_2+r_5)$ (see Figure 7) are non-zero matrices, but their product is equal to zero matrix. These genetic 8-dimensional hypercomplex numbers are different from Cayley's octonions (http://en.wikipedia.org/wiki/Octonion). The algebra of Cayley's octonions is non-associative algebra and correspondingly it does not possess a matrix form of its representation (each of matrix algebras is an associative algebra). The known term "octonions" is not appropriate for the case of the multiplication table on Figure 8 because this mathematical term is usually used for members of normed division non-associative algebra (http://en.wikipedia.org/wiki/Octonion).

For this reason we term these hypercomplex numbers, which are revealed in matrix genetics, as "dyadic-shift genetic octetons" (or briefly "genooctetons" or simply "octetons"). In addition we term such kinds of matrix algebras, which are connected with dyadic-shift decompositions, as dyadic-shift algebras (or briefly DS-algebras). The author supposes that DS-algebras are very important for genetic systems. It is interesting that all the basic matrices $r_0, r_1, \ldots, r_7$ are disposed in the multiplication table (Figure 8) in accordance with the structure of the dyadic-shift matrix (Figure 1). The numeric system of dyadic-shift genooctetons differs cardinally from the system of genetic 8-dimensional numbers which have been described by the author in matrix genetics previously [Petoukhov, 2008a,c; Petoukhov, He, 2010, Section 3] and which are termed "8-dimensional bipolars" or "8-dimensional Yin-Yang genonumbers".

Below we will describe another variant of genooctetons which is connected with Hadamard genomatrices (H-octetons). For this reason we term the first type of genooctetons (Figures 6-8) as $R_{123}$-octetons (here R is the first letter of the name Rademacher; the index 123 means the order 1-2-3 of positions in triplets).

A general form of $R_{123}$-octetons (Figure 7) is the following:

$$R_{123} = x_0*\mathbf{1} + x_1*\mathbf{r_1} + x_2*\mathbf{r_2} + x_3*\mathbf{r_3} + x_4*\mathbf{r_4} + x_5*\mathbf{r_5} + x_6*\mathbf{r_6} + x_7*\mathbf{r_7} \quad (4)$$

where coefficients $x_0, x_1, \ldots, x_7$ are real numbers. Here the first component $x_0$ is a scalar. Other 7 components $x_1*\mathbf{r_1}$, $x_2*\mathbf{r_2}$, $x_3*\mathbf{r_3}$, $x_4*\mathbf{r_4}$, $x_5*\mathbf{r_5}$, $x_6*\mathbf{r_6}$, $x_7*\mathbf{r_7}$ are non-scalar units but imaginary units.

The multiplication table (Figure 8) shows that $R_{123}$-octetons contain 2 sub-algebras of two-dimensional complex numbers z in their (8*8)-matrix forms of representation: $z=x_0*\mathbf{r_0}+x_2*\mathbf{r_2}$ and $z=x_0*\mathbf{r_0}+ x_3*\mathbf{r_3}$. It is interesting because, as science knows, two-dimensional complex numbers, which are a sum of a real item and imaginary item, have appeared as outstanding instruments for the development of theories and calculations in the field of physical problems of heat, light, sounds, vibrations, elasticity, gravitation, magnetism, electricity, liquid streams, and phenomena of the micro-world. Our results in the field of matrix genetics reveal a bunch of (8*8)-matrix forms of representation of two-dimensional complex numbers in 8-dimensional vector spaces of $R_{123}$-octetons (below we will describe cases of positional permutations in triplets which lead to additional (8*8)-matrix forms of representation of complex numbers and of double numbers). From a formal viewpoint, the mentioned branches of theoretical physics can be interpreted in a frame of the 8-dimensional vector space of octetons. Simultaneously the multiplication table (Figure 8) shows that $R_{123}$-octetons contain 5 sub-algebras of two-dimensional double numbers d in their (8*8)-matrix forms of representation:

$d=x_0*\mathbf{r_0}+x_1*\mathbf{r_1}$, $d=x_0*\mathbf{r_0}+x_4*\mathbf{r_4}$, $d=x_0*\mathbf{r_0}+x_5*\mathbf{r_5}$, $d=x_0*\mathbf{r_0}+x_6*\mathbf{r_6}$, $d=x_0*\mathbf{r_0}+x_7*\mathbf{r_7}$. It is known that each complex number can be interpreted in three ways: as a point, or a vector or an operator in a two-dimensional vector space. The same is true for the case of each complex numbers in 8-dimensional vector spaces of octetons.

The multiplication table (Figure 8) shows as well that the product of two different members from the set $r_1, r_2, \ldots, r_7$ is equal to its third member (with the sign "+" or "-"), and such triad of members forms a closed group under multiplication. In this way the following system of 7 triads appears (Figure 9, left column):

| Triads of basis elements | Modulo-2 addition of binary indexes | Decimal notations of these binary indexes |
|---|---|---|
| $\mathbf{r_1,r_2,r_3}$ | 001+010=011; 010+011=001; 011+001=010 | 001=1; 010=2; 011=3 |
| $\mathbf{r_1,r_4,r_5}$ | 001+100=101; 100+101=001; 101+001=100 | 001=1; 100=4; 101=5 |
| $\mathbf{r_1,r_6,r_7}$ | 001+110=111; 110+111=001; 111+001=110 | 001=1; 110=6; 111=7 |
| $\mathbf{r_2,r_4,r_6}$ | 010+100=110; 100+110=010; 110+010=100 | 010=2; 100=4; 110=6 |
| $\mathbf{r_2,r_5,r_7}$ | 010+101=111; 101+111=010; 111+010=101 | 010=2; 101=5; 111=7 |
| $\mathbf{r_3,r_4,r_7}$ | 011+100=111; 100+111=011; 011+111=100 | 011=3; 100=4; 111=7 |
| $\mathbf{r_3,r_5,r_6}$ | 011+101=110; 101+110=011; 011+110=101 | 011=3; 101=5; 110=6 |

Figure 9. Triads of basis elements of $R_{123}$-octetons from Figure 7. Their indexes form closed sets in relation to modulo-2 addition for binary numbers.

This set of triads reveals a new connection of genooctetons with mathematical operation of modulo-2 addition (1). Binary notations of decimal indexes 1, 2, 3, 4, 5, 6, 7 of the basis units $r_1, r_2, \ldots, r_7$ are 001, 010, 011, 100, 101, 110, 111 correspondingly. Figure 9 (two right columns) shows that three indexes of basis units in each triad interact with each other in accordance with the rules of the modulo-2 addition (1). These triads together with the identity matrix $r_0$ define seven 4-dimensional DS-algebras with their multiplication tables presented on Figure 10. One can see that the first three tables define commutative DS-algebras because they are symmetrical; the other four multiplication tables define non-commutative DS-algebras because they are asymmetrical.

| | 1 | $r_1$ | $r_2$ | $r_3$ |
|---|---|---|---|---|
| **1** | **1** | $r_1$ | $r_2$ | $r_3$ |
| $r_1$ | $r_1$ | **1** | $r_3$ | $r_2$ |
| $r_2$ | $r_2$ | $r_3$ | **-1** | $-r_1$ |
| $r_3$ | $r_3$ | $r_2$ | $-r_1$ | **-1** |

| | 1 | $r_1$ | $r_4$ | $r_5$ |
|---|---|---|---|---|
| **1** | **1** | $r_1$ | $r_4$ | $r_5$ |
| $r_1$ | $r_1$ | **1** | $r_5$ | $r_4$ |
| $r_4$ | $r_4$ | $r_5$ | **1** | $r_1$ |
| $r_5$ | $r_5$ | $r_4$ | $r_1$ | **1** |

| | 1 | $r_1$ | $r_6$ | $r_7$ |
|---|---|---|---|---|
| **1** | **1** | $r_1$ | $r_6$ | $r_7$ |
| $r_1$ | $r_1$ | **1** | $r_7$ | $r_6$ |
| $r_6$ | $r_6$ | $r_7$ | **1** | $r_1$ |
| $r_7$ | $r_7$ | $r_6$ | $r_1$ | **1** |

| | 1 | $r_2$ | $r_4$ | $r_6$ |
|---|---|---|---|---|
| **1** | **1** | $r_2$ | $r_4$ | $r_6$ |
| $r_2$ | $r_2$ | **-1** | $-r_6$ | $r_4$ |
| $r_4$ | $r_4$ | $r_6$ | **1** | $r_2$ |
| $r_6$ | $r_6$ | $-r_4$ | $-r_2$ | **1** |

| | 1 | $r_2$ | $r_5$ | $r_7$ |
|---|---|---|---|---|
| **1** | **1** | $r_2$ | $r_5$ | $r_7$ |
| $r_2$ | $r_2$ | **-1** | $-r_7$ | $r_5$ |
| $r_5$ | $r_5$ | $r_7$ | **1** | $r_2$ |
| $r_7$ | $r_7$ | $-r_5$ | $-r_2$ | **1** |

| | 1 | $r_3$ | $r_4$ | $r_7$ |
|---|---|---|---|---|
| **1** | **1** | $r_3$ | $r_4$ | $r_7$ |
| $r_3$ | $r_3$ | **-1** | $-r_7$ | $r_4$ |
| $r_4$ | $r_4$ | $r_7$ | **1** | $r_3$ |
| $r_7$ | $r_7$ | $-r_4$ | $-r_3$ | **1** |

| | 1 | $r_3$ | $r_5$ | $r_6$ |
|---|---|---|---|---|
| **1** | **1** | $r_3$ | $r_5$ | $r_6$ |
| $r_3$ | $r_3$ | **-1** | $-r_6$ | $r_5$ |
| $r_5$ | $r_5$ | $r_6$ | **1** | $r_3$ |
| $r_6$ | $r_6$ | $-r_5$ | $-r_3$ | **1** |

Figure 10. Multiplication tables of 4-dimensional DS-algebras which correspond to triads of $R_{123}$-octetons from Figure 9. The disposition of indexes of elements in each of the tables corresponds to the principle of diadic shifts.

Sum of basic elements of each of these 4-dimensional DS-algebras possesses the following property of its doubling and quadrupling under exponentiation:

$(r_0+r_2+r_4+r_6)^2 = 2*(r_0+r_2+r_4+r_6)$; $(r_0+r_2+r_5+r_7)^2 = 2*(r_0+r_2+r_5+r_7)$; $(r_0+r_3+r_4+r_7)^2 = 2*(r_0+r_3+r_4+r_7)$; $(r_0+r_3+r_5+r_6)^2 = 2*(r_0+r_3+r_5+r_6)$; $(r_0+r_1+r_4+r_5)^2 = 4*(r_0+r_1+r_4+r_5)$; $(r_0+r_1+r_6+r_7)^2 = 4*(r_0+r_1+r_6+r_7)$; $(r_0+r_1)^2 = 2*(r_0+r_1)$; $(r_2+r_3)^2 = -2*(r_0+r_1)$; $(r_1+r_2)^2 = 2*r_3$.

The inversion of signs (plus by minus) of terms of these equations leads, as a rule, to new equations of the same type. For example, in the case of the first equation we have new 7 equations of doubling: $(r_0-r_2+r_4+r_6)^2 = 2*(r_0-r_2+r_4+r_6)$; $(r_0+r_2-r_4+r_6)^2 = 2*(r_0+r_2-r_4+r_6)$; $(r_0+r_2+r_4-r_6)^2 = 2*(r_0+r_2+r_4-r_6)$; $(r_0-r_2-r_4+r_6)^2 = 2*(r_0-r_2-r_4+r_6)$; $(r_0-r_2+r_4-r_6)^2 = 2*(r_0-r_2+r_4-r_6)$; $(r_0+r_2-r_4-r_6)^2 = 2*(r_0+r_2-r_4-r_6)$; $(r_0-r_2-r_4-r_6)^2 = 2*(r_0-r_2-r_4-r_6)$.

The first three tables on Figure 10 are symmetrical and they correspond to commutative algebras. The other four tables correspond to the non-commutative algebra which has been already described in our works in connection with genetic Yin-Yang numbers (or bipolar numbers) [Petoukhov, 2008c, Fig. 14 on the right side].

Let us mention else one interesting aspect of $R_{123}$-octetons. One can replace the symbolic genomatrix $[C\ A;\ U\ G]^{(3)}$ (see Figure 4) by a special numeric representation where each of the triplets is replaced by twice the number of hydrogen bonds in its suffix (it is a letter on the third position of the triplet). For example, all the triplets with suffixes C and G (which possess 3 hydrogen bonds in their complementary pairs in DNA) are replaced by number 2*3=6, and all the triplets with suffixes A and U (which possess 2 hydrogen bonds in their complementary pairs) are replaced by number 2*2=4. These numbers should be taken with the sign "+" for triplets in black cells of the genomatrix on Figure 4 and with the sign "-" for triplets in white cells. One can check that in this case the final numeric matrix is the matrix form of representation of the following $R_{123}$-octetons (4) with a separation of coordinate axes with odd and even indexes:

$$R_{123} = 6*(r_0 + r_2 + r_4 + r_6) + 4*(r_1 + r_3 + r_5 + r_7) \qquad (4a)$$

This special $R_{123}$-octeton (4a) with integer values of all the coordinates is interesting because its square root gives the "golden" $R_{123}$-octeton whose coordinates are equal to irrational numbers of the golden section $f = (1+5^{0.5})/2 = 1.618...$ and of its inverse value $f^{-1} = 0,618...$:

$$R_{123} = f*(r_0 + r_2 + r_4 + r_6) + f^{-1}*(r_1 + r_3 + r_5 + r_7) \qquad (4b)$$

The theme of the golden section and Fibonacci numbers in structural properties of genetic alphabets appears unexpectedly in many cases [Petoukhov, 2008a; Petoukhov, He, 2010]. This theme has attracted an increased attention of scientific community because it is related with quasi-cristalls by D.Shechtman (1984) and his Nobel Prize-2011.

### 5. $R_{123}$-octetons, their norms and metric $R_{123}$-quaternions

Now let us consider some geometric questions concerning $R_{123}$-octetons. By analogy with the case of Cayley's octonions (http://en.wikipedia.org/wiki/Octonion), the conjugation of any 8-dimensional hypercomplex numbers

$$V = x_0*\mathbf{1} + x_1*\mathbf{v_1} + x_2*\mathbf{v_2} + x_3*\mathbf{v_3} + x_4*\mathbf{v_4} + x_5*\mathbf{v_5} + x_6*\mathbf{v_6} + x_7*\mathbf{v_7} \qquad (5)$$

is given by

$$V^{\wedge} = x_0*\mathbf{1} - x_1*\mathbf{v_1} - x_2*\mathbf{v_2} - x_3*\mathbf{v_3} - x_4*\mathbf{v_4} - x_5*\mathbf{v_5} - x_6*\mathbf{v_6} - x_7*\mathbf{v_7} \qquad (6)$$

The *norm* of hypercomplex numbers V is defined as

$$||V|| = (V^{\wedge} * V)^{0.5} \qquad (7)$$

For the case of a $R_{123}$-octeton (Figure 8) we have the following expression (8) for its norm:

$$|| R_{123}||^2 = R_{123}^{\wedge} * R_{123} = ( x_0^2-x_1^2+x_2^2+x_3^2-x_4^2-x_5^2-x_6^2-x_7^2) + $$
$$+ 2*(x_2*x_3-x_4*x_5-x_6*x_7)*r_1 + 2*x_1*(x_3*r_2 + x_2*r_3 - x_5*r_4- x_4*r_5 + x_7*r_6 + x_6*r_7) \qquad (8)$$

The right part of the expression (8) contains the scalar element ( $x_0^2-x_1^2+x_2^2+x_3^2-x_4^2-x_5^2-x_6^2-x_7^2$) and imaginary elements: $2*(x_2*x_3-x_4*x_5-x_6*x_7)*r_1 + 2*x_1*(x_3*r_2+x_2*r_3-x_5*r_4-x_4*r_5+x_7*r_6+x_6*r_7)$. But all these imaginary elements are equal to zero in the following 8 cases corresponding to different 4-dimensional sub-algebras of the 8-dimensional DS-algebra of $R_{123}$-octetons (Figure 11). Each of these 8 types of relevant $R_{123}$-quaternions has a scalar norm and defines a metric space with the signature (+ + - -) of a binary-oppositional kind. Figuratively speaking, the vector space of $R_{123}$-octetons contains a colony of metric sub-spaces of $R_{123}$-quaternions; it is reminiscent of a structure of multicellular organisms in a form of a colony of single-cellular organisms.

| № | Coefficients of zero value | $R_{123}$-quaternions with scalar norms | Squared norms of relevant $R_{123}$-quaternions |
|---|---|---|---|
| 0 | $x_1=x_2=x_4=x_6=0$ | $x_0+x_3*\mathbf{r_3}+x_5*\mathbf{r_5}+x_7*\mathbf{r_7}$ | $x_0^2+x_3^2-x_5^2-x_7^2$ |
| 1 | $x_1=x_2=x_4=x_7=0$ | $x_0+x_3*\mathbf{r_3}+x_5*\mathbf{r_5}+x_6*\mathbf{r_6}$ | $x_0^2+x_3^2-x_5^2-x_6^2$ |
| 2 | $x_1=x_2=x_5=x_6=0$ | $x_0+x_3*\mathbf{r_3}+x_4*\mathbf{r_4}+x_7*\mathbf{r_7}$ | $x_0^2+x_3^2-x_4^2-x_7^2$ |
| 3 | $x_1=x_2=x_5=x_7=0$ | $x_0+x_3*\mathbf{r_3}+x_4*\mathbf{r_4}+x_6*\mathbf{r_6}$ | $x_0^2+x_3^2-x_4^2-x_6^2$ |
| 4 | $x_1=x_3=x_4=x_6=0$ | $x_0+x_2*\mathbf{r_2}+x_5*\mathbf{r_5}+x_7*\mathbf{r_7}$ | $x_0^2+x_2^2-x_5^2-x_7^2$ |
| 5 | $x_1=x_3=x_4=x_7=0$ | $x_0+x_2*\mathbf{r_2}+x_5*\mathbf{r_5}+x_6*\mathbf{r_6}$ | $x_0^2+x_2^2-x_5^2-x_6^2$ |
| 6 | $x_1=x_3=x_5=x_6=0$ | $x_0+x_2*\mathbf{r_2}+x_4*\mathbf{r_4}+x_7*\mathbf{r_7}$ | $x_0^2+x_2^2-x_4^2-x_7^2$ |
| 7 | $x_1=x_3=x_5=x_7=0$ | $x_0+x_2*\mathbf{r_2}+x_4*\mathbf{r_4}+x_6*\mathbf{r_6}$ | $x_0^2+x_2^2-x_4^2-x_6^2$ |

Figure 11. The set of 8 types of $R_{123}$-quaternions with scalar values of squared norms

This set of 8 types of genetic $R_{123}$-quaternions (Figure 11) possesses interesting mathematical properties, for example:

1) Each of these types of $R_{123}$-quaternions is closed in relation to multiplication: the product of any two of its $R_{123}$-quaternions gives a $R_{123}$-quaternion of the same type;
2) For each of these types of $R_{123}$-quaternions with unit values of coordinates ($x_0=x_2=x_3=x_4=x_5=x_6=x_7=1$), the multiplication of such $R_{123}$-quaternion with itself (the square of such $R_{123}$-quaternion) gives a phenomenon of its doubling; for instance $(r_0+r_3+r_5+r_7)^2 = 2*(r_0+r_3+r_5+r_7)$ (see below Figure 15). Correspondingly $(r_0+r_3+r_5+r_7)^n = 2^{n-1}*(r_0+r_3+r_5+r_7)$, where n = 2, 3, 4, …. . This property of doubling is also true for any $R_{123}$-quaternion when each of coordinates of its imaginary part has absolute value equal to 1: $x_2=x_3=x_4=x_5=x_6=x_7=\pm1$. For example, $(r_0-r_3+r_5-r_7)^2 = 2*(r_0-r_3+r_5-r_7)$. This property of doubling seems to be important for two reasons. Firstly, it generates some associations with phenomena of dichotomous reproductions of DNA and somatic cells in a course of mitosis in biological organisms. Secondly, this phenomenon leads to the idea that for a system of genetic coding the main significance belong not to the entire set of possible real values of coordinates of genetic 8-dimensional hypercomplex numbers but only to the subset of numbers $2^0, 2^1, 2^2,…, 2^n$,.. . In other words, in the study of algebraic properties of genetic code systems the emphasis should be placed on the DC-algebras over this subset of numbers $2^0, 2^1, 2^2,…, 2^n$, which is long used in discrete signal processing in some aspects. It seems very probably that for the genetic system DS-algebras are algebras of dichotomous biological processes.
3) The set of 8 types of genetic $R_{123}$-quaternions with unit coordinates ($x_0=x_2=x_3=x_4=x_5=x_6=x_7=1$) is divided into 4 subsets with a pair of "complementary" $R_{123}$-quaternions in each: a) $(r_0+r_3+r_5+r_7)$ and $(r_0+r_2+r_4+r_6)$; b) $(r_0+r_3+r_5+r_6)$ and $(r_0+r_2+r_4+r_7)$; c) $(r_0+r_3+r_4+r_7)$ and $(r_0+r_2+r_5+r_6)$; d) $(r_0+r_3+r_4+r_6)$ and $(r_0+r_2+r_5+r_7)$. The "complementarity" means here that product of both members of any of these pairs gives the Rademacher form R (Figure 6) of representation of the genomatrix [C A; U G](3): $(r_0+r_3+r_5+r_7)*(r_0+r_2+r_4+r_6) = (r_0+r_3+r_5+r_6)*(r_0+r_2+r_4+r_7) = (r_0+r_3+r_4+r_7)*(r_0+r_2+r_5+r_6) =$ $(r_0+r_3+r_4+r_6)*(r_0+r_2+r_5+r_7) = R$. This property of complementarity of $R_{123}$-quaternions is vaguely reminiscent of the genetic phenomenon of complementarity of nitrogenous bases in DNA and RNA. (By the way, the multiplication of this matrix R (Figure 6) with itself gives its tetra-reproduction ($R^2=4*R$), which is reminiscent of the genetic phenomenon of tetra-reproductions of germ cells and genetic materials in meiosis).

It is obvious that for such $R_{123}$-quaternions one can receive a special vector calculation by analogy with the case of Hamilton quaternions where a product of the vector parts of quaternions with their conjugations are used to receive formulas of the vector calculation for the 3-dimensional physical space.

## 6. Dyadic shifts of binary numerations of columns and rows in the initial genomatrix

A disposition of each of the triplets in the initial genomatrix $[C\ A;\ U\ G]^{(3)}$ (Figure 4) is defined by binary numerations of its column and its row (these numerations reflect objective properties of triplets from viewpoints of the first two binary sub-alphabets on Figure 3). If the initial order of positions of binary-numerated columns and rows (Figure 4) is changed, then dispositions of triplets are also changed that lead to a new mosaic genomatrix. Let us study genomatrices which arise if the binary numeration of columns and rows is changed by means of modulo-2 addition of their binary numerations with one of binary numbers from the dyadic group (2). In these cases the initial order of binary numerations of columns and rows on Figure 4 (that is 000, 001, 010, 011, 100, 101, 110, 111) is replaced by the following orders of numerations of columns and rows:

№1) if 001 is taken for modulo-2 addition, then 001, 000, 011, 010, 101, 100, 111, 110;
№2) if 010 is taken for modulo-2 addition, then 010, 011, 000, 001, 110, 111, 100, 101;
№3) if 011 is taken for modulo-2 addition, then 011, 010, 001, 000, 111, 110, 101, 100;
№4) if 100 is taken for modulo-2 addition, then 100, 101, 110, 111, 000, 001, 010, 011;
№5) if 101 is taken for modulo-2 addition, then 101, 100, 111, 110, 001, 000, 011, 010;
№6) if 110 is taken for modulo-2 addition, then 110, 111, 100, 101, 010, 011, 000, 001;
№7) if 111 is taken for modulo-2 addition, then 111, 110, 101, 100, 011, 010, 001, 000.

| | 100 (4) | 101 (5) | 110 (6) | 111 (7) | 000 (0) | 001 (1) | 010 (2) | 011 (3) | |
|---|---|---|---|---|---|---|---|---|---|
| 100(4) | CGG 0 | CGU 1 | CUG 2 | CUU 3 | AGG 4 | AGU 5 | AUG 6 | AUU 7 | + + + + - - - - |
| 101(5) | CGA 1 | CGC 0 | CUA 3 | CUC 2 | AGA 5 | AGC 4 | AUA 7 | AUC 6 | + + + + - - - - |
| 110(6) | CAG 2 | CAU 3 | CCG 0 | CCU 1 | AAG 6 | AAU 7 | ACG 4 | ACU 5 | - - + + - - + + |
| 111(7) | CAA 3 | CAC 2 | CCA 1 | CCC 0 | AAA 7 | AAC 6 | ACA 5 | ACC 4 | - - + + - - + + |
| 000(0) | UGG 4 | UGU 5 | UUG 6 | UUU 7 | GGG 0 | GGU 1 | GUG 2 | GUU 3 | - - - - + + + + |
| 001(1) | UGA 5 | UGC 4 | UUA 7 | UUC 6 | GGA 1 | GGC 0 | GUA 3 | GUC 2 | - - - - + + + + |
| 010(2) | UAG 6 | UAU 7 | UCG 4 | UCU 5 | GAG 2 | GAU 3 | GCG 0 | GCU 1 | - - + + - - + + |
| 011(3) | UAA 7 | UAC 6 | UCA 5 | UCC 4 | GAA 3 | GAC 2 | GCA 1 | GCC 0 | - - + + - - + + |

Figure 12. One of examples of genomatrices received from the initial genomatrix $[C\ A;\ U\ G]^{(3)}$ by means of the dyadic-shift permutation of its columns and rows on the base of modulo-2 addition of the binary number 100 with binary numerations of columns and rows from Figure 4. The corresponding Rademacher form of representation of this genomatrix is shown at the right side (where "+" means "+1", and "-" means "-1"). Dyadic-shifts numerations of triplets are shown.

One can check easily that all these cases of dyadic-shift permutations of columns and rows in $[C\ A;\ U\ G]^{(3)}$ lead only to two variants of Rademacher forms of genomatrices with permutated triplets, columns and rows:

a) The first variant is identical to the Rademacher form on Figure 6; the mentioned cases №№ 1, 4, 5 correspond to this variant;

b) The second variant relates the cases №№ 2, 3, 6, 7 which lead to new genomatrices with a new black-and-white mosaics (one of their examples is shown on Figure 12); this new mosaics is the same for all these cases and it corresponds to a new Rademacher form of their representation. But the dyadic-shift decomposition of this new Rademacher form leads to a new set of 8 basic matrices, which is closed in relation to multiplication and which generates again the multiplication table presented on Figure 8. It means that two different matrix forms of representation of the same genetic DS-algebra of $R_{123}$-octetons are received here by means of described dyadic-shift permutations of column and rows. This DS-algebra is invariant for such kind of dyadic-shift transformations of genomatrices in their Rademacher forms.

## 7. Permutations of positions inside triplets in the genomatrix $[C\ A;\ U\ G]^{(3)}$

The theory of discrete signal processing pays a special attention to permutations of information elements. This section shows that all the possible permutations of positions inside all the triplets lead to new mosaic genomatrices whose Rademacher forms of representation are connected with the same DS-algebra (Figure 8).

It is obvious that a simultaneous permutation of positions in triplets transforms the most of the triplets in cells of the initial genomatrix $[C\ A;\ U\ G]^{(3)}$. For example, in the case of the cyclic transformation of the order 1-2-3 of positions into the order 2-3-1, the triplet CAG is transformed into the triplet AGC, etc. Because each of the triplets is connected with the binary numeration of its column and row, these binary numerations are also transformed correspondingly; for example, the binary numeration 011 is transformed into 110.

Six variants of the order of positions inside triplets are possible: 1-2-3, 2-3-1, 3-1-2, 3-2-1, 2-1-3, 1-3-2. The initial genomatrix $[C\ A;\ U\ G]_{123}^{(3)}$ is related with the first of these orders (Figure 4). Other five genomatrices $[C\ A;\ U\ G]_{231}^{(3)}$, $[C\ A;\ U\ G]_{312}^{(3)}$, $[C\ A;\ U\ G]_{321}^{(3)}$, $[C\ A;\ U\ G]_{213}^{(3)}$, $[C\ A;\ U\ G]_{132}^{(3)}$, which correspond to other five orders, are shown on Figure 13 (subscripts indicate the order of positions in triplets). In these genomatrices, black-and-white mosaic of each row corresponds again to one of Rademacher functions. The replacement of all the triplets with strong and weak roots by entries "+1" and "-1" correspondingly transforms these genomatrices into their Rademacher forms $R_{231}$, $R_{312}$, $R_{321}$, $R_{213}$, $R_{132}$ (Figure 14).

$[C\ A;\ U\ G]_{231}^{(3)} =$

| | 000 (0) | 010 (2) | 100 (4) | 110 (6) | 001 (1) | 011 (3) | 101 (5) | 111 (7) |
|---|---|---|---|---|---|---|---|---|
| 000 (0) | CCC 000 (0) | CAC 010 (2) | ACC 100 (4) | AAC 110 (6) | CCA 001 (1) | CAA 011 (3) | ACA 101 (5) | AAA 111 (7) |
| 010 (2) | CUC 010 (2) | CGC 000 (0) | AUC 110 (6) | AGC 100 (4) | CUA 011 (3) | CGA 001 (1) | AUA 111 (7) | AGA 101 (5) |
| 100 (4) | UCC 100 (4) | UAC 110 (6) | GCC 000 (0) | GAC 010 (2) | UCA 101 (5) | UAA 111 (7) | GCA 001 (1) | GAA 011 (3) |
| 110 (6) | UUC 110 (6) | UGC 100 (4) | GUC 010 (2) | GGC 000 (0) | UUA 111 (7) | UGA 101 (5) | GUA 011 (3) | GGC 001 (1) |
| 001 (1) | CCU 001 (1) | CAU 011 (3) | ACU 101 (5) | AAU 111 (7) | CCG 000 (0) | CAU 010 (2) | ACG 100 (4) | AAG 110 (6) |
| 011 (3) | CUU 011 (3) | CGU 001 (1) | AUU 111 (7) | AGU 101 (5) | CUG 010 (2) | CGG 000 (0) | AUG 110 (6) | AGG 100 (4) |
| 101 (5) | UCU 101 (5) | UAU 111 (7) | GCU 001 (1) | GAU 011 (3) | UCG 100 (4) | UAG 110 (6) | GCG 000 (0) | GAG 010 (2) |
| 111 (7) | UUU 111 (7) | UGU 101 (5) | GUU 011 (3) | GGU 001 (1) | UUG 110 (6) | UGG 100 (4) | GUG 010 (2) | GGG 000 (0) |

$[C\ A;\ U\ G]_{312}^{(3)} =$

| | 000 (0) | 100 (4) | 001 (1) | 101 (5) | 010 (2) | 110 (6) | 011 (3) | 111 (7) |
|---|---|---|---|---|---|---|---|---|
| 000 (0) | CCC 000 (0) | ACC 100 (4) | CCA 001 (1) | ACA 101 (5) | CAC 010 (2) | AAC 110 (6) | CAA 011 (3) | AAA 111 (7) |
| 100 (4) | UCC 100 (4) | GCC 000 (0) | UCA 101 (5) | GCA 001 (1) | UAC 110 (6) | GAC 010 (2) | UAA 111 (7) | GAA 011 (3) |
| 001 (1) | CCU 001 (1) | ACU 101 (5) | CCG 000 (0) | ACG 100 (4) | CAU 011 (3) | AAU 111 (7) | CAU 010 (2) | AAG 110 (6) |
| 101 (5) | UCU 101 (5) | GCU 001 (1) | UCG 100 (4) | GCG 000 (0) | UAU 111 (7) | GAU 011 (3) | UAG 110 (6) | GAG 010 (2) |
| 010 (6) | CUC 010 (2) | AUC 110 (6) | CUA 011 (3) | AUA 111 (7) | CGC 000 (0) | AGC 100 (4) | CGA 001 (1) | AGA 101 (5) |
| 011 (3) | UUC 110 (6) | GUC 010 (2) | UUA 111 (7) | GUA 011 (3) | UGC 100 (4) | GGC 000 (0) | UGA 101 (5) | GGC 001 (1) |
| 001 (3) | CUU 011 (3) | AUU 111 (7) | CUG 010 (2) | AUG 110 (6) | CGU 001 (1) | AGU 101 (5) | CGG 000 (0) | AGG 100 (4) |
| 111 (7) | UUU 111 (7) | GUU 011 (3) | UUG 110 (6) | GUG 010 (2) | UGU 101 (5) | GGU 001 (1) | UGG 100 (4) | GGG 000 (0) |

$[C\ A;\ U\ G]_{321}^{(3)} =$

| | 000 (0) | 100 (4) | 010 (2) | 110 (6) | 001 (1) | 101 (5) | 011 (3) | 111 (7) |
|---|---|---|---|---|---|---|---|---|
| 000 (0) | CCC 000 (0) | ACC 100 (4) | CAC 010 (2) | AAC 110 (6) | CCA 001 (1) | ACA 101 (5) | CAA 011 (3) | AAA 111 (7) |
| 100 (4) | UCC 100 (4) | GCC 000 (0) | UAC 110 (6) | GAC 010 (2) | UCA 101 (5) | GCA 001 (1) | UAA 111 (7) | GAA 011 (3) |
| 010 (2) | CUC 010 (2) | AUC 110 (6) | CGC 000 (0) | AGC 100 (4) | CUA 011 (3) | AUA 111 (7) | CGA 001 (1) | AGA 101 (5) |
| 110 (6) | UUC 110 (6) | GUC 010 (2) | UGC 100 (4) | GGC 000 (0) | UUA 111 (7) | GUA 011 (3) | UGA 101 (5) | GGC 001 (1) |
| 001 (1) | CCU 001 (1) | ACU 101 (5) | CAU 011 (3) | AAU 111 (7) | CCG 000 (0) | ACG 100 (4) | CAU 010 (2) | AAG 110 (6) |
| 101 (5) | UCU 101 (5) | GCU 001 (1) | UAU 111 (7) | GAU 011 (3) | UCG 100 (4) | GCG 000 (0) | UAG 110 (6) | GAG 010 (2) |
| 011 (3) | CUU 011 (3) | AUU 111 (7) | CGU 001 (1) | AGU 101 (5) | CUG 010 (2) | AUG 110 (6) | CGG 000 (0) | AGG 100 (4) |
| 111 (7) | UUU 111 (7) | GUU 011 (3) | UGU 101 (5) | GGU 001 (1) | UUG 110 (6) | GUG 010 (2) | UGG 100 (4) | GGG 000 (0) |

$[C\ A;\ U\ G]_{213}^{(3)} =$

| | 000 (0) | 001 (1) | 100 (4) | 101 (5) | 010 (2) | 011 (3) | 110 (6) | 111 (7) |
|---|---|---|---|---|---|---|---|---|
| 000 (0) | CCC 000 (0) | CCA 001 (1) | ACC 100 (4) | ACA 101 (5) | CAC 010 (2) | CAA 011 (3) | AAC 110 (6) | AAA 111 (7) |
| 001 (1) | CCU 001 (1) | CCG 000 (0) | ACU 101 (5) | ACG 100 (4) | CAU 011 (3) | CAU 010 (2) | AAU 111 (7) | AAG 110 (6) |
| 100 (4) | UCC 100 (4) | UCA 101 (5) | GCC 000 (0) | GCA 001 (1) | UAC 110 (6) | UAA 111 (7) | GAC 010 (2) | GAA 011 (3) |
| 101 (5) | UCU 101 (5) | UCG 100 (4) | GCU 001 (1) | GCG 000 (0) | UAU 111 (7) | UAG 110 (6) | GAU 011 (3) | GAG 010 (2) |
| 010 (2) | CUC 010 (2) | CUA 011 (3) | AUC 110 (6) | AUA 111 (7) | CGC 000 (0) | CGA 001 (1) | AGC 100 (4) | AGA 101 (5) |
| 011 (3) | CUU 011 (3) | CUG 010 (2) | AUU 111 (7) | AUG 110 (6) | CGU 001 (1) | CGG 000 (0) | AGU 101 (5) | AGG 100 (4) |
| 110 (6) | UUC 110 (6) | UUA 111 (7) | GUC 010 (2) | GUA 011 (3) | UGC 100 (4) | UGA 101 (5) | GGC 000 (0) | GGC 001 (1) |
| 111 (7) | UUU 111 (7) | UUG 110 (6) | GUU 011 (3) | GUG 010 (2) | UGU 101 (5) | UGG 100 (4) | GGU 001 (1) | GGG 000 (0) |

$[C\ A;\ U\ G]_{132}^{(3)}$ =

| | 000 (0) | 010 (2) | 001 (1) | 011 (3) | 100 (4) | 110 (6) | 101 (5) | 111 (7) |
|---|---|---|---|---|---|---|---|---|
| 000 (0) | CCC 000 (0) | CAC 010 (2) | CCA 001 (1) | CAA 011 (3) | ACC 100 (4) | AAC 110 (6) | ACA 101 (5) | AAA 111 (7) |
| 010 (2) | CUC 010 (2) | CGC 000 (0) | CUA 011 (3) | CGA 001 (1) | AUC 110 (6) | AGC 100 (4) | AUA 111 (7) | AGA 101 (5) |
| 001 (1) | CCU 001 (1) | CAU 011 (3) | CCG 000 (0) | CAG 010 (2) | ACU 101 (5) | AAU 111 (7) | ACG 100 (4) | AAG 110 (6) |
| 011 (3) | CUU 011 (3) | CGU 001 (1) | CUG 010 (2) | CGG 000 (0) | AUU 111 (7) | AGU 101 (5) | AUG 110 (6) | AGG 100 (4) |
| 100 (4) | UCC 100 (4) | UAC 110 (6) | UCA 101 (5) | UAA 111 (7) | GCC 000 (0) | GAC 010 (2) | GCA 001 (1) | GAA 011 (3) |
| 110 (6) | UUC 110 (6) | UGC 100 (4) | UUA 111 (7) | UGA 101 (5) | GUC 010 (2) | GGC 000 (0) | GUA 011 (3) | GGA 001 (1) |
| 101 (5) | UCU 101 (5) | UAU 111 (7) | UCG 100 (4) | UAG 110 (6) | GCU 001 (1) | GAU 011 (3) | GCG 000 (0) | GAG 010 (2) |
| 111 (7) | UUU 111 (7) | UGU 101 (5) | UUG 110 (6) | UGG 100 (4) | GUU 011 (3) | GGU 001 (1) | GUG 010 (2) | GGG 000 (0) |

Figure 13. Five genomatrices $[C\ A;\ U\ G]_{231}^{(3)}$, $[C\ A;\ U\ G]_{312}^{(3)}$, $[C\ A;\ U\ G]_{321}^{(3)}$, $[C\ A;\ U\ G]_{213}^{(3)}$, $[C\ A;\ U\ G]_{132}^{(3)}$, which correspond to orders of positions in triplets 2-3-1, 3-1-2, 3-2-1, 2-1-3, 1-3-2 in relation to the genomatrix $[C\ A;\ U\ G]_{123}^{(3)}$ on Figure 4. Black and white cells contain triplets with strong and weak roots correspondingly. Binary numerations of columns and rows are shown.

$R_{231}$ =

| | | | | | | | |
|---|---|---|---|---|---|---|---|
| +1 (0) | -1 (2) | +1 (4) | -1 (6) | +1 (1) | -1 (3) | +1 (5) | -1 (7) |
| +1 (2) | +1 (0) | -1 (6) | -1 (4) | +1 (3) | +1 (1) | -1 (7) | -1 (5) |
| +1 (4) | -1 (6) | +1 (0) | -1 (2) | +1 (5) | -1 (7) | +1 (1) | -1 (3) |
| -1 (6) | -1 (4) | +1 (2) | +1 (0) | -1 (7) | -1 (5) | +1 (3) | +1 (0) |
| +1 (1) | -1 (3) | +1 (5) | -1 (7) | +1 (0) | -1 (2) | +1 (4) | -1 (6) |
| +1 (3) | +1 (1) | -1 (7) | -1 (5) | +1 (2) | +1 (0) | -1 (6) | -1 (4) |
| +1 (5) | -1 (7) | +1 (1) | -1 (3) | +1 (4) | -1 (6) | +1 (0) | -1 (2) |
| -1 (7) | -1 (5) | +1 (3) | +1 (1) | -1 (6) | -1 (4) | +1 (2) | +1 (0) |

$R_{312}$ =

| | | | | | | | |
|---|---|---|---|---|---|---|---|
| +1 (0) | +1 (4) | +1 (1) | +1 (5) | -1 (2) | -1 (6) | -1 (3) | -1 (7) |
| +1 (4) | +1 (0) | +1 (5) | +1 (1) | -1 (6) | -1 (2) | -1 (7) | -1 (3) |
| +1 (1) | +1 (5) | +1 (0) | +1 (4) | -1 (3) | -1 (7) | -1 (2) | -1 (6) |
| +1 (5) | +1 (1) | +1 (4) | +1 (0) | -1 (7) | -1 (3) | -1 (6) | -1 (2) |
| +1 (2) | -1 (6) | +1 (3) | -1 (7) | +1 (0) | -1 (4) | +1 (1) | -1 (5) |
| -1 (6) | +1 (2) | -1 (7) | +1 (3) | -1 (4) | +1 (0) | -1 (5) | +1 (1) |
| +1 (3) | -1 (7) | +1 (2) | -1 (6) | +1 (1) | -1 (5) | +1 (0) | -1 (4) |
| -1 (7) | +1 (3) | -1 (6) | +1 (2) | -1 (5) | +1 (1) | -1 (4) | +1 (0) |

$R_{321}$=

| | | | | | | | |
|---|---|---|---|---|---|---|---|
| +1 (0) | +1 (4) | -1 (2) | -1 (6) | +1 (1) | +1 (5) | -1 (3) | -1 (7) |
| +1 (4) | +1 (0) | -1 (6) | -1 (2) | +1 (5) | +1 (1) | -1 (7) | -1 (3) |
| +1 (2) | -1 (6) | +1 (0) | -1 (4) | +1 (3) | -1 (7) | +1 (1) | -1 (5) |
| -1 (6) | +1 (2) | -1 (4) | +1 (0) | -1 (7) | +1 (3) | -1 (5) | +1 (1) |
| +1 (1) | +1 (5) | -1 (3) | -1 (7) | +1 (0) | +1 (4) | -1 (2) | -1 (6) |
| +1 (5) | +1 (1) | -1 (7) | -1 (3) | +1 (4) | +1 (0) | -1 (6) | -1 (2) |
| +1 (3) | -1 (7) | +1 (1) | -1 (5) | +1 (2) | -1 (6) | +1 (0) | -1 (4) |
| -1 (7) | +1 (3) | -1 (5) | +1 (1) | -1 (6) | +1 (2) | -1 (4) | +1 (0) |

$R_{213}$=

| | | | | | | | |
|---|---|---|---|---|---|---|---|
| +1 (0) | +1 (1) | +1 (4) | +1 (5) | -1 (2) | -1 (3) | -1 (6) | -1 (7) |
| +1 (1) | +1 (0) | +1 (5) | +1 (4) | -1 (3) | -1 (2) | -1 (7) | -1 (6) |
| +1 (4) | +1 (5) | +1 (0) | +1 (1) | -1 (6) | -1 (7) | -1 (2) | -1 (3) |
| +1 (5) | +1 (4) | +1 (1) | +1 (0) | -1 (7) | -1 (6) | -1 (3) | -1 (2) |
| +1 (2) | +1 (3) | -1 (6) | -1 (7) | +1 (0) | +1 (1) | -1 (4) | -1 (5) |
| +1 (3) | +1 (2) | -1 (7) | -1 (6) | +1 (1) | +1 (0) | -1 (5) | -1 (4) |
| -1 (6) | -1 (7) | +1 (2) | +1 (3) | -1 (4) | -1 (5) | +1 (0) | +1 (1) |
| -1 (7) | -1 (6) | +1 (3) | +1 (2) | -1 (5) | -1 (4) | +1 (1) | +1 (0) |

$R_{132}$=

| | | | | | | | |
|---|---|---|---|---|---|---|---|
| +1 (0) | -1 (2) | +1 (1) | -1 (3) | +1 (4) | -1 (6) | +1 (5) | -1 (7) |
| +1 (2) | +1 (0) | +1 (3) | +1 (1) | -1 (6) | -1 (4) | -1 (7) | -1 (5) |
| +1 (1) | -1 (3) | +1 (0) | -1 (2) | +1 (5) | -1 (7) | +1 (4) | -1 (6) |
| +1 (3) | +1 (1) | +1 (2) | +1 (0) | -1 (7) | -1 (5) | -1 (6) | -1 (4) |
| +1 (4) | -1 (6) | +1 (5) | -1 (7) | +1 (0) | -1 (2) | +1 (1) | -1 (3) |
| -1 (6) | -1 (4) | -1 (7) | -1 (5) | +1 (2) | +1 (0) | +1 (3) | +1 (1) |
| +1 (5) | -1 (7) | +1 (4) | -1 (6) | +1 (1) | -1 (3) | +1 (0) | -1 (2) |
| -1 (7) | -1 (5) | -1 (6) | -1 (4) | +1 (3) | +1 (1) | +1 (2) | +1 (0) |

Figure 14. The Rademacher forms $R_{231}$, $R_{312}$, $R_{321}$, $R_{213}$, $R_{132}$ of representation of genomatrices $[C\ A;\ U\ G]_{231}^{(3)}$, $[C\ A;\ U\ G]_{312}^{(3)}$, $[C\ A;\ U\ G]_{321}^{(3)}$, $[C\ A;\ U\ G]_{213}^{(3)}$, $[C\ A;\ U\ G]_{132}^{(3)}$ from Figure 13. Brackets contain dyadic numerations of cells from Figure 13.

Each of the Rademacher forms $R_{231}$, $R_{312}$, $R_{321}$, $R_{213}$, $R_{132}$ can be again decomposed into the sum of 8 sparse matrices $r_0$, $r_1$, $r_2$, $r_3$, $r_4$, $r_5$, $r_6$, $r_7$ in accordance with dyadic numerations of its cells on Figure 14. More precisely any sparse matrix $r_k$ (k=0, 1, …, 7) of such decomposition contains entries "+1" or "-1" from the matrix on Figure 14 only in those cells which correspond to cells with the same dyadic numeration "k"; all the other cells of the matrix $r_k$ contain zero. Figure 15 shows all the sets of sparse matrices $r_k$ of such dyadic-shift decompositions.

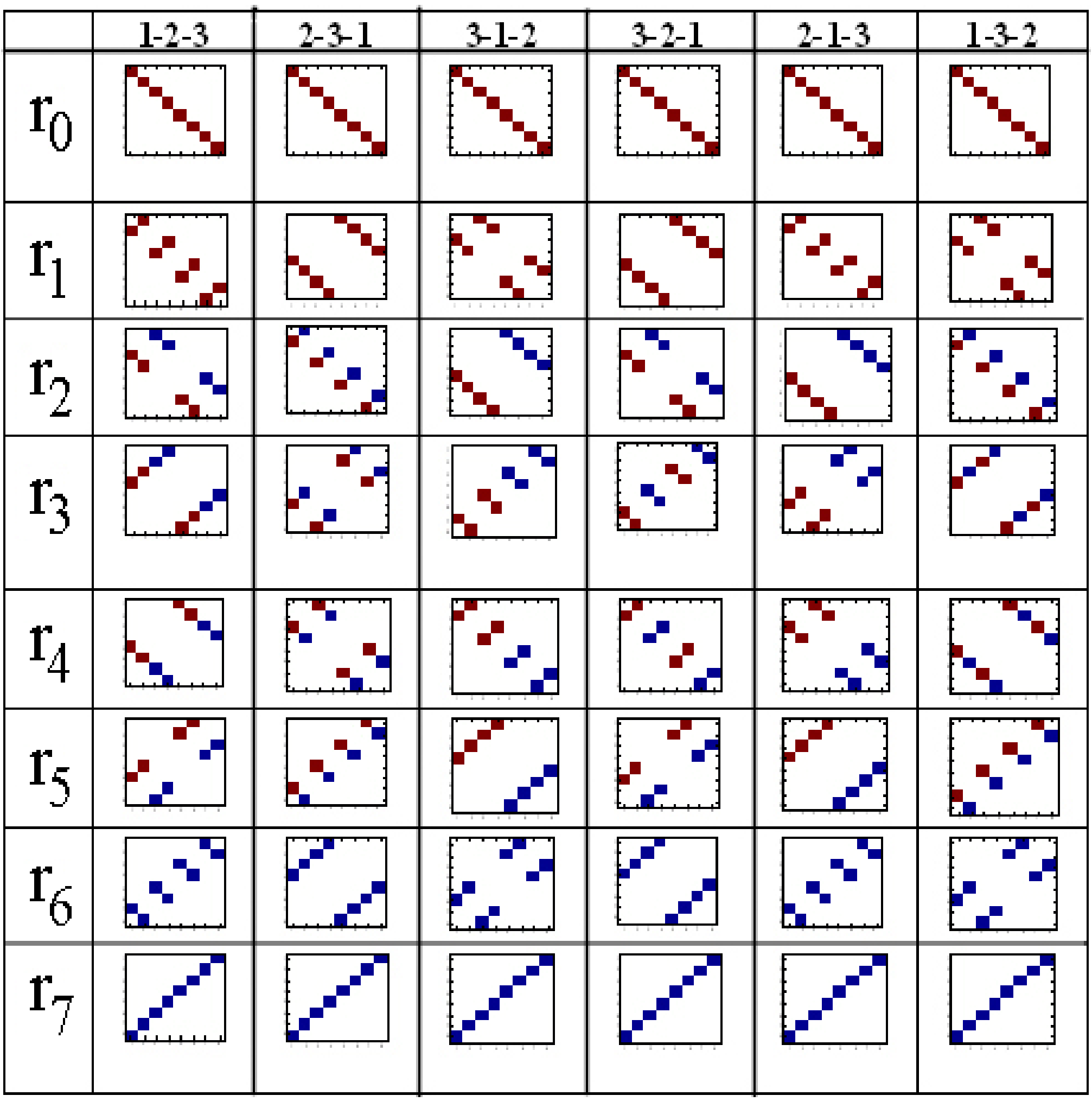

Figure 15. The sets of sparse (8x8)-matrices $r_0$, $r_1$, $r_2$, $r_3$, $r_4$, $r_5$, $r_6$, $r_7$ of the dyadic-shift decomposition of Rademacher forms $R_{123}$, $R_{231}$, $R_{312}$, $R_{321}$, $R_{213}$, $R_{132}$ (Figures 6 and 14). Red and blue cells contain entries "+1" and "-1" correspondingly. White cells contain zero. Each of the columns of the table is denoted by the index of the corresponding matrix.

One can see from Figure 15 that each of the 6 sets with eight sparse matrices $r_0$, $r_1$, $r_2$, $r_3$, $r_4$, $r_5$, $r_6$, $r_7$ is unique and different from other sets ($r_0$ is the identity matrix in all the sets). Unexpected facts are that, firstly, each of these sets is closed in relation to multiplication and, secondly, each of these sets corresponds to the same multiplication table from Figure 8. It means that this genetic DS-algebra of 8-dimensional hypercomplex numbers possesses at least 5 additional forms of its matrix representation (taking into account the result of the previous section about its additional form of matrix representation on Figure 12). In particular the Rademacher forms $R_{123}$, $R_{231}$, $R_{312}$, $R_{321}$, $R_{213}$, $R_{132}$ are different forms of matrix representation of the same R-octeton whose coordinates are equal to 1 ($x_0=x_1=\ldots=x_7=1$). Our results demonstrate that this DS-algebra of genetic octetons possesses a wonderful stability (invariance) in relation to many of the variants of positional permutations in triplets and in relation to the dyadic shifts on the base of modulo-2 addition. All the properties of R-octetons from the section 5 (including metric quaternions) are true in the cases of different forms of matrix representation of R-octetons with the same multiplication table (Figure 8).

One should mention that some other permutations of columns and rows of the initial genomatrix $[C\ A;\ U\ G]^{(3)}$ (Figure 4) lead to the same DS-algebra with the multiplication table (Figure 8) in the case of such dyadic-shift decompositions. For example, permutations of column and rows in $[C\ A;\ U\ G]^{(3)}$ (Figure 4), where they are arranged in the order of binary numerations 000, 001, 010, 011, 100, 101, 110, 111, into the new order 100, 101, 111, 110, 010, 011, 001, 000 lead to a new mosaic genomatrix, whose dyadic-shift decomposition gives a new set of 8 basic matrices with the same multiplication table (Figure 8). This special sequence of numerations 100, 101, 111, 110, 010, 011, 001, 000 is very interesting because this sequence is identical to the ordered sequence 7, 6, 5, 4, 3, 2, 1, 0 in decimal notation from the viewpoint of Gray code (which is widely used in discrete signal processing). The work [Kappraff, Adamson, 2009] was the first article in the field of matrix genetics where a relation of Gray code with genetic matrices was studied initially on the base of numeration of columns and rows of genomatrices in accordance with Gray code sequences. A little later the work [Kappraff, Petoukhov, 2009] has demonstrated a connection of genomatrices with so termed bipolar 8-dimensional algebras on the base of a special type of decompositions of genomatrices ("bipolar decompositions"). This article describes another type of genomatrix decompositions (dyadic-shift decompositions) which lead to DS-algebras of genetic octetons.

### 8. Genetic 8-dimensional hypercomplex numbers and evolution of the genetic code

Beginning with the level of the correspondence between 64 triplets and 20 amino acids, some evolutional changes take place in the general scheme of the genetic code, which lead to some different variants (or dialects) of the genetic code. Such dialects exist in different kinds of organisms or of their subsystems (first of all, in mitochondria, which play a role of factories of energy in biological cells). For this article all initial data about the dialects of the genetic code were taken from the NCBI's website http://www.ncbi.nlm.nih.gov/Taxonomy/Utils/wprintgc.cgi). Today the science knows 17 dialects of the genetic code, which differ one from another by changing the correspondence among some triplets and amino acids; it means that in different dialects of the genetic code changeable triplets encode different amino acids and stop-codons. Not all the triplets, but only a few of them are changeable in the set of the dialects. Figure 16 shows these 17 dialects together with all the changeable triplets in comparison with the Vertebrate Mitochondria Genetic Code which is the most symmetrical among all the dialects; start-codons of each of the dialects are shown as well. For each of triplets its dyadic-shift numeration (Figure 4) is presented on Figure 16 to analyze the phenomenon of evolution (or diversity) of the dialects of the genetic code from the viewpoint of the DS-algebra of genetic octetons (Figure 8).

One can see from the second column of Figure 16 that all the changeable triplets correspond only to four kinds of dyadic-shift numerations 4, 5, 6, 7 except in the special case of Yeast in dialects №15 and №16 where dyadic-shift numerations 2 and 3 exist. But Yeast is unicellular chemoorganoheterotrophic mushrooms, which are reproduced by a vegetative cloning (an asexual reproduction). We think that the genetic-code specificity of the yeast is connected with their asexual reproduction and heterotrophy (our previous works have noted that the dialects of the genetic code of the heterotrophs, which feed on ready living substance, can have some deviations from the canonical forms of the genetic dialects of autotrophic organisms [Petoukhov, 2008a, http:// arXiv:0805.4692]). The additional evidence of molecular-genetic singularity of Yeast is the fact that the histone H1 is not discovered in their genetic system at all (http://drosophila.narod.ru/Review/histone.html).

| **Dialects of the genetic code** | **Changeable triplets and their dyadic-shift numerations in [C A; U G]$^{(3)}$** | **Start-codons and their dyadic-shift numerations in [C A; U G]$^{(3)}$** |
|---|---|---|
| 1) The Vertebrate Mitochondrial Code | | AUU, **7**<br>AUC, 6<br>AUA, 7<br>AUG, 6<br>GUG, 2 |
| 2) The Standart Code | UGA, Stop (Trp), 5<br>AGG, Arg (Stop), 4<br>AGA, Arg (Stop), 5<br>AUA, Ile (Met), 7 | UUG, 6<br>CUG, 2<br>AUG, 6 |
| 3) The Mold, Protozoan, and Coelenterate Mitochondrial Code and the Mycoplasma/Spiroplasma Code | AGG, Arg (Stop), 4<br>AGA, Arg (Stop), 5<br>AUA, Ile (Met), 7 | UUG, 6<br>UUA, 7<br>CUG, 2<br>AUC, 6<br>AUU, 7<br>AUG, 6<br>AUA, 7<br>GUG, 2 |
| 4) The Invertebrate Mitochondrial Code | AGG, Ser (Stop), 4<br>AGA, Ser (Stop), 5 | UUG, 6<br>AUU, 7<br>AUC, 6<br>AUA, 7<br>AUG, 6<br>GUG, 2 |
| 5) The Echinoderm and Flatworm Mitochondrial Code | AGG, Ser (Stop), 4<br>AGA, Ser (Stop), 5<br>AUA, Ile (Met), 7<br>AAA, Asn (Lys), 7 | AUG, 6<br>GUG, 2 |
| 6) The Euplotid Nuclear Code | UGA, Cys (Trp), 5<br>AGG, Arg (Stop), 4<br>AGA, Arg (Stop), 5<br>AUA, Ile (Met), 7 | AUG, 6 |
| 7) The Bacterial and Plant Plastid Code | UGA, Stop (Trp), 5<br>AGG, Arg (Stop), 4<br>AGA, Arg (Stop), 5 | UUG, 6<br>CUG, 2<br>AUC, 6 |
| | AUA, Ile (Met), 7 | AUU, 7<br>AUA, 7<br>AUG, 6 |
| 8) The Ascidian Mitochondrial Code | AGG, Gly (Stop), 4<br>AGA, Gly (Stop), 5 | UUG, 6<br>AUA, 7<br>AUG, 6<br>GUG, 2 |
| 9) The Alternative Flatworm Mitochondrial Code | UAA, Tyr (Stop), 7<br>AGG, Ser (Stop), 4<br>AGA, Ser (Stop), 5<br>AUA, Ile (Met), 7<br>AAA, Asn (Lys), 7 | AUG, 6 |
| 10) Blepharisma Nuclear Code | UGA, Stop (Trp), 5<br>UAG, Gln (Stop), 6<br>AGG, Arg (Stop), 4<br>AGA, Arg (Stop), 5<br>AUA, Ile (Met), 7 | AUG, 6 |
| 11) Chlorophycean Mitochondrial Code | UGA, Stop (Trp), 5<br>UAG, Leu (Stop), 6<br>AGG, Arg (Stop), 4<br>AGA, Arg (Stop), 5<br>AUA, Ile (Met), 7 | AUG, 6 |
| 12) Trematode Mitochondrial Code | AGG, Ser (Stop), 4<br>AGA, Ser (Stop), 5<br>AAA, Asn (Lys), 7 | AUG, 6<br>GUG, 2 |
| 13) Scenedesmus obliquus mitochondrial Code | UGA, Stop (Trp), 5<br>UAG, Leu (Stop), 6<br>UCA, Stop (Ser), 5<br>AGG, Arg (Stop), 5<br>AGA, Arg (Stop), 5<br>AUA, Ile (Met), 7 | AUG, 6 |
| 14) Thraustochytrium Mitochondrial Code | UGA, Stop (Trp), 5<br>UUA, Stop (Leu), 7<br>AGG, Arg (Stop), 4<br>AGA, Arg (Stop), 5<br>AUA, Ile (Met), 7 | AUU, 7<br>AUG, 6<br>GUG, 2 |
| 15) The Alternative Yeast Nuclear Code | UGA, Stop (Trp), 5<br>AGG, Arg (Stop), 4<br>AGA, Arg (Stop), 5<br>AUA, Ile (Met), 7<br>CUG, Ser (Leu), 2 | CUG, 2<br>AUG, 6 |
| 16) The Yeast Mitochondrial Code | AGG, Arg (Stop), 4<br>AGA, Arg (Stop), 5<br>CUG, Thr (Leu), 2<br>CUU, Thr (Leu), 3<br>CUA, Thr (Leu), 3<br>CUC, Thr (Leu), 2 | AUA, 7<br>AUG, 6 |
| 17) The Ciliate, Dasycladacean and Hexamita Nuclear Code | UGA, Stop (Trp), 5<br>UAG, Gln (Stop), 6<br>UAA, Gln (Stop), 7 | AUG, 6 |
| | AGG, Arg (Stop), 4<br>AGA, Arg (Stop), 5<br>AUA, Ile (Met), 7 | |

Figure 16. The 17 dialects of the genetic code and all their changeable triplets (in the second column) together with their dyadic-shift numerations in the genomatrix $[C\ A;\ U\ G]^{(3)}$ (Figure 4). Brackets in each row of the second column show an amino acid or stop-codon which is encoded by this triplet in the dialect № 1. The third column contains start-codons for each dialect together with their dyadic-shift numerations in the genomatrix $[C\ A;\ U\ G]^{(3)}$ (Figure 4). Initial data are taken from http://www.ncbi.nlm.nih.gov/Taxonomy/Utils/wprintgc.cgi.

It is known that Mendel's laws hold true only for bisexual organisms (organisms with a sexual reproduction). Taking into account the asexual reproduction of Yeast, the data of the table on Figure 16 allow formulating the following phenomenologic rule which holds true for all organisms with sexual reproduction:

**Rule 1**:

- The absolute rule for all bisexual organisms is that only those triplets can be involved in the evolutionary changing their correspondence to amino acids or to stop-signals, which possess dyadic-shift numerations 4, 5, 6, 7 in the genomatrix $[C\ A;\ U\ G]^{(3)}$ (Figure 4); in other words, only those triplets can be involved which are connected with the basic matrices $r_4$, $r_5$, $r_6$, $r_7$ (Figure 7) of genetic R-octetons (all these changeable triplets are disposed only in two quadrants along the second diagonal of the matrix $[C\ A;\ U\ G]^{(3)}$).

One can see from the multiplication table on Figure 8 that algebraic properties of the basic matrices $r_4$, $r_5$, $r_6$ and $r_7$ differ from properties of the basic matrices $r_0$, $r_1$, $r_2$ and $r_3$. Really multiplication of any pair of matrices from the set $r_4$, $r_5$, $r_6$ and $r_7$ gives a matrix from another set $r_0$, $r_1$, $r_2$ and $r_3$, but multiplication of any pair of matrices from the set $r_0$, $r_1$, $r_2$ and $r_3$ gives a matrix from the same set. This article shows below that such specific of the disposition of the changeable triplets correlates with the structure of the matrix representation of biquaternions. In addition, it is interesting that the set of changeable triplets (Figure 16) includes only triplets with suffixes A and G (one exception is again the Yeast's dialect № 16 with the changeable triplet CUU).

An additional analysis of the table on Figure 16 allows formulating the second phenomenologic rule.

**Rule 2**:

- The absolute rule for all bisexual organisms is that only triplets with weak roots can be involved in the evolutionary changing their correspondence to amino acids or to stop-signals.

The third column on Figure 16 allows formulating the third phenomenologic rule from the viewpoint of the genetic DS-algebra.

**Rule 3**:

- In all the dialects of the genetic code only triplets with dyadic-shift numerations 2, 6, 7 can be start-codons. In other words, only those triplets can be start-codons, which are connected with the basic matrices $r_2$, $r_6$, $r_7$ (Figure 7) of genetic R-octetons.

It is interesting from the algebraic viewpoint that these three basic matrices $r_2$, $r_6$, $r_7$ define all other basic matrices $r_0$, $r_1$, $r_3$, $r_4$, $r_5$ (Figure 7) by means of multiplication. Really, $r_0 = r_6*r_6$, $r_1= r_6*r_7$, $r_3= r_2*r_6*r_7$, $r_4 = r_2*r_6$, $r_5 = r_2*r_7$.

These initial data and rules testify in a favor of that genetic DS-algebras of 8-dimensional hypercomplex numbers can be useful to understand genetic systems from the mathematical viewpoint more and more.

## 9. Hadamard genomatrices and 8-dimensional hypercomplex numbers

By definition a Hadamard matrix of dimension "n" is the (n*n)-matrix H(n) with elements "+1" and "-1". It satisfies the condition $H(n)*H(n)^T = n*I_n$, where $H(n)^T$ is the transposed matrix and $I_n$ is the identity (n*n)-matrix. Rows of Hadamard matrices are termed Walsh functions. Hadamard matrices are widely used in error-correcting codes such as the Reed-Muller code and Hadamard codes; in the theory of compression of signals and images; in spectral analysis and multi-channel spectrometers with Hadamard transformations; in quantum computers with Hadamard gates; in a realization of Boolean functions by means of spectral methods; in the theory of planning of multiple-factor experiments and in many other branches of science and technology. The works [Petoukhov, 2005b, 2008a,b] have revealed that Kronecker families of genetic matrices are related to some kinds of Hadamard matrices ("Hadamard genomatrices") by means of so termed U-algorithm.

This section describes that the dyadic-shift decompositions of Hadamard genomatrices lead to special 8-dimensional hypercomplex numbers. More precisely Hadamard genomatrices are the sum of basic matrices of some DS-algebras of 8-dimensional hypercomplex numbers; or, in other words, Hadamard (8x8)-genomatrices are 8-dimensional hypercomplex numbers whose coordinates are equal to 1.

Let us begin the description of relations of Hadamard matrices with the system of genetic alphabets. For the U-algorithm, phenomenological facts are essential that the letter U in RNA (and correspondingly the letter T in DNA) is a very special letter in the 4-letter alphabet of nitrogenous bases in the following two aspects:

- Each of three nitrogenous bases A, C, G has one amino-group $NH_2$, but the fourth basis U/T does not have it. From the viewpoint of the existence of the amino-group (which is very important for genetic functions) the letters A, C, G are identical to each other and the letter U is opposite to them;
- The letter U is a single letter in RNA, which is replaced in DNA by another letter, the T.

This uniqueness of the letter U can be utilized in genetic computers of organisms. Taking into account the unique status of the letter U in this genetic alphabet, the author has shown the existence of the following formal "U-algorithm", which demonstrates the close connection between Hadamard matrices and the matrix mosaic of the degeneracy of the genetic code [Petoukhov, 2005b, 2008a,b].

By definition the U-algorithm contains two steps: 1) on the first step, each of the triplets in the black-and-white genomatrix (for example, in the genomatrix $[C\ A;\ U\ G]^{(3)}$ on Figure 4) should change its own color into opposite color each time when the letter U stands in an odd position (in the first or in the third position) inside the triplet; 2) on the second step, black triplets and white triples are interpreted as entries "+1" and "-1" correspondingly. For example, the white triplet UUA (see Figure 4) should become the black triplet (and its matrix cell should be marked by black color) because of the letter U in its first position; for this reason the triplet UUA is interpreted finally as "+1". Or the white triplet UUU should not change its color because of the letter U in its first and third positions (the color of this triplet is changed twice according to the described algorithm); for this reason the triplet UUU is interpreted finally as "-1". The triplet ACG does not change its color because the letter U is absent in this triplet.

By means of the U-algorithm, all the genomatrices $[C\ A;\ U\ G]_{123}^{(3)}$, $[C\ A;\ U\ G]_{231}^{(3)}$, $[C\ A;\ U\ G]_{312}^{(3)}$, $[C\ A;\ U\ G]_{321}^{(3)}$, $[C\ A;\ U\ G]_{213}^{(3)}$, $[C\ A;\ U\ G]_{132}^{(3)}$ (Figures 4 and 13) are transformed into relevant numeric genomatrices $H_{123}$, $H_{231}$, $H_{312}$, $H_{321}$, $H_{213}$, $H_{132}$ on Figure 17. Determinants of all these genomatrices are equal to $2^{12} = 4096$.

$H_{123}$ =

| | | | | | | | |
|---|---|---|---|---|---|---|---|
| (0) | (1) | (2) | (3) | (4) | (5) | (6) | (7) |
| (1) | (0) | (3) | (2) | (5) | (4) | (7) | (6) |
| (2) | (3) | (0) | (1) | (6) | (7) | (4) | (5) |
| (3) | (2) | (1) | (0) | (7) | (6) | (5) | (4) |
| (4) | (5)) | (6) | (7) | (0) | (1) | (2) | (3) |
| (5) | (4) | (7) | (6) | (1) | (0) | (3) | (2) |
| (6) | (7) | (4) | (5) | (2) | (3) | (0) | (1) |
| (7) | (6) | (5) | (4) | (3) | (2) | (1) | (0) |

$H_{231}$ =

| | | | | | | | |
|---|---|---|---|---|---|---|---|
| (0) | (2) | (4) | (6) | (1) | (3) | (5) | (7) |
| (2) | (0) | (6) | (4) | (3) | (1) | (7) | (5) |
| (4) | (6) | (0) | (2) | (5) | (7) | (1) | (3) |
| (6) | (4) | (2) | (0) | (7) | (5) | (3) | (1) |
| (1) | (3) | (5) | (7) | (0) | (2) | (4) | (6) |
| (3) | (1) | (7) | (5) | (2) | (0) | (6) | (4) |
| (5) | (7) | (1) | (3) | (4) | (6) | (0) | (2) |
| (7) | (5) | (3) | (1) | (6) | (4) | (2) | (0) |

$H_{312}$ =

| | | | | | | | |
|---|---|---|---|---|---|---|---|
| (0) | (4) | (1) | (5) | (2) | (6) | (3) | (7) |
| (4) | (0) | (5) | (1) | (6) | (2) | (7) | (3) |
| (1) | (5) | (0) | (4) | (3) | (7) | (2) | (6) |
| (5) | (1) | (4) | (0) | (7) | (3) | (6) | (2) |
| (2) | (6) | (3) | (7) | (0) | (4) | (1) | (5) |
| (6) | (2) | (7) | (3) | (4) | (0) | (5) | (1) |
| (3) | (7) | (2) | (6) | (1) | (5) | (0) | (4) |
| (7) | (3) | (6) | (2) | (5) | (1) | (4) | (0) |

$H_{321}$ =

| | | | | | | | |
|---|---|---|---|---|---|---|---|
| (0) | (4) | (2) | (6) | (1) | (5) | (3) | (7) |
| (4) | (0) | (6) | (2) | (5) | (1) | (7) | (3) |
| (2) | (6) | (0) | (4) | (3) | (7) | (1) | (5) |
| (6) | (2) | (4) | (0) | (7) | (3) | (5) | (1) |
| (1) | (5) | (3) | (7) | (0) | (4) | (2) | (6) |
| (5) | (1) | (7) | (3) | (4) | (0) | (6) | (2) |
| (3) | (7) | (1) | (5) | (2) | (6) | (0) | (4) |
| (7) | (3) | (5) | (1) | (6) | (2) | (4) | (0) |

$H_{213}$ =

| | | | | | | | |
|---|---|---|---|---|---|---|---|
| (0) | (1) | (4) | (5) | (2) | (3) | (6) | (7) |
| (1) | (0) | (5) | (4) | (3) | (2) | (7) | (6) |
| (4) | (5) | (0) | (1) | (6) | (7) | (2) | (3) |
| (5) | (4) | (1) | (0) | (7) | (6) | (3) | (2) |
| (2) | (3) | (6) | (7) | (0) | (1) | (4) | (5) |
| (3) | (2) | (7) | (6) | (1) | (0) | (5) | (4) |
| (6) | (7) | (2) | (3) | (4) | (5) | (0) | (1) |
| (7) | (6) | (3) | (2) | (5) | (4) | (1) | (0) |

$H_{132}$ =

| | | | | | | | |
|---|---|---|---|---|---|---|---|
| (0) | (2) | (1) | (3) | (4) | (6) | (5) | (7) |
| (2) | (0) | (3) | (1) | (6) | (4) | (7) | (5) |
| (1) | (3) | (0) | (2) | (5) | (7) | (4) | (6) |
| (3) | (1) | (2) | (0) | (7) | (5) | (6) | (4) |
| (4) | (6) | (5) | (7) | (0) | (2) | (1) | (3) |
| (6) | (4) | (7) | (5) | (2) | (0) | (3) | (1) |
| (5) | (7) | (4) | (6) | (1) | (3) | (0) | (2) |
| (7) | (5) | (6) | (4) | (3) | (1) | (2) | (0) |

Figure 17. The Hadamard genomatrices $H_{123}$, $H_{231}$, $H_{312}$, $H_{321}$, $H_{213}$, $H_{132}$ which are received from the genomatrices $[C\ A;\ U\ G]_{123}^{(3)}$, $[C\ A;\ U\ G]_{231}^{(3)}$, $[C\ A;\ U\ G]_{312}^{(3)}$, $[C\ A;\ U\ G]_{321}^{(3)}$, $[C\ A;\ U\ G]_{213}^{(3)}$, $[C\ A;\ U\ G]_{132}^{(3)}$ (Figures 4 and 13) by means of the U-algorithm. Brackets

contain dyadic numerations of cells by analogy with matrices on Figures 4, 13 and 14. Black color and white color of cells mean entries “+1” and “-1” in these cells correspondingly.

One can make the dyadic-shift decomposition of each of these six Hadamard genomatrices $H_{123}$, $H_{231}$, $H_{312}$, $H_{321}$, $H_{213}$, $H_{132}$ (Figure 17) by analogy with the described decompositions of the relevant genomatrices $R_{123}$, $R_{231}$, $R_{312}$, $R_{321}$, $R_{213}$, $R_{132}$ on Figures 7, 14 and 15. In the result, six new different sets of basic sparse matrices $h_0$, $h_1$, $h_2$, $h_3$, $h_4$, $h_5$, $h_6$, $h_7$ arise (where $h_0$ is the identity matrix). All their determinants are equal to 1. It is unexpected but each of these six sets of Hadamard genomatrices is closed in relation to multiplication. Moreover each of these sets $h_0$, $h_1$, $h_2$, $h_3$, $h_4$, $h_5$, $h_6$, $h_7$ corresponds to the same multiplication table on Figure 19. For example, the dyadic-shift decomposition of $H_{123}$ gives the following set of basic matrices (Figure 18) which corresponds to the multiplication table on Figure 19:

$$H_{123} = h_0+h_1+h_2+h_3+h_4+h_5+h_6+h_7 =$$

$$\begin{bmatrix} 1&0&0&0&0&0&0&0\\ 0&1&0&0&0&0&0&0\\ 0&0&1&0&0&0&0&0\\ 0&0&0&1&0&0&0&0\\ 0&0&0&0&1&0&0&0\\ 0&0&0&0&0&1&0&0\\ 0&0&0&0&0&0&1&0\\ 0&0&0&0&0&0&0&1 \end{bmatrix} + \begin{bmatrix} 0&1&0&0&0&0&0&0\\ -1&0&0&0&0&0&0&0\\ 0&0&0&1&0&0&0&0\\ 0&0&-1&0&0&0&0&0\\ 0&0&0&0&0&1&0&0\\ 0&0&0&0&-1&0&0&0\\ 0&0&0&0&0&0&0&1\\ 0&0&0&0&0&0&-1&0 \end{bmatrix} + \begin{bmatrix} 0&0&-1&0&0&0&0&0\\ 0&0&0&-1&0&0&0&0\\ 1&0&0&0&0&0&0&0\\ 0&1&0&0&0&0&0&0\\ 0&0&0&0&0&0&-1&0\\ 0&0&0&0&0&0&0&-1\\ 0&0&0&0&1&0&0&0\\ 0&0&0&0&0&1&0&0 \end{bmatrix} + \begin{bmatrix} 0&0&0&-1&0&0&0&0\\ 0&0&1&0&0&0&0&0\\ 0&1&0&0&0&0&0&0\\ -1&0&0&0&0&0&0&0\\ 0&0&0&0&0&0&0&-1\\ 0&0&0&0&0&0&1&0\\ 0&0&0&0&0&1&0&0\\ 0&0&0&0&-1&0&0&0 \end{bmatrix} +$$

$$\begin{bmatrix} 0&0&0&0&1&0&0&0\\ 0&0&0&0&0&1&0&0\\ 0&0&0&0&0&0&-1&0\\ 0&0&0&0&0&0&0&-1\\ -1&0&0&0&0&0&0&0\\ 0&-1&0&0&0&0&0&0\\ 0&0&1&0&0&0&0&0\\ 0&0&0&1&0&0&0&0 \end{bmatrix} + \begin{bmatrix} 0&0&0&0&0&1&0&0\\ 0&0&0&0&-1&0&0&0\\ 0&0&0&0&0&0&0&-1\\ 0&0&0&0&0&0&1&0\\ 0&-1&0&0&0&0&0&0\\ 1&0&0&0&0&0&0&0\\ 0&0&0&1&0&0&0&0\\ 0&0&-1&0&0&0&0&0 \end{bmatrix} + \begin{bmatrix} 0&0&0&0&0&0&-1&0\\ 0&0&0&0&0&0&0&-1\\ 0&0&0&0&-1&0&0&0\\ 0&0&0&0&0&-1&0&0\\ 0&0&1&0&0&0&0&0\\ 0&0&0&1&0&0&0&0\\ 1&0&0&0&0&0&0&0\\ 0&1&0&0&0&0&0&0 \end{bmatrix} + \begin{bmatrix} 0&0&0&0&0&0&0&-1\\ 0&0&0&0&0&0&1&0\\ 0&0&0&0&0&-1&0&0\\ 0&0&0&0&1&0&0&0\\ 0&0&0&1&0&0&0&0\\ 0&0&-1&0&0&0&0&0\\ 0&1&0&0&0&0&0&0\\ -1&0&0&0&0&0&0&0 \end{bmatrix}$$

Figure 18. The dyadic-shift decomposition of the Hadamard genomatrix $H_{123}$ (Figure 17) into the sum of 8 sparse matrices $h_0$, $h_1$,…, $h_7$.

| | **1** | **$h_1$** | **$h_2$** | **$h_3$** | **$h_4$** | **$h_5$** | **$h_6$** | **$h_7$** |
|---|---|---|---|---|---|---|---|---|
| **1** | **1** | **$h_1$** | **$h_2$** | **$h_3$** | **$h_4$** | **$h_5$** | **$h_6$** | **$h_7$** |
| **$h_1$** | **$h_1$** | **-1** | **$h_3$** | **$-h_2$** | **$h_5$** | **$-h_4$** | **$h_7$** | **$-h_6$** |
| **$h_2$** | **$h_2$** | **$h_3$** | **-1** | **$-h_1$** | **$-h_6$** | **$-h_7$** | **$h_4$** | **$h_5$** |
| **$h_3$** | **$h_3$** | **$-h_2$** | **$-h_1$** | **1** | **$-h_7$** | **$h_6$** | **$h_5$** | **$-h_4$** |
| **$h_4$** | **$h_4$** | **$h_5$** | **$h_6$** | **$h_7$** | **-1** | **$-h_1$** | **$-h_2$** | **$-h_3$** |
| **$h_5$** | **$h_5$** | **$-h_4$** | **$h_7$** | **$-h_6$** | **$-h_1$** | **1** | **$-h_3$** | **$h_2$** |
| **$h_6$** | **$h_6$** | **$h_7$** | **$-h_4$** | **$-h_5$** | **$h_2$** | **$h_3$** | **-1** | **$-h_1$** |
| **$h_7$** | **$h_7$** | **$-h_6$** | **$-h_5$** | **$h_4$** | **$h_3$** | **$-h_2$** | **$-h_1$** | **1** |

Figure 19. The multiplication table for the dyadic-shift decompositions of Hadamard genomatrices $H_{123}$, $H_{231}$, $H_{312}$, $H_{321}$, $H_{213}$, $H_{132}$ (Figure 17). Here the identity matrix $h_0$ is replaced by the symbol 1.

The existence of the multiplication table (Figure 19) means that a new 8-dimensional DS-algebra or a new system of 8-dimensional hypercomplex numbers (9) exists on the base of these Hadamard genomatrices which are connected with six different forms of matrix representation of this hypercomplex system:

$$H = x_0*1 + x_1*h_1 + x_2*h_2 + x_3*h_3 + x_4*h_4 + x_5*h_5 + x_6*h_6 + x_7*h_7 \qquad (9)$$

We term these new 8-dimensional hypercomplex numbers as H-octetons (here "H" is the first letter in the name Hadamard) because they differ from R-octetons (Figure 8) and from Cayley's octonions. The six Hadamard genomatrices $H_{123}$, $H_{231}$, $H_{312}$, $H_{321}$, $H_{213}$, $H_{132}$ are different forms of matrix representation of the same H-octeton whose coordinates are equal to 1 ($x_0=x_1=\ldots=x_7=1$).

The DS-algebra of H-octetons (Figure 19) is non-commutative associative non-division algebra. It has zero divisors: for example ($h_3+h_4$) and ($h_2-h_5$) are non-zero H-octetons, but their product is equal to zero. It is interesting that the quantity and the disposition of signs "+" and "-" in the multiplication table on Figure 19 are identical to their quantity and disposition in a Hadamard matrix. In addition, indexes of basic matrices are again disposed in the multiplication table (Figure 19) in accordance with the dyadic-shift matrix on Figure 1. The tables of triads of basic matrices for the case of H-octetons and for the case of R-octetons (Figure 9) are identical.
For each of these Hadamard genomatrices, the dyadic-shift decomposition of the transposed Hadamard genomatrix leads to a DS-algebra of 8-dimensional hypercomplex numbers with the multiplication table, which is the transposed analogue of the multiplication table on Fig. 19.

One can note that H-octetons (9) contain four (8*8)-matrix forms of representation s of two-dimensional complex numbers z ($z=x_0*h_0+x_1*h_1$; $z=x_0*h_0+x_2*h_2$; $z=x_0*h_0+x_4*h_4$; $z=x_0*h_0+x_6*h_6$) and three (8*8)-matrix forms of representation of two-dimensional double numbers d ($d=y_0*h_0+y_3*h_3$; $d=y_0*h_0+y_5*h_5$; $d=y_0*h_0+y_7*h_7$).

It should be noted that Hadamard matrices play important roles in many tasks of discrete signal processing; tens of thousands of publications devoted to them (see reviews in [Seberry, 2005]). For example, codes based on Hadamard matrices have been used on spacecrafts «Mariner» and «Voyager», which allowed obtaining high-quality photos of Mars, Jupiter, Saturn, Uranus and Neptune in spite of the distortion and weakening of the incoming signals.

Only a few symmetrical Hadamard matrices (Figure 20) are usually used in the field of discrete signal processing. But, as we have checked, dyadic-shift decompositions of these "engineering" Hadamard matrices do not lead to any 8-dimensional hypercomplex numbers in contrast to the asymmetrical Hadamard genomatrices described in our article (in these "engineering" Hadamard matrices their main diagonals contain entries "+1" and "-1" and for this reason decompositions of these matrices do not generate identity matrix which should exist in the set of basic elements of hypercomplex numbers). Moreover the author knows no publications about the facts that Hadamard matrices can be the base for matrix forms of representation of 8-dimensional hypercomplex numbers. It seems that the genetic code has led the author to discovering the new interesting fact in the field of theory of Hadamard matrices about the unexpected relation of some Hadamard matrices with multidimensional DS-algebras of hypercomplex numbers. This fact can be useful for many applications of Hadamard genomatrices for simulating of bioinformation phenomena, for technology of discrete signal processing, etc. A great number of Hadamard (8x8)-matrices exists (according to some experts, their quantity is equal to approximately 5 billion). Perhaps, only the genetic Hadamard matrices, which represent a small subset of a great set of all the Hadamard matrices, are related with multidimensional DS-algebras but it is an open question now.

| | | | | | | | |
|---|---|---|---|---|---|---|---|
| + | + | + | + | + | + | + | + |
| + | - | + | - | + | - | + | - |
| + | + | - | - | + | + | - | - |
| + | - | - | + | + | - | - | + |
| + | + | + | + | - | - | - | - |
| + | - | + | - | - | + | - | + |
| + | + | - | - | - | - | + | + |
| + | - | - | + | - | + | + | - |
| + | + | + | + | + | + | + | + |
| + | + | + | + | - | - | - | - |
| + | + | - | - | + | + | - | - |
| + | + | - | - | - | - | + | + |
| + | - | + | - | + | - | + | - |
| + | - | + | - | - | + | - | + |
| + | - | - | + | + | - | - | + |
| + | - | - | + | - | + | + | - |
| + | + | + | + | + | + | + | + |
| + | + | + | + | - | - | - | - |
| + | + | - | - | - | - | + | + |
| + | + | - | - | + | + | - | - |
| + | - | - | + | + | - | - | + |
| + | - | - | + | - | + | + | - |
| + | - | + | - | - | + | - | + |
| + | - | + | - | + | - | + | - |

Figure 20. Three symmetrical Hadamard matrices, which are frequently used in the technology of digital signal processing [Trahtman & Trahtman, 1975]. On the left side: the Hadamard matrix with the Walsh-Hadamard system of functions in its rows. In the middle: the Hadamard matrix with the Walsh-Paley system of functions in its rows. On the right side: the Hadamard matrix with a monotone sequence of increasing of numbers of reciprocal replacements of entries +1 and -1 in rows

Why living nature uses just such the genetic code that is associated with Hadamard genomatrices? It is an open question also. We suppose that its reason is related with solving in biological organisms the same information tasks which lead to a wide using of Hadamard matrices in digital signal processing and in physics. The orthogonal systems of Walsh functions in Hadamard genomatrices can be a natural basis to organize storage and transfer of genetic information with noise-immunity properties, etc. The described Hadamard genomatrices can be considered as a patent of living nature which was discovered through matrix genetics. It should be emphasized that a few authors have applied orthogonal systems of Walsh functions from "engineering" Hadamard matrices to study macro-physiological systems [Shiozaki, 1980; Carl, 1974; Ginsburg et all, 1974]. It seems that applications of "genetic" Walsh functions from Hadamard genomatrices will be more relevant for such macro-physiological researches.

One can mentioned additionally that orthogonal systems of Walsh functions are connected with modulo-2 addition and with dyadic groups [Harmuth, 1977, Section 1.1.4]. In an orthogonal system of Walsh functions each of functions can be numerated by a number of its intersections of zero (which is termed as "the sequency") [Harmuth, 1977, Section 1.1.4; http://fourier.eng.hmc.edu/e161/lectures/wht/node3.html]. For example in the case of Hadamard genomatrix $H_{123}$ (Figure 17) its Walsh functions have the following binary numerations from the upper row to the bottom row: 011, 100, 001, 110, 010, 101, 000 and 111 (or in decimal numeration 3, 4, 1, 6, 2, 5, 0, 7). Product of two Walsh functions provides a new Walsh function, the binary numeration of which is equal to the result of modulo-2 addition of binary numerations of initial factors. For example in the case of the genomatrix $H_{123}$ (Figure 17) the product of two Walsh functions with their numerations 011 and 100 (see the first two rows) gives the Walsh function with its numeration 111 (the bottom row). This situation is the analogue with the situation of triads of basic matrices of genetic octetons (see Figure 9). Walsh functions form a commutative group in relation to multiplication and this group is isomorphic to discrete dyadic group [Harmuth, 1977, Section 1.1.4].

### 10. H-octetons, their norms and metric H-quaternions

Let us study the question about a norm of H-octetons with the multiplication table on Figure 19. By analogy with the section 5 and taking into account the multiplication table (Figure 19) and the expression (7) for the norm, we have the following product of a H-octeton (9) and its conjugation H^:

$$|| H||^2 = H\^{} * H = (x_0^2+x_1^2+x_2^2-x_3^2+x_4^2-x_5^2+x_6^2-x_7^2)*h_0 + 2*(x_2*x_3+x_4*x_5+x_6*x_7)*h_1 + \\ + 2*x_1*(-x_3*h_2 + x_2*h_3 + x_5*h_4 - x_4*h_5 - x_7*h_6 + x_6*h_7) \qquad (10)$$

Here the right part contains the scalar element $(x_0^2+x_1^2+x_2^2-x_3^2+x_4^2-x_5^2+x_6^2-x_7^2)$ and imaginary elements: $2*(x_2*x_3+x_4*x_5+x_6*x_7)*h_1 + 2*x_1*(-x_3*h_2 + x_2*h_3 + x_5*h_4 - x_4*h_5 - x_7*h_6 + x_6*h_7)$. All these imaginary elements are equal to zero in the following 8 cases corresponding to different 4-dimensional sub-algebras of the 8-dimensional DS-algebra of H-octetons (Figure 21). Each of these 8 types of relevant H-quaternions has a scalar value of its squared norm and defines a metric space. Figuratively speaking, the vector space of H-octetons contains a colony of metric spaces of H-quaternions; it is reminiscent of a structure of multicellular organisms in a form of a colony of single-cellular organisms.

| № | Coefficients of zero value | H-quaternions with scalar values of squared norms | Squared norms of relevant H-quaternions |
|---|---|---|---|
| 0 | $x_1=x_2=x_4=x_6=0$ | $x_0+x_3*h_3+x_5*h_5+x_7*h_7$ | $x_0^2-x_3^2-x_5^2-x_7^2$ |
| 1 | $x_1=x_2=x_4=x_7=0$ | $x_0 +x_3*h_3+x_5*h_5+x_6*h_6$ | $x_0^2-x_3^2-x_5^2+x_6^2$ |
| 2 | $x_1=x_2=x_5=x_6=0$ | $x_0 +x_3*h_3+x_4*h_4+x_7*h_7$ | $x_0^2-x_3^2+x_4^2-x_7^2$ |
| 3 | $x_1=x_2=x_5=x_7=0$ | $x_0 +x_3*h_3+x_4*h_4+x_6*h_6$ | $x_0^2-x_3^2+x_4^2+x_6^2$ |
| 4 | $x_1=x_3=x_4=x_6=0$ | $x_0 +x_2*h_2+x_5*h_5+x_7*h_7$ | $x_0^2+x_2^2-x_5^2-x_7^2$ |
| 5 | $x_1=x_3=x_4=x_7=0$ | $x_0 +x_2*h_2+x_5*h_5+x_6*h_6$ | $x_0^2+x_2^2-x_5^2+x_6^2$ |
| 6 | $x_1=x_3=x_5=x_6=0$ | $x_0 +x_2*h_2+x_4*h_4+x_7*h_7$ | $x_0^2+x_2^2+x_4^2-x_7^2$ |
| 7 | $x_1=x_3=x_5=x_7=0$ | $x_0 +x_2*h_2+x_4*h_4+x_6*h_6$ | $x_0^2+x_2^2+x_4^2+x_6^2$ |

Figure 21. The set of 8 types of H-quaternions with scalar values of their squared norms

## 11. The spectral analysis on the base of the dyadic-shift decomposition of Hadamard genomatrices. Permutation genomatrix operators

In discrete signal processing, Hadamard (n*n)-matrices are widely used for spectral analysis of n-dimensional vectors. A transform of a vector ā by means of a Hadamard matrix H gives the vector ū = H*ā, which is termed "Hadamard spectrum" and which is based on Walsh functions in rows of the matrix H [Ahmed, Rao, 1975]. But in the case of Hadamard genomatrices we have additional possibilities of spectrum representation of the same vector ā by means of basic matrices $h_0, h_1,\ldots, h_7$ for each of Hadamard genomatrices $H_{123}$, $H_{231}$, $H_{312}$, $H_{321}$, $H_{213}$, $H_{132}$. Let us describe it in more detail.

For example in the case of a Hadamard genomatrix $H_{123}$ (Figure 17) and a 8-dimensional vector $\bar{a}=[a_0; a_1; a_2; a_3; a_4; a_5; a_6; a_7]$, whose coordinates $a_k$ are real numbers, the corresponding Hadamard spectrum ū is presented by the following 8-dimensional vector:

$$\bar{u} = H_{123}* \bar{a} = [a_0+a_1-a_2-a_3+a_4+a_5-a_6-a_7; \quad -a_0+a_1+a_2-a_3-a_4+a_5+a_6-a_7; \quad a_0+a_1+a_2+a_3-a_4-a_5-a_6-a_7; \quad -a_0+a_1-a_2+a_3+a_4-a_5+a_6-a_7; \quad -a_0-a_1+a_2+a_3+a_4+a_5-a_6-a_7; \quad a_0-a_1-a_2+a_3-a_4+a_5+a_6-a_7; \quad a_0+a_1+a_2+a_3+a_4+a_5+a_6+a_7; \quad -a_0+a_1-a_2+a_3-a_4+a_5-a_6+a_7] \qquad (11)$$

It means that if numerated rows 0, 1, 2, …, 7 of the matrix $H_{123}$ correspond to Walsh functions $w_0, w_1, w_2,\ldots, w_7$ (each of them is a 8-dimensional vector with entries +1 and -1) then the initial vector ā is expressed by means of the following expression (12) on the base of its spectral representation (11):

$$\bar{a} = (a_0+a_1-a_2-a_3+a_4+a_5-a_6-a_7)*w_0 + (-a_0+a_1+a_2-a_3-a_4+a_5+a_6-a_7)*w_1 + (a_0+a_1+a_2+a_3-a_4-a_5-a_6-a_7)*w_2 + (-a_0+a_1-a_2+a_3+a_4-a_5+a_6-a_7)*w_3 + (-a_0-a_1+a_2+a_3+a_4+a_5-a_6-a_7)*w_4 + (a_0-a_1-a_2+a_3-a_4+a_5+a_6-a_7)*w_5 + (a_0+a_1+a_2+a_3+a_4+a_5+a_6+a_7)*w_6 + (-a_0+a_1-a_2+a_3-a_4+a_5-a_6+a_7)*w_7 \qquad (12)$$

Taking into account that the Hadamard genomatrix $H_{123}$ can be interpreted as sum of basic matrices of genetic H-octetons that is $H_{123} = h_0+h_1+h_2+h_3+h_4+h_5+h_6+h_7$ (Figure 18), one can represent the same Hadamard spectrum ū in a new form without the direct using of Walsh functions:

$$\bar{u} = H_{123}* \bar{a} = h_0* \bar{a} +h_1* \bar{a} +h_2* \bar{a} +h_3* \bar{a} +h_4* \bar{a} +h_5* \bar{a} +h_6* \bar{a} +h_7* \bar{a} \qquad (13)$$

The set of basic matrices $h_k$ (k=0, 1, 2, …, 7) of $H_{123}$-octetons (Figure 18) is characterized by the interesting property (14):

$$h_k * h_k^T = h_k^T * h_k = E \qquad (14)$$

where $h_k^T$ is a transposed matrix and E is identical matrix. The property (14) holds true also for each of basic matrices in cases of other H-octetons $H_{231}$, $H_{312}$, $H_{321}$, $H_{213}$, $H_{132}$ (Figure 17) and in cases of R-octetons $R_{123}$, $R_{231}$, $R_{312}$, $R_{321}$, $R_{213}$, $R_{132}$ (Figures 7 and 14). In other words, it is a general property.

Each of the basic matrices $h_k$ of $H_{123}$-octetons possesses a system of orthogonal rows which can be used for a new spectral representation of any 8-dimensional vector $\bar{a}=[a_0; a_1; a_2; a_3; a_4; a_5; a_6; a_7]$. We will call the expression $\bar{u}_k = h_k*\bar{a}$ as Hadamard $h_k$-spectrum of the vector $\bar{a}$. Expressions (15) show Hadamard $h_k$-spectrums of the vector $\bar{a}$ for different matrix $h_k$ from Figure 18:

$$\begin{aligned}
\bar{u}_0 &= h_0*\bar{a} = [\ a_0,\ a_1,\ a_2,\ a_3,\ a_4,\ a_5,\ a_6,\ a_7] \\
\bar{u}_1 &= h_1*\bar{a} = [\ a_1, -a_0,\ a_3, -a_2,\ a_5, -a_4,\ a_7, -a_6] \\
\bar{u}_2 &= h_2*\bar{a} = [\ -a_2, -a_3,\ a_0,\ a_1, -a_6, -a_7,\ a_4,\ a_5] \\
\bar{u}_3 &= h_3*\bar{a} = [\ -a_3,\ a_2,\ a_1, -a_0, -a_7,\ a_6,\ a_5, -a_4] \\
\bar{u}_4 &= h_4*\bar{a} = [\ a_4,\ a_5, -a_6, -a_7, -a_0, -a_1,\ a_2,\ a_3] \\
\bar{u}_5 &= h_5*\bar{a} = [\ a_5, -a_4, -a_7,\ a_6, -a_1,\ a_0,\ a_3, -a_2] \\
\bar{u}_6 &= h_6*\bar{a} = [\ -a_6, -a_7, -a_4, -a_5,\ a_2,\ a_3,\ a_0,\ a_1] \\
\bar{u}_7 &= h_7*\bar{a} = [\ -a_7,\ a_6, -a_5,\ a_4,\ a_3, -a_2,\ a_1, -a_0]
\end{aligned} \qquad (15)$$

Taking into account the property (14), one can restore the initial vector $\bar{a}$ from its $h_k$-spectrum $\bar{u}_k$ by means of multiplication of the $h_k$-spectrum $\bar{u}_k$ with the transposed matrix $h_k^T$:

$$h_k^T*\bar{u}_k = h_k^T*(h_k*\bar{a}) = \bar{a} \qquad (16)$$

In a general case, numeric coordinates of the Hadamard spectrum $\bar{u}$ (see expression (11)) are quite different from numeric coordinates of the initial vector $\bar{a} = [a_0; a_1; a_2; a_3; a_4; a_5; a_6; a_7]$. In contrast to this case, coordinates of $h_k$-spectrums $\bar{u}_k$ (15) are always taken from the same set of coordinates $a_0; a_1; a_2; a_3; a_4; a_5; a_6; a_7$ of the initial vector $\bar{a}$ (only their signs "+" and "–" can be changed). In other words, all the $h_k$-spectral operations are special cases of permutation operators which not only make a permutation of coordinates of 8-dimensional vectors but also invert sign of some coordinates. Such kind of spectral analysis can be conditionally called the "genetic permutation spectral analysis". In a general case of such analysis, 8-dimensional vectors $\bar{a}$ can contain not only real numbers but also many other objects if their sequences are presented in a form of 8-dimensional vectors: genetic triplets and their associations, letters, phrases, complex and hypercomplex numbers, matrices, photos, musical notes and fragments of musical works, etc. From the linguistic viewpoint, $h_k$-spectrums are anagrams to within plus or minus sign of entries. To illustrate this we show below two lingustics examples of transformation of 8-dimensional vectors by means of the basic matrices $h_0, h_1, \ldots, h_7$ of $H_{123}$-octeton (Figure 18). In these examples the word "adultery" and the phrase "I also love to run over long distances" are used as initial 8-dimensional vectors:

$h_0$*[ a; d; u; l; t; e; r; y] = [ a; d; u; l; t; e; r; y]
$h_1$*[ a; d; u; l; t; e; r; y] = [ d; -a; l; -u; e; -t; y; -r]
$h_2$*[ a; d; u; l; t; e; r; y] = [-u; -l; a; d; -r; -y; t; e]
$h_3$*[ a; d; u; l; t; e; r; y] = [-l; u; d; -a; -y; r; e; -t]
$h_4$*[ a; d; u; l; t; e; r; y] = [ t; e; -r; -y; -a; -d; u; l]
$h_5$*[ a; d; u; l; t; e; r; y] = [ e; -t; -y; r; -d; a; l; -u]
$h_6$*[ a; d; u; l; t; e; r; y] = [-r; -y; -t; -e; u; l; a; d]
$h_7$*[ a; d; u; l; t; e; r; y] = [-y; r; -e; t; l; -u; d; -a]

$h_0$*[I;also;love;to;run;over;long;distances] = [I; also; love; to; run; over; long; distances]
$h_1$*[I;also;love;to;run;over;long;distances] = [also; -I; to; -love; over; -run; distances; -long]
$h_2$*[I;also;love;to;run;over;long;distances] = [-love; -to; I; also; -long; -distances; run; over]
$h_3$*[I;also;love;to;run;over;long;distances] = [-to; love; also; -I; -distances; long; over; -run]
$h_4$*[I;also;love;to;run;over;long;distances] = [run; over; -long; -distances; -I; -also; love; to]
$h_5$*[I;also;love;to;run;over;long;distances] = [over; -run; -distances; long; -also; I; to; -love]
$h_6$*[I;also;love;to;run;over;long;distances] = [-long; -distances; -run; -over; love; to; I; also]
$h_7$*[I;also;love;to;run;over;long;distances] = [-distances; long; -over; run; to; -love; also; -I]

The basis matrices of DS-algebras of other Hadamard genomatrices $H_{231}$, $H_{312}$, $H_{321}$, $H_{213}$, $H_{132}$ (Рис. 17) transform the same 8-dimensional vectors otherwise.

Why the discovery of the connection of the genetic code with permutation matrix operators is interesting for molecular genetics? One of the main reasons is that genes demonstrate phenomena of rearrangements of their pieces, presented in facts of transposons and splicing (see the websites http://en.wikipedia.org/wiki/Transposon and http://en.wikipedia.org/wiki/RNA_splicing with thematic references). For example, transposons are sequences of DNA that can move or transpose themselves to new positions within the genome of a single cell. Barbara McClintock's discovery of these jumping genes early in her career earned her a Nobel Prize in 1983. Now jumping genes are used to create new medicines, etc. Genetic permutation operators, which are described in our article, can be useful for mathematical modeling and analysis of such phenomena of rearrangements of parts of a whole gene or of ensembles of genes. If the quantity of vector elements in a genetic sequence is not divisible by 8, the remaining short vector can be extended to 8-dimensional vector by adding to it at the end of the required number of zeros by analogy with methods of digital signal processing. There is a separate question about a way to interpret binary-oppositional signs "+" and "-" in such genetic permutations (15). It seems that facts of existence of binary-oppositional objects or characteristics can be used here: codon-anticodon, purine-pyrimidine, etc. (see Figure 3).

Not only molecular genetics but many other fields of heritable physiological systems for biological communications deal with permutation structures and processes. A special attention should be paid to problems of anagrams in linguistics because impressive discoveries in the field of the genetic code have been described by its researchers using the terminology borrowed from linguistics and the theory of communications. As experts in molecular genetics remark, "*the more we understand laws of coding of the genetic information, the more strongly we are surprised by their similarity to principles of linguistics of human and computer languages*" [Ratner, 2002, p. 203].

Problems of anagrams in linguistics have a long history concerning with culture (poetry, lingustcs games, musics, semiotics, etc.; anagrams were widely used in ancient epics: the "Mahabharata" and "Ramayana", the "Iliad" and the "Odyssey", as well as in the Bible), religions (anagrammatical words, phrases and symbols which were interpreted from the ancient time as magic tools) and medicine (anagrams in speech and writing are typical for some human diseases: aphasia motoria, aphasia sensoria, literal paraphasia, etc.). The website http://en.wikipedia.org/wiki/Anagram gives many data and references about anagrams.

By analogy with the genetic code, linguistics is one of the significant examples of existence and importance of ensembles of binary oppositions in information physiology. Leading experts on structural linguistics have believed for a long time that languages of human dialogue were formed not from an empty place, but they are continuation of genetic language or, somehow, are closely connected with it, confirming the idea of information commonality of organisms. Analogies between systems of genetic and linguistic information are contents of a wide and important scientific sphere, which can be illustrated here in short only. We reproduce below some thematic thoughts by R. Jakobson [1985, 1987, 1999], who is one of the most famous experts and the author of a deep theory of binary oppositions in linguistics. Jakobson and some other scientists believe that we possess a genetic language which is as old as life and which is the most alive language among all languages. Among all systems of information transfer, only the genetic code and linguistic codes are based on the use of discrete components, which themselves make no sense, but serve for the construction of the minimum units which make sense after their special grouping. (By the way, one can note here that matrix genetics deals with matrix forms of groupings of elements of genetic language). As Jakobson stated, the genetic code system is the basic simulator, which underlines all verbal codes of human languages. "*The heredity in itself is the fundamental form of communications … Perhaps, the bases of language structures, which are superimposed on molecular communication, have been constructed by means of its structural principles directly*" [Jakobson, 1985, p. 396]. These questions have arisen to Jakobson as a consequence of its long-term researches of connections between linguistics, biology and physics. Such connections were considered at a united seminar of physicists and linguists, which was organized by Niels Bohr and Roman Jakobson jointly at the Massachusetts Institute of Technology.

"*Jakobson reveals distinctly a binary opposition of sound attributes as underlying each system of phonemes... The subject of phonology has changed by him: the phonology considered phonemes (as the main subject) earlier, but now Yakobson has offered that distinctive attributes should be considered as "quantums" (or elementary units of language)… . Jakobson was interested especially in the general analogies of language structures with the genetic code, and he considered these analogies as indubitable*" [Ivanov, 1985]. One can remind also of the title of the monograph "On the Yin and Yang nature of language" [Baily, 1982], which is characteristic for the theme of binary oppositions in linguistics (see more detail in [Petoukhov, 2008a; Petoukhov, He, 2010]).

Similar questions about a connection of linguistics with the genetic code excite many researchers. In addition a linguistic language is perceived by many researchers as a living organism. The book "Linguistic genetics" [Makovskiy, 1992] states: "*The opinion about language as about a living organism, which is submitted to the laws of a nature, ascends to a deep antiquity … Research of a nature, of disposition and of reasons of isomorphism between genetic and linguistic regularities is one of the most important fundamental problems for linguistics of our time*".

We believe that genetic DS-algebras of hypercomplex numbers, genetic permutation operators and other achievements of matrix genetics will be useful for an additional revealing deep connections between genetic and linguistic languages. Below we show that the Hadamard genomatrices of duplets and triplets are matrix representations of the Hamilton quaternion and biquaternion with unit coordinates. Taking this fact into account, the materials of this Section about spectral representations can be interpreted as the materials about spectral representations on the base of the basic Hamilton quaternions and biquaternions.

## 12. The DS-algebras of the genomatrices of 16 duplets and coquaternions by Cockle

By analogy with the considered case of the (8*8)-genomatrices of triplers, one can study the question about of DS-algebras of the (4*4)-genomatrices of duplets. This section describes some results of such studying for the cases of the genomatrix $[C\ A;\ U\ G]^{(2)}$ (Figure 4) and of the genomatrix $[C\ U;\ A\ G]^{(2)}$ (Figure 24). Each of them contains 8 duplets with strong roots (they are marked by black color) and 8 duplets with weak roots (white color).

$[C\ A;\ U\ G]_{12}^{(2)}$ =

| | 00(0) | 01(1) | 10(2) | 11(3) |
|---|---|---|---|---|
| 00 (0) | CC 00(0) | CA 01(1) | AC 10(2) | AA 11(3) |
| 01 (1) | CU 01(1) | CG 00(0) | AU 11(3) | AG 10(2) |
| 10 (2) | UC 10(2) | UA 11(3) | GC 00(0) | GA 01(1) |
| 11 (3) | UU 11(3) | UG 10(2) | GU 01(1) | GG 00(0) |

; $[C\ A;\ U\ G]_{21}^{(2)}$ =

| | 00(0) | 10(2) | 01(1) | 11(3) |
|---|---|---|---|---|
| 00 (0) | CC 00(0) | AC 10(2) | CA 01(1) | AA 11(3) |
| 10 (2) | UC 10(2) | GC 00(0) | UA 11(3) | GA 01(1) |
| 01 (1) | CU 01(1) | AU 11(3) | CG 00(0) | AG 10(2) |
| 11 (3) | UU 11(3) | GU 01(1) | UG 10 (2) | GG 00 (0) |

$R_{12}$ =

| | 00 | 01 | 10 | 11 |
|---|---|---|---|---|
| 00 | +1(0) | -1(1) | +1(2) | -1(3) |
| 01 | +1(1) | +1(0) | -1(3) | -1(2) |
| 10 | +1(2) | -1(3) | +1 (0) | -1(1) |
| 11 | -1(3) | -1(2) | +1 (1) | +1(0) |

; $R_{21}$ =

| | 00 | 10 | 01 | 11 |
|---|---|---|---|---|
| 00 | +1(0) | +1(2) | -1(1) | -1(3) |
| 10 | +1(2) | +1(0) | -1(3) | -1(1) |
| 01 | +1(1) | -1(3) | +1(0) | -1(2) |
| 11 | -1(3) | +1(1) | -1 (2) | +1(0) |

**$R0_{12}$ =[1 0 0 0; 0 1 0 0; 0 0 1 0; 0 0 0 1]**
**$R1_{12}$ =[0 -1 0 0; 1 0 0 0; 0 0 0 -1; 0 0 1 0]**
**$R2_{12}$ =[0 0 1 0; 0 0 0 -1; 1 0 0 0; 0 -1 0 0]**
**$R3_{12}$ =[0 0 0 -1; 0 0 -1 0; 0 -1 0 0; -1 0 0 0]**

**$R0_{21}$ =[1 0 0 0; 0 1 0 0; 0 0 1 0; 0 0 0 1]**
**$R1_{21}$ =[0 0 -1 0; 0 0 0 -1; 1 0 0 0; 0 1 0 0]**
**$R2_{21}$ =[0 1 0 0; 1 0 0 0; 0 0 0 -1; 0 0 -1 0]**
**$R3_{21}$ =[0 0 0 -1; 0 0 -1 0; 0 -1 0 0; -1 0 0 0]**

Figure 22. Top row: the genomatrices $[C\ A;\ U\ G]_{12}^{(2)}$ (from Figure 4) and $[C\ A;\ U\ G]_{21}^{(2)}$ which differ from each other by means of the order of positions in doublets (1-2 and 2-1); black and white cells contain duplets with strong and weak roots correspondingly; brackets contain dyadic-shift numeration of each duplet. Middle row: the Rademacher forms $R_{12}$ and $R_{21}$ of representation of these genomatrices. Bottom row: two sets of four sparse matrices which arise in the result of dyadic-shift decompositions of these Rademacher forms $R_{12}$ and $R_{21}$. The matrices $R0_{12}$ and $R0_{21}$ are identity matrices.

Figure 22 shows this genomatrix together with its transformation by means of changing the initial order of positions 1-2 in dublets into the order 2-1. Each row of these two genomatrices corresponds to one of Rademacher functions. For this reason the Rademacher forms of these two genomatrices are also presented on Figure 22 together with the result of their dyadic-shift decompositions.

Each of two sets of four sparse matrices (bottom row on Figure 22), which arise in the result of dyadic-shift decompositions of these Rademacher forms $R_{12}$ and $R_{21}$, is closed in relation to multiplication and satisfies the same multiplication table (Figure 23, left). In other words, these two sets are different forms of representation of the DS-algebra of 4-dimensional hypercomplex numbers with such multiplication table (on Figure 23 the indices 12 and 21 of the elements are omitted). We term these 4-dimensional numbers as R-genoquaternions. It is interesting that this multiplication table is the transposed analogue of the known multiplication table of the coquaternions (or split-quaternions) by J.Cockle [http://en.wikipedia.org/wiki/Split-quaternion] (Figure 23, right). (In the field of matrix genetics this multiplication table was firstly published in the author's book [Petoukhov, 2008, Figures 5.2/2 and 5.2/3] with a mention about Cockle co-quaternions).

| | 1 | $R_1$ | $R_2$ | $R_3$ |
|---|---|---|---|---|
| 1 | 1 | $R_1$ | $R_2$ | $R_3$ |
| $R_1$ | $R_1$ | -1 | $-R_3$ | $R_2$ |
| $R_2$ | $R_2$ | $R_3$ | 1 | $R_1$ |
| $R_3$ | $R_3$ | $-R_2$ | $-R_1$ | 1 |

| | 1 | i | j | k |
|---|---|---|---|---|
| 1 | 1 | i | j | k |
| i | i | -1 | k | -j |
| j | j | -k | 1 | -i |
| k | k | j | i | 1 |

Figure 23. The left table: the multiplication table for the sparse matrices of dyadic-shift decompositions of Rademacher forms of the genomatrices $[C\ A;\ U\ G]_{12}^{(2)}$ and $[C\ A;\ U\ G]_{21}^{(2)}$ from Figure 22; this table defines the 4-dimensional algebra of R-genoquaternions. The right

table: the multiplication table of coquaternions (or split-quaternions) by J.Cockle [http://en.wikipedia.org/wiki/Split-quaternion]

To get the classical multiplication table of the 4-dimensional algebra of coquaternions by Cockle one can use one of the following two ways:

1) One can take the new order of basic matrices of this hypercomplex number R0+R1+R3+R2 (instead of the order R0+R1+R2+R3 on Figure 22, bottom);
2) One can take from the very beginning the genomatrix [C U; A G]$^{(2)}$ (Figure 24, left) which is the transposed analogue of the considered genomatrix [C A; U G]$^{(2)}$. The DS-decomposition of the Rademacher form of the mosaic genomatrix [C U; A G]$^{(2)}$ gives the set of 4 sparse matrices $R0^T$, $R1^T$, $R2^T$, $R3^T$ which are transposed analogues of the basic matrices R0, R1, R2, R3 (Figure 22, bottom) and which give the multiplication table of Cockle coquaternions (Figure 23, right). It is known that the coquaternion basis can be identified as the basis elements of either the Clifford algebra $C\ell_{1,1}(R)$ or the algebra $C\ell_{2,0}(R)$ [http://en.wikipedia.org/wiki/Split-quaternion].

History of science knows many examples of how abstract mathematical structures created by mathematicians, later turned out to be miraculously realized inside complex natural systems. The realization of Cockled coquaternions in the system of molecular-genetic alphabets can be considered as a new example of such type.

| | | | |
|---|---|---|---|
| CC | CU | UC | UU |
| CA | CG | UA | UG |
| AC | AU | GC | GU |
| AA | AG | GA | GG |

| | | | | | | | |
|---|---|---|---|---|---|---|---|
| **CCC** | **CCU** | **CUC** | **CUU** | **UCC** | **UCU** | **UUC** | **UUU** |
| **CCA** | **CCG** | **CUA** | **CUG** | **UCA** | **UCG** | **UUA** | **UUG** |
| **CAC** | **CAU** | **CGC** | **CGU** | **UAC** | **UAU** | **UGC** | **UGU** |
| **CAA** | **CAG** | **CGA** | **CGG** | **UAA** | **UAG** | **UGA** | **UGG** |
| **ACC** | **ACU** | **AUC** | **AUU** | **GCC** | **GCU** | **GUC** | **GUU** |
| **ACA** | **ACG** | **AUA** | **AUG** | **GCA** | **GCG** | **GUA** | **GUG** |
| **AAC** | **AAU** | **AGC** | **AGU** | **GAC** | **GAU** | **GGC** | **GGU** |
| **AAA** | **AAG** | **AGA** | **AGG** | **GAA** | **GAG** | **GGA** | **GGG** |

Figure 24. The left side: the genomatrix of duplets [C U; A G]$^{(2)}$. The right side: the genomatrix of triplets [C U; A G]$^{(3)}$. These genomatrices are received by means of the transposition of genomatrices [C A; U G]$^{(2)}$ and [C A; U G]$^{(3)}$ from Figure 4. The mosaic of their columns corresponds to Rademacher functions.

Figure 25 shows that the Rademacher forms of the (4*4)-genomatrix of duplets [C U; A G]$^{(2)}$ and the (8*8)-genomatrix of triplets [C U; A G]$^{(3)}$ are interconnected by the simple operation: the Rademacher form of the [C U; A G]$^{(3)}$ is received by means of the Kronecker product of the Rademacher form of the [C U; A G]$^{(2)}$ with the matrix form of the double numbers [x y; y x] where x=y=1 (see about double numbers or split-complex numbers http://en.wikipedia.org/wiki/Split-complex_number ).

| | | | |
|---|---|---|---|
| 1 | 1 | 1 | -1 |
| -1 | 1 | -1 | -1 |
| 1 | -1 | 1 | 1 |
| -1 | -1 | -1 | 1 |

⊗

| | |
|---|---|
| 1 | 1 |
| 1 | 1 |

=

| | | | | | | | |
|---|---|---|---|---|---|---|---|
| 1 | 1 | 1 | 1 | 1 | 1 | -1 | -1 |
| 1 | 1 | 1 | 1 | 1 | 1 | -1 | -1 |
| -1 | -1 | 1 | 1 | -1 | -1 | -1 | -1 |
| -1 | -1 | 1 | 1 | -1 | -1 | -1 | -1 |
| 1 | 1 | -1 | -1 | 1 | 1 | 1 | 1 |
| 1 | 1 | -1 | -1 | 1 | 1 | 1 | 1 |
| -1 | -1 | -1 | -1 | -1 | -1 | 1 | 1 |
| -1 | -1 | -1 | -1 | -1 | -1 | 1 | 1 |

Figure 25. The interconnection of the Rademacher forms of the genomatrices [C U; A G]$^{(2)}$ and [C U; A G]$^{(3)}$ on the base of the Kronecker product with the double number [1 1; 1 1].

The DS-decomposition of this Rademacher form of the [C U; A G]$^{(3)}$ (Figure 25, right) gives the set of 8 sparse matrices $r_0^T$, $r_1^T$,…, $r_7^T$ which are transposed analogues of the basic matrices $r_0$, $r_1$,…, $r_7$ of R-octetons (Figure 7) and which produce the transposed analogue of the multiplication table of R-octetons (Figure 8).

Let us turn now to the question about DS-algebras of relevant Hadamard genomatrices $H_{12}$ and $H_{21}$ which are received from these genomatrices [C A; U G]$_{12}^{(2)}$ and [C A; U G]$_{21}^{(2)}$ by means of the same U-algorithm (see § 9). Figure 26 shows these Hadamard genomatrices together with the results of their dyadic-shift decompositions.

The case of the order 1-2:

| | 00(0) | 01(1) | 10(2) | 11(3) |
|---|---|---|---|---|
| 00 (0) | CC 00(0) | CA 01(1) | AC 10(2) | AA 11(3) |
| 01 (1) | CU 01(1) | CG 00(0) | AU 11(3) | AG 10(2) |
| 10 (2) | UC 10(2) | UA 11(3) | GC 00(0) | GA 01(1) |
| 11 (3) | UU 11(3) | UG 10(2) | GU 01(1) | GG 00(0) |

The case of the order 2-1:

| | 00(0) | 10(2) | 01(1) | 11(3) |
|---|---|---|---|---|
| 00 (0) | CC 00(0) | AC 10(2) | CA 01(1) | AA 11(3) |
| 10 (2) | UC 10(2) | GC 00(0) | UA 11(3) | GA 01(1) |
| 01 (1) | CU 01(1) | AU 11(3) | CG 00(0) | AG 10(2) |
| 11 (3) | UU 11(3) | GU 01(1) | UG 10 (2) | GG 00(0) |

$H_{12}$ =

| | 00 | 01 | 10 | 11 |
|---|---|---|---|---|
| 00 | +1(0) | -1(1) | +1(2) | -1(3) |
| 01 | +1(1) | +1(0) | -1(3) | -1(2) |
| 10 | -1(2) | +1(3) | +1 (0) | -1(1) |
| 11 | +1(3) | +1(2) | +1 (1) | +1(0) |

$H_{21}$ =

| | 00 | 10 | 01 | 11 |
|---|---|---|---|---|
| 00 | +1(0) | +1(2) | -1(1) | -1(3) |
| 10 | -1(2) | +1(0) | 1(3) | -1(1) |
| 01 | +1(1) | -1(3) | +1(0) | -1(2) |
| 11 | +1(3) | +1(1) | +1 (2) | +1(0) |

**$h0_{12}$ =[1 0 0 0; 0 1 0 0; 0 0 1 0; 0 0 0 1]**
**$h1_{12}$ =[0 -1 0 0; 1 0 0 0; 0 0 0 -1; 0 0 1 0]**
**$h2_{12}$ =[0 0 1 0; 0 0 0 -1; -1 0 0 0; 0 1 0 0]**
**$h3_{12}$ =[0 0 0 -1; 0 0 -1 0; 0 1 0 0; 1 0 0 0]**

**$h0_{21}$ =[1 0 0 0; 0 1 0 0; 0 0 1 0; 0 0 0 1]**
**$h1_{21}$ =[0 0 -1 0; 0 0 0 -1; 1 0 0 0; 0 1 0 0]**
**$h2_{21}$ =[0 1 0 0; -1 0 0 0; 0 0 0 -1; 0 0 1 0]**
**$h3_{21}$ =[0 0 0 -1; 0 0 1 0; 0 -1 0 0; 1 0 0 0]**

Figure 26. Top and middle rows: the results of transformation of the genomatrices [C A; U G]$_{12}^{(2)}$ and [C A; U G]$_{21}^{(2)}$ (Figure 22) by means of the U-algorithm into Hadamard matrices $H_{12}$ and $H_{21}$. Bottom row: two sets of four sparse matrices which arise in the result of dyadic-shift decompositions of these Hadamard genomatrices $H_{12}$ and $H_{21}$. The matrices $H0_{12}$ and $H0_{21}$ are identity matrices.

Each of two sets of four sparse matrices (bottom row on Figure 26), which arise in the result of dyadic-shift decompositions of these Hadamard forms $H_{12}$ and $H_{21}$, is closed in relation to multiplication and satisfies the same multiplication table on Figure 27. In other words, these two sets are different forms of representation of the matrix DS-algebra of 4-dimensional hypercomplex numbers with such multiplication table (on Figure 27 the indices 12 and 21 of the elements are omitted). We term these 4-dimensional numbers as H-genoquaternions.

| | 1 | h1 | h2 | h3 |
|---|---|---|---|---|
| 1 | 1 | h1 | h2 | h3 |
| h1 | h1 | -1 | -h3 | h2 |
| h2 | h2 | h3 | -1 | -h1 |
| h3 | h3 | -h2 | h1 | -1 |

| | 1 | i | j | k |
|---|---|---|---|---|
| 1 | 1 | i | j | k |
| i | i | -1 | k | -j |
| j | j | -k | -1 | i |
| k | k | j | -i | -1 |

Figure 27. The multiplication table (the left side) for the dyadic-shift decompositions of Hadamard genomatrices $H_{12}$ and $H_{21}$ from Figure 26. This table defines the 4-dimensional algebra of H-genoquaternions which is simply connected with the algebra of Hamilton quaternions because this table is the transposed analogue of the multiplication table of Hamilton quaternions (the right side). From the viewpoint of a disposition of signs "+" and "-" in these multiplication tables, both tables correspond additionally to relavant Hadamard matrices.

### 13. The Hadamard genomatrices, Hamilton quaternions and biquaternions

The multiplication table (Figure 27, left) for the dyadic-shift decompositions of Hadamard genomatrices $H_{12}$ and $H_{21}$ is the transposed analogue of the multiplication table of Hamilton quaternions (Figure 27, right).

By the way, such transposed variant of the multiplication table is known in the study of the quaternion's structure of the operators which describe vibrational-rotational molecular states [Lobodenko, 2009]; it is only one of a few evidences in a favor of the our supposition that the origin of such hypercomplex structure of the system of genetic alphabets is connected with quanto-mechanical properties of genetic molecules (see additionally below about our spin-invariant approaches to the system of genetic alphabets).

To get the classical multiplication table of Hamilton quaternions one can use one of the following two ways:

3) One can take the new order of basic matrices of H-genooctetons h0+h3+h2+h1 (instead of the order h0+h1+h2+h3 on Figure 26, bottom);
4) One can use from the very beginning the genomatrix [C U; A G]$^{(2)}$ (Figure 28, left) which is the transposed analogue of the considered genomatrix [C A; U G]$^{(2)}$ (Figure 4). The DS-decomposition of the Hadamard form of the mosaic genomatrix [C U; A G]$^{(2)}$ gives the set of 4 sparse matrices $h0^T$, $h1^T$, $h2^T$, $h3^T$ which are transposed analogues of the basic matrices h0, h1, h2, h3 (Figure 26, bottom) and which give the multiplication table of Hamilton quaternions (Figure 27, right).

| | | | |
|---|---|---|---|
| CC | CU | UC | UU |
| CA | CG | UA | UG |
| AC | AU | GC | GU |
| AA | AG | GA | GG |

| | | | | | | | |
|---|---|---|---|---|---|---|---|
| CCC | CCU | CUC | CUU | UCC | UCU | UUC | UUU |
| CCA | CCG | CUA | CUG | UCA | UCG | UUA | UUG |
| CAC | CAU | CGC | CGU | UAC | UAU | UGC | UGU |
| CAA | CAG | CGA | CGG | UAA | UAG | UGA | UGG |
| ACC | ACU | AUC | AUU | GCC | GCU | GUC | GUU |
| ACA | ACG | AUA | AUG | GCA | GCG | GUA | GUG |
| AAC | AAU | AGC | AGU | GAC | GAU | GGC | GGU |
| AAA | AAG | AGA | AGG | GAA | GAG | GGA | GGG |

Figure 28. The result of the transformation of genomatrices [C U; A G]$^{(2)}$ and [C U; A G]$^{(3)}$ from Figure 24 by means of the U-algorithm. The represented matrices are the Hadamard matrices if each of black components is replaced by the element "+1" and each of white components is replaced by the element "-1".

Below we will term the Hamilton quaternion with unit coordinates as "the frame Hamilton quaternion" and the complex number with unit coordinates Z=1+i as "the frame complex number".The Hadamard form of the genomatrix [C U; A G]$^{(3)}$ possesses a block structure phenomenologically: each of its (2*2)-subquadrants is the matrix representation [1 -1; 1 1] of the complex number Z=1+i (Figure 29, the top row). In the result the Hadamard form of the (8*8)-genomatrix [C U; A G]$^{(3)}$ can be represented as a result of the Kronecker product of the frame Hamilton quaternion (or the Hadamard form of the (4*4)-genomatrix [C U; A G]$^{(2)}$) with the matrix representation [1 -1; 1 1] of the complex number Z=1+i (Figure 29, the bottom row).

| | | | | | | | |
|---|---|---|---|---|---|---|---|
| 1 | -1 | 1 | -1 | -1 | 1 | 1 | -1 |
| 1 | 1 | 1 | 1 | -1 | -1 | 1 | 1 |
| -1 | 1 | 1 | -1 | 1 | -1 | 1 | -1 |
| -1 | -1 | 1 | 1 | 1 | 1 | 1 | 1 |
| 1 | -1 | -1 | 1 | 1 | -1 | 1 | -1 |
| 1 | 1 | -1 | -1 | 1 | 1 | 1 | 1 |
| -1 | 1 | -1 | 1 | -1 | 1 | 1 | -1 |
| -1 | -1 | -1 | -1 | -1 | -1 | 1 | 1 |

=

| | | | |
|---|---|---|---|
| Z | Z | -Z | Z |
| -Z | Z | Z | Z |
| Z | -Z | Z | Z |
| -Z | -Z | -Z | Z |
| 1 | 1 | -1 | 1 |
| -1 | 1 | 1 | 1 |
| 1 | -1 | 1 | 1 |
| -1 | -1 | -1 | 1 |

$\otimes$

| | |
|---|---|
| 1 | -1 |
| 1 | 1 |

=

| | | | | | | | |
|---|---|---|---|---|---|---|---|
| 1 | -1 | 1 | -1 | -1 | 1 | 1 | -1 |
| 1 | 1 | 1 | 1 | -1 | -1 | 1 | 1 |
| -1 | 1 | 1 | -1 | 1 | -1 | 1 | -1 |
| -1 | -1 | 1 | 1 | 1 | 1 | 1 | 1 |
| 1 | -1 | -1 | 1 | 1 | -1 | 1 | -1 |
| 1 | 1 | -1 | -1 | 1 | 1 | 1 | 1 |
| -1 | 1 | -1 | 1 | -1 | 1 | 1 | -1 |
| -1 | -1 | -1 | -1 | -1 | -1 | 1 | 1 |

Figure 29. The top row: the block structure of the Hadamard form of the genomatrix $[C\ U;\ A\ G]^{(3)}$ includes in its (2*2)-sub-quadrants the matrix representation [1 -1; 1 1] of the complex number Z=1+i (with the sign plus or minus). The bottom row: the Hadamard form B of the (8*8)-genomatrix $[C\ U;\ A\ G]^{(3)}$ is the result of the Kronecker product of the Hamilton quaternion H with unit coordinates (or Hadamard form of the (4*4)-genomatrix $[C\ U;\ A\ G]^{(2)}$) with this (2*2)-matrix representation of the complex number Z=1+i.

Usually the result of the Kronecker product of Hamilton quaternions with the matrix representation [x -y; y x] of the complex numbers x+i*y is termed "complexification". The result of such complexification of Hamilton quaternions is termed "biquaternions"; in abstract algebra, the biquaternions are the numbers w+i*x+j*y+k*z, where w, x, y and z are complex numbers and the elements of {1, i, j, k} multiply as in the quaternion group [http://en.wikipedia.org/wiki/Biquaternion]. In other words, the Hadamard (8*8)-genomatrix $H_{123}$ is produced by means of complexification of the Hamilton quaternion $H_{12}$. In this article we will use the symbols $q_0= h0_{12}^T=1$, $q_1= h1_{12}^T$, $q_2= h2_{12}^T$, $q_3= h3_{12}^T$ (here the matrices $h0_{12}$, $h1_{12}$, $h2_{12}$, $h3_{12}$ are taken from the left side of Figure 26, bottom; T – the symbol of matrix transposition) for basic quaternions instead of {1, i, j, k}. If all coordinates of the mentioned complex numbers are equal to 1 (w=x=y=z=1+i) and the Hamilton quaternion with the unit coordinates ($1+q_1+q_2+q_3$) is used, then the following biquaternion Q with the unit coordinates arises (where $i=q_4$, $q_0*i=q_4$, $q_1*i=q_5$, $q_2*i=q_6$, $q_3*i=q_7$ from the multiplication table on Figure 30):

$$Q = q_0*(1+i) + q_1*(1+i) + q_2*(1+i) + q_3*(1+i) = \\ = (q_0+q_1+q_2+q_3)+(q_0+q_1+q_2+q_3)*i = (q_0+q_1+q_2+q_3)+(q_4+q_5+q_6+q_7) \qquad (17)$$

It is essential that this biquaternion contains two different halves: the first half is the "real" part of biquaternion ($q_0+q_1+q_2+q_3$) and the second half is the "imaginery" part of the biquaternion ($q_4+q_5+q_6+q_7$). Now we will show that these two halves possess a specific connection with the phenomenon of evolution (or diversity) of dialects of the genetic code (see Figure 16).

The DS-decomposition of the (8*8)-matrix from Figure 29 gives the set of 8 sparse matrices $q_0$, $q_1$, $q_2$, $q_3$, $q_4$, $q_5$, $q_6$, $q_7$ which is shown on Figure 30:

$$Q = (q_0+q_1+q_2+q_3)+(q_4+q_5+q_6+q_7) =$$

$$\begin{bmatrix} 1&0&0&0&0&0&0&0\\ 0&1&0&0&0&0&0&0\\ 0&0&1&0&0&0&0&0\\ 0&0&0&1&0&0&0&0\\ 0&0&0&0&1&0&0&0\\ 0&0&0&0&0&1&0&0\\ 0&0&0&0&0&0&1&0\\ 0&0&0&0&0&0&0&1 \end{bmatrix} + \begin{bmatrix} 0&-1&0&0&0&0&0&0\\ 1&0&0&0&0&0&0&0\\ 0&0&0&-1&0&0&0&0\\ 0&0&1&0&0&0&0&0\\ 0&0&0&0&0&-1&0&0\\ 0&0&0&0&1&0&0&0\\ 0&0&0&0&0&0&0&-1\\ 0&0&0&0&0&0&1&0 \end{bmatrix} + \begin{bmatrix} 0&0&1&0&0&0&0&0\\ 0&0&0&1&0&0&0&0\\ -1&0&0&0&0&0&0&0\\ 0&-1&0&0&0&0&0&0\\ 0&0&0&0&0&0&1&0\\ 0&0&0&0&0&0&0&1\\ 0&0&0&0&-1&0&0&0\\ 0&0&0&0&0&-1&0&0 \end{bmatrix} + \begin{bmatrix} 0&0&0&-1&0&0&0&0\\ 0&0&1&0&0&0&0&0\\ 0&1&0&0&0&0&0&0\\ -1&0&0&0&0&0&0&0\\ 0&0&0&0&0&0&0&-1\\ 0&0&0&0&0&0&1&0\\ 0&0&0&0&0&1&0&0\\ 0&0&0&0&-1&0&0&0 \end{bmatrix} +$$

$$\begin{bmatrix} 0&0&0&0&-1&0&0&0\\ 0&0&0&0&0&-1&0&0\\ 0&0&0&0&0&0&1&0\\ 0&0&0&0&0&0&0&1\\ 1&0&0&0&0&0&0&0\\ 0&1&0&0&0&0&0&0\\ 0&0&-1&0&0&0&0&0\\ 0&0&0&-1&0&0&0&0 \end{bmatrix} + \begin{bmatrix} 0&0&0&0&0&1&0&0\\ 0&0&0&0&-1&0&0&0\\ 0&0&0&0&0&0&0&-1\\ 0&0&0&0&0&0&1&0\\ 0&-1&0&0&0&0&0&0\\ 1&0&0&0&0&0&0&0\\ 0&0&0&1&0&0&0&0\\ 0&0&-1&0&0&0&0&0 \end{bmatrix} + \begin{bmatrix} 0&0&0&0&0&0&1&0\\ 0&0&0&0&0&0&0&1\\ 0&0&0&0&1&0&0&0\\ 0&0&0&0&0&1&0&0\\ 0&0&-1&0&0&0&0&0\\ 0&0&0&-1&0&0&0&0\\ -1&0&0&0&0&0&0&0\\ 0&-1&0&0&0&0&0&0 \end{bmatrix} + \begin{bmatrix} 0&0&0&0&0&0&0&-1\\ 0&0&0&0&0&0&1&0\\ 0&0&0&0&0&-1&0&0\\ 0&0&0&0&1&0&0&0\\ 0&0&0&1&0&0&0&0\\ 0&0&-1&0&0&0&0&0\\ 0&1&0&0&0&0&0&0\\ -1&0&0&0&0&0&0&0 \end{bmatrix}$$

| | 1 | $q_1$ | $q_2$ | $q_3$ | $q_4$ | $q_5$ | $q_6$ | $q_7$ |
|---|---|---|---|---|---|---|---|---|
| 1 | 1 | $q_1$ | $q_2$ | $q_3$ | $q_4$ | $q_5$ | $q_6$ | $q_7$ |
| $q_1$ | $q_1$ | -1 | $q_3$ | $-q_2$ | $q_5$ | $-q_4$ | $q_7$ | $-q_6$ |
| $q_2$ | $q_2$ | $q_3$ | -1 | $-q_1$ | $q_6$ | $q_7$ | $-q_4$ | $-q_5$ |
| $q_3$ | $q_3$ | $-q_2$ | $-q_1$ | 1 | $q_7$ | $-q_6$ | $-q_5$ | $q_4$ |
| $q_4$ | $q_4$ | $q_5$ | $-q_6$ | $-q_7$ | -1 | $-q_1$ | $q_2$ | $q_3$ |
| $q_5$ | $q_5$ | $-q_4$ | $-q_7$ | $q_6$ | $-q_1$ | 1 | $q_3$ | $-q_2$ |
| $q_6$ | $q_6$ | $q_7$ | $q_4$ | $q_5$ | $-q_2$ | $-q_3$ | -1 | $-q_1$ |
| $q_7$ | $q_7$ | $-q_6$ | $q_5$ | $-q_4$ | $-q_3$ | $q_2$ | $-q_1$ | 1 |

Figure 30. The top part: the set of 8 sparse matrices $q_0$, $q_1$, $q_2$, $q_3$, $q_4$, $q_5$, $q_6$, $q_7$ from the DS-decomposition of the (8*8)-matrix representation of the biquaternion from Figure 29. Here $q_0$ is the unity matrix. The bottom part: the multiplication table of the basic biquaternions.

This set of the 8 sparse matrices is closed in relation to multiplication and defines the multiplication table of the algebra of biquaternions (Figure 30, bottom). One can check that non-zero entries of the first four basic biquaternions $q_0$, $q_1$, $q_2$ and $q_3$ ocupy only two (4*4)-quadrants along the main diagonal of the matrix representation of the biquaternion on Figure a2. And non-zero entries of the last four basic biquaternions $q_4$, $q_5$, $q_6$ and $q_7$ ocupy only two (4*4)-quadrants along the second diagonal of the matrix representation of the biquaternion on Figure 29.

In accordance with the phenomenologic rule № 1 (see Section 8 and Figure 16) about evolutionary changes of coding meaning of triplets in different dialects of the genetic code in organisms with sexual reproduction, one can see that all the changeable triplets are disposed in two quadrants along the second diagonal of the matrix $[C\ U;\ A\ G]^{(3)}$ (Figure 28). Consequently all the changeable triplets are connected by means of their non-zero entries only with the "imaginery" basic biquaternions $q_4$, $q_5$, $q_6$ and $q_7$ (Figures 29 and 30). The "real" basic biquaternions $q_0$, $q_1$, $q_2$ and $q_3$ are connected by means of their non-zero entries only with non-changeable triplets. A hope appears to understand the evolutionary changes in dialects of the genetic code from the viewpoint of Hamilton quaternions (or of Cockle coquaternions) and their interconnections with many phenomena in the field of atomic physics and quanto-mechanical theory of chemical valency.

| | | | |
|---|---|---|---|
| A | B | C | D |
| -B | A | -D | C |
| -C | D | A | -B |
| -D | -C | B | A |

➔ F =

| | | | |
|---|---|---|---|
| 1 | 1 | 1 | 1 |
| -1 | 1 | -1 | 1 |
| -1 | 1 | 1 | -1 |
| -1 | -1 | 1 | 1 |

; $F*F^T=4*E$

Z⊗F =

| | | | | | | | |
|---|---|---|---|---|---|---|---|
| 1 | 1 | 1 | 1 | 1 | 1 | 1 | 1 |
| -1 | 1 | -1 | 1 | -1 | 1 | -1 | 1 |
| -1 | 1 | 1 | -1 | -1 | 1 | 1 | -1 |
| -1 | -1 | 1 | 1 | -1 | -1 | 1 | 1 |
| -1 | -1 | -1 | -1 | 1 | 1 | 1 | 1 |
| 1 | -1 | 1 | -1 | -1 | 1 | -1 | 1 |
| 1 | -1 | -1 | 1 | -1 | 1 | 1 | -1 |
| 1 | 1 | -1 | -1 | -1 | -1 | +1 | 1 |

Figure 31. Top row, left: the traditional form of the (4*4)-matrix representation of Hamilton quaternions (from http://en.wikipedia.org/wiki/Quaternion). Right: the matrix F presents the Hamilton quaternion with unit coordinates ("the frame Hamilton quaternion") and the matrix F is the Hadamard matrix. The entries with the sign "+" are marked by black color. E is the unity matrix. Bottom row: the Hamilton biquaternion with unit coordinates (the frame Hamilton biquaternion) as the result of the Kronecker product of Z=[1 1; -1 1], which is the matrix representation of the complex number Z=1+1, with the frame Hamilton quaternion F. Cells with entries "+1" are marked by black color.

Hamilton quaternions are well-known in mathematics and theoretical physics (see, for example http://en.wikipedia.org/wiki/Quaternion). Hamilton quaternions were first described by Irish mathematician Sir William Rowan Hamilton in 1843 and applied to mechanics in three-dimensional space. Their historical impact on physics is described in many works (see a review in http://en.wikipedia.org/wiki/Quaternion). In our study Hamilton quaternions and their complexification show themselves in the structural organization of information basics of living matter: in the multi-level system of molecular-genetic alphabets. Hamilton quaternions in their matrix form are usually written in the way shown on Figure 31 where the scheme of the dyadic-shift decomposition is laid inside as a natural part of this matrix form. One can check that the matrix form of the Hamilton quaternion with unit coordinates is the Hadamard (4*4)-matrix F which satisfies the criterial condition $F*F^T=4*E$, where E is the unity matrix (author knows of no publications in which this interesting fact of the interconnection of Hamilton quaternions with Hadamard matrices was noted). The complex number with unit coordinates Z=1+i is also the Hadamard (2*2)-matrix [1 1; - 1 1]. Such "frame complexification" of the frame Hamilton quaternion by means of the frame complex number gives a Hadamard matrix again because of the rule: Kronecker product of two Hadamard matrices gives a new Hadamard matrix. All the forms of (4*4)-matrix representation of the frame Hamilton quaternion, which are known for the author, are also Hadamard matrices. Since the frame Hamilton quaternion and its frame complexification are expressed by Hadamard matrices, Hamilton quaternions and their complexification are interesting not only for physics but also for informatics including noise-immunity coding where Hadamard matrices play an important role long ago. On the base of Hamilton quaternions and their complexification mathematical biology is docked (or joined) with theoretical physics.

One should note specially a known connection of basic quaternions with the Pauli matrices which play an important role in quantum mechanics and which were used by W.Pauli in his study of spin in quantum mechanics [http://en.wikipedia.org/wiki/Pauli_matrices; http://en.wikipedia.org/wiki/Spin_(physics)]. In quantum mechanics and particle physics, spin is a fundamental characteristic property of elementary particles, composite particles (hadrons), and atomic nuclei. Now the connection of genetic alphabets with quaternions and Pauli matrices allows using in biological research many approaches, data and formalisms from quantum mechanics, atomic physics, quanto-mechanical theory of chemical valency, theory of spin-invariants, spin chemistry, theory of atomic memory [Brewer, Hahn, 1984], theory of spinors and spinor fields, theory of quantum information (where the Pauli matrices are some of the most important single-qubit operations), theory of screws, Clifford algebra, etc. In other words, this connection of the genetic system with Hamilton quaternions and Pauli matrices can be considered as one of bases to develop "quanto-mechanical bioinformatics". Some of these perspective problems are studied now in Moscow, Russia in the laboratory of biomechanical systems headed by the author.

In matrix genetics from a huge family of Hadamard matrices the author distinguishes as the most important the subfamily of Hadamard matrices, which is associated with the Hamilton quaternions and their complexifications. One should mark else that Hamilton quaternions and their complexification are usually considered over the field of real numbers. From the viewpoint of matrix genetics, the author supposes that the most appropriate variant to model mathematically the genetic system is provided by means of Hamilton quaternions and their complexification over the field of algebraic numbers (see below the Section 16). Mathematics knows the complexification of Hamilton quaternions long ago in a connection with the 8-dimensional hypercomplex algebra of biquaternions which also plays an important role in theoretical physics.

What one can say about algebraic interconnections of Hamilton quaternions and Cockle coquaternions? It is known that both of these algebras are Clifford algebras for the case n=2 [http://www.mi.ras.ru/noc/11_12/cllifalg04.12.11.pdf ]. Perhaps this fact will be useful in further algebraic study of genetic systems.

How new variants of matrix representations of Hamilton quaternion and its complexification can be algorithmically received if only one of their possible variants is known, for example, the variant of matrix representation of the Hamilton quaternion on Figure 31 and the variant of its complexification? At first glance, it is not simple task. But matrix genetics reveals that this task can be solved by means of the following simple algorithm which uses the system of hidden parameters for these matrices in a form of symbolic duplets and triplets. We term this algorithm as "the algorithm of hidden parameters". Let us take the frame Hamilton quaternion and its frame complexification whose matrix representations are Hadamard matrices of the $4^{th}$ and $8^{th}$ orders (their entries "+1" and "-1" are disposed in black and white cells correspondingly as usual in this article). On the second step let us take a four-letter alphabet K, L, M, N and two matrices $[K\ L;\ M\ N]^{(2)}$ for 16 duplets and $[K\ L;\ M\ N]^{(3)}$ for 64 triplets on the base of this alphabet. Cells of these two matrices are marked by black and white colors by analogy with black-and-white mosaics of the traditional matrix representations of the frame Hamilton quaternion F and the frame biquaternion $[1\ 1;\ -1\ 1]\otimes F$ (from Figure 31). All possible variants of positional permutations of letters inside duplets (two variants exist: 1-2 and 2-1) and inside triplets (six variants exist: 1-2-3, 2-3-1, 3-1-2, 3-2-1, 2-1-3, 1-3-2) generate two variants of (4*4)-matrix of duplets (Figure 32) and six variants of (8*8)-matrix of triplets (Figure 33).

The interesting result is the following. The replacement of the black (white) doublets and triplets in the matrices on Figures 32 and 33 by entries "+1" (or "-1" respectively) gives rise to two matrix representations of the frame Hamilton quaternion (or the sums of the two relevant sets of 4 basic quaternions) and six matrix representations of its frame biquaternion (or the sums of the six relevant sets of 8 basic biquaternions) on the base of the DS-decompositions of these matrices. Knowledge of these two matrix sets of 4 basic quaternions allows presenting Hamilton quaternions in two different matrix forms. Knowledge of these two matrix sets of 4 basic quaternions allows presenting Hamilton quaternions in two different matrix forms. And knowledge of these six matrix sets of 8 basic biquaternions allows presenting Hamilton biquaternions in six different matrix forms. One can conclude that the proposed "algorithm of hidden parameters" works effectively. The situation with this algorithm resembles in some extend the situation with quarks which form the system of hidden parameters in physics of elementary particles.

**$F_{12}$ =**

| | | | |
|---|---|---|---|
| 1 | 1 | 1 | 1 |
| -1 | 1 | -1 | 1 |
| -1 | 1 | 1 | -1 |
| -1 | -1 | 1 | 1 |

➔ **$[K\ L;\ M\ N]^{(2)}_{12}$ =**

| | | | |
|---|---|---|---|
| KK | KL | LK | LL |
| -KM | KN | -LM | LN |
| -MK | ML | NK | -NL |
| -MM | -MN | NM | NN |

**$[K\ L;\ M\ N]^{(2)}_{21}$ =**

| | | | |
|---|---|---|---|
| **KK** | **LK** | **KL** | **LL** |
| **-MK** | **KN** | **ML** | **-NL** |
| **-KM** | **-LM** | **KN** | **LN** |
| **-MM** | **NM** | **-MN** | **NN** |

➔ **$F_{21}$=**

| | | | |
|---|---|---|---|
| **1** | **1** | **1** | **1** |
| **-1** | **1** | **1** | **-1** |
| **-1** | **-1** | **1** | **1** |
| **-1** | **1** | **-1** | **1** |

Figure 32. The producing a new matrix representation of Hamilton quaternions by means of the algorithm of positional permutations inside 16 duplets in the matrix $[K\ L;\ M\ N]^{(2)}$ which plays a role of the system of hidden parameters in a form of duplets. The matrix $F_{12}$ is identical to the matrix F on Figure 31. The final matrix $F_{12}$ is the sum of basic quaternions in their new matrix representations. Indexes 12 and 21 correspond to variants of positional orders 1-2 and 2-1 inside duplets in the matrix $[K\ L;\ M\ N]^{(2)}$.

$[K\ L;\ M\ N]^{(3)}_{123}$ =

| | | | | | | | |
|---|---|---|---|---|---|---|---|
| KKK | KKL | KLK | KLL | LKK | LKL | LLK | LLL |
| KKM | KKN | KLM | KLN | LKM | LKN | LLM | LLN |
| KMK | KML | KNK | KNL | LMK | LML | LNK | LNL |
| KMM | KMN | KNM | KNN | LMM | LMN | LNM | LNN |
| MKK | MKL | MLK | MLL | NKK | NKL | NLK | NLL |
| MKM | MKN | MLM | MLN | NKM | NKN | NLM | NLN |
| MMK | MML | MNK | MNL | NMK | NML | NNK | NNL |
| MMM | MMN | MNM | MNN | NMM | NMN | NNM | NNN |

$[K\ L;\ M\ N]^{(3)}_{231}$ =

| | | | | | | | |
|---|---|---|---|---|---|---|---|
| KKK | KLK | LKK | LLK | KKL | KLL | LKL | LLL |
| KMK | KNK | LMK | LNK | KML | KNL | LML | LNL |
| MKK | MLK | NKK | NLK | MKL | MLL | NKL | NLL |
| MMK | MNK | NMK | NNK | MML | MNL | NML | NNL |
| KKM | KLM | LKM | LLM | KKN | KLN | LKN | LLN |
| KMM | KNM | LMM | LNM | KMN | KNN | LMN | LNN |
| MKM | MLM | NKM | NLM | MKN | MLN | NKN | NLN |
| MMM | MNM | NMM | NNM | MMN | MNN | NMN | NNN |

$[K\ L;\ M\ N]^{(3)}_{213}$ =

| | | | | | | | |
|---|---|---|---|---|---|---|---|
| KKK | KKL | LKK | LKL | KLK | KLL | LLK | LLL |
| KKM | KKN | LKM | LKN | KLM | KLN | LLM | LLN |
| MKK | MKL | NKK | NKL | MLK | MLL | NLK | NLL |
| MKM | MKN | NKM | NKN | MLM | MLN | NLM | NLN |
| KMK | KML | LMK | LML | KNK | KNL | LNK | LNL |
| KMM | KMN | LMM | LMN | KNM | KNN | LNM | LNN |
| MMK | MML | NMK | NML | MNK | MNL | NNK | NNL |
| MMM | MMN | NMM | NMN | MNM | MNN | NNM | NNN |

$[K\ L;\ M\ N]^{(3)}{}_{321}$ =

| | | | | | | | |
|---|---|---|---|---|---|---|---|
| KKK | LKK | KLK | LLK | KKL | LKL | KLL | LLL |
| MKK | NKK | MLK | NLK | MKL | NKL | MLL | NLL |
| KMK | LMK | KNK | LNK | KML | LML | KNL | LNL |
| MMK | NMK | MNK | NNK | MML | NML | MNL | NNL |
| KKM | LKM | KLM | LLM | KKN | LKN | KLN | LLN |
| MKM | NKM | MLM | NLM | MKN | NKN | MLN | NLN |
| KMM | LMM | KNM | LNM | KMN | LMN | KNN | LNN |
| MMM | NMM | MNM | NNM | MMN | NMN | MNN | NNN |

$[K\ L;\ M\ N]^{(3)}{}_{312}$ =

| | | | | | | | |
|---|---|---|---|---|---|---|---|
| KKK | LKK | KKL | LKL | KLK | LLK | KLL | LLL |
| MKK | NKK | MKL | NKL | MLK | NLK | MLL | NLL |
| KKM | LKM | KKN | LKN | KLM | LLM | KLN | LLN |
| MKM | NKM | MKN | NKN | MLM | NLM | MLN | NLN |
| KMK | LMK | KML | LML | KNK | LNK | KNL | LNL |
| MMK | NMK | MML | NML | MNK | NNK | MNL | NNL |
| KMM | LMM | KMN | LMN | KNM | LNM | KNN | LNN |
| MMM | NMM | MMN | NMN | MNM | NNM | MNN | NNN |

$[K\ L;\ M\ N]^{(3)}{}_{132}$ =

| | | | | | | | |
|---|---|---|---|---|---|---|---|
| KKK | KLK | KKL | KLL | LKK | LLK | LKL | LLL |
| KMK | KNK | KML | KNL | LMK | LNK | LML | LNL |
| KKM | KLM | KKN | KLN | LKM | LLM | LKN | LLN |
| KMM | KNM | KMN | KNN | LMM | LNM | LMN | LNN |
| MKK | MLK | MKL | MLL | NKK | NLK | NKL | NLL |
| MMK | MNK | MML | MNL | NMK | NNK | NML | NNL |
| MKM | MLM | MKN | MLN | NKM | NLM | NKN | NLN |
| MMM | MNM | MMN | MNN | NMM | NNM | NMN | NNN |

Figure 33. The producing of new matrix representations of Hamilton biquaternions by means of "the algorithm of hidden parameters" which is based on positional permutations inside 64 triplets in the matrix $[K\ L;\ M\ N]^{(3)}$. Here the set of triplets plays a role of the system of hidden parameters. The black-and-white mosaic of upper matrix $[K\ L;\ M\ N]^{(3)}{}_{123}$ is identical to the mosaic of the matrix representation of the frame Hamilton biquaternion on Figure 31. Indexes correspond to variants of positional orders 1-2-3, 2-3-1, etc. of symbols inside the triplets in $[K\ L;\ M\ N]^{(3)}$. Each of triplets possesses the same color in all matrices.

As Hamilton quaternions describe the properties of three-dimensional physical space, the discovery of the connection of the genetic system with Hamilton quaternions attracts our attention to some fundamental questions. It is, for example, the question about innate spatial representations in humans and animals [Russell, 1956; Petoukhov, 1981]. Or the question about a development of physical theories, in which the concept of space is not primary, but derived from more fundamental concepts of special mathematical systems [Kulakov, 2004; Vladimirov, 1998]. The described effectiveness of the algorithm of hidden parameters allows thinking about the systems of hidden parameters as about a possible base for additional development of these theories.

Molecular biology and bioinformatics possess their own problems where Hamilton quaternions and their complexification can be used. For example some approaches are known about algorithmic constructions of fractal patterns in biological structures including fractals in genetic molecules (see [Pellionisz et al, 2011; http://www.junkdna.com/the_genome_is_fractal.html]). A development of geometric algorythms for such approaches needs those geometrical operations inside physical 3D-space which are connected with the molecular-genetic system and which can be used as a basis of these geometric algorithms. Hamilton quaternions and their complexifications, which are connected with the system of genetic alphabets and which correspond to geometric properties of our physical space, seem to be promising candidates for this purpose.

Our article shows that now Hamilton quaternions and their complexifications are connected not only with theoretical physics but also with molecular genetics and bioinformatics. The discovery of the relation between the system of molecular-genetic alphabets and Hamilton quaternions together with their complexification provides a bridge between theoretical physics and biology for their mutual enrichment. It can be considered as a next step to discover the mathematical unity of nature.

## 14. Encrypted matrices and permutation genomatrices

In this section we pay special attention to an interesting general feature of the described DS-algebras which were revealed in a course of studying the multilevel system of genetic alphabets. All the basic matrices (with the exception of the identity matrices) of the described genetic octetons and 4-dimensional hypercomplex numbers are not only permutation matrices but they satisfy the following special conditions:

1) Their main diagonal has zero entries only;

2) Their square is equal to the identity matrix that is $P^2 = I$ (not all permutation matrices possess such property; for example, the permutation matrix [0 1 0 0; 0 0 1 0; 0 0 0 1; 1 0 0 0] does not possess it).

In informatics such special permutation matrices are termed as "*encryption matrices*" and they are used for effective encryption of text-based messages. The work [Martinelli, 2003] shows that such encryption matrices are effectively used in communication technologies and they are bases of algorithms with the following properties:

1) all characters are swapped;
2) a second encryption returns the original message.

If P is an encryption matrix of dimension N, then N is an even integer. In other words, an alphabetic vector must contain an even number N of characters to be encrypted. Our basic encryption matrices of genetic octetons (see for example Figures 7 and 18) are differ from encryption matrices in the work [Martinelli, 2003] in that aspect that they can contain not only non-zero entries "+1" but also entries "-1"; it gives additional abilities for encryption in cases of binary-oppositional alphabets (for example in the case of 64 genetic triplets when each codon has its anticodon).

One can remind that in informatics, the electronic transmission of text-based information is widespread today. Many situations arise in which some degree of privacy is desired for the transmitted message. Encrypted matrices provide a complete rearrangement of the elements of encrypted messages (if a permutation matrix has non-zero entries on its diagonals, then not all elements of messages are rearranged under its influence). In addition they provide a comfortable possibility of using the same technological device of a permutation for both encrypting and decrypting a message.

But what about a possible biological meaning of encryption genomatrices such as those described in our article? Firstly, they can be used in living matter to provide a principle of molecular economy in genetic coding [Petoukhov, 2008; Petoukhov, He, 2010, Chapter 6]. The speach is that a single genetic sequence can be read by a biological computer of an organism by many ways on the base of its transformation by appropriate algorithms of encryption matrices. Or in some situations this initial genetic sequence can generate a set of new genetic sequences with rearrangements of its components by means of algorithms of encryption matrices. It means that in the genetic system not all genetic sequencies should exist in real molecular forms from the very beginning and in each moment of time, but a set of genetic sequences can be generated in real forms or in virtual forms some later or from time to time in process of ontogenesis on the base of genetic encryption matrices. Such a principle gives a general economy of molecular materials and an additional possibility of increasing the diversity of genetic material at various stages of development of organisms.

Secondly the encryption genomatrices and algorithms on their bases can be useful for existence of biological organisms as symbiotic entities. Modern symbio-genetics (or inter-organismal genetics) shows that each complex macroorganism is a genetic symbiosis with a great number of microorganisms.

## 15. Punnett squares: their Kronecker products and dyadic-shift decompositions

The structural features of different levels of genetic systems must be coordinated to provide the unity of the organism. In view of this, methods of matrix genetics may also be useful in their applying to different levels of organization of the genetic system.

Till this place we said about alphabets of genetic n-plets and about their subalphabets. This section is devoted to a new class of objects of matrix genetics at a higher biological level where new terms are used as central: “alleles”, “gametes”, “genotype”, “zygote”, etc. We will analyze Punnett squares, which represent alphabets of genotypes or, more precisely, alphabets of possible combinations of male and female gametes in Mendelian crosses of organisms from a viewpoint of a certain amount of inherited traits taken into account. We will show a possibilty of interpreting a set of Punnett squares for polyhybrid crosses not as tables but as square matrices (we term them “Punnett matrices”) of Kronecker products of Punnett (2*2)-matrices for monohybrid crosses. Based on Kronecker multiplication of matrices this approach gives a simple algebraic method to construct Punnett squares for the cases of multi-hybrid crosses. In addition we show that dyadic-shift decompositions of these “Punnett matrices” lead in some cases to a certain method of classification of different sub-sets of combinations of alleles from male and female gametes. Some of these results of matrix genetics at the level of genotypes are formally identical to results described above concerning to the lower level of genetic multiplets. These results demonstrate the generality of the algebraic structure of the genetic organization at different levels.

Punnetts squares are one well-known tool of genetics (http://en.wikipedia.org/wiki/Punnett_square). It was introduced by the British geneticist R. C. Punnett in 1905 to help in prediction of traits of offspring. It is a quite popular method among most text-books of genetics.

Heredity is the passing of traits from parent to offspring. Traits are controlled by genes. The different forms of a gene for a certain trait are called alleles. There are two alleles for every trait. Alleles can be dominant or recessive. Each cell in an organism's body contains two alleles for every trait. One allele is inherited from the female parent and one allele is inherited from the male parent. Punnett square is a simple method for predicting the ways in which alleles can be combined. The Punnett square is a summary of every possible combination of one maternal allele with one paternal allele for each gene being studied in the cross. In a Punnett square, dominant and recessive alleles are usually represented by letters. An uppercase letter represents a dominant allele (we will use for dominant alleles symbols H, B, C,…), and a lowercase letter represents a recessive allele (we will use for recessive alleles symbols h, b, c,…). An organism is homozygous if it has identical alleles for a particular trait, for example HH or hh. An organism is heterozygous if it has non-identical alleles for a particular trait, for example Hh. There are three possible combinations of alleles of an organism for a particular trait: homozygous dominant (HH), heterozygous (Hh), and homozygous recessive (hh). In the classic way of constructing Punnett square, alleles of a maternal gamete are put on one side of the square (usually on the top) and alleles of a paternal gamete are put on the left side; the possible progeny are produced by filling the squares with one allele from the top and one allele from the left to produce the progeny genotypes. By tradition an uppercase letter of dominant allele stands before a lowercase letter of a recessive allele (for example Hh, but not hH).

If only one trait is being considered in a genetic cross, the cross is called monohybrid. If two or three traits are being considered in a genetic cross, the cross is called dihybrid or trihybrid correspondingly. Figure 34 shows an example of Punnett squares for one of cases of trihybrid cross of parents with the maternal genotype HhbbCc and the paternal genotype HhBbCc (this particular example of initial genotypes is taken from a site about Punnett squares http://www.changbioscience.com/genetics/punnett.html). In this example each set of gametes include 8 gametes; the set of female gametes is different from the set of male gametes. One can note that this is not the easiest thing to construct Punnett squares for different cases of multi-hydrid crosses by means of the classical way.

| | | maternal gametes | | | | | | | |
|---|---|---|---|---|---|---|---|---|---|
| | | HbC | Hbc | HbC | Hbc | hbC | hbc | hbC | hbc |
| paternal gametes | HBC | HHBbCC | HHBbCc | HHBbCC | HHBbCc | HhBbCC | HhBbCc | HhBbCC | HhBbCc |
| | HBc | HHBbCc | HHBbcc | HHBbCc | HHBbcc | HhBbCc | HhBbcc | HhBbCc | HhBbcc |
| | HbC | HHbbCC | HHbbCc | HHbbCC | HHbbCc | HhbbCC | HhbbCc | HhbbCC | HhbbCc |
| | Hbc | HHbbCc | HHbbcc | HHbbCc | HHbbcc | HhbbCc | Hhbbcc | HhbbCc | Hhbbcc |
| | hBC | HhBbCC | HhBbCc | HhBbCC | HhBbCc | hhBbCC | hhBbCc | hhBbCC | hhBbCc |
| | hBc | HhBbCc | HhBbcc | HhBbCc | HhBbcc | hhBbCc | hhBbcc | hhBbCc | hhBbcc |
| | hbC | HhbbCC | HhbbCc | HhbbCC | HhbbCc | hhbbCC | hhbbCc | hhbbCC | hhbbCc |
| | hbc | HhbbCc | Hhbbcc | HhbbCc | Hhbbcc | hhbbCc | hhbbcc | hhbbCc | hhbbcc |

Figure 34. A typical Punnett square for a trihybrid cross; the maternal genotype is HhbbCc and the paternal genotype is HhBbCc (from http://www.changbioscience.com/genetics/punnett.html).

But we pay attention that this Punnett (8*8)-matrix (in the bold frame on Figure 34) is identical to a result of Kronecker product of three Punnett (2*2)-matrices of monohybrid crosses in relation to each of the traits (with a preciseness up to the order of the elements). Figure 35 shows three Punnett squares of monohybrid crosses for the considered case of the maternal genotype HhbbCc and the paternal genotype HhBbCc.

| | maternal gametes | |
|---|---|---|
| | H | h |
| H | HH | Hh |
| h | Hh | hh |

;

| | maternal gametes | |
|---|---|---|
| | b | b |
| B | Bb | Bb |
| b | bb | bb |

;

| | maternal gametes | |
|---|---|---|
| | C | c |
| C | CC | Cc |
| c | Cc | cc |

Figure 35. Three Punnett squares of monohybrid cross for the case of the Punnett square of the trihybrid cross on Figure 34.

Let us show a connection of the Punnett matrices with matrices of dyadic shifts on the example of the trihybrid cross with identical maternal and paternal genotypes HhBbCc (Figure 36). This Punnett matrix can be presented (with a preciseness up to the order of the elements) in a form of Kronecker product of three Punnett matrices for monohybrid crosses (Figure 36): [HH Hh; Hh hh]⊗ [BB Bb; Bb bb]⊗[CC Cc; Cc cc].

| | maternal gametes | |
|---|---|---|
| | H | h |
| H | HH | Hh |
| h | Hh | hh |

;

| | maternal gametes | |
|---|---|---|
| | B | b |
| B | BB | Bb |
| b | Bb | bb |

;

| | maternal gametes | |
|---|---|---|
| | C | c |
| C | CC | Cc |
| c | Cc | cc |

Figure 36. Three Punnett matrices of monohybrid crosses. Their Kronecker multiplication gives the Punnett matrix of the trihybrid cross on Figure 37.

One of simple ways to show the mentioned connection with dyadic shifts is the following. In the three-letter numerations of columns and rows of the Punnett matrix (Figure 37), each of the uppercase letters H, B, C can be denoted by means of the number 0, and each of the lowercase letters h, b, c can be denoted by means of the number 1. In this case for example the letter numerations of columns and rows HBC, HBc, HbC, Hbc, hBC, hBc, hbC, hbc are becoming numerical numerations 000, 001, 010, 011, 100, 101, 110, 111 correspondingly (Figure 37). Using of modulo-2 addition of these binary numeration of columns and rows of the Punnett matrix by analogy with the matrix of dyadic shifts on Figure 1 gives the following binary numeration of cells of the Punnett matrix (or combinations of alleles inside the matrix): 000, 001, …, 111 (in decimal notation 0, 1,…, 7) (Figure 37).

| | HBC(000) | HBc(001) | HbC(010) | Hbc(011) | hBC(100) | hBc(101) | hbC(110) | hbc(111) |
|---|---|---|---|---|---|---|---|---|
| HBC (000) | HHBBCC (000) | HHBBCc (001) | HHBbCC (010) | HHBbCc (011) | HhBBCC (100) | HhBBCc (101) | HhBbCC (110) | HhBbCc (111) |
| HBc (001) | HHBBCc (001) | HHBBcc (000) | HHBbCc (011) | HHBbcc (010) | HhBBCc (101) | HhBBcc (100) | HhBbCc (111) | HhBbcc (110) |
| HbC (010) | HHBbCC (010) | HHBbCc (011) | HHbbCC (000) | HHbbCc (001) | HhBbCC (110) | HhBbCc (111) | HhbbCC (100) | HhbbCc (101) |
| Hbc (011) | HHBbCc (011) | HHBbcc (010) | HHbbCc (001) | HHbbcc (000) | HhBbCc (111) | HhBbcc (110) | HhbbCc (101) | Hhbbcc (100) |
| hBC (100) | HhBBCC (100) | HhBBCc (101) | HhBbCC (110) | HhBbCc (111) | hhBBCC (000) | hhBBCc (001) | hhBbCC (010) | hhBbCc (011) |
| hBc (101) | HhBBCc (101) | HhBBcc (100) | HhBbCc (111) | HhBbcc (110) | hhBBCc (001) | hhBBcc (000) | hhBbCc (011) | hhBbcc (010) |
| hbC (110) | HhBbCC (110) | HhBbCc (111) | HhbbCC (100) | HhbbCc (101) | hhBbCC (010) | hhBbCc (011) | hhbbCC (000) | hhbbCc (001) |
| hbc (111) | HhBbCc (111) | HhBbcc (110) | HhbbCc (101) | Hhbbcc (100) | hhBbCc (011) | hhBbcc (010) | hhbbCc (001) | hhbbcc (000) |

Figure 37. The Punnett matrix of the trihybrid cross in the case of identical maternal and paternal genotypes HhBbCc. Numeration of columns and rows is presented not only in the letter form but also in numeric bitwise manner by means of the denotation of each of uppercase letters H, B, C as number 0 and each of the lowercase letters h, b, c as number 1. Binary numeration of each cell (or each of combinations of alleles) is received by means of modulo-2 addition of binary numerations of its column and row by analogy with the matrix of dyadic shifts (Figure 1).

One can see on Figure 37 that such dyadic-shift numeration devides the set of combination of alleles in the Punnett matrix into 8 different sub-sets of homozygous or heterozygous combinations:

- the binary numeration 000 denotes all the homozygous combinations of alleles HHBBCC, HHBBcc, HHbbCC, HHbbcc, hhBBCC, hhBBcc, hhbbCC, hhbbcc (they stand on the main diagonal of the matrix);
- the binary numeration 001 denotes all the combinations of alleles where only the third trait is heterozygous: HHBBCc, HHbbCc, hhBBCc, hhbbCc;
- the binary numeration 010 denotes all the combinations of alleles where only the second trait is heterozygous: HHBbCC, HHBbcc, hhBbCC, hhBbcc;
- the binary numeration 011 denotes all the combinations of alleles where only the first trait is homozygous: HHBbCc, hhBbCc;
- the binary numeration 100 denotes all the combinations of alleles where only the first trait is heterozygous: HhBBCC, HhBBcc, HhbbCC, Hhbbcc;

- the binary numeration 101 denotes all the combinations of alleles where only the second trait is homozygous: HhBBCc, HhbbCc;
- the binary numeration 110 denotes all the combinations of alleles where only the third trait is homozygous: HhBbCC, HhBbcc;
- the binary numeration 111 denotes all the combinations of alleles where only heterozygous traits exist: HhBbCc.

It is obvious that by analogy with the DS-decompositions of the genetic (8*8)-matrices (Figures 7, 8, 18, 19), this Punnett matrix can be decomposed on the base of DS-decomposition into the sum of 8 sparse matrices each of which contains only one of these 8 sub-sets of combinations of alleles. This example illustrates that by means of DS-decompositions of Punnett matrices we get the algorithmic method of classification of different sub-sets of homozygous or heterozygous organisms in cases of multi-hybrid crosses. (By analogy this method is applicable for the table of 64 hexagrams in Fu-Xi's order which is discussed below in Appendix A; this method separates 8 sub-sets, each of which contains 8 hexagrams with the same „homozygous" or "heterozygous" type from the viewpoint of a combination of its three digrams).

Why we have described analogies between matrices of the genetic alphabets and Punnett matrices? In our opinion these analogies are included as a particular part in our conception of "alphabetic-molecular Mendelism" [Petoukhov, 2004]. This conception argues that supramolecular phenomena of inheritance of traits in holistic biological organisms, which are described by the Mendel's laws, did not arise at an empty place, but they are a continuation of molecular-structural phenomena which are defined by deeper laws of the alphabetic-molecular level. These deeper laws possess important analogies with Mendel's laws. A set of inherited traits in biological organisms can be considered as a special alphabetic system of a high level. From such viewpoint, biological evolution can be interpreted in some extend as an evolution of multilevel systems of inherited and interconnected biological alphabets beginning at least from the molecular-genetic level. We suppose an existence of universal bio-algorithms of evolutionary producing of interrelated biological alphabets at higher and higher levels of organisation. Kronecker multiplications of genetic matrices, algorithms of dyadic shifts and some other algorithmic operations described in our works can be used to simulate this evolutionary process of a complication of a multilevel system of biological alphabets.

## 16. Octetons, symbolic roots of triplets and mathematical fourth roots of unity

This section pays a special attention to the interesting fact that all the basic matrices of R-octetons $r_0, r_1,\ldots, r_7$ (Figure 7) and all the basic matrices of H-octetons $h_0, h_1,\ldots, h_7$ (Figure 18) satisfy the following equations (18, 19):

$$(\pm r_k)^4 = 1, \quad (18)$$
$$(\pm h_k)^4 = 1, \quad (19)$$

where k = 0, 1, 2,…, 7. In other words, all of the basic matrices of octetons are the fourth roots of unity. Of course they are also the $(2^p*4)^{th}$ roots of unity (where p is a positive integer number) but the degree 4 is the minimal degree here. The equations (18) and (19) are connected with the fact that each of the basic matrices $r_0, r_1,\ldots, r_7$ and $h_0, h_1,\ldots, h_7$ is a form of (8*8)-matrix representation of the imaginery unit $i^2 = -1$ of complex numbers or the imaginery unit i=+1 of double numbers (their another name is Lorentz numbers) in accordance with the multiplication tables on Figures 8 and 19. It is known that these imaginary units can possess a few variants of their forms of matrix representation in a multidimensional algebra as it was mentioned above (Figures 14, 17, 22 and 26). The equations (18) and (19) are particular cases of matrix polynomials (http://en.wikipedia.org/wiki/Matrix_polynomial).

By definition the *n*th root of a number x is a number r which, when raised to the power of n, equals x:

$$r^n = x, \quad (20)$$

where n is the degree of the root (http://en.wikipedia.org/wiki/Nth_root).

Another fact is that the basic matrices of 4-dimensional algebra of R-genoquaternions (Figures 22, 23) and of H-genoquaternions (Figures 26 and 27) are also the 4$^{th}$ roots of unity:

$$R0_{12}^4 = R1_{12}^4 = R2_{12}^4 = R3_{12}^4 = R0_{21}^4 = R1_{21}^4 = R2_{21}^4 = R3_{21}^4 = 1 \quad (21)$$

$$H0_{12}^4 = H1_{12}^4 = H2_{12}^4 = H3_{12}^4 = H0_{21}^4 = H1_{21}^4 = H2_{21}^4 = H3_{21}^4 = 1 \quad (22)$$

In general case one can ask, what are the roots of $z^4 = 1$? The correct answer on this question depends on what type of algebra in question and on what type of matrix representations of this algebra in question, for example:

- in the case of 1-dimensional algebra of real numbers the correct answer shows the 2 roots: $z_{1,2} = \pm 1$;
- in the case of 2-dimensional algebra of complex numbers, the correct answer shows the 4 roots: $z_{1,2} = \pm 1$, $z_{3,4} = \pm i$;
- in the case of 4-dimensional algebra of Hamilton quaternions the 8 roots exist: $z_{1,2} = \pm 1$; $z_{3,4} = \pm$[0 -1 0 0; 1 0 0 0; 0 0 0 -1; 0 0 1 0]; $z_{5,6} = \pm$[0 0 -1 0; 0 0 0 1; 1 0 0 0; 0 -1 0 0]; $z_{7,8} = \pm$[0 0 0 -1; 0 0 -1 0; 0 1 0 0; 1 0 0 0];
- in the case of the 4-dimensional algebra of R-genoquaternions (Figure 22, 23) another 8 roots exist: $z_{1,2} = \pm 1$; $z_{3,4} = \pm$[0 -1 0 0; 1 0 0 0; 0 0 0 -1; 0 0 1 0]; $z_{5,6} = \pm$[0 0 1 0; 0 0 0 -1; 1 0 0 0; 0 -1 0 0]; $z_{7,8} = \pm$[0 0 0 -1; 0 0 -1 0; 0 -1 0 0; -1 0 0 0];
- in the case of the 4-dimensional algebra of H-genoquaternions, which are another form of matrix representations of Hamilton quaternion, the following 8 roots exist: $z_{1,2} = \pm 1$; $z_{3,4} = \pm$[0 -1 0 0; 1 0 0 0; 0 0 0 -1; 0 0 1 0]; $z_{5,6} = \pm$[0 0 1 0; 0 0 0 -1; -1 0 0 0; 0 1 0 0]; $z_{7,8} = \pm$[0 0 0 -1; 0 0 -1 0; 0 1 0 0; 1 0 0 0];
- in the case of the 8-dimensional algebra of genetic R-octetons 16 roots exist: $\pm r_0$, $\pm r_1$, $\pm r_2$, $\pm r_3$, $\pm r_4$, $\pm r_5$, $\pm r_6$, $\pm r_7$ (Figure 7);
- in the case of the 8-dimensional algebra of genetic H-octetons another 16 roots exist: $\pm h_0$, $\pm h_1$, $\pm h_2$, $\pm h_3$, $\pm h_4$, $\pm h_5$, $\pm h_6$, $\pm h_7$ (Figure 18).

One of the fundamental questions about the system of molecular-genetic alphabets is the following: what is that mathematical law that defines the phenomenon of symmetric distributions of strong and weak roots of triplets in the (8*8)-genomatrix [C A; U G]$^{(3)}$ of 64 triplets and in the (4*4)-genomatrix [C A; U G]$^{(2)}$ of 16 duplets on Figure 4 (these roots are marked by black and white colors correspondingly)? What is that meaningful mathematical language, that allows explaining an origin of these phenomenological mosaics in the genetic matrices, which are constructed formally by means of the second and third Kronecker powers of the simplest alphabetic (2*2)-genomatrix [C A; U G]? These phenomenological mosaics reflect a hidden natural interrelation among the genetic roots which form the holistic system.

The author's results discover that the phenomenological set of the strong and weak roots of triplets is connected with the deep mathematical notions of Hamilton quaternions, Cockle coquaternions and 4$^{th}$ roots of unity. Moreover, the genetic strong roots of triplets are expressed as the positive components of these mathematical roots of unity (in forms of their matrix representation on Figures 7, 18, 22 and 26). And the genetic weak roots of triplets are expressed by negative components of these mathematical roots of unity.

The genetic matrices [C A; U G]$^{(2)}$ and [C A; U G]$^{(3)}$ in their Rademacher and Hadamard representations are simply the sum of the appropriate sets of the 4$^{th}$ roots of unity from relevant hypercomplex vector spaces. Each of the 4th roots is interrelated with the other roots of the same set by means of its product with the imaginary unities in their forms of matrix representations (see multiplication tables on Figures 8, 19, 22 and 26). Using the interrelations among the 4th roots of unity in each of the considered cases allows deducing the complete set of the 4th roots of unity from a few of these roots (one can also describe such interrelations by using the language of basic Hamilton quaternions or basic Cockle coquaternions).

For example let us take the Rademacher form of the (8*8)-genomatrix [C A; U G]$^{(3)}$ on Figure 6 (or the R-octeton on Figure 6: R = $r_0+r_1+r_2+r_3+r_4+r_5+r_6+r_7$). One can present this R-octeton as the sum of the two sparse matrices, one of which contains only zero components in both quadrants along the main diagonal (Figure 38, left) and the second one contains only zero components in both quadrants along the second diagonal (Figure 38, right). The first of these sparse matrices is the sum $r_0+r_1+r_2+r_3$ of the first four basic matrices of R-octetons (or the sum of the four members of the set (18) of 4th roots of unity). The second matrix is the sum $r_4+r_5+r_6+r_7$ of the last four basic matrices of R-octetons (or the sum of another four members of the set (18) of 4th roots of unity). The first of these sparse matrices is transformed precisely into the second sparse matrix by means of its multiplication with the basic matrix $r_7$ (Figure 38).

| | | | | | | | |
|---|---|---|---|---|---|---|---|
| 0 | 0 | 0 | 0 | 1 | 1 | −1 | −1 |
| 0 | 0 | 0 | 0 | 1 | 1 | −1 | −1 |
| 0 | 0 | 0 | 0 | −1 | −1 | −1 | −1 |
| 0 | 0 | 0 | 0 | −1 | −1 | −1 | −1 |
| 1 | 1 | −1 | −1 | 0 | 0 | 0 | 0 |
| 1 | 1 | −1 | −1 | 0 | 0 | 0 | 0 |
| −1 | −1 | −1 | −1 | 0 | 0 | 0 | 0 |
| −1 | −1 | −1 | −1 | 0 | 0 | 0 | 0 |

*

| | | | | | | | |
|---|---|---|---|---|---|---|---|
| 0 | 0 | 0 | 0 | 0 | 0 | 0 | −1 |
| 0 | 0 | 0 | 0 | 0 | 0 | −1 | 0 |
| 0 | 0 | 0 | 0 | 0 | −1 | 0 | 0 |
| 0 | 0 | 0 | 0 | −1 | 0 | 0 | 0 |
| 0 | 0 | 0 | −1 | 0 | 0 | 0 | 0 |
| 00 | 0 | −1 | 0 | 0 | 0 | 0 | 0 |
| 0 | −1 | 0 | 0 | 0 | 0 | 0 | 0 |
| −1 | 0 | 0 | 0 | 0 | 0 | 0 | 0 |

=

| | | | | | | | |
|---|---|---|---|---|---|---|---|
| 1 | 1 | −1 | −1 | 0 | 0 | 0 | 0 |
| 1 | 1 | −1 | −1 | 0 | 0 | 0 | 0 |
| 1 | 1 | 1 | 1 | 0 | 0 | 0 | 0 |
| 1 | 1 | 1 | 1 | 0 | 0 | 0 | 0 |
| 0 | 0 | 0 | 0 | 1 | 1 | −1 | −1 |
| 0 | 0 | 0 | 0 | 1 | 1 | −1 | −1 |
| 0 | 0 | 0 | 0 | 1 | 1 | 1 | 1 |
| 0 | 0 | 0 | 0 | 1 | 1 | 1 | 1 |

Figure 38. The transformation of one half of the mosaic Rademacher genomatrix (Figure 6) into its second half by means of its multiplication with the basic matrix $r_7$ of R-octetons. Cells with zero components are marked by yellow. The first half is the sum $r_0+r_1+r_2+r_3$ of the first four basic matrices of R-octetons and the second half is the sum $r_4+r_5+r_6+$r7 of the last four basic matrices of R-octetons (Figure 7).

Taking into account our results, the mathematical language of *n*th roots of unity seems to be the adequate language for the multilevel system of genetic alphabets in addition to the language of multidimensional algebras and Kronecker products of matrices.

One should specially note that each of the considered 4th roots of unity belongs to algebraic numbers. Any superposition of algebraic numbers on the base of arithmetic operations (of summation, subtraction, multiplication and division) is an algebraic number again. It is obvious that genetic octetons and genoquaternions allow constructing polynomials. The author's results provide some evidences that the algebraic modeling of the molecular-genetic system should be based on the field of algebraic numbers (that is on roots of polynomials with integer coefficients).

The theme of *n*th roots of numbers and polynomials are studied in mathematics and its applications long ago (see http://en.wikipedia.org/wiki/Nth_root). This theme is connected with algebraic numbers, polynomial equations, abstract algebra, algebraic geometry, theory of coding and noise-immunity communications, cyclic groups, cyclotomic fields, spectral analysis, automatic control theory, optimization theory, approximation theory and many others. In history of mathematics the most attention was paid to the sub-theme of nth roots of unity (see for example http://en.wikipedia.org/wiki/Root_of_unity;

http://en.wikipedia.org/wiki/Cyclotomic_polynomial;
http://en.wikipedia.org/wiki/Root_of_unity_modulo_n). Roots of algebraic polynomials of arbitrary degree are ultimately determined by solving a linear or quadratic equation. And these roots define the so-termed free component of the solution of a linear differential equation. But the most of dynamic objects and systems around us, which can perceive external stimuli and respond to changes in their states, are described by means of such equations under conditions of sufficiently small influences. Thus, the solution of polynomial equations is a part of the description of real objects and systems, which enables to study them and to build on this basis useful devices for managing them properly. Many results of the theme of nth roots of polynomials are used in many fields from physics and chemistry to economics and social science. One can hope that now on the base of matrix genetics the theme of *n*th roots of unity will be also useful for genetics.

One can mention here about group properties of the roots of unity. In the case of R-octetons the set of the 16 roots $\pm r_0, \pm r_1, \pm r_2, \pm r_3, \pm r_4, \pm r_5, \pm r_6, \pm r_7$ (Figure 7) forms a group in relation to multiplication. In this group the matrix $r_0$ is the identity matrix. Each of the 16 elements has its inverse element from this set (Figure 8): $\pm r_0 = \pm r_0^{-1}$; $\pm r_1 = \pm r_1^{-1}$; $r_2^{-1} = -r_2$; $r_3^{-1} = -r_3$; $\pm r_4^{-1} = \pm r_4$; $\pm r_5^{-1} = \pm r_5$; $\pm r_6^{-1} = \pm r_6$; $\pm r_7^{-1} = \pm r_7$. The condition of associativity holds true because matrix algebras are associative algebras.

In the case of R-octetons the set of the 16 roots $\pm h_0, \pm h_1, \pm h_2, \pm h_3, \pm h_4, \pm h_5, \pm h_6, \pm h_7$ (Figure 18) forms a group in relation to multiplication also. In this group the matrix $h_0$ is the identity matrix. Each of the 16 elements has its inverse element also (see Figure 19).

In the cases of R-genoquaternions (Figure 22) and H-genoquaternions (Figure 26) their sets of the 8 roots form two individual groups also.

Our results about 4$^{th}$ roots of unity generate some associations with known noise-immunity "quaternary codes" which are codes over an alphabet of four elements [Zinoviev, Sole, 2004; Nechaev, 1989; Hammons et al., 1994; Stepanov, 2006]. These quaternary codes use 4$^{th}$ roots of unity on the complex plane. Perhaps, a generalization of these quaternary codes on the base of the algebras of genetic octetons and genoquaternions will provide a new progress in bioinformatics and noise-immunity coding.

For further study of the genetic system, the mentioned groups of the 4$^{th}$ roots of unity can be analyzed from a viewpoint of theory of group representations in the nearest future.

## 17. Cyclic groups of the Hamilton frame quaternion and biquaternion and their applications to heritable biological phenomena

Let us return to the Hamilton quaternion and the Hamilton biquaternion with unit coordinates on Figure 29 but in their normalized forms with unit determinants (Figure 39):

$$H = 0.5 * \begin{bmatrix} 1 & 1 & -1 & 1 \\ -1 & 1 & 1 & 1 \\ 1 & -1 & 1 & 1 \\ -1 & -1 & -1 & 1 \end{bmatrix}; \quad Q = 2^{-1.5} * \begin{bmatrix} 1 & -1 & 1 & -1 & -1 & 1 & 1 & -1 \\ 1 & 1 & 1 & 1 & -1 & -1 & 1 & 1 \\ -1 & 1 & 1 & -1 & 1 & -1 & 1 & -1 \\ -1 & -1 & 1 & 1 & 1 & 1 & 1 & 1 \\ 1 & -1 & -1 & 1 & 1 & -1 & 1 & -1 \\ 1 & 1 & -1 & -1 & 1 & 1 & 1 & 1 \\ -1 & 1 & -1 & 1 & -1 & 1 & 1 & -1 \\ -1 & -1 & -1 & -1 & -1 & -1 & 1 & 1 \end{bmatrix}$$

Figure 39. The Hamilton quaternion H and the Hamilton biquaternion Q with unit determinants.

The expression $H^n$ represents the cyclic group whose period is equal to 6, that is $H^{n+6} = H^n$ (n = 1, 2, 3,…). Each member of this group is a Hamilton quaternion. This group, which is connected with genetic system, is connected also with some genetically inherited physiological phenomena.

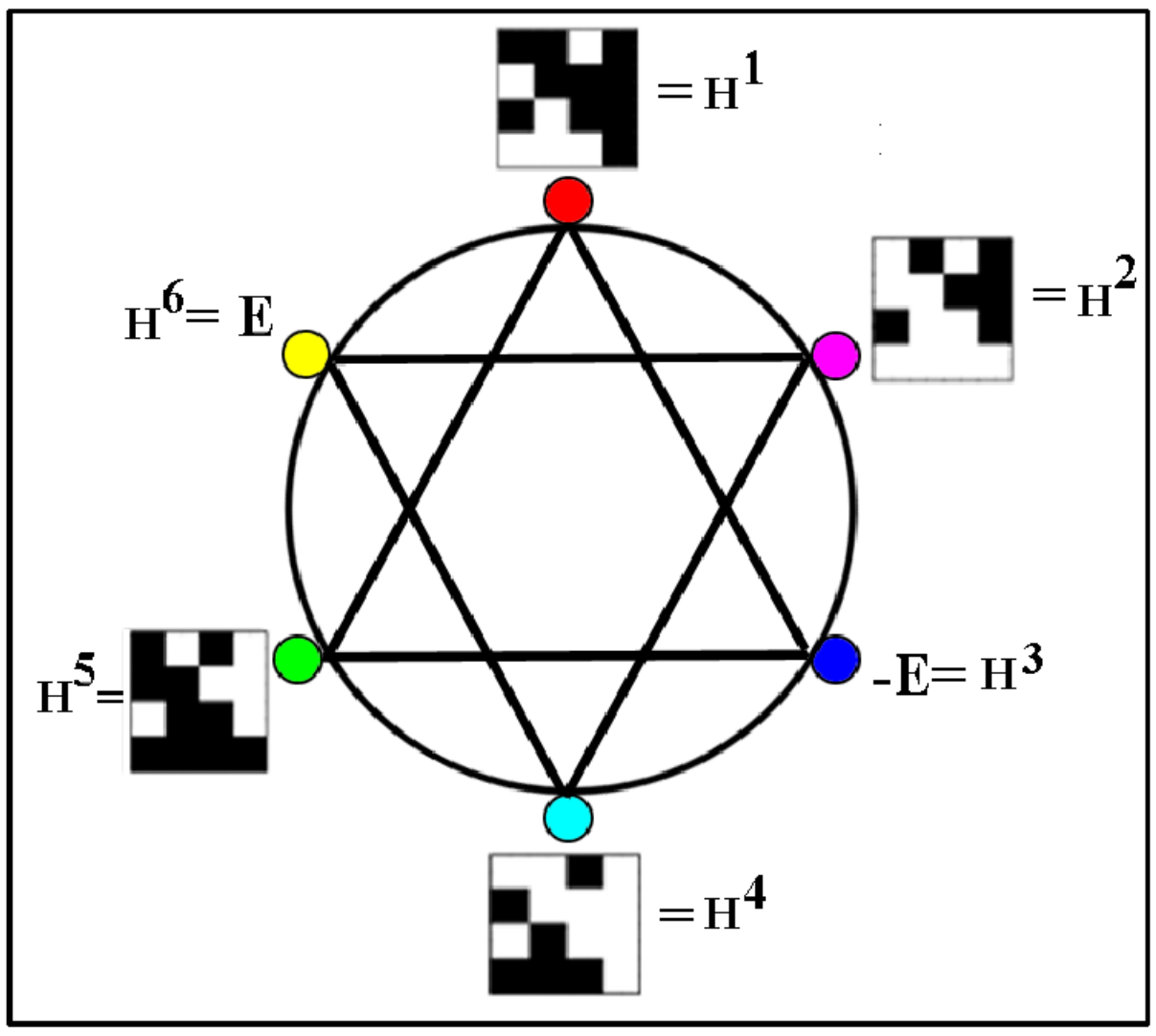

Figure 40. The circular disposition of the members $H^1$, $H^2$, $H^3$, $H^4$, $H^5$, $H^6$ of the cyclic group $H^n$ of the normalized quaternion by Hamilton. Here E is identity matrix. This "genetic" cyclic group posseses a deep analogy with the known circle of color perception, which represents one of genetically inherited phenomena.

The members $H^1$, $H^2$, $H^3$, $H^4$, $H^5$, $H^6$ of the cyclic group $H^n$ can be disposed on a circle (Figure 40) to represent in a clearer form the following properties of this group:
- any two diagonal quaternions on the circle differ from each other only in the opposite sign, that is their sum is equal to zero: $H^n = H^{n+3}$ (in other words, these quaternions are complementary);
- each quaternion of this cyclic group is equal to the sum of its adjacent quaternions on the circle: $H^k = H^{k-1} + H^{k+1}$;
- the sum of the three quaternions with odd degrees "n" is equal to zero (these three quaternions lie at the verticies of a triangle of the "star of David" on Figure 40) and the sum of the three quaternions with even degrees is equal to zero also (these three quaternions lie at the verticies of another triangle of the "star of David" on Figure 40).

But these properties of the cyclic group of the normalized Hamilton quaternion are similar of known properties of human color perception, which is one of genetically inherited physiological phenomena. Properties of color perception is traditionally represented in a form of the Newton's color circle [http://hyperphysics.phy-astr.gsu.edu/hbase/vision/newtcol.html] which is a convenient way to summarize the additive mixing properties of colors. The three colors - red, blue and green - are thought of as the primary colors, and their complementary (or secondary) colors - cyan, yellow and magenta correspondingly - are placed across from them on the circle (see Figure 40). The center of the circle is said to be "achromatic" or without color.

The known properties of color perception on the Newton's color circle are the following:
- complementary colors, which are opposite each other on this circle, neutralize each other in their summation;
- each color on the circle is the sum of two adjacent colors;
- the three primary colors neutralize each other in their summation; the three secondary colors neutralize each other in their summation also.

In accordance with Figure 40 one can assume that the red, magenta, blue, cyan, green and yellow colors are formally expressed by means of the quaternions $H^1$, $H^2$, $H^3$, $H^4$, $H^5$, $H^6$ correspondingly. In this case the problem of mixing of colors can now be solved in terms of the cyclic group of Hamilton quaternions $H^n$. For example, what color is obtained by mixing 3 shares of red, 2 shares of yellow and 5 shares of blue? The answer can be obtained by means of the summation of corresponding Hamilton quaternions: $3H^1 + 2H^6 + 5H^3 = 3H^2$ that is 3 shares of magenta.

It is known that colors are not physical properties of objects but they reflect a genetically inherited human reaction on light influences with taking into account a general light environment. For example a man in color glasses percepts color objects adequately. In other words, "color perception is not in the eye of the beholder: it's in the brain" [http://www.sciencedaily.com/releases/2005/10/051026082313.htm]. Why does our color perception is arranged in accordance with the Newton's color circle and with the cyclic group of the Hamilton quaternions? The possible answer is: because the human organism processes light information on the basis of heritable general algorithms of processing of genetic and physical information with a participation of the genetically inherited cyclic group of the Hamilton quaternions which represents essential aspects of the system of the genetic alphabets. The author believes that biological principles of processing of sensory information from physical surroundings are working in close similarity with the principles of storing and processing of genetic information: principles of processing sensory information have appeared not at an empty place but they are a continuation of principles of processing genetic information.

Now let us turn to the (8*8)-matrix representation of the Hamilton biquaternion Q on Figure 39. The expression $Q^n$ represents the cyclic group whose period is equal to 24, that is $Q^{n+24} = Q^n$ (n = 1, 2, 3,…). Each member of this cyclic group is a biquaternion. This "genetic" cyclic group can be used in algebraic modeling of heritable physiological cycles and biorhythms, which are connected with the physiological division of the diurnal cycle on 24 parts (or 24 hours) from the ancient time around the world. It should be explained in more details.

The statement that biological organisms exist in accordance with cyclic processes of environment and with their own cyclic physiological processes is one of the most classical statements of biology and medicine from ancient times (see for example (Dubrov, 1989; Wright, 2006)). Many branches of medicine take into account the time of day specially, when diagnostic, pharmacological and therapeutic actions should be made for individuals. The set of this medical and biological knowledge is usually united under names of chrono-medicine and chrono-biology. Many diseases are connected with disturbances of natural biological rhythms in organisms. The problem of internal clocks of organisms, which participate in coordination of all interrelated processes of any organism, is one of the main physiological problems.

Molecular biology deals with this problem of physiological rhythms and of cyclic re-combinations of molecular ensembles on the molecular level as well. Really, it is the well-known fact that in biological organisms proteins are disintegrated into amino acids and then they are re-built (are re-created) from amino acids again in a cyclic manner systematically. A half-life period (a duration of renovation of half of a set of molecules) for proteins of human organisms is approximately equal to 80 days in most cases; for proteins of the liver and blood plasma – 10 days; for the mucilaginous cover of bowels – 3-4 days; for insulin – 6-9 minutes (Aksenova, 1998, v. 2, p. 19). Such permanent rebuilding of proteins provides a permanent cyclic renovation of human organisms. Such cyclic processes at the molecular-genetic level are one of the parts of a hierarchical system of a huge number of interelated cycles in organisms. The phenomenon of repeated recombinations of molecular ensembles, which are carried out inside separate cycles, is one of the main problems of biological self-organization.

So our organism is a huge chorus of rythmic processes. Phases of activity and passivity of many physiological systems agree with each other; in accordance with ancient and modern medicine their coordination defines a traditional division of the diurnal cycle to 24 equal parts which are termed "24 hours" (see the figure about the circadian clock in [Wright, 2006]). If a person is placed in a dark cave for a long time, his idea about the length of the diurnal cycle may change, but the duration of the "new" diurnal cycle will be physiologically divided on 24 parts again. The bio-rythmic organization of heritable processes is distributed throughout the body of organisms of all kinds.

The cyclic "genetic" groups of biquaternions $Q^n$ can be used to model phenomena of 24-hour rythms of the diurnal cycle especially as biquaternions in physics are used to model spatio-temporal relations. Figure 41 shows a 24-hour dial where numbers correspond to the exponent of this biquaternion Q. The biquaternions $Q^n$ and $Q^{n+12}$, which are disposed opposite each other on this dial, differ only in the opposite sign ($Q^n = -Q^{n+12}$) by analogy with the opposite of daytime and night time hours (for example, in ordinary life, often say instead of "14 hours" simply "2 o'clock", calling the number which lies opposite on this circle).

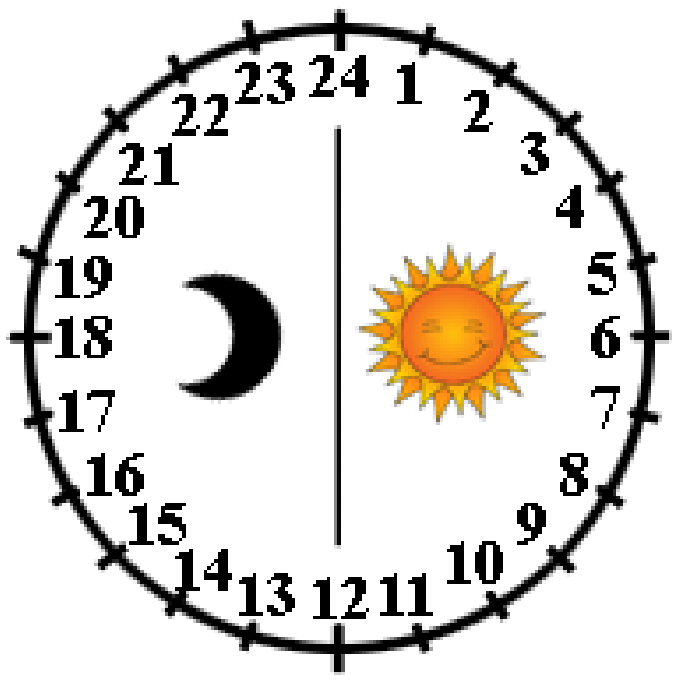

Figure 41. The 24-hour dial where numbers correspond to exponentiations of members of the cyclic group of biquaternions $Q^n$.

This cyclic group of biquaternions $Q^n$ possesses also other interestion properties, for example, $Q^n = Q^{n-4}+Q^{n+4}$; $Q^n = -(Q^{n-8}+Q^{n+8})$; $Q^n = 2^{-0.5}*(Q^{n-3}+Q^{n+3})$. One should mention that the expression $Q^{(24/k)*n}$ (where k and n are integer numbers) corresponds to the cyclic group whose period is equal to k. In the case of the Hamilton quaternion H from Figure 39, the same holds true for the expression $H^{(6/k)*n}$ which corresponds to the cyclic group whose period is equal to k. Various polynoms can be constructed on the base of these cyclic groups for different applications some of which are been developing in the author's laboratory.

A model of an ensemble of circadian bio-rythms (and a model of biological clock) can be proposed on the base of an ensemble of cyclic groups of biquaternions $Q^{(24/k)*n}$ (or cyclic groups of quaternions $H^{(6/k)*n}$). In such model, an ensemble of many circles exists which contains different pairs of circles with mutual agreed sizes; these circles are rolling over each other without sliding with opposite directions of rotation. This ensemble of many rotating circles returns to its initial state in a cyclic manner after certain quantities of tacts of time.

The author proposes additionally a special approach to generalized crystallography on the base of cyclic groups of hypercomplex numbers (first of all, the Hamilton quaternions, biquaternions and complex numbers). The speech is about "a principle of nodal points" (instead of the crystallographic principle of close packing of modules). These nodal points are disposed in n-dimensional space (in particular in 2-dimensional plane or in 3-dimensional space) on the base of ensembles of n-dimensional spheres where each sphere contains the so termed nodal points: the points with coordinates of members of the cyclic group (in a general case different spheres can contain the different nodal points). Adjacent spheres intersect each other in such way that some of their nodal points are coincident. In a result, a net of nodal points arises in such n-dimensional space. If initial cyclic groups and their nodal points on spheres are changed in some spheres, then the whole net of nodal points is changed. The proposed approach can be useful to model nanomaterial structures and their changes in different conditions and also to model quasi-crystalls by Shechtman.

## 18. About fractal patterns of genetic sequences on the base of the CGR-method and the dyadic-shift sub-alphabets of the genetic alphabets.

This section describes a receiving of fractal patterns of genetic sequences by means of a method Chaos Game Representation (CGR-method) on the base of dyadic-shift decompositions of the alphabet of 64 triplets into 4 sub-alphabets with 16 triplets in each. In the end of the section a special attention is paid to a comparative analysis of quantities of members of these sub-alphabets in genetic sequences.

The CGR-method is well-known in bioinformatics and molecular biology. This iterative mapping method allows converting a nucleotide sequence into a scale-independent and unique visual image. It was introduced in genome analysis in the work (Jeffrey, 1990). In this work J.Jeffrey proposed a visualization of a non-randomness character of DNA sequences by means of a chaos game algorithm for four points. Such approach permits the representation and investigation of patterns in sequences, visually revealing previously unknown structures. Based on a technique from chaotic dynamics, the method produces a picture of a gene sequence which displays both local and global patterns. The pictures have a complex structure which varies depending on the sequence. CGR raises a new set of questions about the structure of DNA sequences, and is a new tool for investigating gene structure. We should reproduce some materials about CGR from the work by Jeffrey because we will describe a new variant of this method for investigations of genetic sequences of DNA on the basis of the mentioned dyadic-shift sub-alphabets.

During the past 15 years a new field of physics has developed, known as 'non-linear dynamics', 'chaotic dynamical systems', or simply 'chaos' (Barnsley, 1988; Devaney, 1989). Central to much of the field are questions of the structure of certain complex curves known as 'fractals'. The Chaos Game is an algorithm which allows one to produce pictures of fractal structures, using a computer. In simplest form, it proceeds as follows:

1. Locate three dots on a piece of paper. They can be anywhere, as long as they are not all on a line. We will call these dots vertices.
2. Label one vertex with the numerals 1 and 2, one of the others with the numerals 3 and 4, and the third with the numerals 5 and 6.
3. Pick a point anywhere on the paper, and mark it. This is the initial point.
4. Roll a 6-sided die. Since in Step 2 the vertices were labelled, the number that comes up on the die is a label on a vertex. Thus, the number rolled on the die picks out a vertex. On the paper, place a mark half way between the previous point and the indicated vertex. (The first time the die is rolled, the 'previous point' is the initial point picked in Step 3.) For example, if 3 is rolled, place a mark on the paper half way between the previous point and the vertex labelled '3'.
5. Continue to roll the die, on each roll marking the paper at the point half way between the previous point and the indicated vertex.

One might expect that this procedure, if repeated many times, would yield a paper covered with random dots or, perhaps, a triangle filled with random dots. Such is not the case. In fact, if the Chaos Game is run for several thousand points, the result is a beautiful fractal figure which is known in mathematics for many years under the name 'Sierpinski triangle', after the mathematician who first defined it. For cases of five points, six, or seven initial points the chaos game produces a figure with visible patterns (pentagons within pentagons, a striated hexagon, or heptagons within heptagons), but for eight or more point the game yields essentially a filled-in polygon, except that the center is empty.

With four initial points, however, the result is different. It is not squares within squares, as one might expect; in fact there is no pattern at all. The chaos game on four points produces a square uniformly and randomly filled with dots in the case of random rolling of a 6-sided die. The picture produced by the chaos game is known as the attractor. Mathematically, the chaos game is described by an iterated function system (IFS).

If a sequence of numbers is used to produce an attractor for an IFS code and that attractor has visually observable then we have, intuitively, revealed some underlying structure in the sequence of numbers. Experiments (Jeffrey, 1990) had shown that the Chaos Game can be used to display certain kinds of non-randomness visually. This led to the following question. Since a genetic sequence can be treated formally as a string composed from the four letters 'a', 'c', 'g', and 't' (or 'u'), what patterns arise if we use for Chaos Game not a series of random numbers but DNA sequences in their relevant numeric form? Instead of 'rolling a 4-sided die', use the next base (a, c, g, t/u) to pick the next point. Each of the four corners of the square is labelled by symbols 'a', 'c', 'g', or 'u/t'. If a 'c', for example, is the next base, then a point is plotted half way between the previous point and the 'c' corner. In such way Jeffrey has receive a set of CGR patterns of different DNA sequences. These patterns had fractal characters and they were used for many tasks of comparative analysis in bioinformatics and for formulating new scientific questions.

After Jeffrey many scientists studied CGR patterns of nucleotide sequences intensively. In the result, "*alignment free methods based on Chaos Game Representation (CGR), also known as sequence signature approaches, have proven of great interest for DNA sequence analysis. Indeed, they have been successfully applied for sequence comparison, phylogeny, detection of horizontal transfers or extraction of representative motifs in regulation sequence*s" (Deschavanne, Tuffey, 2008, p. 615).

Previous sections of our article described connections of the genetic system with the hypercomplex numerical systems (Hamilton quaternions and biquaternions, Cockle split-quaternions and bi-split-quaternions, complex and double numbers) which were revealed by means of dyadic-shift decompositions of the genetic matrices in their Rademacher and Hadamard representations. Let us return to the genetic (4*4)-matrix $[C\ U;\ A\ G]^{(2)}$ which contains 16 duplets (Figures 28 and 42) and whose DS-decomposition contains the 4 sparse matrices $H_0$, $H_1$, $H_2$, $H_3$ (Figure 42). The four $H_k$-sub-alphabets of the alphabet of 64 triplets exist on the base of the following principle: each of $H_k$-sub-alphabets contains 16 triplets whose roots belong to the appropriate sparse matrix $H_k$ (Figure 42). These $H_k$-sub-alphabets can be termed "sub-alphabets of duplets" because of the main role of the duplets in their definition. Taking into acount that the Hadamard representation of the (4*4)-matrix $[C\ U;\ A\ G]^{(2)}$ is the matrix representation of the Hamilton quaternion with unit coordinates (Figure 29), one can also term these $H_k$-sub-alphabets conditionally "sub-alphabets of the quaternion type" (below we will consider also $B_s$-sub-alphabets of a biquaternion type). Each of the $H_k$-sub-alphabets contains only those triplets whose roots (that is the first two positions) have the same binary dyadic-shift numeration. These numerations are defined by means of the replacement of each letter in the first two positions of each triplet by its dyadic-shift numeration from Figure 4: C=G=0 and A=U/T=1. For example the triplet CAG has its two first positions with their dyadic-shift numeration 01. Through this numerical criterion one can easily identify to which sub-alphabet any triplet belongs.

$$[C\ U;\ A\ G]^{(2)} = \begin{bmatrix} CC & CU & UC & UU \\ CA & CG & UA & UG \\ AC & AU & GC & GU \\ AA & AG & GA & GG \end{bmatrix} = \begin{bmatrix} CC & 0 & 0 & 0 \\ 0 & CG & 0 & 0 \\ 0 & 0 & GC & 0 \\ 0 & 0 & 0 & GG \end{bmatrix} + \begin{bmatrix} 0 & CU & 0 & 0 \\ CA & 0 & 0 & 0 \\ 0 & 0 & 0 & GU \\ 0 & 0 & GA & 0 \end{bmatrix} + \begin{bmatrix} 0 & 0 & UC & 0 \\ 0 & 0 & 0 & UG \\ AC & 0 & 0 & 0 \\ 0 & AG & 0 & 0 \end{bmatrix} + \begin{bmatrix} 0 & 0 & 0 & UU \\ 0 & 0 & UA & 0 \\ 0 & AU & 0 & 0 \\ AA & 0 & 0 & 0 \end{bmatrix}$$

| | |
|---|---|
| $H_0$-sub-alphabet (binary DS-numeration of each root in these triplets is 00) | **CC**C, **CC**U, **CC**A, **CC**G, **CG**C, **CG**U, **CG**A, **CG**G, **GC**C, **GC**U, **GC**A, **GC**G, **GG**C, **GG**U, **GG**A, **GG**G |
| $H_1$-sub-alphabet (binary DS-numeration of each root of triplets is 01) | **CU**C, **CU**U, **CU**A, **CU**G, **CA**C, **CA**U, **CA**A, **CA**G, **GU**C, **GU**U, **GU**A, **GU**G, **GA**C, **GA**U, **GA**A, **GA**G |
| $H_2$-sub-alphabet (binary DS-numeration of each root in these triplets is 10) | **UC**C, **UC**U, **UC**A, **UC**G, **UG**C, **UG**U, **UG**A, **UG**G, **AC**C, **AC**U, **AC**A, **AC**G, **AG**C, **AG**U, **AG**A, **AG**G |
| $H_3$-sub-alphabet (binary DS-numeration of each root in these triplets is 11) | **UU**C, **UU**U, **UU**A, **UU**G, **UA**C, **UA**U,**UA**A, **UA**G, **AU**C, **AU**U, **AU**A, **AU**G, **AA**C, **AA**U, **AA**A, **AA**G |

Figure 42. Upper row: the DS-decomposition of the genetic matrix [C U; A G]$^{(2)}$ in the 4 sparse matrices: [C U; A G]$^{(2)}$ = $H_0$+$H_1$ +$H_2$+$H_3$. Bottom: the table of the four $H_k$-sub-alphabets of the alphabet of 64 triplets, where each of the $H_i$-sub-alphabets contains only triplets with those roots which belong to the appropriate sparse matrix $H_k$ (k = 0, 1, 2, 3). Roots of triplets are marked specially. The binary DS-numeration of roots of triplets is defined by means of the replacing of each of the letters by its DS-numeration from Figure 4: C=G=0, A=U=1.

Each of the $H_k$-sub-alphabets contains 8 pairs of “codon-anticodon”. Now each of symbolic genetic sequences of triplets can be transformed in numeric sequence of the four binary numbers 00, 01, 10, 11 (or their analoques in decimal notation 0, 1, 2, 3 correspondingly) if each of triplets is replaced by its binary number from the table on Figure 42. For example, the sequence UUC, ACG, AAG, CUC GGC is transformed in the sequence 3, 2, 3, 1, 0 (in decimal notation). If one has such numeric representation of a long genetic sequence of triplets, then CGR-method can be applied to receive a CGR-fractal pattern for this sequence. The first point of a future CGR pattern is placed half way between the center of the square and the vertex corresponding to the first member of a considered numeric sequence; its i-th point is then placed half way between the (i-1)-th point and the vertex corresponding to the i-th letter.

Figure 43 shows an example of CGR-fractal patterns which was received in such way by the author for the qenetic sequence of Homo sapiens BAC clone CH17-186K1 (http://www.ncbi.nlm.nih.gov/nuccore/AC239584.4) which contains 62974 triplets.

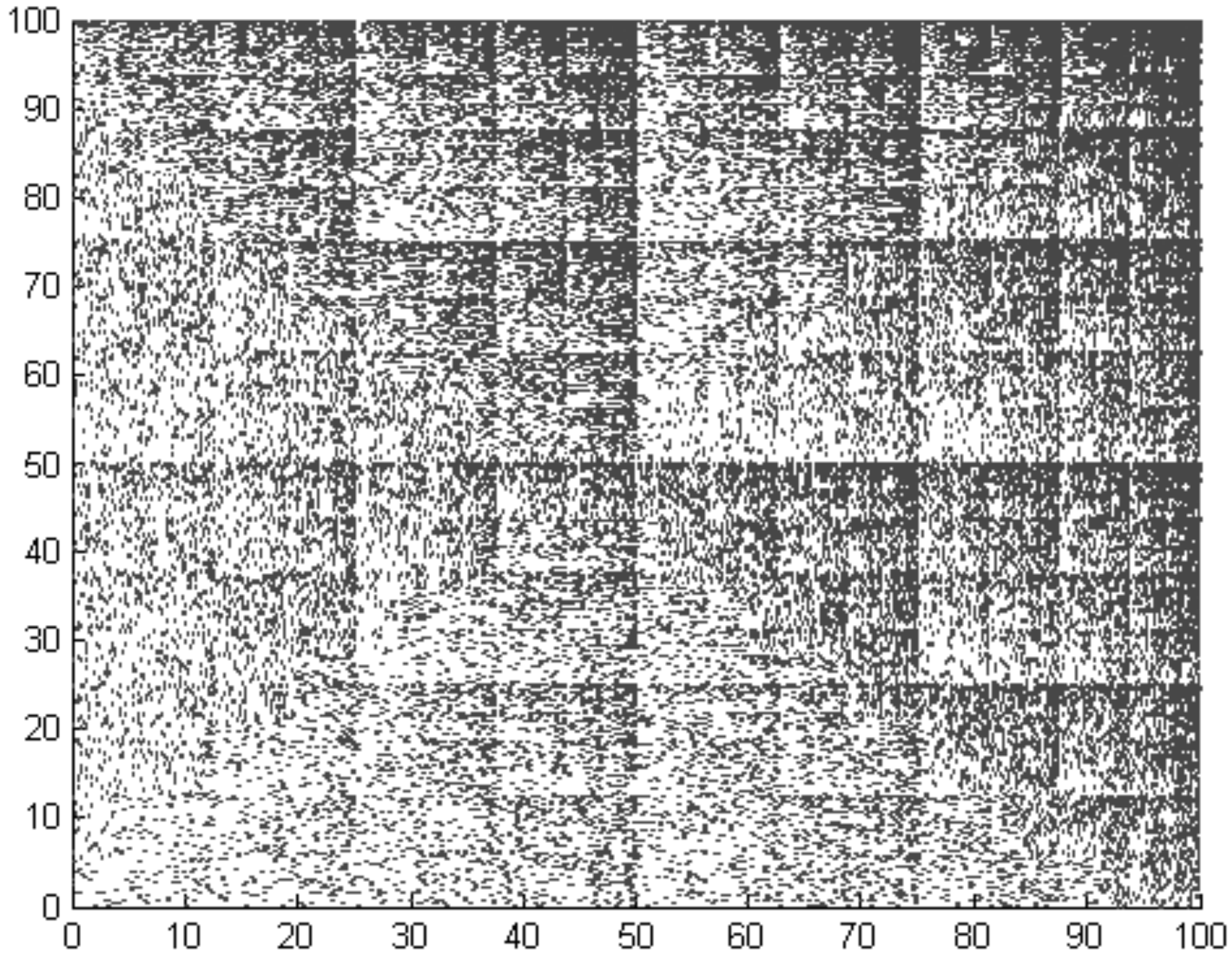

Figure 43. The CGR-fractal pattern of the sequence of Homo sapiens BAC clone CH17-186K1 (http://www.ncbi.nlm.nih.gov/nuccore/AC239584.4) on the base of the four $H_k$-sub-alphabets of the quaternion type from Figure 42.

The produced CGR pattern has a fractal structure or a self-similarity character (Figure 43). Really one can see that the CGR pattern of the whole square is reproduced in each quadrant of the square, and in each subquadrant, etc. Identifying chaos in experimental data such as biological sequences, a researcher can include searching for a strange attractor in the special dynamics, identified by its fractal structure. Having found such an attractor, one can try to estimate its dimension, which is a measure of the number of active variables and hence the complexity of the equations required to model the dynamics. Fractals are to chaos what geometry is to algebra. They are the useful geometric manifestation of the chaotic dynamics. They are called "the fingerprints of chaos" sometimes. The revealing the fractal structure of such CGR patterns of genetic sequences shows hidden connections of genetic structures with non-linear dynamics or chaotic dynamical systems as the new field of physics. In our case it reveals also some relations between the genetic DS-algebras and non-linear dynamics. The disclosure of fractals which are connected with genetic sequences in the described variant of the CGR method, allows using this new knowledge in different branches of bioinformatics, molecular and evolutionary biology by analogy with many known applications of CGR techniques made by different authors before. For example, fractal properties of CGR images such as fractal dimension and multifractal spectra are the tools for genomic sequences comparison and phylogeny studies.

Now let us return to the symbolic (8*8)-matrix [C U; A G]$^{(3)}$. Figure 44 shows its DS-decomposition in the eight sparse matrices $B_0$, $B_1$, $B_2$, $B_3$, $B_4$, $B_5$, $B_6$, $B_7$.

**[C U; A G]$^{(3)}$ = $B_0+B_1+B_2+B_3+B_4+B_5+B_6+B_7$ =**

| | | | | | | | |
|---|---|---|---|---|---|---|---|
| CCC | CCU | CUC | CUU | UCC | UCU | UUC | UUU |
| CCA | CCG | CUA | CUG | UCA | UCG | UUA | UUG |
| CAC | CAU | CGC | CGU | UAC | UAU | UGC | UGU |
| CAA | CAG | CGA | CGG | UAA | UAG | UGA | UGG |
| ACC | ACU | AUC | AUU | GCC | GCU | GUC | GUU |
| ACA | ACG | AUA | AUG | GCA | GCG | GUA | GUG |
| AAC | AAU | AGC | AGU | GAC | GAU | GGC | GGU |
| AAA | AAG | AGA | AGG | GAA | GAG | GGA | GGG |

=

| | | | | | | | |
|---|---|---|---|---|---|---|---|
| CCC | 0 | 0 | 0 | 0 | 0 | 0 | 0 |
| 0 | CCG | 0 | 0 | 0 | 0 | 0 | 0 |
| 0 | 0 | CGC | 0 | 0 | 0 | 0 | 0 |
| 0 | 0 | 0 | CGG | 0 | 0 | 0 | 0 |
| 0 | 0 | 0 | 0 | GCC | 0 | 0 | 0 |
| 0 | 0 | 0 | 0 | 0 | GCG | 0 | 0 |
| 0 | 0 | 0 | 0 | 0 | 0 | GGC | 0 |
| 0 | 0 | 0 | 0 | 0 | 0 | 0 | GGG |

+

| | | | | | | | |
|---|---|---|---|---|---|---|---|
| 0 | CCU | 0 | 0 | 0 | 0 | 0 | 0 |
| CCA | 0 | 0 | 0 | 0 | 0 | 0 | 0 |
| 0 | 0 | 0 | CGU | 0 | 0 | 0 | 0 |
| 0 | 0 | CGA | 0 | 0 | 0 | 0 | 0 |
| 0 | 0 | 0 | 0 | 0 | GCU | 0 | 0 |
| 0 | 0 | 0 | 0 | GCA | 0 | 0 | 0 |
| 0 | 0 | 0 | 0 | 0 | 0 | 0 | GGU |
| 0 | 0 | 0 | 0 | 0 | 0 | GGA | 0 |

+

| | | | | | | | |
|---|---|---|---|---|---|---|---|
| 0 | 0 | CUC | 0 | 0 | 0 | 0 | 0 |
| 0 | 0 | 0 | CUG | 0 | 0 | 0 | 0 |
| CAC | 0 | 0 | 0 | 0 | 0 | 0 | 0 |
| 0 | CAG | 0 | 0 | 0 | 0 | 0 | 0 |
| 0 | 0 | 0 | 0 | 0 | 0 | GUC | 0 |
| 0 | 0 | 0 | 0 | 0 | 0 | 0 | GUG |
| 0 | 0 | 0 | 0 | GAC | 0 | 0 | 0 |
| 0 | 0 | 0 | 0 | 0 | GAG | 0 | 0 |

+

| | | | | | | | |
|---|---|---|---|---|---|---|---|
| 0 | 0 | 0 | CUU | 0 | 0 | 0 | 0 |
| 0 | 0 | CUA | 0 | 0 | 0 | 0 | 0 |
| 0 | CAU | 0 | 0 | 0 | 0 | 0 | 0 |
| CAA | 0 | 0 | 0 | 0 | 0 | 0 | 0 |
| 0 | 0 | 0 | 0 | 0 | 0 | 0 | GUU |
| 0 | 0 | 0 | 0 | 0 | 0 | GUA | 0 |
| 0 | 0 | 0 | 0 | 0 | GAU | 0 | 0 |
| 0 | 0 | 0 | 0 | GAA | 0 | 0 | 0 |

+

| | | | | | | | |
|---|---|---|---|---|---|---|---|
| 0 | 0 | 0 | 0 | UCC | 0 | 0 | 0 |
| 0 | 0 | 0 | 0 | 0 | UCG | 0 | 0 |
| 0 | 0 | 0 | 0 | 0 | 0 | UGC | 0 |
| 0 | 0 | 0 | 0 | 0 | 0 | 0 | UGG |
| ACC | 0 | 0 | 0 | 0 | 0 | 0 | 0 |
| 0 | ACG | 0 | 0 | 0 | 0 | 0 | 0 |
| 0 | 0 | AGC | 0 | 0 | 0 | 0 | 0 |
| 0 | 0 | 0 | AGG | 0000 | | | |

+

| | | | | | | | |
|---|---|---|---|---|---|---|---|
| 0 | 0 | 0 | 0 | 0 | UCU | 0 | 0 |
| 0 | 0 | 0 | 0 | UCA | 0 | 0 | 0 |
| 0 | 0 | 0 | 0 | 0 | 0 | 0 | UGU |
| 0 | 0 | 0 | 0 | 0 | 0 | UGA | 0 |
| 0 | ACU | 0 | 0 | 0 | 0 | 0 | 0 |
| ACA | 0 | 0 | 0 | 0 | 0 | 0 | 0 |
| 0 | 0 | 0 | AGU | 0 | 0 | 0 | 0 |
| 0 | 0 | AGA | 0 | 0 | 0 | 0 | 0 |

+

| | | | | | | | |
|---|---|---|---|---|---|---|---|
| 0 | 0 | 0 | 0 | 0 | 0 | UUC | 0 |
| 0 | 0 | 0 | 0 | 0 | 0 | 0 | UUG |
| 0 | 0 | 0 | 0 | UAC | 0 | 0 | 0 |
| 0 | 0 | 0 | 0 | 0 | UAG | 0 | 0 |
| 0 | 0 | AUC | 0 | 0 | 0 | 0 | 0 |
| 0 | 0 | 0 | AUG | 0 | 0 | 0 | 0 |
| AAC | 0 | 0 | 0 | 0 | 0 | 0 | 0 |
| 0 | AAG | 0 | 0 | 0 | 0 | 0 | 0 |

+

| | | | | | | | |
|---|---|---|---|---|---|---|---|
| 0 | 0 | 0 | 0 | 0 | 0 | 0 | UUU |
| 0 | 0 | 0 | 0 | 0 | 0 | UUA | 0 |
| 0 | 0 | 0 | 0 | 0 | UAU | 0 | 0 |
| 0 | 0 | 0 | 0 | UAA | 0 | 0 | 0 |
| 0 | 0 | 0 | AUU | 0 | 0 | 0 | 0 |
| 0 | 0 | AUA | 0 | 0 | 0 | 0 | 0 |
| 0 | AAU | 0 | 0 | 0 | 0 | 0 | 0 |
| AAA | 0 | 0 | 0 | 0 | 0 | 0 | 0 |

Figure 44. The dyadic-shift decomposition of the genetic matrix $[C\ U;\ A\ G]^{(3)}$.

The Hadamard representation of the genomatrix $[C\ U;\ A\ G]^{(3)}$ have led us to the Hamilton biquaternion $Q = q_0+q_1+q_2+q_3+q_4+q_5+q_6+q_7$ with the unit coordinates (see Figures 28-30). It is known that Hamilton biquaternions are Hamilton quaternions over the complex number field. The Hamilton biquaternion with unit coordinates can be represented in the following form (taking into account the expression (17) and the expressions $i=q_4$, $q_0*i = q_4$, $q_1*i=q_5$, $q_2*i=q_6$, $q_3*i=q_7$ from the multiplication table on Figure 30):

$$Q = q_0*(1+i) + q_1*(1+i) + q_2*(1+i) + q_3*(1+i) = (q_0+q_4) + (q_1+q_5) + (q_2+q_6) + (q_3+q_7) \qquad (23)$$

The expression (23) shows the natural division of the biquaternion Q into the four parts $(q_0+q_4)$, $(q_1+q_5)$, $(q_2+q_6)$ and $(q_3+q_7)$. By analogy one can represent the genomatrix $[C\ U;\ A\ G]^{(3)}$ of the 64 triplets as the sum of the four pairs (24) from Figure 44:

$$[C\ U;\ A\ G]^{(3)} = (B_0+B_4)+(B_1+B_5)+(B_2+B_6)+(B_3+B_7) \qquad (24)$$

So the four sub-alphabets $(B_0+B_4)=B_{04}$, $(B_1+B_5)=B_{15}$, $(B_2+B_6)=B_{26}$, $(B_3+B_7)=B_{37}$ of the alphabet of 64 triplets arise. Each of the matrices $B_{04}$, $B_{15}$, $B_{26}$, $B_{37}$ possesses the identical dispositions of its non-zero entries in all of its four sub-quadrants. Each of these sub-alphabets contains 8 pairs of "codon-anticodon". These sub-alphabets $B_{04}$, $B_{15}$, $B_{26}$, $B_{37}$ can conditionally be termed sub-alphabets of the biquaternion type. Figure 45 shows these four sub-alphabets, each of which contains 16 triplets, all of which possess two last positions with the same binary dyadic-shift numeration. These numerations are defined by means of the replacement of each letter in the two last positions of each triplet by its dyadic-shift numeration from Figure 4: C=G=0 and A=U/T=1. For example the triplet CAG has its two last positions with their dyadic-shift numeration 10. Through this numerical criterion one can easily identify to which sub-alphabet any triplet belongs.

| | |
|---|---|
| $B_{04}$-sub-alphabet (the binary DS-numeration of two last positions in each triplet is 00) | CCC, CCG, CGC, CGG, GCC, GCG, GGC, GGG, UCC, UCG, UGC, UGG, ACC, ACG, AGC, AGG |
| $B_{15}$-sub-alphabet (the binary DS-numeration of two last positions in each triplet is 01) | CCU, CCA, CGU, CGA, GCU, GCA, GGU, GGA, UCU, UCA, UGU, UGA, ACU, ACA, AGU, AGA |
| $B_{26}$-sub-alphabet (the binary DS-numeration of two last positions in each triplet is 10) | CUC, CUG, CAC, CAG, GUC, GUG, GAC, GAG, UUC, UUG, UAC, UAG, AUC, AUG, AAC, AAG |
| $B_{37}$-sub-alphabet (the binary DS-numeration of two last positions in each triplet is 11) | CUU, CUA, CAU, CAA, GUU, GUA, GAU, GAA, UUU, UUA, UAU, UAA, AUU, AUA, AAU, AAA |

Figure 45. The four sub-alphabets of the biquaternion type (explanations in the text).

So we have received the new set of four sub-alphabets of triplets to construct CGR-patterns of long genetic sequences. One should emphasized that each of the sub-alphabets of the biquaternion type (Figure 45) differs essentially from each of the sub-alphabets of the quaternion type (Figure 42). To construct CGR-patterns of sequences of triplets one can mark each of triplets of the $B_{04}$-sub-alphabet by number 0, each of triplets of the $B_{15}$-sub-alphabet - by number 1, each of triplets of the $B_{26}$-sub-alphabet - by number 2 and each of triplets of the $B_{37}$-sub-alphabet - by number 3. In this case any symbolic sequence of triplets is transformed into a relevant sequence of numbers 0, 1, 2, 3.

Figure 46 shows the CGR-pattern which was received by the author for the same sequence of Homo sapiens BAC clone CH17-186K1 (http://www.ncbi.nlm.nih.gov/nuccore/AC239584.4) but now on the base of the four sub-alphabets of the biquaternion type from Figure 45. This pattern has again a fractal structure or a self-similarity character because the CGR pattern of the whole square is reproduced in each quadrant of the square, and in each subquadrant, etc. One can see that this new CGR-fractal is very similar to the CGR-fractal obtained early on the basis of the $H_k$-sub-alphabets of the quaternion type (Figure 43).

Now our laboratory investigates CGR-fractal patterns of different genetic sequences on the base of the sub-alphabets of quaternion and biquaternion types. CGR-fractal patterns indicate non-random but law-governed nature of the organization of genetic sequences. The similarity of CGR-fractals on Figures 43 and 46 allows putting forward a hypothesis that the organization of genetic sequences is connected with Hamilton quaternions and biquaternions.

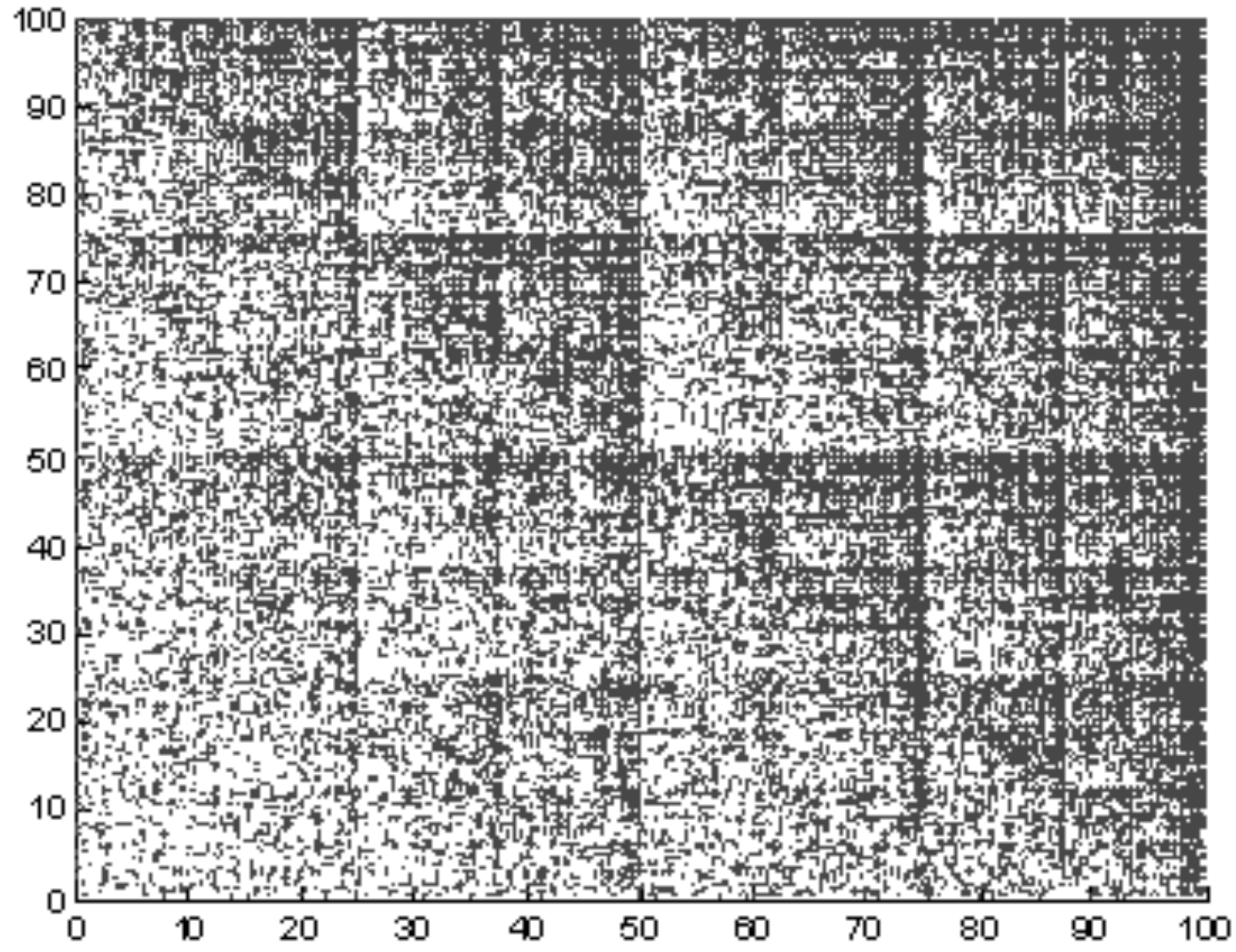

Figure 46. The CGR-fractal pattern of the sequence of Homo sapiens BAC clone CH17-186K1 (http://www.ncbi.nlm.nih.gov/nuccore/AC239584.4) on the base of the four sub-alphabets of the biquaternion type from Figure 45.

To check this hypothesis, the author makes a comparative analysis of quantities of representatives of the different sub-alphabets of quaternion and biquaternion types in various genetic sequences. Figure 47 shows some initial results of such comparative analysis for a set of long genetic sequences that was chosen randomly. More precisely the author has used a possibility of a choice of nucleotide sequences of arbitrary lengths on the site of http://www.ncbi.nlm.nih.gov/sites/entrez. On this site the author has asked the search of sequences of nucleotides with lenghts between 200000 and 200010 by means of the following typical instruction: 200000:200010[SLEN]. In the result the site has shown 54 variants of sequences. The most of these variants include gaps inside sequences or they possess a fragmentary character and, by this reason, they were uncomfortable for the initial testing. In this situation the author has chosen only those 11 variants of the sequences which were devoid such defects.

Each of these 11 symbolic sequences of nucleotides was transformed into an appropriate sequence of triplets and then was transformed into two different sequences of the four numbers 0, 1, 2 and 3 as we did it above for sequences of nucleotides to construct CGR-patterns. The first numeric sequence is based on the four sub-alphabets of the quaternion type from Figure 42; one can conditionally term this sequence "the quaternion representation of the genetic sequence". The second numeric sequence is based on the four sub-alphabets of the biquaternion type from Figure 45; one can conditionally term this sequence "the biquaternion representation of the genetic sequence". Figure 47 shows quantities of numbers 0, 1, 2 and 3 separately inside the quaternion representation and inside the biquaternion representation for the mentioned 11 long sequences.

| **Title of the sequence and its number of nucleotides (in brackets)** | **The quaternion and biquaternion representations of the sequence of triplets** | **Quantity of numbers 0** | **Quantity of numbers 1** | **Quantity of numbers 2** | **Quantity of numbers 3** |
|---|---|---|---|---|---|
| 1. Arabidopsis thaliana DNA chromosome 4, contig fragment No. 13 (200001 bp) GenBank: AL161501.2 | in the quaternion representation: | 7297 | 15597 | 15598 | 28175 |
| | in the biquaternion representation: | 7261 **Δ=-0,49%** | 15634 **Δ=0,24%** | 15431 **Δ=-1,07%** | 28341 **Δ= 0,59%** |
| 2.Arabidopsis thaliana DNA chromosome 4, contig fragment No. 71 (200001 bp) GenBank: AL161575.2 | in the quaternion representation: | 8181 | 16162 | 16198 | 26126 |
| | in the biquaternion representation: | 8196 **Δ= 0,18%** | 16183 **Δ=0,13%** | 16410 **Δ= 1,31%** | 25878 **Δ= -0,95%** |
| 3.Glycine max strain Williams 82 clone GM_WBb0017A17, complete sequence (200003 bp) GenBank: AC235203.1 | in the quaternion representation: | 10725 | 15991 | 16315 | 23636 |
| | in the biquaternion representation: | 10750 **Δ= 0,23%** | 16290 **Δ=0,87%** | 16114 **Δ=-1,23%** | 23513 **Δ= -0,52%** |
| 4.Glycine max strain Williams 82 clone GM_WBb0077L15, complete sequence (200010 bp) | in the quaternion representation: | 12152 | 16867 | 16282 | 21369 |
| | in the biquaternion representation: | 12075 **Δ=-0,63%** | 16359 **Δ=-3,01%** | 16346 **Δ= 0,39%** | 21890 **Δ= 2,44%** |
| GenBank: AC235320.1 | | | | | |
| 5.Homo sapiens DNA, immunoglobulin heavy-chain variable region, complete sequence, 3 of 5 (200000 bp) GenBank: AB019439.1 | in the quaternion representation: | 10223 | 17285 | 17480 | 21678 |
| | in the biquaternion representation: | 10164<br>**Δ=-0,58%** | 17539<br>**Δ=1,47%** | 17256<br>**Δ=-1,28%** | 21707<br>**Δ=0,13%** |
| 6.Homo sapiens DNA, immunoglobulin heavy-chain variable region, complete sequence, 1 of 5 (200000 bp) GenBank: AB019437.1 | in the quaternion representation: | 10108 | 17311 | 17447 | 21800 |
| | in the biquaternion representation: | 10388<br>**Δ=2,77%** | 17167<br>**Δ=-0,83%** | 17461<br>**Δ=0,08%** | 21650<br>**Δ=-0,69%** |
| 7.Homo sapiens DNA, immunoglobulin heavy-chain variable region, complete sequence, 4 of 5, (200000 bp) GenBank: AB019440.1 | in the quaternion representation: | 9799 | 17308 | 17312 | 22247 |
| | in the biquaternion representation: | 9670<br>**Δ=-1,37%** | 17441<br>**Δ=0,77%** | 17266<br>**Δ=-0,27%** | 22289<br>**Δ=0,19%** |
| 8. Mus musculus BAC clone RP23-449L24 from chromosome 12, complete sequence (200004 bp). GenBank: AC122367.4 | in the quaternion representation: | 9993 | 17645 | 17351 | 21679 |
| | in the biquaternion representation: | 10024<br>**Δ=0,31%** | 17320<br>**Δ=-1,84%** | 17511<br>**Δ=0,92%** | 21813<br>**Δ=0,62%** |
| 9.Mustela putorius furo contig100519, whole genome shotgun sequence (200008 bp). GenBank: AEYP01100519.1 | in the quaternion representation: | 12061 | 17632 | 17493 | 19483 |
| | in the biquaternion representation: | 11946<br>**Δ=-0,95%** | 17608<br>**Δ=-0,17%** | 17504<br>**Δ=0,06%** | 19611<br>**Δ=0,66%** |
| 10.Neofelis nebulosa clone CH87-108K8, working draft sequence, 10 ordered pieces (200004 bp) GenBank: AC225874.2 | in the quaternion representation: | 9322 | 16545 | 16768 | 23742 |
| | in the biquaternion representation: | 9472<br>**Δ= 1,61%** | 16618<br>**Δ=0,44%** | 16739<br>**Δ=-0,17%** | 23548<br>**Δ=-0,82%** |
| 11. Pongo abelii BAC clone CH276-516L3 from chromosome 21, complete sequence (200001 bp). GenBank: AC197588.3 | in the quaternion representation: | 12483 | 17162 | 17576 | 19446 |
| | in the biquaternion representation: | 12526<br>**Δ=0,34%** | 17533<br>**Δ=2,16%** | 17410<br>**Δ=-0,94%** | 19198<br>**Δ=-1,27%** |

Figure 47. The initial data for a comparative analysis of the quaternion representation and the biquaternion representation of long sequences of triplets in a form of appropriate sequences of the four numbers 0, 1, 2 and 3. The symbol Δ means the percentage difference of quantities of the numbers 0, 1, 2 and 3 in the biquaternion representation in comparison with their quantities in the quaternion representation where these quantities are taken as 100%.

The data on Figure 47 show that in the numerical relation the quaternion representation and the biquaternion representation of all of these long sequences are practically identical to each other: in the most cases the difference Δ of their quantities is equal to 1-2% approximately or much lesser. In addition, in each of these sequences the following regularities exist in the both representations:

1) the quantity of numbers 3 is more than quantities of each of other numbers;
2) the quantity of numbers 0 is less than quantities of each of other numbers;
3) the quantities of numbers 1 and 2 are approximately equal to each other;
4) the total quantities of numbers 1 and 2 is approximately equal to the total quantities of numbers 0 and 3.

The initial analysis of short sequences has revealed examples where their quaternion representation and the biquaternion representation differ each from another in numeric relations significantly. At this initial stage of investigation it seems that the longer the genetic sequences the more these regularities reveal themselfs. This situation resembles the situation with attractors in the theory of dynamical chaos [http://en.wikipedia.org/wiki/Attractor].

Our initial data allows putting forward the working hypothesis about an existence of the following phenomenological rule.

**The rule** of the quaternion and biquaternion representations of long genetic sequences:

- In long sequences their quaternion and biquaternion representations are similar to each other in numeric relations; the longer the sequences the more this numeric similarity reveals itself.

Of course, this hypothesis should be tested in a great number of genetic sequences and for a variety of taxonomic groups. Now relevant works are conducted in our laboratory.

It should be emphasized that, for working with these numeric representations of genetic sequences, biologists do not need to know the mathematical notions of quaternions and biquaternions by the following reason. The sub-alphabets of the quaternion and biquaternion types (Figures 42 and 45) can be represented as the sub-alphabets of the first two positions of triplets and as the sub-alphabets of the last two positions of triplets correspondingly without any mention of the words "quaternions" and "biquaternions"(it was described above). By this reason the sub-alphabets of the quaternion type can be conditionally termed “the root sub-alphabets” (because the two first positions in any triplet is its root) and the sub-alphabets of the biquaternion type can be conditionally termed “the suffix sub-alphabets” (because the last two positions of any triplet contain its suffix that is the third position). Correspondingly the hypothetical rule formulated above can be also formulated in terms of the root and suffix sub-alphabets.

A connection of two first positions of triplets with two last positions in triplets is known in the phenomenology of the molecular-genetic system: on one filament of DNA two first positions of any codon are connected in the complementary relation with the last two positions of its anti-codon which is disposed on the second filament and which is readed in the inverse direction. One can remind here that each of the sub-alphabets on Figures 42 and 45 contains pairs of „codon-anti-codon“ as it was mentioned above.

Our results in the field of matrix genetics testify that one of the effective ways to understanding genetic system is based on the studying the system of genetic alphabets and on revealing new alphabets which are important for the functional organization of the genetic system. It seems that some of important genetic alphabets are not yet known or have not been studied till now. Any communication system "transmitter-receiver" possesses the feature that the receiver always knows the alphabet of signals that the transmitter uses to send information. Taking this into account, the matrix genetics pays a great attention to study the multi-level system of genetic alphabets.

## 18. Discussion

The author has revealed a close relation of the genetic code with 8-dimensional hypercomplex numbers (first of all, R-octetons and H-octetons) and with dyadic shifts and Hadamard matrices. This relation is interesting in many aspects. Some of them are the following.

Numeric representations of genetic sequences are useful to study hidden genetic regularities [Cristea, 2002, 2010; Petoukhov, He, 2010; etc.]. On the base of the described results, new approaches of numeric representations of genetic sequences can be proposed for such aims. It seems appropriate to interpret genetic sequences as sequences of 8-dimensional vectors where genetic elements are replaced by their special numeric representations which are connected with the described DS-algebras. Then Hadamard spectrums, dyadic distances and some other characteristics of these vector sequences can be studied. If the quantity of vector elements in a genetic sequence is not divisible by 8, the remaining short vector can be extended to an 8-dimensional vector by adding to its end of the required number of zeros by analogy with methods of digital signal processing.

For example, the α-chain of insulin, which has the genetic sequence with 21 triplets, can be interpreted in a few ways as the sequence of 8-dimensional vectors: (GGC, ATC, GTT, GAA, CAG, TGT, TGC, ACT), (TCT, ATC, TGC, TCT, CTT, TAC, CAG, CTT), (GAG, AAC, TAC, TGT, AAC, 0, 0, 0). Firstly, taking into account the representation of triplets with strong and weak roots as elements +1 and -1 in the Rademacher form (Fig. 4 and 6), this chain can be interpreted as the vector sequence (+1, -1, +1, -1, -1, -1, -1, -1), (+1, -1, -1, +1, +1, -1, -1, +1), (-1, -1, -1, -1, -1, 0, 0, 0). Secondly, taking into account the representation of triplets as elements +1 and -1 in Hadamard forms (Fig. 17), this chain can be interpreted as another vector sequence (+1, -1, -1, -1, -1, -1, +1, +1), (+1, -1, +1, +1, -1, +1, -1, -1), (-1, -1, +1, -1, -1, 0, 0, 0). Thirdly, taking into account the representation of triplets by means of their dyadic-shift numerations (Fig. 2), this chain can be interpreted as the sequence (000, 110, 011, 011, 010, 101, 100, 101), (101, 110, 100, 101, 011, 110, 010, 011), (010, 110, 110, 101, 110, 000, 000, 000). Some results of studying such numeric representations will be published separately.

Molecular genetics possesses examples of a special phenomenological role of number 8. For instance, in eukaryote cells, filaments of DNA are coiled around nucleosomes, each of which is an octamer shank consisting of the histones of the four types: H2A, H2B, H3 and H4. More precisely, a single nucleosome contains the ensemble of eight histones, where two histones of each of the four types H2A, H2B, H3 and H4 are included. The DNA molecule is reeled up on this octamer shank. The octamer structure of nucleosome plays the main role in the packing of DNA on all levels. The wide-known concept of the histone code exists in molecular genetics (Jenuwein, Allis, 2001, etc.). One can hope that mathematical formalisms of genetic octetons can be used to simulate some phenomenological facts in this fundamental concept.

Let us go further. The following question arises additionally: why the molecular-genetic system is connected with associative algebras of genetic octetons but not with non-associative algebras of Cayley's octonions and split-octonions which are very popular in theoretical physics of non-living matter? The possible reason is that associativity is very important in the field of noise-immunity coding, where many types of codes (for example cyclic codes) are widely used which are based on the concept of algebraic groups and of Galois fields. But the notions of a group and Galois fields contain conditions of an associativity of their elements. Consequently, associative algebras of octetons have the fundamental advantage in the field of noise-immunity coding in comparison with the non-associative algebras of Cayley's octonions and split-octonions.

The wide using of Hadamard matrices and their Walsh functions in digital technologies is based on their properties. Firstly, stair-step functions by Walsh are implemented technologically much simpler than trigonometric functions and many others. Secondly, the using of Walsh functions allows providing the processing of digital signals by means of operations of addition and subtraction only without using of multiplication and division [Ahmed, Rao, 1975]. Since division operations are not required here, digital informatics can use algebras without dividing, for example the described DS-algebras of octetons. This distinguishes computer informatics from theoretical physics, where multiplication and division are essential, and therefore attention of theorists is drawn to the Cayley algebra of octonions which includes a division operation. It is essential that operations of additions and subtractions for molecular-genetic systems can be organized simply by means of interconnections or disconnections of molecular elements which is easier than molecular organization of multiplication operations.

Thirdly, Hadamard transforms are well suited to dyadic-invariant signal processing [Regalia, Mitra, 1989]. "*The concept of a dyadic shift is very important in the study of Hadamard transforms, as it plays the analogous role of a circular shift to a Discrete Fourier Transform. For example, the power spectrum of a Hadamard transform is invariant to a dyadic shift of the input data. Likewise, dyadic convolution and dyadic correlation results for the Hadamard transform parallel those of circular convolution and circular correlation for the DFT*" [Regalia, Mitra, 1989, p. 608]. Taking into account the described relations of the genetic system with Hadamard matrices and dyadic shifts, one can suppose that dyadic-invariant signal processing is essential for genetic informatics. This supposition should be studied attentively on the base of genetic data. It should be additionally mentioned that some works about applications of Walsh functions for spectral analysis of genetic sequences and genetic algorithms are known [Forrest, Mitchell, 1991; Geadah Corinthios 1977; Goldberg, 1989; Lee, Kaveh, 1986; Shiozaki, 1980; Vose, Wright, 1998; Waterman, 1999]. The work [Kargupta, 2001] uses the Walsh basis to analyse the gene expression process, which plays a key role in evaluating the fitness of DNA through the production of different proteins in different cells; this work shows that genetic code-like transformations introduce interesting properties to the representation of a genetic fitness function.

It is known that multi-dimensional numeric systems are used in the field of multi-dimensional digital signal processing in cases of multi-channel communication to provide some functional advantages. For example 2-dimensional complex numbers are used in digital processing of 2-dimensional signals [Lyons, 2004, Chapter 8]; in this case the real part of complex numbers corresponds to a signal of the first channel and the imaginary part of complex numbers corresponds to a signal of the second channel. But biological organisms are systems of multi-channel informatics. For example a retina of eyes possesses a heritable set of receptors that define a multi-channel transmission of information about a projected image from a great number of separate receptors into the nervous system. But a cooperative principle is needed for systems with a great number of independent channels to operate efficiently with multi-channel flows of signals; such cooperative principle can be realized in a form of an algebraic system of multi-dimensional hypercomplex numbers.

We suppose that the described genetic octetons are used in heritable biological multi-channel informatics for this purpose. One can add that in the field of molecular genetics whole families of proteins should work to provide physiological processes in an active and living context (a separate protein is not a living substance). But these families should contain individual quantities of proteins of different types. The genetic system succesfully solves this numeric task of a genetic determination of amounts of proteins of each type inside separate families of proteins for various physiological processes. The described genetic octetons can be additionally used to construct mathematical models of this genetic phenomenon. Generally speaking, we suppose that in the phenomenological field of molecular genetics and heritable physiological systems, the 8-dimensional genetic algebras (which have been revealed in our works) can be a natural genetic basis to simulate numeric regularities in heritable families of information elements. A possible reason for the variety of different genetic algebras is connected with multi-channel character of biological informatics where, for example, the same 8-dimensional vector can be read for different physiological sub-systems as a representative of various multi-dimensional numeric systems and vector spaces.

In the 6th and 7th versions of our article we have shown the close connection of the system of molecular-genetic alphabets with Hamilton quaternions and Cockle coquaternions. It seems that this result is one of the most important results in the field of matrix genetics. This result was already discussed briefly in Sections 12, 13, 17 and 18. In physics of XX century thousands of papers (see the bibliographic review [http://arxiv.org/abs/math-ph/0511092]) were devoted to Hamilton quaternions which appear now in genetics and algebraic biology together with Cockle split-quaternions. From this time the theme of Hamilton quaternions and Cockle split-quaternions will be one of central themes in our further publications.

In the study of the multi-level system of genetic alphabets we have obtained the multiplication tables of genetic octetons and 8-dimensional Yin-Yang numbers (or bipolar numbers) [Petoukhov, 2008c]. However, it is possible that the value of the 8-dimensional genetic algebras is not limited only by the system of genetic alphabets. Let's try to look at the described 8-dimensional numerical systems hypothetically as one of algebraic basises of living matter. In this case we can recall the following facts for their discussing.

The notion "number" is the main notion of mathematics. In accordance with the known thesis, the complexity of civilization is reflected in complexity of numbers which are utilized by the civilization. "*Number is one of the most fundamental concepts not only in mathematics, but also in all natural sciences. Perhaps, it is the more primary concept than such global categories, as time, space, substance or a field*" [from the editorial article of the journal "Hypercomplex numbers in geometry and physics", 2004, №1, in Russian].

In the history of development of the notion "number", after establishment of real numbers, complex numbers have appeared. These 2-dimensional numbers have played the role of the magic tool for development of theories and calculations in the field of problems of heat, light, sounds, fluctuations, elasticity, gravitation, magnetism, electricity, current of liquids, the quantum-mechanical phenomena. In modern theoretical physics systems of 4-dimensional Hamilton quaternions and 8-dimensional hypercomplex numbers (mainly, Cayley's octonions and split-octonions) are one of important mathematical objects.

Pythagoras has formulated the famous idea: "All the things are numbers". This idea had a great influence on mankind. B.Russell wrote that he did not know any other person who has influenced with such power on other people in the field of thought [Russell, 1967]. In this relation, the world does not know the more fundamental scientific idea than the idea by Pythagoras (it should be mentioned that the notion "number" was perfected after Pythagoras in the direction of generalized numbers such as hypercomplex numbers). Our results give additional materials to the great idea by Pythagoras.

In the beginning of the XIX century the following opinion existed: the world possesses the single real geometry (Euclidean geometry) and the single arithmetic. But this opinion was rejected after the discovery of non-Euclidean geometries and quaternions by Hamilton. The science understood that different natural systems can possess their own individual geometries and their own individual algebras (see this theme in the book [Kline, 1980]). The example of Hamilton, who has wasted 10 years in his attempts to solve the task of description of transformations of 3D space by means of 3-dimensional algebras without a success, is the very demonstrative. This example says that if a scientist does not guess right what type of algebras is appropriate for a natural system, he can waste many years to study this system without result by analogy with Hamilton. One can add that geometrical and physical-geometrical properties of concrete natural systems (including laws of conservations, theories of oscillations and waves, theories of potentials and fields, etc.) can depend on the type of algebras, which are adequate for them.

The fact, that the genetic code has led us to DS-algebras of 8-dimensional hypercomplex numbers (first of all, R-octenons and H-octetons), testifies in favor of the importance of these algebras for organisms. It seems that many difficulties of modern science to understand genetic and biological systems are determined by approaches to these systems from the viewpoint of non-adequate algebras, which were developed entirely for other systems. In particular, the classical vector calculation, which plays the role of the important tool in classical mechanics and which corresponds to geometrical properties of our physical space, can be inappropriate for many biological phenomena. One can think that living substance lives in its own biological space which has specific algebraic and geometric properties. Genetic octetons, which are described in this article, can correspond to many aspects in such biological space. In comparison with our physical space and its vector calculation, vector spaces of genetic octetons are related with another type of vector calculations, etc.

Here it can be remind that E. Schrodinger thought that gaining knowledge about a "stream of order" in living matter is a critical task. He wrote in his book [Schrodinger, 1992, Chapter VII]: "*What I wish to make clear in this last chapter is, in short, that from all we have learnt about the structure of living matter, we must be prepared to find it working in a manner that cannot be reduced to the ordinary laws of physics. And that not on the ground that there is any "new force" or what not, directing the behavior of the single atoms within a living organism, but because the construction is different from anything we have yet tested in the physical laboratory... The unfolding of events in the life cycle of an organism exhibits an admirable regularity and orderliness, unrivalled by anything we meet with in inanimate matter… To put it briefly, we witness the event that existing order displays the power of maintaining itself and of producing orderly events… We must be prepared to find a new type of physical law prevailing in it /living matter/*".

The hypothesis about a non-Euclidean geometry of living nature exists long ago [Vernadsky, 1965] but without a concrete definition of the type of such geometry. Our results draw attention to concrete multidimensional vector spaces which have arised in researches of the genetic code. All the described facts provoke the high interest to the question: what is life from the viewpoint of algebra and geometry? This question exists now in parallel with the old question from the famous book by E.Schrodinger: what is life from the viewpoint of physics? One can add that attempts are known in modern theoretical physics to reveal information bases of physics; in these attempts information principles are considered as the most fundamental.

Mathematics has deals not only with algebras of numbers but with algebras of operators also (see historical remarks in the book [Kline,1980, Chapter VIII]). G.Boole has published in 1854 year his brilliant work about investigations of laws of thinking. He has proposed the Boole's algebra of logics (or logical operators). Boole tried to construct such operator algebra which would reflect basic properties of human thinking. The Boole's algebra plays a great role in the modern science because of its connections with many scientific branches: mathematical logic, the problem of artificial intelligence, computer technologies, bases of theory of probability, etc.

In our opinion, the genetic DS-algebra of octetons can be considered not only as the algebras of the numeric systems but as the algebra of proper logical operators of genetic systems also. This direction of thoughts can lead us to deeper understanding of logic of biological systems including an advanced variant of the idea by Boole (and by some other scientists) about development of algebraic theory of laws of thinking. One can add that biological organisms possess known possibilities to utilize the same structures for multi-purpose destinations. The genetic DS-algebras can be utilized by biological organisms in different purposes also.

In our article a significant attention was paid to the binary and dichotomic properties of the system of genetic alphabets and of the 8-dimensional numerical systems which remind dichotomic bioprocesses and binary bio-objects like meiosis, double helix of DNA, complementary pairs of nitrogenous bases, male and female gametes and beginnings, pairs of chromosomes, etc. Here one can add else a binary bio-factt which was used by R. Penrose and S. Hameroff in the known theory about microtubules of cytoskeletons as a place of quantum consciousness (see the book [Penrose, 1996]). These nano-sized microtubules exist practically in all the cells. The elements that make up these microtubules are termed tubulins. The significan fact is that tubulins have only two possible states. These nanoscale tubulins possess only two possible states. Switching between these two states takes nanoseconds that are extremely fast in comparison with many others biological processes. In other words biological organisms possess binary "cellular automata". It seems that genetic octetons and bipolar 8-dimensional numbers, which were described by us, can be useful to develop a theory of such cellular automata.

Many practical tasks are connected with hypercomplex numbers in the field of informatics. For example, different hypercomplex numbers are used to encode information in digital communication [Bulow, 1999, 2001; Chernov, 2002; Felberg, 2001; Furman et al., 2003; McCarthy, 2000; Sin'kov, 2010; Toyoshima, 1999, 2002; etc.]. Genetic 8-dimensional hypercomplex numbers and their metric 4-dimensional subspaces can be used to construct new genetic algorithms and new decisions in the field of artificial intellect, robotics, etc. Some works try to analyze genetic sequences on the base of complex and hypercomplex numbers but in these attempts a difficult problem exists: what kind of hypercomplex numbers should be chosen for the analysis from a great set of all the types of hypercomplex numbers [Cristea, 2002, 2010; Shu, Li, 2010; Shu, Ouw, 2004]? It seems obviously that hypercomplex numbers, which are described in our article, should be actively used in analyzing of genetic sequences.

Concerning Hadamard matrices and their DS-algebras one can say the following. We suppose that Hadamard genomatrices can be used in genetic systems by analogy with applications of Hadamard matrices in different fields of science and technology: signal processing, error-correcting and other codes, spectral analysis, multi-channel spectrometers, etc. It should be emphasized that a few authors have already revealed the importance of orthogonal systems of Walsh functions in macro-physiological systems, which should be agreed with genetic structures for their transferring along a chain of generations [Shiozaki, 1980; Carl, 1974; Ginsburg et all, 1974].

Genetic molecules are objects of quantum mechanics, where normalized Hadamard matrices play important role as unitary operators (it is known that an evolution of closed quantum system is described by unitary transformations). In particular, quantum computers use these matrices in a role of Hadamard gates [Nielsen, Chuang, 2001]. In view of this, some new theoretical possibilities are revealed to transmit achievements of quantum computer conceptions into the field of molecular genetics and to consider the genetic system as a quantum computer. From the viewpoint of quantum mechanics and its unitary operators, first of all, Hadamard operators, a possible answer on the fundamental question about reasons for the nature to choose the four-letter genetic alphabet is the following one: the possible reason is that Hadamard (2x2)-matrices, which are the simplest unitary matrices in two-dimensional space, consist of four elements. It seems probably that principles of quantum mechanics and quantum computers underlie many structural peculiarities of the genetic code.

Orthogonal systems of Walsh functions play the main role in the fruitful sequency theory by Harmuth for signal processing [Harmuth, 1970, 1977, 1981, 1989]. Rows of Hadamard genomatrices correspond to very special kinds of Walsh functions ("genosystems" of Walsh functions), which define the special ("genetic") variants of sequency analysis in signal processing. The author believes that this "genetic" sequency analysis, whose bases have been revealed in the field of matrix genetics, can be a key to understand important features not only of genetic informatics but also of many other heritable physiological systems (morphogenetic, sensori-motor, etc.). In comparison with spectral analysis by means of sine waves, which is applicable to linear time-invariant systems, the sequency analysis is based on non-sinusoidal waves and it is used to study systems, which are changed in time (biological systems belong to such systems) [Harmuth, 1977, 1989]. The author believes that mechanisms of biological morphogenesis are closely associated with spatial and temporal filters from the field of sequency analysis for genetic systems (the general theory of spatial and temporary filters of sequency theory has been described by Harmuth). From this viewpoint, many inherited morphogenetic abnormalities are consequences of disruptions in physiological spatial and temporal filters of sequency types. Taking into account the sequency theory by Harmuth together with our data about Hadamard genomatrices and genetic H-octetons, one can assume that biological evolution can be interpreted largely like the evolution of physiological spatial and temporal filters of sequency types. In this direction of thoughts, the author develops in his Laboratory new approaches in bioinformatics, bioengineering and medicine.

In the 5$^{th}$ version of the article the author describes at the first time his results about the important connection of the system of genetic alphabets with Hamilton quaternions and their complexification (biquaternions) which are widely applied in theoretical physics. These author's results show new promising ways to build a bridge between theoretical physics and mathematical biology. They can be considered as a new step to knowledge of a mathematical unity of the nature. A continuation of the study of the mentioned physical-biological bridge will be published in the nearest future.

Bioinformatics should solve many problems about heritable properties of biological bodies:

- Noise-immunity property of genetic coding;
- Management and synchronization of a huge number of heritable cyclic processes;
- Compression of heritable biological data;
- Spatial and temporal filtering of genetic information;
- Primary structure of proteins;
- Multi-channel informatics;
- Hidden rules of structural interrelations among parts of genetic systems;
- genetic programming the phenomenon of doubling of bio-information (mitosis, etc);
- genetic programming the phenomenon of complication of the body in the ontogeny from a sexual fertilized cell (zygote) to a multi-cellular organism (increasing the multidimensionality of a phase space of the body);
- Laws of evolution of dialects of the genetic code, etc.

The principle of dyadic shifts and DS-algebras can be useful for many of these problems. This article contains only initial results of researches about DS-algebras and genetic hypercomplex numbers. It is obvious that many other researches of genetic matrices and genetic phenomena should be made from the viewpoint of DS-algebras step by step, for example:

1) An existence of DS-algebras not only for the genomatrix [C A; U G]$^{(3)}$ but also for all other variants of genomatrices and their transposed matrices ([C G; U A]$^{(3)}$, [C U; G A]$^{(3)}$, etc.);
2) An existence of DS-algebras in cases when all other variants of binary sub-alphabets (Figure 3) are used for binary numeration of genetic letters and multiplets;
3) An existence of DS-algebras in different cases of all the possible permutations of columns and rows in genomatrices;
4) Different cases of possible dyadic-shift decompositions of genomatrices;
5) Cases with different values "n" (n = 2, 3, 4, …) in Kronecker families of genomatrices [C A; U G]$^{(n)}$ and others.

## Appendix A. The genetic code, dyadic shifts and „I Ching"

G. Stent [1969, p. 64] has published a hypothesis about a possible connection between genetic code structures and a symbolic system of the Ancient Chinese "The Book of Changes" (or "I Ching"). He is the famous expert in molecular genetics, and his thematic textbooks for students were translated into many countries including Russia. A few authors have supported him and his hypothesis later. For example, the Nobel Prize winner in molecular genetics F. Jacob [1974, p. 205] wrote as well: *« C'est peut-être I Ching qu'il faudrait étudier pour saisir les* *relations entre hérédité et langage»* (it means in English: perhaps, for revealing of relations between genetics and language it would be necessary to study them through the Ancient Chinese "I Ching"). In whole, a position about the necessity of the profound analysis of named parallels and their possible expansion exists in molecular genetics for 40 years. Our researches on matrix genetics have given additional materials to this area and to the hypothesis by Stent [Petoukhov, 2001, 2005a, 2008; Petoukhov, He, 2010]. "I Ching", which is devoted to the binary-oppositional system of "Yin and Yang", declares a universality of a cyclic principle of organization in nature. The Ancient Chinese culture has stated about close relations between heritable physiological systems and the system of "I Ching", and for this reason traditional Oriental medicine (including acupuncture, Tibetan pulse diagnostics, etc.) was based on positions of this book. A great number of literature sources are devoted to "I Ching" which was written some thousand years ago. Many of these sources label "I Ching" as one of the greatest and most mysterious human creations. Symbols and principles "I Ching" penetrated into all spheres of life of traditional China – from theoretical conceptions and high art to household subjects and decorations.

Many western scientists studied and used "I Ching". For example, the creator of analytical psychology C. Jung has developed his doctrine about a collective unconscious in connection with this book. According to Jung, the trigrams and the hexagrams of "I Ching" "fix a universal set of archetypes (innate psychic structures)" [Shchutskii, 1997, p. 12]. Niels Bohr has chosen the symbol Yin-Yang as his personal emblem. Many modern physicists, who feel unity of the world, connect their theories with ideas of traditional Oriental culture, which unite all nature. For example, it has been reflected in the title "the eightfold way" of the famous book [Gell-Mann, Ne'eman, 1964]. Intensive development of modern sciences about self-organizing and nonlinear dynamics of complex systems (synergetrics) promotes strengthening attention of western scientists to traditional eastern world-view (for example see [Capra, 2000]). Special groups study "Book of Changes" in many eastern and western universities. The great numbers of web-sites in Internet are devoted to the similar studies. The influence of "I Ching" is widely presented in the modern life of the countries of the East. For example, the national flag of South Korea bears symbols of trigrams of "I Ching".

Figure 39 shows the most famous tables of 64 hexagrams of "I Ching". Some of them (first of all, the table of 64 hexagrams in Fu-Xi's order) were considered a universal natural archetype in Chinese tradition. Ancient Chinese culture knew nothing about the genetic code of living organisms but the genetic code with its 64 triplets, etc. possesses many unexpected parallels with the symbolic system of "I Ching" (see historical reviews and some results of studying in [Petoukhov, 2001a,b, 2008a; Petoukhov, He, 2010]).

Each hexagram is a pile of six broken and unbroken (solid) lines. Each broken line symbolizes Yin and each unbroken line symbolizes Yang. According western tradition, these broken and unbroken lines are shown in the form of the binary symbols "0" and "1" and each hexagram is shown as a sequence of such six binary symbols. Each position in all the hexagrams has its own individual number: in the Chinese graphical representation, a numbering of the lines of each hexagram is read in the sequence bottom-up; in the western numeric representation of a hexagram, positions of its binary symbols are numbered left-to-right by numbers from 1 to 6. In Chinese tradition, each hexagram is considered constructed from two independent trigrams (piles of three lines or three-digit binary numbers): a bottom trigram and a top trigram. For example, the book [Shchutskii, 1997, p. 86] states: "*The theory of "I Ching" considers that a bottom trigram concerns an internal life… and a top trigram concerns to an external world… . Similar positions in a top trigram and in a bottom trigram have the nearest relation to each other. In view of this, the first position relates by analogy to the fourth position, the second position – to the fifth position, and the third position relates by analogy to the sixth position*". From the Chinese viewpoint, "*trigrams, hexagrams and their components in all possible combinatory combinations form a universal hierarchy of classification schemes. These schemes in visual* *patterns embrace any aspects of reality – spatial parts, time intervals, the elements, numbers, colors, body organs, social and family conditions, etc.*" [Shchutskii, 1997, p. 10].

Let us mention one interesting historical moment. We live in an era of computers.The creator of the first computer G. Leibniz, who had ideas of a universal language, was amazed by this table of 64 hexagrams when he became acquainted with it because he considered himself as the originator of the binary numeration system, which was presented in this ancient table already. Really, he saw the following fact. If each hexagram is presented as the six-digit binary number (by replacement of each broken line with the binary symbol "0" and by replacement of each unbroken line with the binary symbol "1"), this ancient sequence of 64 hexagrams in Fu-Xi's order was identical to the ordinal series of numbers from 63 to 0 in decimal notation. By analogy, a sequence of 8 trigrams in Fu-Xi's order is identical to the ordinal series of numbers from 7 to 0 in decimal notation. "*Leibniz has seen in this similarity the evidence of the pre-established harmony and unity of the divine plan for all epochs and for all people*" [Shchutskii, 1997, p. 12].

But Leibnitz would be surprised even more if he had been told that the ancient Chinese knew not only binary notation, but also a logical modulo-2 addition (1), which is one of the main operations in modern computer technology and which leads to dyadic shifts. Below the author gives some evidences of this knowledge and a relevant realization of the principle of modulo-2 addition and dyadic shifts in tables of hexagrams of "I Ching".

The connections of the genetic code with dyadic shifts, which are described in this article, lead to the idea that dyadic shifts can also be important for famous tables of "I Ching". It is obvious that each of the Chinese hexagrams can be characterized by means of its dyadic-shift numeration on the base of modulo-2 addition of its bottom and top trigrams. For example, the hexagram 110101 is characterized by means of its dyadic-shift numeration 011 on the base of modulo-2 addition of its trigrams: 110+101=011. If each of the hexagrams is replaced by its dyadic-shift numeration, then each of the tables of 64 hexagrams (Figure 48) is transformed in another table of 64 dyadic-shift numerations. We term such new table as dyadic-shift representation of the initial table. Let us analyze dyadic-shift representations of all the famous tables of 64 hexagrams on Figure 48.

| | *111* ***CHYAN*** | *110* ***TUI*** | *101* ***LI*** | *100* ***CHEN*** | *011* ***HSUN*** | *010* ***KAN*** | *001* ***KEN*** | *000* ***KUN*** |
|---|---|---|---|---|---|---|---|---|
| 111 ***CHYAN*** | 111*111* | 111*110* | 111*101* | 111*100* | 111*011* | 111*010* | 111*001* | 111*000* |
| 110 ***TUI*** | 110*111* | 110*110* | 110*101* | 110*100* | 110*011* | 110*010* | 110*001* | 110*000* |
| 101 ***LI*** | 101*111* | 101*110* | 101*101* | 101*100* | 101*011* | 101*010* | 101*001* | 101*000* |
| 100 *CHEN* | 100*111* | 100*110* | 100*101* | 100*100* | 100*011* | 100*010* | 100*001* | 100*000* |
| 011 ***HSUN*** | 011*111* | 011*110* | 011*101* | 011*100* | 011*011* | 011*010* | 011*001* | 011*000* |
| 010 *KAN* | 010*111* | 010*110* | 010*101* | 010*100* | 010*011* | 010*010* | 010*001* | 010*000* |
| 001 *KEN* | 001*111* | 001*110* | 001*101* | 001*100* | 001*011* | 001*010* | 001*001* | 001*000* |
| 000 *KUN* | 000*111* | 000*110* | 000*101* | 000*100* | 000*011* | 000*010* | 000*001* | 000*000* |

A.1) The table of 64 hexagrams in Fu-Xi's order

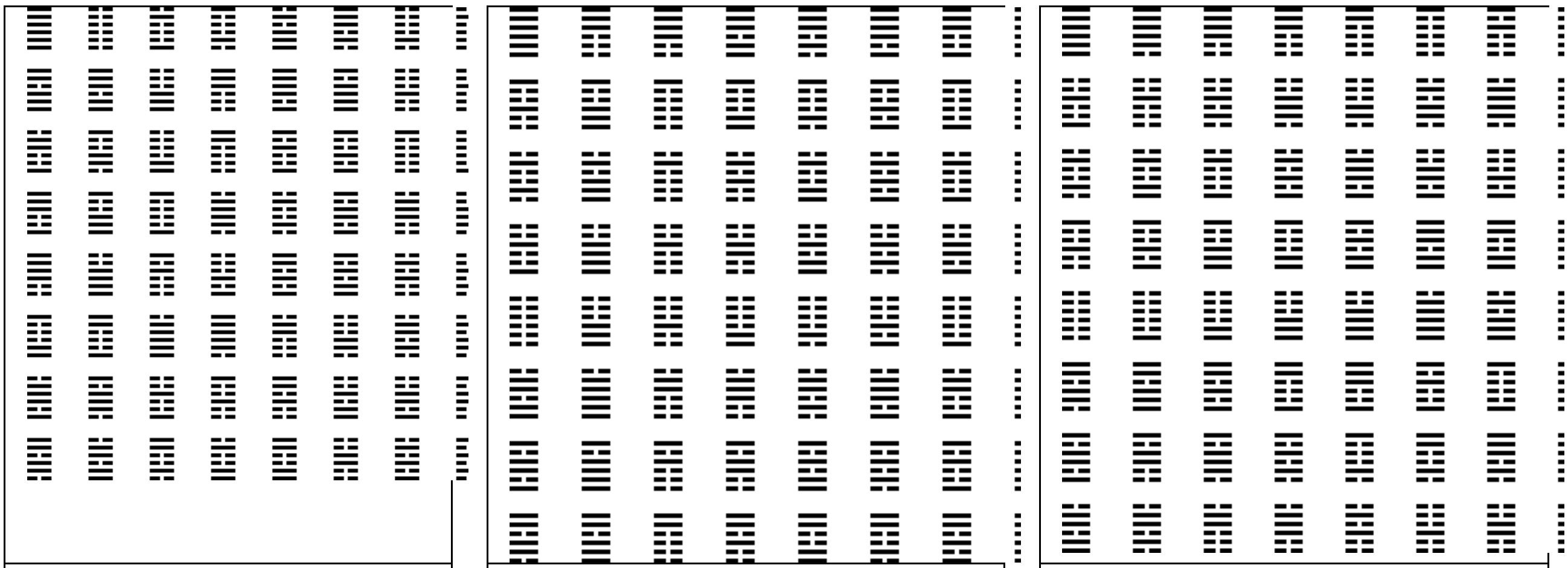

A.2) The King Wen's order

A.3) The Mawangdui table

A.4) Jing Fang's 'Eight Palaces'

Figure 48. The famous tables of 64 hexagrams from "I Ching". The tables A.2-A.4 are taken from http://www.biroco.com/yijing/sequence.htm and [Shchutskii, 1997].

**A.1)** The dyadic-shift representation of the table of 64 hexagrams in Fu-xi's order (Figure 48) is shown on Figure 49.

| | *111* | *110* | *101* | *100* | *011* | *010* | *001* | *000* |
|---|---|---|---|---|---|---|---|---|
| **111** | **000(0)** | **001(1)** | **010(2)** | **011(3)** | **100(4)** | **101(5)** | **110(6)** | **111(7)** |
| **110** | **001(1)** | **000(0)** | **011(3)** | **010(2)** | **101(5)** | **100(4)** | **111(7)** | **110(6)** |
| **101** | **010(2)** | **011(3)** | **000(0)** | **001(1)** | **110(6)** | **111(7)** | **100(4)** | **101(5)** |
| **100** | **011(3)** | **010(2)** | **001(1)** | **000(0)** | **111(7)** | **110(6)** | **101(5)** | **100(4)** |
| **011** | **100(4)** | **101(5)** | **110(6)** | **111(7)** | **000(0)** | **001(1)** | **010(2)** | **011(3)** |
| **010** | **101(5)** | **100(4)** | **111(7)** | **110(6)** | **001(1)** | **000(0)** | **011(3)** | **010(2)** |
| **001** | **110(6)** | **111(7)** | **100(4)** | **101(5)** | **010(2)** | **011(3)** | **000(0)** | **001(1)** |
| **000** | **111(7)** | **110(6)** | **101(5)** | **100(4)** | **011(3)** | **010(2)** | **001(1)** | **000(0)** |

Figure 49. The dyadic-shift representation of the table of 64 hexagrams in Fu-Xi's order (A.1 on Figure 48). Dyadic-shift characteristics of hexagrams are shown; brackets contain values of these binary numbers in decimal notation. Cells with hexagrams, which are characterized by odd dyadic-shift numbers, are marked by black color.

One can see that this dyadic-shift representation of the table of Fu-xi's order is identical to the matrix of diadic shifts in the beginning of the article (Figure 1). Such kind of representation of the table of Fu-xi's order shows some symmetrical patterns of the block character of the table. For example, two quadrants along the main diagonal contain identical block elements, which are (4x4)-matrices of dyadic shifts. Two quadrants along the second diagonal are identical blocks in a form of (4x4)-matrices of diadic shifts also. If tabular cells with odd dyadic-shift numerations are painted in black, then the well-known pattern of 64 cells of chess-board appears in this (8x8)-matrix of dyadic shifts (Figure 49); one can think that the popularity of many games on such chess-boards is connected with the archetypal significance of this pattern.

The table of 64 hexagrams in Fu-xi's order possesses essential analogies with Punnett squares for the case of a tri-hybrid cross (see the section about Punnett squares above). By analogy with genetic matrices this table can be considered as a representation of a special system of alphabets. The ancient Chinese claimed that the system of "I Ching" is a universal classification system or a universal archetype of nature where 8 kinds of energy exist. From viewpoint of data of our article, it leads to the conclusion that our world is constructed on algebraic multilevel systems of interrelated alphabets. In view of this one can suppose that future development of science will be based on a profound theory of such interrelated alphabets.

**A.2)** Concerning the principle of organization of the table of 64 hexagrams in King Wen's order, it is known that "*the reasoning, if any, that informs this sequence is unknown. The hexagrams proceed in pairs, each the inverse of the other, except for the eight hexagrams that are the same both ways up, when the hexagrams of the pair complement each other (yin where yang was and vice-versa) .... There are also four pairs that are both the inverse and complement of each other*" [http://www.biroco.com/yijing/sequence.htm].

The dyadic-shift representation of the table of 64 hexagrams in King Wen's order (Figure 50) reveals a hidden block principle of organization of the table on the base of modulo-2 addition. Each of the rows on Figure 50 consists of pairs of adjacent cells 1-2, 3-4, 5-6, 7-8, which contain binary numbers which differ only in the direction they are read (pairs of mirror-symmetrical numbers). For example the cells 3 and 4 of the first row contain dyadic-shift numbers 110 and 011 which are mirror-symmetrical to each other. If you add up binary numbers in relevant cells of different rows, then new rows appear, in which the same pairs of cells 1-2, 3-4, 5-6, 7-8 contain again mirror-symmetric binary numbers. Figure 51 gives two of possible examples of such modulo-2 addition of rows from Figure 50.

| | | | | | | | | |
|---|---|---|---|---|---|---|---|---|
| №1 | 000 (0) | 000 (0) | 110 (6) | 011 (3) | 101 (5) | 101 (5) | 010 (2) | 010 (2) |
| №2 | 100 (4) | 001 (1) | 111 (7) | 111 (7) | 010 (2) | 010 (2) | 001 (1) | 100 (4) |
| №3 | 010 (2) | 010 (2) | 110 (6) | 011 (3) | 001 (1) | 100 (4) | 001 (1) | 100 (4) |
| №4 | 011 (3) | 110 (6) | 101 (5) | 101 (5) | 000 (0) | 000 (0) | 111 (7) | 111 (7) |
| №5 | 110 (6) | 011 (3) | 101 (5) | 101 (5) | 110 (6) | 011 (3) | 011 (3) | 110 (6) |
| №6 | 111 (7) | 111 (7) | 001 (1) | 100 (4) | 110 (6) | 011 (3) | 100 (4) | 001 (1) |
| №7 | 011 (3) | 110 (6) | 000 (2) | 000 (0) | 010 (2) | 010 (2) | 001 (1) | 100 (4) |
| №8 | 000 (0) | 000 (0) | 001 (1) | 100 (4) | 101 (5) | 101 (5) | 111 (7) | 111 (7) |

Figure 50. The dyadic-shift representation of the table of 64 hexagrams in King Wen's order (A.2 on Figure 48). Dyadic-shift characteristics of hexagrams are shown; brackets contain values of these binary numbers in decimal notation.

| | | | | | | | | |
|---|---|---|---|---|---|---|---|---|
| $\Sigma_{12}$ | 100(4) | 001(1) | 001(1) | 100(4) | 111(7) | 111(7) | 011(3) | 110(6) |
| $\Sigma_{34}$ | 001(1) | 100(4) | 111(7) | 111(7) | 001(1) | 100(4) | 110(6) | 011(3) |
| $\Sigma_{56}$ | 001(1) | 100(4) | 100(4) | 001(1) | 000(0) | 000(0) | 111(7) | 111(7) |
| $\Sigma_{78}$ | 011(3) | 110(6) | 001(1) | 100(4) | 111(7) | 111(7) | 110(6) | 011(3) |
| $\Sigma_{1234}$ | 101(5) | 101(5) | 110(6) | 011(3) | 110(6) | 011(3) | 101(5) | 101(5) |
| $\Sigma_{5678}$ | 010(2) | 010(2) | 101(5) | 101(5) | 111(7) | 111(7) | 001(1) | 100(4) |
| $\Sigma_{12345678}$ | 111(7) | 111(7) | 011(3) | 110(6) | 001(1) | 100(4) | 100(4) | 001(1) |

Figure 51. Some results of summing of rows from Figure 50. In the symbol Σ indices indicate the numbers of summed rows.

<u>**A.3)**</u> The dyadic-shift representation of the Mawangdui table of 64 hexagrams order is shown on Figure 52. To describe the properties of this table, we will distinguish among heterogeneous and homogeneous numbers from the viewpoint of a mix of 0 and 1 in a binary number: each of heterogeneous numbers contains 0 and 1 simultaneously (for example, 011 or 010); each of homogeneous numbers contains only 0 or only 1 (only two homogenous numbers exist here: 000 and 111). Two binary numbers we call "inverse numbers" if their appropriate positions are occupied by opposite elements 0 and 1 (for example, 100 and 011 are inverse binary numbers).

| | | | | | | | | | $\Sigma_L / \Sigma_R$ |
|---|---|---|---|---|---|---|---|---|---|
| | 000(0) | 111(7) | 110(6) | 001(1) | 101(5) | 010(2) | 011(3) | 100(4) | 000/000 |
| | 000(0) | 110(6) | 001(1) | 111(7) | 011(3) | 100(4) | 101(5) | 010(2) | 000/000 |
| | 000(0) | 101(5) | 010(2) | 011(3) | 100(4) | 111(7) | 110(6) | 001(1) | 100/100 |
| | 000(0) | 011(3) | 100(4) | 101(5) | 010(2) | 110(6) | 001(1) | 111(7) | 010/010 |
| | 000(0) | 111(7) | 001(1) | 110(6) | 010(2) | 101(5) | 100(4) | 011(3) | 000/000 |
| | 000(0) | 001(1) | 110(6) | 111(7) | 100(4) | 011(3) | 010(2) | 101(5) | 000/000 |
| | 000(0) | 010(2) | 101(5) | 100(4) | 011(3) | 111(7) | 001(1) | 110(6) | 011/011 |
| | 000(0) | 100(4) | 011(3) | 010(2) | 101(5) | 001(1) | 110(6) | 111(7) | 101/101 |
| $\Sigma_U/\Sigma_B$ | 000/000 | 111/000 | 001/001 | 000/111 | 000/000 | 111/000 | 001/001 | 000/111 | |

Figure 52. The dyadic-shift representation of the Mawangdui table of 64 hexagrams. Identical numerical blocks are marked by similar color. The right column shows the result of modulo-2 addition of binary numbers in the left half of the table ($\Sigma_L$) and in the right half ($\Sigma_R$). The lower row shows the result of modulo-2 addition of binary numbers in the four cells of the upper half of the table ($\Sigma_U$) and in the four cells of the lower half ($\Sigma_B$).

This dyadic-shift representation of the Mawangdui table demonstrates the following symmetrical features in a connection with modulo-2 addition (Figure 52):

- the table consists of a repetition of identical numerical blocks which are marked in colors;
- the total configuration of the colored blocks in the upper half of the table is identical to the total configuration of the colored blocks of its lower half;
- all the heterogenous binary numbers in the upper half of the table correspond to their inverse binary numbers in the lower half;
- all the homogenous binary numbers occupy the identical positions in the upper and lower halves of the tables;
- each of the rows contains all the numbers 000, 001, …, 111 of the dyadic group (2);
- for each of the rows sum of the numbers shown in the right column is the same for quadruples of cells in the left and right halves of the table, that is $\Sigma_L = \Sigma_R$ for each row;
- for each of the columns binary numbers shown in the lower row are the same or are inverse for quadruples of cells in the upper and lower halves of the table;
- the numerical contents of the left and right halves of the lower row are identical;
- for each of the tabular halves (upper, lower, left and right), modulo-2 addition of the relevant four binary numbers of each kind $\Sigma_U$, $\Sigma_B$, $\Sigma_L$ or $\Sigma_R$ gives the same sum 110 (in decimal notation this binary number is equal to 6 which symbolized Yin in Ancient China). For example, modulo-2 addition of the four binary numbers $\Sigma_U$ in the left half of the table gives the value: 000+111+001+000=110.

**A.4)** The dyadic-shift representation of the Jing Fang's 'Eight Palaces' of 64 hexagrams is shown on Figure 53.

| | | | | | | | |
|---|---|---|---|---|---|---|---|
| 000(0) | 100(4) | 110(6) | 111(7) | 011(3) | 001(1) | 101(5) | 010(2) |
| 000(0) | 100(4) | 110(6) | 111(7) | 011(3) | 001(1) | 101(5) | 010(2) |
| 000(0) | 100(4) | 110(6) | 111(7) | 011(3) | 001(1) | 101(5) | 010(2) |
| 000(0) | 100(4) | 110(6) | 111(7) | 011(3) | 001(1) | 101(5) | 010(2) |
| 000(0) | 100(4) | 110(6) | 111(7) | 011(3) | 001(1) | 101(5) | 010(2) |
| 000(0) | 100(4) | 110(6) | 111(7) | 011(3) | 001(1) | 101(5) | 010(2) |
| 000(0) | 100(4) | 110(6) | 111(7) | 011(3) | 001(1) | 101(5) | 010(2) |
| 000(0) | 100(4) | 110(6) | 111(7) | 011(3) | 001(1) | 101(5) | 010(2) |

Figure 53. The dyadic-shift representation of the Jing Fang's 'Eight Palaces' of 64 hexagrams.

This table demonstrates symmetrical features again. Each of the columns contains the same binary number in its cells and all the rows are identical from the viewpoint of their binary numbers. The result of modulo-2 addition of binary numbers in the four cells of each row in the left tabular half and in the right half is the same: 000+100+110+111 = 011+001+101+010 =101 (that is 5 in decimal notation). The results of modulo-2 addition of binary numbers in the eight cells of each row and in the eight cells of each column give the same binary number 000.

All the described facts of this Appendix give evidences that the notions of modulo-2 addition and dyadic-shifts were known in Ancient Chinese culture. They were used in constructions of Chinese tables. It seems that these mathematical notions have played a significant role in theoretical understanding of ancient Chinese sages about the world. The idea of dyadic shifts can be a part of the doctrine of "I Ching" about laws of cyclic changes in the nature. Results of matrix genetics can be useful for deeper understanding of the system of "I Ching". One should emphasize specially that modulo-2 addition and dyadic shifts are notions and operations from the field of informatics but not from physics, chemistry, etc. If the world is created in accordance with the doctrine of "I Ching" where these informatics notions are used in the construction of the main tables then the world is created in accordance with informational principles.

One can add that the traditional Chinese calendar is connected with a division of a circle into 12 equal parts each of which corresponds to one of 12 animals on Figure 54 (see more details in [Shelly H. Wu. 2005; http://en.wikipedia.org/wiki/chinese_zodiac]). The set of these 12 animals is divided into the four trines (triads) each of which includes 3 animals in the vertices of one of the four triangles on Figure 54. Two animals which lie opposite each another on this Chinese circle are antagonist. But the cyclic group of the Hamilton biquaternions $B^{2n}$ (where n=1, 2, 3,.., and the "genetic" biquaternion B is shown on Figure 48), whose period is equal to 12, posseses the similar properties:

1) each of its appropriate triads is characterized by the property that the sum of its members is equal to zero: $B^{2*n} + B^{2*(n+4)} + B^{2*(n+8)} = 0$;

2) two members $B^{2*n}$ and $B^{2*(n+6)}$, which lie opposite inside this cyclic group, differ each another only by the opposite sign (that is they are antagonist; their sum is equal to zero).

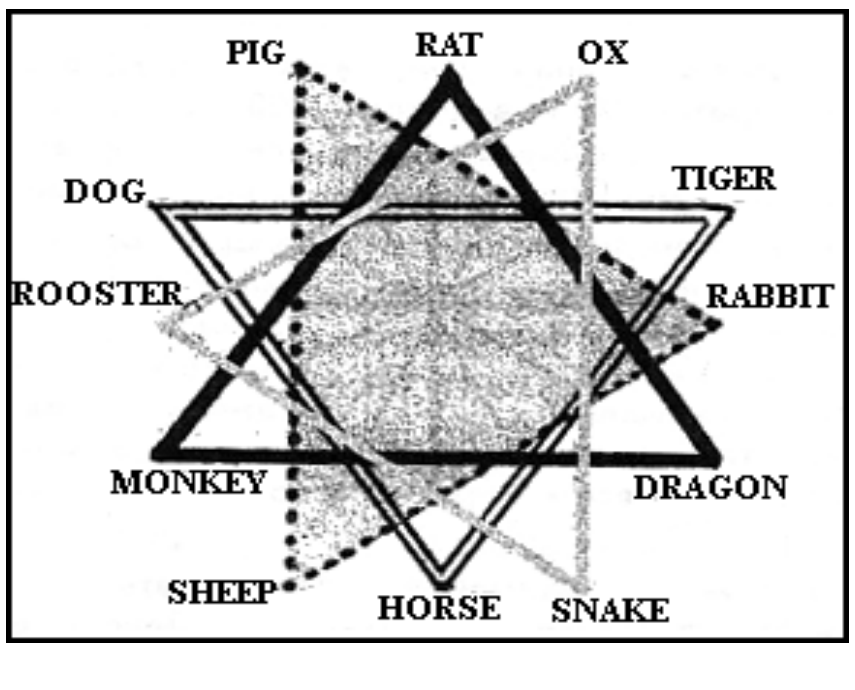

Figure 54. The triangles of a compatibility of 12 animals of the Chinese calendar (from [L.Too, 1993]).

In connection with the cyclic group of the Hamilton biquaternions $B^n$, whose period is equal to 12, one can mention that the traditional Chinese agricultural calendar is divided into 24 two week segments known as Solar Terms; the Chinese zodiac is also used to label times of day, with each sign corresponding to a "large-hour", which is a two-hour period (24 divided by 12 animals) [http://en.wikipedia.org/wiki/chinese_zodiac].

## Appendix B. The Kronecker product of matrices and the conception of resonance genomes

In mathematics, the Kronecker (or tensor) product of matrices, denoted by ⊗, is an operation on two matrices of arbitrary size resulting in a block matrix [http://en.wikipedia.org/wiki/Kronecker_product]. In cases of square matrices considered in our article, the Kronecker multiplication of m-by-m matrix $S=(s_{ij})$ and n-by-n matrix $P=(p_{kr})$ is mn-by-mn matrix $S \otimes P$:

$$S \otimes P = [s_{11}P \quad s_{12}P \quad \ldots \quad s_{1m}P \\ \ldots\ldots\ldots\ldots\ldots\ldots \\ s_{1m}P \quad s_{2m}P \quad \ldots \quad s_{mm}P] \qquad (25)$$

The Kronecker product of matrices is endowed with the property of "inheritance" of their eigenvalues: if the original matrix V and W have the eigenvalues $\lambda_i$ and $\mu_j$ respectively, then all the eigenvalues of their Kronecker product $Q = V \otimes W$ are equal to $\lambda_i * \mu_j$ (figuratively speaking, $\lambda_i$ and $\mu_j$ are inherited in this tensor way) [Bellman, 1960]. For example, let us take a (2*2)-matrix $S_1$, which has two eigenvalues H and h, and a (2*2)-matrix $S_2$, which has two eigenvalues B and b, then their Kronecker product $S_1 \otimes S_2$ is the (4*4)-matrix with its 4 eigenvalues H*B, H*b, h*B, h*b.

Features of such tensor inheritance of eigenvalues of the original matrices (or "parental" matrices) in the result of their Kronecker product can be conveniently represented in the form of "tables of inheritance". The top row of Fig. 55 shows the example of two simplest cases, conventionally referred to as monohybrid and dihybrid cases of the Kronecker hybridization of eigenvalues of parental matrices. In the first case, the Kronecker product of two (2*2)-matrices V and W, which have the same set of eigenvalues A and a, gives the (4*4)-matrix $Q=V \otimes W$ with its 4 eigenvalues A*A, A*a, A*a, a*a. In the second case, the Kronecker product of (4*4)-matrices having the same set of eigenvalues AB, Ab, aB, ab gives a (16*16)-matrix with 16 eigenvalues, representated in the tabular form.

| | | maternal spectrum | |
|---|---|---|---|
| | | A | a |
| paternal spectrum | A | AA | a |
| | a | aA | aa |

| | | maternal spectrum | | | |
|---|---|---|---|---|---|
| | | AB | Ab | aB | ab |
| pat. sp. | AB | AABB | AABb | AaBB | AaBb |
| | Ab | AABb | AAbb | AaBb | Aabb |
| | aB | A BB | AaBb | aaBB | aaBb |
| | ab | AaB | Aabb | aaBb | aabb |

| | | maternal gametes | |
|---|---|---|---|
| | | A | a |
| paternal gametes | A | AA | Aa |
| | a | aA | aa |

| | | maternal gametes | | | |
|---|---|---|---|---|---|
| | | AB | Ab | aB | ab |
| pat. gam. | AB | AABB | AABb | AaBB | AaBb |
| | Ab | AABb | AAbb | AaBb | Aabb |
| | aB | AaBB | AaBb | aaBB | aaBb |
| | ab | AaBb | Aabb | aaBb | aabb |

Figure 55. Top row: the tables of inheritance of eigenvalues of $(2^n*2^n)$-matrices present the result of their Kronecker product (shown cases of monohybrid and dihybrid «Kronecker hybridizations» of sets of eigenvalues of parental matrices). Bottom row: examples of Punnet squares for monohybrid and dihybrid hybridizations of organisms under the laws of Mendel. Abbreviations «pat.sp.» and «pat. gam.» mean «paternal spectrum» and «paternal gametes».

The author notes that these tables of the Kronecker inheritance of eigenvalues of parental matrices are formally identical to Punnet squares for poly-hybrid crosses of organisms (Fig. 55, bottom row), which were already considered in the Section 15. A small difference between them is that Punnet squares represent combinations of dominant and recessive alleles of genes from parent reproductive cells (gametes) instead of the representation of combinations of eigenvalues of parental matrices in the case of the Kronecker inheritance.

It is known that eigenvalues of matrices play an important role in theory of resonances of oscillatory systems with many degrees of freedom (see for example [Gladwell, 2004]). Matrices possess a wonderful property to express resonances, which sometimes is called as their main quality [Balonin, p. 21, 26]. Physical resonance phenomenon is familiar to everyone. The expression y=A*S simulates a transmition of a signal S via an acoustic system represented by its matrix A. If an input signal “y” is a resonant tone, then the output signal will repeat it with a precision up to a scale factor y = λ*S by analogy with a situation when a string sounds in unison with the neighboring vibrating string. In the case of a matrix A, the number of resonant tones $S_i$ corresponds to its size. They are called its eigenvectors, and the scale factors $\lambda_i$ with them are called its eigenvalues or, briefly, a spectrum A. Frequencies $\omega_i = \lambda_i^{0.5}$ are called as natural frequencies of the system, and the corresponding eigenvectors are called as its own forms of oscillations (or simply, natural oscillations).

On the base of the noted formal analogy - between Punnet squares of combinations of alleles and tables of the Kronecker inheritance of eigenvalues of matrices of oscillatory systems -, the author puts forward the following idea:

- alleles of genes and their combinations can be interpreted as eigenvalues of $(2^n*2^n)$-matrices from tensor families of matrices of oscillatory systems. For genetic systems, this model approach focuses an attention on the possible importance of special classes of mutually related resonances from Kronecker families of matrices, which play the role of biological "matrix archetypes."

This idea seems to be a perspective for further study taking into account that living organisms are endowed with an innate ability to tune in to resonances, and they can use resonant frequencies as carriers of information, for example, in human vocal singing and speech communication. One might think that this inherited ability is based on a fundamental fact of existence of a living organism in the form of a huge choir of coordinated oscillatory processes (mechanical, electromagnetic, etc.); this choir is capable to a significant complication in the course of the ontogenetic development of the organism. Since ancient times chronomedicine claims that all diseases are caused by disorders of this harmony in the choir. From a formal point of view, any living organism is an oscillating system with a large number of degrees of freedom. Onthogenetic development of an organism from embryo to adult is accompanied by a great increasing of number of its degrees of freedom while maintaining the coherence of oscillatory processes at each stage of development.

The mentioned idea together with some additional materials has led the author to the conception of resonance genomes [Petoukhov, 2015a,b,c]. This conception proposes the following points of view:

- molecular alphabets of molecular-genetic systems are interpreted as alphabets of resonances of appropriate oscillatory systems; respectively, the genetic code can be interpreted as the code of special systems of resonances, and genetic texts on the basis of these alphabets can be interpreted as texts written in the language of these resonances;
- heritable biological processes are associated with phenomena of vibrational mechanics.

- In connection with the last point it should be added that amazing phenomena of vibrational mechanics are widely used in engineering technologies for vibrational separation and structurization of multiphase media, vibro-transportations, vibro-transmitting energy from one part of the system to another, etc. [Blekhman, 2000]. Familiarity with some of these phenomena creates the impression of staying in the world with other physical laws. Practically invisible vibrations, which may have high frequency and small amplitudes, can provide, for example, the following phenomena: a) the upper position of the inverted pendulum becomes stable; b) a heavy metal ball "pops" in a layer of sand; c) rotation of the rotor of the electro-motor, which is disconnected from the power supply, can be steadily supported by means of vibro-transfer of energy from another motor, which is connected to the power supply and which stands on a general vibro-platform with the first motor; d) movements of metronomes, which stand on a general movable platform, become synchronized, etc. In living organisms, numerous phenomena exist, which are associated with the vibrations at different levels and which occur as if they were produced by means of mysterious forces. We hope that appropriate mathematical models of such biological processes will be developed on the base of formalisms of vibrational mechanics and the conception of resonance genomes, where the theory of inherited systems of biological resonances is under development. Some additional details of this theory will be published in the nearest future. In particularly, the conception of resonance genomes leads to interesting models of molecular ensembles of genetic elements in a form of a set of special oscillatory systems. Taking into account that eigenvalues of self-adjoint matrices play an essential role in foundations of quantum physics, one can additionally think about quantum-physical models in the conception of resonance genomes, which is also based on mathematics of eigenvalues of matrices. Our study of the conception of resonance genomes has a number of practical aspects related to using of wave and vibration processes in medicine, biotechnology, artificial intelligence systems, etc.

All natural objects (living and non-living) can be considered as oscillatory systems with many degrees of freedom. But in view of the concept of resonance genomes, which studies inherited systems of mutually coordinated bio-resonances, living organisms are distinguished by special inherited tensor-matrix systems of resonances in them. These biological systems of resonances represent a relatively narrow class of all systems of conceivable resonances. In our opinion, the conception of resonance genomes and its matrix mathematics of simulations can be useful for understanding many biological problems.

The conception of resonace genomes also connects physical phenomena of resonances of oscillatory systems with abstract binary-numeric systems of computer technology and mathematics, including dyadic groups of binary numbers, which were noted above in this article in relation with the genetic binary sub-alphabets (Fig. 3 and 4). Really, each of attributes of nitrogenous bases A, C, G, U/T in Fig. 3 can be interpreted as connected with its own resonance characteristics. For example, it is obvious that purines may have resonance characteristics that differ from the resonance characteristics of pyrimidines due to differences in the structure of the purine and pyrimidine molecules. In this light, each of pairs of binary-oppositional attributes in Fig. 3 can be treated as a pair of binary-oppositional kinds of resonance characteristics. In this case, numeric symbols 0 and 1 in each of binary sub-alphabets in Fig. 3 are representations of binary-oppositional kinds of resonance characteristics. For comparison, we recall that in computer technology binary elements 0 and 1 are physically realized through using two types of signal amplitudes (eg oppositional in polarity) or two kinds of laser beams, etc. But in the considered genetic case, the binary opposition of resonance characteristics gives an opportunity of considering genetic systems as computers on the base of the same binary numbers. In other words, binary-oppositional properties of the DNA (and RNA) alphabet A, C, G, T/U, which are treated as oppositions of appropriate resonance characteristics, define its binary sub-alphabets and provide a possibility of binary-numeric representations of the alphabets and n-plets of DNA (and RNA) for studying the living bodies as genetic computers on the base of oscillatory bio-systems.

In the past century, science has discovered that molecular-genetic bases of all living organisms are the same (alphabets of DNA, RNA, etc.) and that they are very simple. A hope arises that the algorithmic foundations of organisms, which are subordinated to genetic laws such as Mendel's laws, are also very simple and are unified for all living things. Identifying these algorithms of living matter is important. We assume that resonance algorithms of matching and ordering subsystems, which are associated with formalisms of tensor products of matrices, play one of key roles in living matter.

**Appendix C. The logic holography by Walsh waves in genetic models**

Living organisms possess properties, which seem to be analogical to properties of holography with its non-local record of information. For example, German embryologist Hans Driesch separated from each other two or four blastomeres of sea urchin eggs or on changing their mutual positions. The main result of Driesch’s experiments was that fairly normal (although proportionally diminished) larvae with all of their organs properly arranged could be obtained from a single embryonic cell (blastomere) containing no more than 1⁄2 (if two first blastomeres were separated) or even 1⁄4 (in the case of four blastomeres separation) of the entire egg’s material. Rather soon these effects (defined by Driesch as “embryonic regulations”) were numerously confirmed and extended to the species belonging to almost all taxonomic groups of metazoans, from sponges to mammalians [Belousov, 2015; Driesch, 1921]. In 1901 Hans Spemann has conducted an experiment on the separation of the amphibian embryo into individual cells, from which quite normal tadpoles have grown (in 1935 he won the Nobel Prize for the discovery of organizing effects in embryonic development). These experimental results testify that complete sets of “causes” required for further development are contained not only within whole eggs/embryos but also in their halves, quarters, etc. The similar properties exist in holograms, where one can restore a whole holographic image of a material object from a part of the hologram. A hologram has such property since each element of the hologram possesses information about all elements of the object (in difference to ordinary photos).

One can mention also known hypothesis about possible connections of holography with brain functions including associative memory, physiological processing visual information, etc. (see for example [Greguss, 1968; Pribram, 1971]). But the brain and the nervous system have appeared at a relatively late stage of biological evolution. A great number of species of organisms lived perfectly up to this, and is now living without neuronal networks. It is clear that the origins of the similarity between holography and nonlocal informatics of living organisms need to look at the level of the genetic system.

Physical holography, which possesses the highest properties of noise-immunity, is based on a record of standing waves from two coherent physical waves of the object beam and of the reference beam [Soroko, 1980]. But physical waves can be modeled digitally. Correspondingly noise-immunity and other properties of optical and acoustical holography can be modeled by digitally, in particular, with using Walsh functions and logic operations concerning dyadic groups of binary numbers. This can be made on the base of discrete electrical or other signals without any application of physical waves. The pioneer work about «holography by Walsh waves» was [Morita, Sakurai, 1973]. The work was devoted to Walsh waves (or Walsh functions), which propagate through electronic circuits - composed of logical and analog elements - by the analogy with the optical Fourier transform holography. In this digital Walsh-holography, objects, whose digital holograms should be made, are represented in forms of $2^n$-dimensional vectors. Each component of these vectors corresponds to one of $2^n$ input channels of appropriate electric circuits; the same is true for $2^n$ output channels, which are related with components of resulting vectors. Examples of electrical circuits for logic holography are shown in the works [Morita, Sakurai,1973; Soroko, 1974]. Due to application of Walsh transformation, information about such vector is written in each component of the appropriate hologram, which is also a $2^n$-dimensional vector, to provide nonlocal character of storing information.

One should specially note that Walsh functions are close related with dyadic groups since Walsh functions are characters of the dyadic groups [Fine, 1949]. Therefore, the Fourier analysis on dyadic groups is defined in terms of Walsh functions. In the same way, the discrete Walsh functions are characters of the finite dyadic groups, on which the switching functions are defined. Therefore, the Fourier analysis for switching functions considered as a subset of complex valued functions is formulated in terms of the discrete Walsh functions [Karpovsky, Stankovic, Astola, 2008].

This digital Walsh-holography under the title «logical holography» was also considered later in [Derzhypolskyy, Melenevskyy, Gnatovskyy, 2007; Golubov, Efimov, Skvortsov, 1987; Soroko, 1974]. Examples of electrical circuits for logic holography are in the works [Morita, Sakurai,1973; Soroko, 1974]. All these and other works about logical Walsh-holography considered possibilities of its apllications in engineering technologies of digital signals processing without any supposition of its application in biology, in particular, in genetics. In our paper and in some other author publications [Petoukhov, 2008a,b,c, 2016], deep connections of the genetic code system with Walsh functions, Hadamard matrices, dyadic groups and logical modulo-2 addition are described (one can note that Rademacher functions, which were mentioned in representations of genetic matrices in Fig. 6, coincide with appropriate Walsh functions). Due to these results and some other materials, the author puts forward the hypothesis that genetic system is based on principles of logical holography and appropriate logic operations.

This hypothesis leads to a new class of mathematical models of genetic structures and phenomena on the base of logical holography and appropriate logical operations. Correspondingly the author develops the theory of «genetic logical holography» (GLH), where mathematics of logical holography and logical operations is used for modeling genetic phenomena. The mathematical basis of this modeling approach is lattice functions, logical operations with them, dyadic spaces, dyadic groups of binary numbers, logic modulo-2 addition, dyadic convolution, dyadic derivatives as well as isomorphic and homomorphic transformations of Walsh in close relation with peculiarities of molecular-genetic systems. In our opinion, a realization of digital-like mechanisms of the logical holography in biological organisms is created by Nature on the base of their binary computers on resonances [Petoukhov, 2015c, 2016]. The new kind of mathematics in modeling genetic phenomena gives possibilities of new heuristic associations and new understanding of the natural phenomena. Let us show a few examples of mathematical models and approaches in this new field of researches.

**The example № 1** concerns the proposed model of the phenomenon of zipper-like reproduction of DNA molecules. A prerequisite for the preservation of existing genetic information in a series of successive generations of cells and organisms is replication of genetic material in DNA. Replication of DNA strands takes place before each division of normal DNA-containing structures in eukaryotes (nuclei, mitochondria and plastids), and also before each division of bacterial cells and before each reproduction of DNA-viruses. Then doubled DNA in the process of segregation is distributed equally between the two daughter cell nuclei or bacterial cells. The process of replication of genetic information and its subsequent dichotomous segregation equally between child objects can be repeated many times as you like. In this scheme the nature provides a very high degree of noise-immunity of the transmitted information, which can be compared with the highest noise-immunity of information in holographic recordings.

The replication of DNA resembles a zipper mechanism with its teeth, which may cling to each other or unzipp from each other step by step. The work [Nie, Kleinermanns, de Vries, 2000] considers four molecular bases of DNA as individual teeth of the zipper and studyes the mechanism of the DNA zipper by means of the whole isolation of these bases so that there is no interaction with anything else. In this work the authors were able to see, in particular, the frequencies of vibrations of hydrogen bonds in molecules. The DNA zipper has a specificity: two pairs of complementary teeth exist in it (complementary bases C-G and A-T). Moreover the pair C-G differs from the pair A-T in two aspects: 1) the pair C-G is characterized by odd number of hydrogen bonds (3 bonds) and the pair A-T is characterized by even number of hydrogen bonds (2 bonds); 2) the families of triplets, whose first letter is C or G, are opposite to the families of triplets, whose first letter is A or T/U, on the criterion of the ratio of amounts of triplets with strong and weak roots (3:1 or 1:3 in Fig. 4).

The following model of a vector zipper with its two pairs of complementary teeth can be proposed from logical holography. Let us take the constant Walsh function in a form of the 4-dimensional vector V, which is sum of 4 spare vectors $2\overline{C}$, $2\overline{A}$, $2\overline{G}$, $2\overline{T}$ (26):

$$V = [1,1,1,1] = [1,0,0,0] + [0,1,0,0] + [0,0,1,0] + [0,0,0,1] = 2\overline{C} + 2\overline{A} + 2\overline{G} + 2\overline{T}, \qquad (26)$$

where $\overline{C} = [0.5, 0, 0, 0]$, $\overline{G} = [0, 0, 0.5, 0]$, $\overline{A} = [0, 0.5, 0, 0]$, $\overline{T} = [0, 0, 0, 0.5]$.

The algorithm of logical holography operates with object vectors and reference vectors by analogy with classical holography, which is based on interference of object waves with reference waves. We will describe this algorithm following the work [Soroko, 1974]. The algorithm consists of two parts: 1) generation of logic hologram of a vector-signal; 2) restoration of the vector-signal from logic hologram.

The generation of logic hologram consists of the following three steps, which are illustrated by us for the case of the vector V (26).

In the first step, an object vector and a reference vector are taken (both of them have identical $2^n$-dimensionality but they differ each from other). Then their sum is considred for further transformations. For our example № 1, we will generate holograms for each of four object vectors $\overline{C}$, $\overline{G}$, $\overline{A}$, $\overline{T}$ from (26). For the case of object vectors $\overline{C}$ and $\overline{G}$, the role of the reference vector will play the vector $\overline{P}_1$=[0,0,0,1]; for the case of object vectors $\overline{A}$ and $\overline{T}$, the role of the reference vector will play the vector $\overline{P}_2$=[1,0,0,0] (the reference vectors is more «intensive» than the object vectors by analogy with the classical holography [Soroko, 1974]). In general case, the non-zero component of the reference vector can be equal to another real number than 1 but for simplicity here we follow numeric examples in the work, where this component is equal to 1 [Soroko, 1974]. In our example № 1 corresponding sums of the object vectors and the reference vectors are equal to:

$$\overline{C}+\overline{P}_1 = [0.5,0,0,1],\ \ \overline{G}+\overline{P}_1 = [0,0,0.5,1],\ \ \overline{A} + \overline{P}_2 = [1,0.5,0,0],\ \ \overline{T} + \overline{P}_2 = [1,0,0,0.5] \qquad (27)$$

In the second step, the sum of the object vector and the reference vector is multiplied with one of Hadamard matrices of the order $2^n$ to receive the Walsh-Hadamard spectrum in relation to this Hadamard matrix. For our example, taking into account the difference between complementary pairs C-G and A-T, we will use the Hadamard (4*4)-matrix $H_{CG} = [1, -1; 1, 1]^{(2)}$ for the vectors $\overline{C}+\overline{P}_1$ and $\overline{G}+\overline{P}_1$ (26); opposite we will use Hadamard (4*4)-matrix $H_{AT} = [1, 1; 1, -1]^{(2)}$ for the vectors $\overline{A} + \overline{P}_2$ and $\overline{T} + \overline{P}_2$ (26). In our example, the Walsh-Hadamard spectra $W_C$, $W_G$, $W_A$ and $W_C$ are equal to:

$$\begin{aligned} W_C &= (\overline{C}+\overline{P}_1) * H_{CG} = [1.5,\ 0.5,\ 0.5,\ 1.5] \\ W_G &= (\overline{G}+\overline{P}_1) * H_{CG} = [1.5,\ 0.5,\ 1.5,\ 0.5] \\ W_A &= (\overline{A}+\overline{P}_2) * H_{AT} = [1.5,\ 0.5,\ 1.5,\ 0.5] \\ W_T &= (\overline{T}+\overline{P}_2) * H_{AT} = [1.5,\ 0.5,\ 0.5,\ 1.5] \end{aligned} \qquad (28)$$

In the third step, a logical hologram is constructed for each of these spectra $W_C$, $W_G$, $W_A$ and $W_T$ (28) by means of exponentiation in square of each of its components:

$$\begin{aligned} Hol_C &= [2.25, 0.25,\ 0.25,\ 2.25] \\ Hol_G &= [2.25, 0.25,\ 2.25,\ 0.25] \\ Hol_A &= [2.25, 0.25,\ 2.25,\ 0.25] \\ Hol_T &= [2.25, 0.25,\ 0.25,\ 2.25] \end{aligned} \qquad (29)$$

The algorithm of restoration of vector-signals from logic holograms also consists of three steps. In the first step, the Walsh-Hadamard spectrum of each reference vector is received in relation to the Hadamard matrix, which was used at the stage of generating the logical hologram. In our example, these spectra are equal to:

$$\overline{P}_1 * H_{CG} = [1, 1, 1, 1]; \quad \overline{P}_2 * H_{AT} = [1, 1, 1, 1] \qquad (30)$$

In the second step, such spectrum should be componentwise multiplied with approprite logic hologram. In our example, the results of these actions are the following:

$$\begin{aligned} Hol_C .* [1, 1, 1, 1] &= [2.25, 0.25,\ 0.25,\ 2.25] \\ Hol_G .* [1, 1, 1, 1] &= [2.25, 0.25,\ 2.25,\ 0.25] \\ Hol_A .* [1, 1, 1, 1] &= [2.25, 0.25,\ 2.25,\ 0.25] \\ Hol_T .* [1, 1, 1, 1] &= [2.25, 0.25,\ 0.25,\ 2.25] \end{aligned} \qquad (31)$$

In the third step, the result of such multiplication should be multiplied by the transposed version of the same Hadamard matrix, and the result must be divided by the numerical order of the matrix. In the result the object vector is restored together with the reference vector but with a new ratio 1/1.25 = 4/5 between their non-zero components by analogy with classical holography [Soroko, 1974]. In our example, the result of these actions with using the same Hadamard matrices are the following:

| | |
|---|---|
| $R_C=(Hol_C.*[1,1,1,1]*(H_{CG}^T)/4=[1,0,0,1.25]=[1, 0, 0, 0]+[0, 0, 0, 1.25]=2*\overline{C} + 1.25*\overline{P}_1$<br>$R_G=(Hol_G.*[1,1,1,1]*(H_{CG}^T)/4=[0,0,1,1.25]=[0, 0, 1, 0]+[0, 0, 0, 1.25]=2*\overline{G} + 1.25*\overline{P}_1$<br>$R_A=(Hol_A.*[1,1,1,1]*(H_{AT}^T)/4=[1.25,1,0,0]=[0, 1, 0, 0]+[1.25, 0, 0, 0]= 2*\overline{A} + 1.25*\overline{P}_2$<br>$R_T=(Hol_T.*[1,1,1,1]*(H_{AT}^T)/4=[1.25, 0, 0, 1]= [0, 0, 0, 1]+[1.25, 0, 0, 1]= 2*\overline{T} + 1.25*\overline{P}_2$ | (32) |

Expressions (32) show that in the result the object vectors $\overline{C}$, $\overline{G}$, $\overline{A}$, $\overline{T}$ and the reference vectors $\overline{P}_1$ and $\overline{P}_2$ are restored though with their additional weight coefficients 2 and 1.25, which are not essential for vectors processing in electric circuites of logical holography. In classical holography the reference beam also exists at the output. In (32) the sum of restored object vectors is equal to the initial Walsh function in (26): $2*\overline{C} + 2*\overline{A} + 2*\overline{G} + 2*\overline{T} = [1, 1, 1, 1]$. The theory of logical holography, including its noise-immunity, is described in the work [Soroko, 1974].

For us the following property of replication in logical Walsh-holography, which was not noted in previous works of other authors, is interesting specially. In classical holography each separate half of the hologram allows restoring the object beam and reference beam with a certain deterioration of their quality. Let us study now what properties are revealed when, for instance, only left half of logical hologram is used for restoring the vector-signal?

The left half (or right half) of each logical hologram from (29) can be denoted by the symbol beginning with the letter L (or R correspondingly). Using such symbols, the left halves and the right halves of the holograms (29) are represented in the following forms:

| | | |
|---|---|---|
| $LHol_C$ = [2.25, 0.25, 0, 0]<br>$LHol_G$ = [2.25, 0.25, 0, 0]<br>$LHol_A$ = [2.25, 0.25, 0, 0]<br>$LHol_T$ = [2.25, 0.25, 0, 0] | $RHol_C$ = [0, 0, 0.25, 2.25]<br>$RHol_G$ = [0, 0, 2.25, 0.25]<br>$RHol_A$ = [0, 0, 2.25, 0.25]<br>$RHol_T$ = [0, 0, 0.25, 2.25] | (33) |

The following expressions (34a-d, left) show vector-signals $SLHol_C$, $SLHol_G$, $SLHol_A$ and $SLHol_T$, which are restored from $LHol_C$, $LHol_G$, $LHol_A$ and $LHol_T$ (33) by means of the algorithm of restoration described above. Expressions (34a-d, right) show vector-signals $SRHol_C$, $SRHol_G$, $SRHol_A$ and $SRHol_T$, which are restored from $RHol_C$, $RHol_G$, $RHol_A$ and $RHol_T$ (33) by means of the same algorithm of restoration. In this expressions D means the matrix of dyadic shifts [0, 0, 1, 0 ; 0, 0, 0, 1; 1, 0, 0, 0; 0, 1, 0, 0], E means the identity matrix [1, 0, 0, 0; 0, 1, 0, 0; 0, 0, 1, 0; 0, 0, 0, 1]. The ratio 0.5/0.625 of non-zero components of the restored object vectors to non-zero components of the restored reference vectots is equal again 4/5.

| | | |
|---|---|---|
| $SLHol_C$ = [0.5, 0.625, 0.5, 0.625]<br>= [0.5, 0, 0.5, 0]+[0, 0.625, 0, 0.625]<br>$= \overline{C} + \overline{G} + 0.625*(D+E)*\overline{P}_1$ | $SRHol_C$ = [0.5, -0.625, -0.5, 0.625] =<br>[0.5, 0, 0, 0]+[0, 0, 0, 0.625]+[0, -0.625, -0.5, 0]<br>$= \overline{C} + 0.625*\overline{P}_1$ + [0, -0.625, -0.5, 0] | (34a) |
| $Hol_G$ = [0.5, 0.625, 0.5, 0.625]<br>= [0.5, 0, 0.5, 0]+[0, 0.625, 0, 0.625]<br>$= \overline{C} + \overline{G} + 0.625*(D+E)*\overline{P}_1$ | $SRHol_G$ = [-0.5, -0.625, 0.5, 0.625] =<br>[0, 0, 0.5, 0]+[0, 0, 0, 0.625]+[-0.5, -0.625, 0, 0]<br>$= \overline{G} + 0.625*\overline{P}_1$ + [-0.5, -0.625, 0, 0] | (34b) |
| $SLHol_A$ = [0.625, 0.5, 0.625, 0.5]<br>= [0, 0.5, 0, 0.5]+[0.625, 0, 0.625, 0]<br>$= \overline{A} + \overline{T} + 0.625*(D+E)*\overline{P}_2$ | $SRHol_A$ = [0.625, 0.5, -0.625, -0.5] =<br>[0, 0.5, 0, 0]+[0.625, 0, 0, 0]+[0, 0, -0.625, -0.5]<br>$= \overline{A} + 0.625*\overline{P}_2$ + [0, 0, -0.625, -0.5] | (34c) |
| $SLHol_T$ = [0.625, 0.5, 0.625, 0.5]<br>= [0, 0.5, 0, 0.5]+[0.625, 0, 0.625, 0]<br>$= \overline{A} + \overline{T} + 0.625*(D+E)*\overline{P}_2$ | $SRHol_T$ = [0.625, -0.5, -0.625, 0.5] =<br>[0, 0, 0, 0.5]+[0.625, 0, 0, 0]+[0, -0.5, -0.625, 0]<br>$= \overline{T} + 0.625*\overline{P}_2$ + [0, -0.5, -0.625, 0] | (34d) |

From the left column (34a-b) one can see the following. From the left half of the logical hologram $Hol_C$ (31) for the object vector $\overline{C}$, this algorithm of logical holography restores the pair of the object vectors $\overline{C} + \overline{G}$ (34a). Also from the left half of the logical hologram $Hol_G$ (31) for the object vector $\overline{G}$ this algorithm restores the same pair of the object vectors $\overline{C} + \overline{G}$ (34b). By this reason the pair of the object vectors $\overline{C}$ and $\overline{G}$ is interpreted as an analogue of the complementary pair of cytosine C and guanine G in this modeling approach.

From the left half of the logical hologram $Hol_A$ (31) for the object vector $\overline{A}$, this algorithm restores the pair of the object vectors $\overline{A} + \overline{T}$ (34c). Also from the left half of the logical hologram $Hol_T$ (31) for the object vector $\overline{T}$ this algorithm restores the same pair of the object vectors $\overline{A} + \overline{T}$ (34d). By this reason the pair of vectors $\overline{A}$ and $\overline{T}$ is interpreted as an analogue of complementary pair of adenine A and thymine T in this modeling approach.

These results allow proposing the following model of the phenomenon of zipper-like reproduction of DNA molecules (Fig. 56). The model supposes that biological mechanisms embody the described algorithm of logical holography in this phenomenon.

In the case of a separate DNA-chain (Fig. 56a), each of nitrogenous bases in the form of appropriate vector ($\overline{C}$, $\overline{G}$, $\overline{A}$ or $\overline{T}$) is read as an appropriate logic hologram (31), from the left half of which (33, left) the described algorithm restores the complementary pair $\overline{C}$+$\overline{G}$ (34b) or $\overline{A}$+$\overline{T}$ (34c). Such stepwise reading of each of next nitrogenous bases on the DNA-chain provides the zipper-like constructing of double chains of DNA (56b).

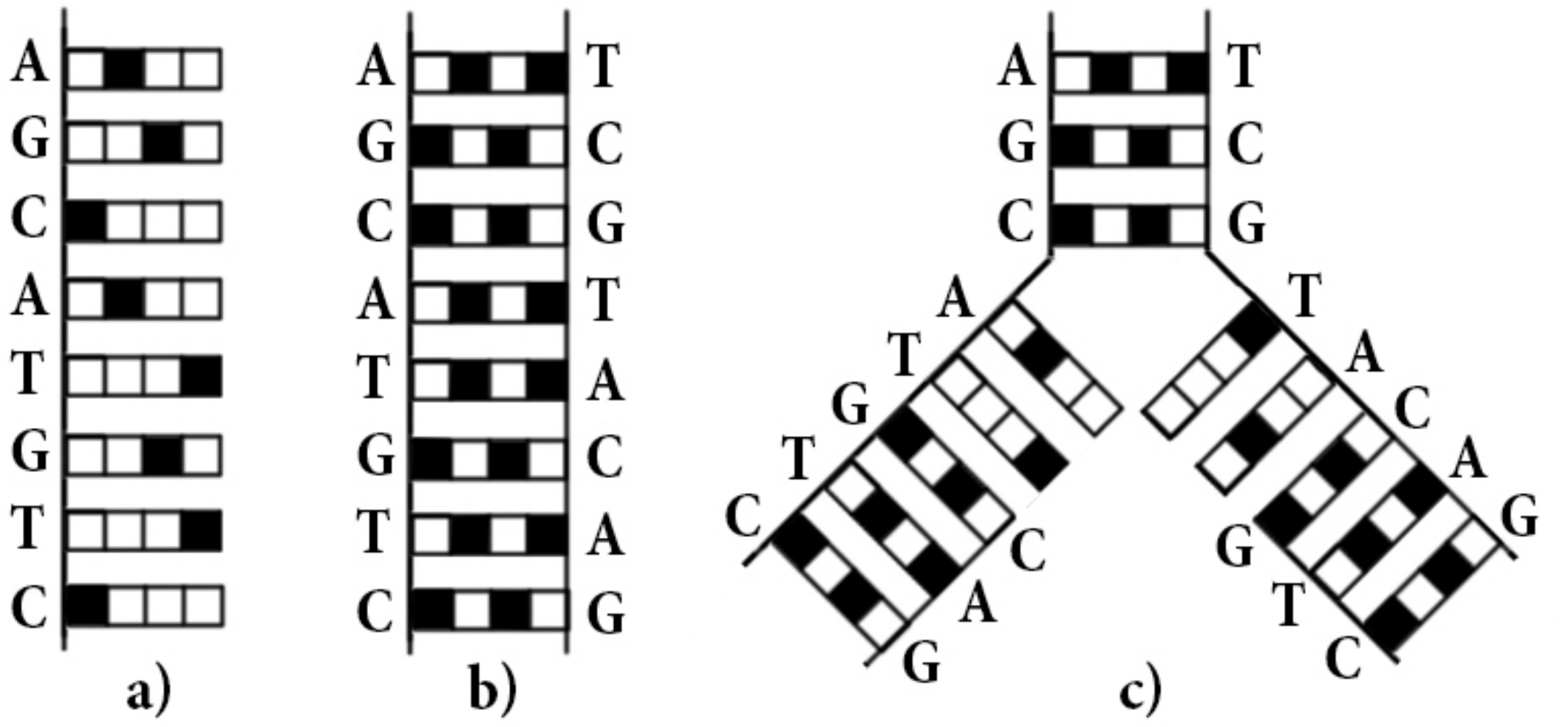

Figure 56. The illustration to the model of the zipper-like reproduction of DNA. Nitrogenous bases of DNA are denoted as informational vectors (26): $\overline{C}$ = [0.5, 0, 0, 0], $\overline{G}$ = [0, 0, 0.5, 0], $\overline{A}$ = [0, 0.5, 0, 0], $\overline{T}$ = [0, 0, 0, 0.5]. These vectors are shown schematically as structures with 4 cells in each, where each black cell means 0.5 and each white cell means 0. a) A separate DNA-chain; b) two complementary chains of DNA; c) unzipping two chains of DNA.

The second part of this model concerns unzipping of two chains of DNA (56c). In this case each of nitrogenous bases on a DNA chain is read again as an appropriate logic hologram (31), from the right half of which (33, right) the described algorithm restores only one of vectors $\overline{C}$, $\overline{G}$, $\overline{A}$ or $\overline{T}$ (34) without its complementary pair, but with additional vectorial "trash" (like [0, -0.625, -0.5, 0] in (34a)), which doesn't correspond to any of considered vectors $\overline{C}$, $\overline{G}$, $\overline{A}$, $\overline{T}$, $\overline{P}_1$ or $\overline{P}_2$. In this model, such result of the restoration means unzipping the complementary pair $\overline{C}$+$\overline{G}$ or $\overline{A}$+$\overline{T}$.

**The example № 2** of models and approaches from the theory of logic holography in genetics is devoted to phenomena of different repetitions in nucleotide sequences of DNA and RNA where a great number of complementary palindromes and simple repetitions exists [Gusfield, 1997; Lehninger, 1982]. For instance, families of repetitive sequences occupy about one-third of the human genome. The importance of the problem of repeats in genetic sequents is reflected in the fact that during 20 years before 1991 on this subject was published 6000 articles [Gribskov, Devereux, 1991]. The author proposes a new approach for modeling and comparative analyses of the families of repetitive sequences. The proposed approach concerns, first of all, to reproduction of fragments of nucleotide sequences but also to some other biological phenomena where similar kinds of repetitions (simple repetitions or with dyadic shifts) are realized.

One should remind about the difference in notions of an ordinary palindrome and a complementary palindrome. By definition, an ordinary palindrome is a string that reads the same from beginning and from the end. By contrast, a complementary palindrome in molecular genetics is a fragment of a chain of DNA or RNA, which becomes an ordinary palindrome, if each symbol in one half of the fragment is replaced by its complementary symbol (A↔T, C↔G) [Gusfield,1997]. For instance, AGCTCGCGAGCT is a complemetary palindrome. In molecular genetics the term «mirror repetition» is sometimes used for ordinary palindromes, which also exist in genetic strings.

One of many applications of classical holography is devoted to the reproduction of numerous copies of flat templates required for the production of printed circuits, etc. [Wenyon, 1978]. In this case a special holographic operation is used, which has its analogue in mathematics under the title "convolution of two functions". For example, if such convolution of a flat image of a triangle with images of three points is made, the result is an image of three triangles [Wenyon, 1978].

In logical holography the analogue of this operation is the dyadic convolution of two $2^n$-dimensional vectors $\overline{X}$ and $\overline{Y}$, which is known in the theory of discrete signal processing and which is calculated by means of the following expression, where ⊕ means logical modulo-2 addition [Ahmed, Rao, 1975]:

$$Z(m) = n^{-1} * \sum_{h=0}^{n-1} X(h)Y(m\oplus h)\ ,\ \ m = 0, 1,\ldots, n\text{-}1 \qquad (35)$$

Our model uses the dyadic convolution (35) of two $2^n$-dimensional vectors $\overline{X}$ and $\overline{Y}$, where some components of $\overline{X}$ represent nitrogenous bases A, C, G and T. Taking into account the opposite character of the bases, which form complementary pairs (A and T, C and G) on two opposite chains of DNA, we assume T = -A, G = -C. Let us consider dyadic convolution between, for instance, the spare vector $\overline{X}$ = [A, G, C, A, 0, 0, 0, 0] and various variants of the reproducting vector $\overline{Y}$ (Fig. 57). We call the symbolic part A,G,C,A in $\overline{X}$ as the "template" (or "word", or "kernel"), which can be reproducted in various variants by means of its dyadic convolution with different variants of the reproducting vector $\overline{Y}$. In general case this genetic template may be very different in its length and in composition and order of its nitrogenous bases. The reproducting vector $\overline{Y}$ is an analogue of the image of points in the mentioned example from classical holography about reproduction of image of a single triangle into its many copies. The results of dyadic convolutions of 8-dimensional vectors $\overline{X}$ and $\overline{Y}$ are shown in Fig. 57.

| № | $\overline{X}$ | $\overline{Y}$ | Result of dyadic convolution |
|---|---|---|---|
| 1 | [A,G,C,A,0,0,0,0] | [8,0,0,0,0,0,0,-8] | [A,G,C,A,-A,-C,-G,-A] = [A,G,C,A,T,G,C,T] |
| 2 | [A,G,C,A,0,0,0,0] | [8,0,0,0,8,0,0,0] | [A,G,C,A,A,G,C,A] |
| 3 | [A,G,C,A,0,0,0,0] | [8,0,0,0,0,0,0,8] | [A,G,C,A,A,C,G,A] |
| 4 | [A,G,C,A,0,0,0,0] | [8,0,0,0,0,0,8,0] | [A,G,C,A,C,A,A,G] |
| 5 | [A,G,C,A,0,0,0,0] | [8,0,0,0,0,8,0,0] | [A,G,C,A,G,A,A,C] |
| 6 | [A,G,C,A,0,0,0,0] | [8,0,0,0,-8,0,0,0] | [A,G,C,A,-A,-G,-C,-A] = [A,G,C,A,T,C,G,T] |
| 7 | [A,G,C,A,0,0,0,0] | [8,0,0,0,0,-8,0,0] | [A,G,C,A,-G,-A,-A,-C] = [A,G,C,A,C,T,T,G] |
| 8 | [A,G,C,A,0,0,0,0] | [8,0,0,0,0,0,-8,0] | [A,G,C,A,-C,-A,-A,-G] = [A,G,C,A,G,T,T,C] |

Figure 57. The dyadic convolution of two vectors $\overline{X}$ and $\overline{Y}$.

These results (Fig. 57) show that the operation of dyadic convolution of the same vector X, which comprises the template AGCT, with different reproducting vectors $\overline{Y}$ generates different types of regularly organized vectors, which correspond to the well-studied types of repetitions in genetic sequences, as well as to their more complex dyadic-shift variants, which previously didn't attract attention of researchers as we can judge:

- the row №1 shows an example of obtaining a complementary palindrome;
- the row №2 shows an example of obtaining a simple repetition of the template;
- the row №3 shows an example of an ordinary palindrome (or a mirror repetition);
- the rows №4,5 show examples of non-trivial repetitions, where the second half of vectors repites the template of the first half but with dyadic-shift permutations of components;
- the row №6 shows an example of a simple complementary repetition, where the second half of the vector contains the complementary repetition of the first half;
- the rows №7,8 show examples of non-trivial complemetary repetitions, where the second half of the vectors contain the complementary repetitions of the first half but with dyadic-shift permutations of components.

Reminding what does it mean «dyadic-shift permutations of vector components», Fig. 58 illustrates such permutations of components in the 4-dimensional vector [$a_1$, $a_2$, $a_3$, $a_4$] by means of the appropriate dyadic-shift matrix (Fig. 1 shows the dyadic-shift matrix for the case of 8-dimensional vectors). Each row of this matrix represents a vector with a dyadic shift of its components in comparison with vectors in other rows (the set of components is identical in all the vectors).

| | | | |
|---|---|---|---|
| $a_1$ | $a_2$ | $a_3$ | $a_4$ |
| $a_2$ | $a_1$ | $a_4$ | $a_3$ |
| $a_3$ | $a_4$ | $a_1$ | $a_2$ |
| $a_4$ | $a_3$ | $a_2$ | $a_1$ |

Figure 58. The dyadic-shift matrix for the case of 4-dimensional vectors.

Complemenary palindromes, simple repetitions, mirror repetitions, example of which are shown in rows №№1-3 in Fig. 57, are known in nucleotide sequences since they were studied previously in works of different authors [Gusfield, 1997]. But more complex kinds of repetitions of templates, which are generated by the dyadic convolution and examples of which are shown in rows №№4-8, were not studied in nucleotide sequences as we can judge. Our theory of genetic logical holography predicts that the last kinds of repetitions also exist in nucleotide sequences since all these known and unknown kinds of repetitions of templates are only different variants of the same mechanism of logical holography and dyadic convolutions. This prediction should be checked in further researches.

It is obvious that similar applications of the method of dyadic convolution for reproduction of templates and for generating variuos repetitions can be used not only for 8-dimensional vectors but also for any $2^n$-dimensional vectors for modeling and analysis of genetic sequences. This method of logical holography allows modeling long sequences, in which an initial template is repited many times in different parts of the sequences. For instance, let us consider the 16-dimensional vector [A,G,C,A,0,0,0,0,0,0,0,0,0,0,0,0] with the template A,G,C,A and the reproducting vector [16,0,0,0,0,0,0,-16,0,0,0,0,16,0,0,0]. The result of their dyadic convolution is the following vector with nitrogenous bases (we assume again that A=-T, C=-G): [A, G, C, A,-A, -C, -G, -A, 0, 0, 0, 0, A, G, C, A] = [A, G, C, A, T, G, C, T, 0, 0, 0, 0, A, G, C, A]. Here in the result we get the new 16-dimensional vector, the first half of which is a complementary palindrome, and the last part is the simple repetition of the template A,G,C,A.

In our opinion, the proposed method of the dyadic convolution can be used in modeling and analysis of phenomena of jumping genes [McClintock, 1950], exons and introns, etc., that is different phenomena related with transpositions and structural interrelations of fragments of DNA and RNA. Also we propose to use the method of dyadic convolution for comparison analysis of different genetic sequences, which contain repetitions and which are realized at different levels and branches of biological evolution.

Holographic methods in engineering allow quickly detecting individual elements in a huge image. The theory of genetic logic holography allows assuming that one of secrets of noise-immunity of genetic informatics is based on the similar possibilities of genetic holography. Each of DNA molecules can be only one of elements of the whole logic hologram of a certain kind in organism (inside organism many different logic holograms can exist for its different organs). In this case, the genetic logic holography – by analogy with classic holography - also allows giving integral estimates of the entire molecule of DNA and promptly compares different molecules of DNA from different parts of organism. If one DNA molecule mutates, its genetic information - in the result of a comparison of this DNA with other DNA molecules by means of holographic methods - found to be incorrect for further using in organism.

**The example № 3** of new approaches related with the theory of logic holography is based on application of the dyadic derivative of J.Gibbs [Gibbs, 1967; Golubov, 2006; Stankovic, Astola, 2008] to study ensembles of genetic structures, first of all, nucleotide sequences. By contrast to ordinary derivative in classical mathematical analysis, the operator of the dyadic derivative is the nonlocal operator. Correspondingly the dyadic derivative at the point is defined by values of a considered function in a few points, which are specially located. Its properties are closely related to the theory of characters on dyadic carriers. Dyadic derivatives save forms of Walsh functions, which are the characters of dyadic groups.

The dyadic derivative for n-dimensional vectors has its matrix representation in a form of a square matrix of order $2^n$ symmetric about both diagonals [Stankovic, Astola, 2008]. For instance, the dyadic derivative for 8-dimensional vectors is represented by the following matrix:

$$D = -0.5* \begin{vmatrix} -7 & 1 & 2 & 0 & 4 & 0 & 0 & 0 \\ 1 & -7 & 0 & 2 & 0 & 4 & 0 & 0 \\ 2 & 0 & -7 & 1 & 0 & 0 & 4 & 0 \\ 0 & 2 & 1 & -7 & 0 & 0 & 0 & 4 \\ 4 & 0 & 0 & 0 & -7 & 1 & 2 & 0 \\ 0 & 4 & 0 & 0 & 1 & -7 & 0 & 2 \\ 0 & 0 & 4 & 0 & 2 & 0 & -7 & 1 \\ 0 & 0 & 0 & 4 & 0 & 2 & 1 & -7 \end{vmatrix} \tag{36}$$

To apply dyadic derivatives for comparison analysis of genetic sequences of n-plets (monoplets, doublets, triplets, etc.), one should represent each of n-plets numerically. One of possibilities of numeric representations of triplets is related with their binary numeration in the genetic (8*8)-matrix of triplets in Fig. 59. In this case each of triplets can be numerated by the concatenation of binary numerations of the row and the column, on crossing of which the triplet is located. In this case the triplet CCC gets number 000000, the triplet CCA gets number 000001, etc. In the result the numeration of triplets in their matrix (Fig. 59) becomes identical to binary meanings of 64 hexagrams in Fu-Xi's order shown above in Fig. A.1 of the Appendix A.

| | 000 (0) | 001 (1) | 010 (2) | 011 (3) | 100 (4) | 101 (5) | 110 (6) | 111 (7) |
|---|---|---|---|---|---|---|---|---|
| 000<br>(0) | CCC<br>000000<br>(0) | CCA<br>000001<br>(1) | CAC<br>000010<br>(2) | CAA<br>000011<br>(3) | ACC<br>000100<br>(4) | ACA<br>000101<br>(5) | AAC<br>000110<br>(6) | AAA<br>000111<br>(7) |
| 001<br>(1) | CCT<br>001000<br>(8) | CCG<br>001001<br>(9) | CAT<br>001010<br>(10) | CAG<br>001011 (11) | ACT<br>001100<br>(12) | ACG<br>001101<br>(13) | AAT<br>001110<br>(14) | AAG<br>001111<br>(15) |
| 010<br>(2) | CTC<br>010000<br>(16) | CTA<br>010001<br>(17) | CGC<br>010010<br>(18) | CGA<br>010011 (19) | ATC<br>010100<br>(20) | ATA<br>010101<br>(21) | AGC<br>010110<br>(22) | AGA<br>010111<br>(23) |
| 011<br>(3) | CTT<br>011000<br>(24) | CTG<br>011001<br>(25) | CGT<br>011010<br>(26) | CGG<br>011011 (27) | ATT<br>011100<br>(28) | ATG<br>011101<br>(29) | AGT<br>011110<br>(30) | AGG<br>011111<br>(31) |
| 100<br>(4) | TCC<br>100000<br>(32) | TCA<br>100001<br>(33) | TAC<br>100010<br>(34) | TAA<br>100011 (35) | GCC<br>100100<br>(36) | GCA<br>100101<br>(37) | GAC<br>100110<br>(38) | GAA<br>100111<br>(39) |
| 101<br>(5) | TCT<br>101000<br>(40) | TCG<br>101001<br>(41) | TAT<br>101010<br>(42) | TAG<br>101011 (43) | GCT<br>101100<br>(44) | GCG<br>101101<br>(45) | GAT<br>101110<br>(46) | GAG<br>101111<br>(47) |
| 110<br>(6) | TTC<br>110000<br>(48) | TTA<br>110001<br>(49) | TGC<br>110010<br>(50) | TGA<br>110011 (51) | GTC<br>110100<br>(52) | GTA<br>110101<br>(53) | GGC<br>110110<br>(54) | GGA<br>110111<br>(55) |
| 111<br>(7) | TTT<br>11100<br>(56) | TTG<br>111001<br>(57) | TGT<br>111010<br>(58) | TGG<br>111011 (59) | GTT<br>111100<br>(60) | GTG<br>111101<br>(61) | GGT<br>111110<br>(62) | GGG<br>111111<br>(63) |

Figure 59. The numeration of each of 64 triplets by means of concatenation of binary numerations of the row and the column (from Fig. 4), on crosing of which the triplet is located. Decimal meanings of binary numerations are shown in brackets.

For instance, using the matrix operator D (36), let us calculate the dyadic derivative of the following sequence of 8 triplets GGC-ATC-GTT-GAA-CAG-TGT-TGC-ACT, which is a fragment of the gene of insulin. Taking into account the decimal numeraton of each of these triplets from Fig. 59, we represent these sequence in the form of the 8-dimensional vector B=[54, 20, 60, 39, 11, 40, 50, 12]. The dyadic derivative of the first order for this vector is equal to B*D =[97, -76, 36.5, 62.5, -139.5, 82.5, 38, -101]. Also the dyadic derivative of the second order can be calculated here to compare them with the dyadic derivatives of other genetic fragments.

Mathematical formalisms of the logic holography and the theory of logic functions, including the dyadic convolution and the dyadic derivative, can be applied for comparative studying not only nucleotide sequences in DNA and RNA but also other biological string-like patterns and repetitions in them, which are under influence of genetic templates. For example such string patterns and different kinds of binary oppositions and repetitions are typical for the following fields of human creativity, where influence of genetic templates and principles seems to be exist:

1) music with its known metric and rhythmic structures, and also with different symmetric permutations in many musical piecies;
2) verse (poetry), where meter and rhyme play important roles;
3) linguistic texts, where binary oppositions exist on the base of vowels and consonants, etc.;
4) multi-block structures in architecture and design with different kinds of symmetry, etc.

The dyadic derivative can be used in medical diagnosis to analize complex biorhythms, for instance, in cases of cardiac arrhythmias. In the last case, components of the vector for taking its dyadic derivatve represent values of time intervals between cardiac pulsations.

Some long known biological phenomena also testify about physiologic importance of logic functions in inherited parts and algorithms of organisms. One can remind here about “all-or-none law" for nerve fibers and muscle fibers (https://en.wikipedia.org/wiki/All-or-none_law). Accoring to this famous law, each neuron - under influence of appropriate stimuluses - generates a nerve impulse with its maximum intensity or does not generate it at all. In other words, it works in the alternative mode - "yes" or "no" - by analogy with Boolean variables and logic functions.

Our data testify in favor that genetics and bioinformatics in wide sense are based on logic functions and logic operations on dyadic groups including binary codes, modulo-2 addition, etc. Due to obtained results the author puts forward a new working hypothesis: 20 amino acids of the genetic code represent 20 logic functions, which are defined on the dyadic group of 6-bit numbers with 64 members in it. Results of the development of this hypothesis in the author's laboratory will be published later.

## Appendix D. About the geno-logic code in DNA and epigenetics

In the middle of XX century a great discovery was made about the genetic code as a code correspondence between the set of 64 triplets and the set of 20 amino acids and stop-signals. From that time the genetic code is understood as a rule, in accordance with which informational fragments of DNA (or RNA) in a form of triplets define corresponding amino acids and stop-codons, that is biochemical elements and punctuation signs of proteins synthesis. This classical biochemical understanding of the genetic code can be named «the first genetic code».

Different authors supposed that other kinds of genetic coding can also exist. For example, a supposition about the histone code is well known (https://en.wikipedia.org/wiki/Histone_code#cite_note-Jenuwein-1). The histone code is a hypothesis that the transcription of genetic information encoded in DNA is in part regulated by chemical modifications to histone proteins, primarily on their unstructured ends. Together with similar modifications such as DNA methylation it is part of the epigenetic code (see for example [Jenuwein, Allis, 2001]). No mathematical approaches to model such additional kinds of genetic code were proposed.

In this Appendix D, the author puts forward his idea that molecules DNA (and RNA) are not only carriers of the genetic code of protein sequences of amino acids but also they are participants of the second type of genetic encoding that we term «geno-logic code». This geno-logic code of DNA and of epigenetic mechanisms is based on systems of Boolean functions, their Walsh-Hadamard spectra and logical operations. DNA is an important part of this integrated coding system, peculiarities of which are reflected in structures of DNA-alphabets and in features of the degeneration of the genetic code of amino acids. The proposed approach is related with mathematics that allows creating a wide class of mathematical models in genetics.

The system of the geno-logic code exists in parallel with the genetic code of amino acids and participates in its functioning, we believe. As known, for a creation of a computer, an usage of material substances for its hardware is not sufficient but logical operations should be also included to provide work of computer. These logical operations can succesfully work in different hardware from very different materials. The same situation is true for living bodies, where genetical systems should provide genetic information not only about material substances (proteins) but also about logic of interrelated operations in biological processes. One can think that the first genetic code defines material aspects of biological bodies and the geno-logic code defines logic rules and functions of their operating work. In the proposed new modeling approach about the geno-logic code, separate molecular-genetic elements (monoplets, doublets, triplets, etc.) are represented as Boolean functions or systems of these functions.

Account of existence of the geno-logic code allows you to interpret the work of genetic systems using analogies with computers, which also operate on the basis of Boolean functions and logic operations. As known, Boolean functions in computers can be realized by means of triggers (the states “yes” or “no”), which can regulate flows of electric current, optical signals, liquids, etc.; flows of different substances can be used for constructing computers to provide systems of Boolean functions playing a key role in them. Achievements of computer technologies and mathematics are used in the proposed direction of bioinformation researches. An important role in these boinformation researches belongs to so called spectral logic on Walsh-Hadamard spectra from the theory of information processing and digital communication [Karpovsky, Moskalev, 1973; Karpovsky, Stankovic, Astola, 2008; Zalmazon, 1989]. As known, spectral methods of development of logical schemes give ways to create devices of manage and communication, which possess by effective resources of self-monitoring, self-correction, adaptability and ability of structural adjustment under changes of external conditions (biological organisms possess similar properties). Spectral methods are easily realized by computers for logical systems of any degree of complexity. Walsh functions and spectra are used in a creation of multi-channel systems of communication, where different signals are simultaneously transmitted via each of communication channels (multiplexing).

In computers, logical systems of Boolean functions provide a huge variety of computer functions. By analogy, the geno-logic code of DNA and RNA provides an inheritance of a huge variety of biological functions. In our opinion, it is the code of providing biological variety in the course of biological evolution, in which systems of genetic Boolean functions are evolving. We believe that knowledge about the geno-logic code will be useful for understanding those genetic phenomena, which can not be explained from knowledge of the code correspondence among triplets and amino acids. In our opinion, the epigenetic code is related with the geno-logic code. Genetics can be additionally developed as a science about genetic systems of logic functions. Results of this development can be used not only for deeper understanding living matter but also for progress in the fields of artificial intellect and artificial life (A-life), where mathematical logic plays a key role. In this case those computer systems and theoretical models should be developed, which are based on genetic systems of logic functions, which constitute a relative small part of the infinite number of possible systems of logic functions. As known, artificial intellect, which possesses an ability of reproducing features of biological intellect, can not be constructed without usage of mathematical logic [Yaglom, 1980], and so the idea of the geno-logic code is very natural.

The author has generated the idea of the geno-logic code due to his studying of structures of the genetic alphabets and some other inherited biological phenomena taking into account the following facts:

1) The basic alphabet of DNA (and RNA) is closely related with binary numbers since it is represented by the set of 4 specific polyatomic molecular constructions, which bears the symmetric system of pairs of binary-oppositional attributes and forms three binary sub-alphabets (Fig. 2 and 3). This peculiarity of the alphabet is associated with thoughts about biological computers on the base of binary-oppositional resonances of genetic molecules [Petoukhov, 2016];
2) the set of alphabets of genetic n-plets is connected with the dyadic groups of 2n-bit binary numbers (n = 1, 2, 3,…). Really each of such dyadic groups contains $2^{2n}$ members by analogy with a corresponding alphabet of genetic n-plets: the alphabet of monoplets (the nitrogenous bases of DNA or RNA) contains $2^2 = 4$ members; the alphabet of doublets contains $2^4 = 16$ members; the alphabet of triplets contains $2^6 = 64$ members; … ;
3) dyadic groups of binary numbers are deeply related with Walsh functions from the standpoint of mathematical theory of characters on dyadic groups: Walsh functions are the characters of dyadic groups; groups of Walsh functions are isomorphic to appropriate dyadic groups [Fine, 1949; Harmuth, 1977; Karpovsky, Stankovic, Astola, 2008]. But phenomenological structures of the alphabets of doublets and triplets in their matrix representations demonstrate their relations with the set of Walsh functions (Fig. 4), which includes Rademacher functions as its particular cases. Additional connections of the genetic alphabets with complete sets of Walsh functions are shown in [Petoukhov, 2016].

4) Binary numbers are the basis of Boolean algebra, on which technical computers with their trigger devices work. Hypothesis about analogies between functioning of living organisms and technical computers exist long ago (see for example [Elsevier, 2014; Hameroff, Penrose, 1996, 2013; Igamberdiev, Shklovskiy-Kordi, 2016; Ji, 2012; Liberman,1972; Penrose, 1996, 1999]). Our results of studying molecular-genetic systems lead to the evidences that genetic systems work on the base of Boolean algebra; it is important since the level of the genetic code is the deeper level than secondary levels of inherited nervous systems or separate kinds of proteins such as tubulin, trigger properties of which were used in the theory of consciousness in publications [Hameroff, Penrose, 1996, 2013; Penrose, 1996, 1999]. According to the concept of the systemic-resonant genetics [Petoukhov, 2016], binary-oppositional kinds of molecular resonances can be the natural basis of biocomputers working with genetic systems of Boolean functions.

Our idea of the geno-logic coding is confirmed, in particular, by the well-known physiological law "all-or-none" for excitable tissues: by analogy with triggers in computers, a nerve cell or muscle fiber give only their answers "yes" or "no" under action of different stimulus. If a stimulus is above a certain threshold, a nerve or muscle fiber will fire with full response. Essentially, there will either be a full response or there will be no response at all [Kaczmarek, Levitan, 1987; Martini, 2005]. A separate muscle, which contains many muscle fibers, can reduce its length in different degree due to the combined work of the plurality of its muscle fibers. Nervous system can also account stimulus of a different force due to the combined action of its many nerve fibers (and also due to the ability to change the frequency of the generation of nerve impulses at their fixed amplitude).

In view of these data, the human body (and other organisms) can be interpreted as an immense network of trigger devices, which provide systems of logic functions and logic operations on them. In light of this it is not so surprising that the genetic system, that provides transmission of corresponding logic networks along the chain of generations, is also built on the principles of Boolean algebra of logic, including the case of the logical holography by Walsh functions (see above the Appendix C).

One can note additionally that from the standpoint of the first genetic code (the table of correspondence between 64 triplets and 20 amino acids) it is unclear why the alphabets of n-plets of DNA and RNA are related with dyadic groups of binary numbers and with Walsh functions. But from the standpoint of the geno-logic code, this relationship plays a key role since dyadic groups are closely related not only with Boolean algebra of logic but also with Walsh functions, which are characters of dyadic groups. By this reason the dyadic-group structure of the set of DNA alphabets connects genetics with Boolean algebra of logic by the natural way.

The first genetic code has a strong degeneracy: 64 triplets encode 20 amino acids and a stop-signal. Correspondingly a few triplets encode each of amino acids. By this reason any fragment of a sequence of amino acids in proteins has many different variants of its encoding by different triplets. For example, let us consider the short sequence of 3 amino acids: Ser-Pro-Leu. As known (see Fig. 5), the amino acid Ser can be encoded by 6 triplets (TCC, TCT, TCA, TCG, AGC, AGT), the amino acid Pro – by 4 triplets (CCC, CCT, CCA, CCG) and the amino acid Leu – by 6 triplets (CTC, CTT, CTA, CTG, TTA, TTG). Due to this, the short sequence Ser-Pro-Leu can be encoded by means of 144 (=6*4*6) different variants of a sequence of 3 triplets: TCC-CCC-CTC, TCT-CCT-CTG, AGT-CCT-TTG, etc. If more and more long sequences of amino acids are considered (some proteins have sequences with many thousands of amino acids in them), the number of variants of encoding increases rapidly to astronomic quantities.

Why living matter needs such a tremendously excessive number of encoding options, which can greatly complicate the work of reliable coding system? This is a difficult question without a satisfactory explanation in modern science.

Taking into account that in the geno-logic code no degeneracy exists (see below), we think that a good answer on this question is the following. DNA (RNA) molecules are carriers at least two different genetic codes: in one of them triplets encode amino acids, and in the second code the same triplets (or n-plets in general case) encode genetic systems of Boolean functions. In this case, existence of different variants of sequences of triplets, which encode the same amino acid sequence, has a sense since any two different sequences of triplets, which encode the same amino acid sequence, simultaneously encode different systems of Boolean functions. One can suppose that the amino acid sequence, encoded by the first set of triplets, is designed to work in cooperation with a special genetic system of Boolean functions (which is encoded by the first set of triplets); by contrast, the same amino acid sequence, encoded by the second set of triplets, is designed to work in cooperation with another genetic system of Boolean functions (which is encoded by the second set of triplets). This implies that identical proteins are intended for use in different genetic systems of Boolean functions, if they are encoded by different sequences of triplets. And the fate of the protein depends on its membership of a particular genetic system of Boolean functions. It leads to a conclusion that genetic systems of Boolean functions dominate over the set of encoded proteins.

Such standpoint gives new abilities to understand some genetic phenomena. For example, the existence of introns, which do not encode proteins (unlike exons in nucleotide sequences), is explained by the fact that introns encode a realization of genetic systems of Boolean functions without adding of amino acids into proteins. The known problem of jumping genes [McClintock, 1950] can also be related with the existence of two kinds of the genetic code in DNA: jumping of DNA fragments need for changing of encoded systems of Boolean functions without changing the amino acid sequence in the encoded protein.

For developing theory of the geno-logic code one needs to study possible variants of natural relations of molecular-genetic alphabets with systems of Boolean functions and their methods of anaylizes developed in mathematics, including spectral logics with using Walsh functions, Kronecker (or tensor) product and Fibonacci transforms [Karpovsky, Stankovic, Astola, 2008; Stankovic, Astola, 2003; Zalmazon, 1989]. These known mathematical methods pay special attention to Walsh-Hadamard spectra and dyadic autocorrelation characteristics of systems of Boolean functions. Different classes of Boolean functions exist: linear, self-dual, anti-self-dual, threshold and other Boolean functions. Knowledge of a class of Boolean functions is important for the synthesis of devices that implement complex logic functions or their systems since for each of such classes specific synthesis methods are known. Dyadic autocorrelation characteristics, which are coinside with dyadic convolutions [Ahmed, Rao, 1975], are traditionally used to get this knowledge. Below we will indicate Walsh-Hadamard spectra and autocorrelation characteristics for each case of Boolean functions or their systems in the representation of genetic elements as elements of algebra of logic.

One can remind that for a $2^n$-dimensional dimensional vector $\overline{V}$ its Walsh-Hadamard spectrum $\overline{S}$ (in relation to a certain Hadamard matrix H of the same order $2^n$) is calculated by means of the expression (37):

$$\overline{S} = \overline{V}*H \qquad (37)$$

The dyadic autocorrelation characteristic (the dyadic convolution) Z(m) of a real-valued function X(m), where m=0, 1, 2,…, N-1, is calculated [Ahmed, Rao,1975] by means of the expression (38):

$$Z(m) = \left(\frac{1}{N}\right) * \sum_{h=0}^{N-1} X(h)X(m \oplus h) \ , \qquad (38)$$

where ⊕ means the logical operation of modulo-2 addition for binary numbers.

Let us describe the first of possible variants of the representation of elements of molecular-genetic alphabets in forms of Boolean functions and their systems for further studying and classifications.

The four nytrogenous bases of DNA (or RNA) correspond to the dyadic group of 2-bit binary numbers $X_0X_1$, which contains four members: 00, 01, 10, 11 (each of symbols $X_0$ and $X_1$ can take only two values – 0 or 1). Previously, in the pair of binary-oppositional attributes "amino-keto" we have identified the attribute "amino" as 0, and the attribute "keto" as 1 (Fig. 3). Similarly, in the pair of binary-oppositional attributes "pyrimidine-purine" we have identified the attribute "pyrimidine" as 0 and the attribute "purine" as 1. In the DNA alphabet, each of the nitrogenous bases is uniquely determined by the indication of its pair of attributes from the set: "amino", "keto", "pyrimidine" or "purine". Each nitrogenous base can be denoted by its individual 2-bit number, in which the first position corresponds to its attribute from the binary-oppositional pair "amino-keto" and the second position corresponds to its attribute from the second binary-oppositional pair "pyrimidine-purine". In this case cytosine C is denoted by number 00 (it bears the attributes "amino" and "pyrimidine"); adenine A – by number 01 ("amino" and "purine"); thymine T – by number 10 ("keto" and "pyrimidine"); guanine G – by number 11 ("keto" and "purine"). The same is true for the nitrogenous bases of RNA. (In our concept of systemic-resonant genetics, these attributes are associated with resonances, which are related with these molecular features (see above Appendix B)).

In the described variant of algebraic-logical representation of molecular elements of the genetic system we will follow the methods of spectral logic, which have been developed for engineering purposes [Karpovsky, Stankovic, Astola, 2008; Zalmazon, 1989]. According to these general methods, in our case the 2-bit inputs $X_0X_1$, the following Boolean functions $f_C(X_0X_1)$, $f_A(X_0X_1)$, $f_T(X_0X_1)$ and $f_G(X_0X_1)$ can be defined, which represent, respectively, cytosine, adenine, thymine and guanine in a form of 4-dimensional Boolean vectors (Fig. 60):

| $X_0X_1$ | $f_C(X_0X_1)$ | $f_A(X_0X_1)$ | $f_T(X_0X_1)$ | $f_G(X_0X_1)$ |
|---|---|---|---|---|
| 00 | 1 | 0 | 0 | 0 |
| 01 | 0 | 1 | 0 | 0 |
| 10 | 0 | 0 | 1 | 0 |
| 11 | 0 | 0 | 0 | 1 |

Figure 60. Boolean functions $f_C(X_0X_1)$, $f_A(X_0X_1)$, $f_T(X_0X_1)$ and $f_G(X_0X_1)$, which represent cytosine, adenine, thymine and guanine

Referring now to the alphabet 16 of doublets, each of which can be represented as an ordered system of the two individual Boolean functions from Fig. 60, which correspond to nitrogenous bases in the first and second positions of the doublet. Systems of Boolean functions are analyzed and synthesized by so-called step functions Φ, which are associated with trigger states of electrical networks [Karpovsky, Stankovic, Astola, 2008, § 1.2]. Step functions Φ are obtained from the lattice functions f(z), representing the studied system of Boolean functions, by means of the following expression:

$$\Phi(z) = f(k), \quad \text{when } z \in [k, k+1),\ k = 0, 1, \ldots, n-1, \qquad (39)$$

where n means dimensionality of Boolean vectors.

Let us explain this with an example of the doublet CG, for which we construct a lattice function $f_{CG}$ (Fig. 61). Each component of the function $f_{CG}$ is equal to the decimal expression of a binary number, composed of individual components of Boolean functions $f_C(X_0X_1)$ and $f_G(X_0X_1)$ from the left side in the same row of the table. The step function $\Phi_{CG}$, which corresponds to the lattice function $f_{CG}$, is shown in the right side of Fig. 61.

| $X=X_0X_1$ | $f_C(X_0X_1)$ | $f_G(X_0X_1)$ | $f_{CG}$ |
|---|---|---|---|
| 00 | 1 | 0 | 2 |
| 01 | 0 | 0 | 0 |
| 10 | 0 | 0 | 0 |
| 11 | 0 | 1 | 1 |

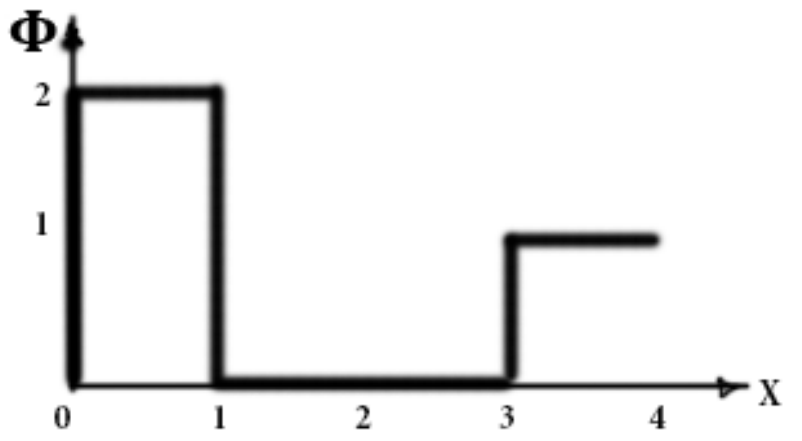

Figure 61. The example of the representation of the doublet CG by means of the lattice function $f_{CG}$ = [2, 0, 0, 1] and the corresponding step function Φ (on the right).

The table in Fig. 62 shows the data about lattice functions, which represent 16 doublets of DNA, their spectra Walsh-Hadamard (with respect to the Hadamard matrix $[1, 1; 1, -1]^{(2)}$) and dyadic autocorrelation characteristics of the corresponding step functions. Dyadic autocorrelations coincide with dyadic convolutions and are calculated according to (38).

| Doublet | Lattice function f | Walsh-Hadamard spectrum | Dyadic autocorrelation |
|---|---|---|---|
| CC | [3, 0, 0, 0] | [3, 3, 3, 3] | [2.25, 0, 0, 0] |
| CA | [2, 1, 0, 0] | [3, 1, 3, 1] | [1.25, 1, 0, 0] |
| AC | [1, 2, 0, 0] | [3, -1, 3, -1] | [1.25, 1, 0, 0] |
| AA | [0, 3, 0, 0] | [3, -3, 3, -3] | [2.25, 0, 0, 0] |
| CT | [2, 0, 1, 0] | [3, 3, 1, 1] | [1.25, 0, 1, 0] |
| CG | [2, 0, 0, 1] | [3, 1, 1, 3] | [1.25, 0, 0, 1] |
| AT | [0, 2, 1, 0] | [3, -1, 1, -3] | [1.25, 0, 0, 1] |
| AG | [0, 2, 0, 1] | [3, -3, 1, -1] | [1.25, 0, 1, 0] |
| TC | [1, 0, 2, 0] | [3, 3, -1, -1] | [1.25, 0, 1, 0] |
| TA | [0, 1, 2, 0] | [3, 1, -1, -3] | [1.25, 0, 0, 1] |
| GC | [1, 0, 0, 2] | [3, -1, -1, 3] | [1.25, 0, 0, 1] |
| GA | [0, 1, 0, 2] | [3, -3, -1, 1] | [1.25, 0, 1, 0] |
| TT | [0, 0, 3, 0] | [3, 3, -3, -3] | [2.25, 0, 0, 0] |
| TG | [0, 0, 2, 1] | [3, 1, -3, -1] | [1.25, 1, 0, 0] |
| GT | [0, 0, 1, 2] | [3, -1, -3, 1] | [1.25, 1, 0, 0] |
| GG | [0, 0, 0, 3] | [3, -3, -3, 3] | [2.25, 0, 0, 0] |

Figure 62. Lattice functions of 16 doublets, their Walsh-Hadamard spectra and dyadic autocorrelations.

By analogy, for triplets and n-plets (n = 4, 5, 6, ...) one can also construct their representation in the form of systems of Boolean functions on the basis of individual Boolean functions of nitrogenous bases. Let us explain this with an example of the triplet CAG, for which we construct a lattice function $f_{CAG}$ (Fig. 63). Each component of the function $f_{CAG}$ is equal to the decimal expression of a binary number, composed of components of Boolean functions $f_C(X_0X_1)$, $f_A(X_0X_1)$ and $f_G(X_0X_1)$ from the left side in the same row of the table. The step function $\Phi_{CAG}$, which corresponds to the lattice function $f_{CAG}$, is shown in the right side of Fig. 63.

| $X=X_0X_1$ | $f_C(X)$ | $f_A(X)$ | $f_G(X)$ | $f_{CAG}$ |
|---|---|---|---|---|
| 00 | 1 | 0 | 0 | 4 |
| 01 | 0 | 1 | 0 | 2 |
| 10 | 0 | 0 | 0 | 0 |
| 11 | 0 | 0 | 1 | 1 |

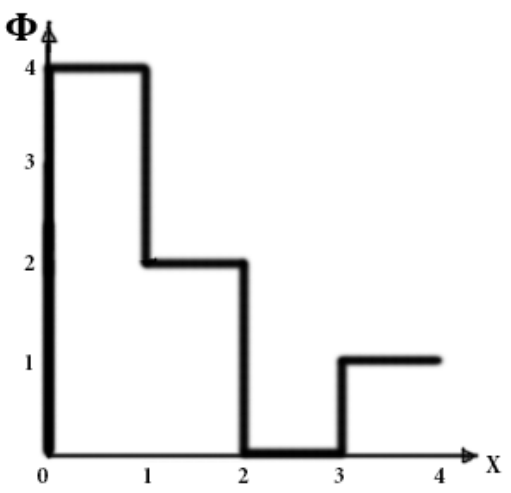

Figre 63. The example of the representation of the triplet CAG by means of the lattice function $f_{CAG}$ = [4, 2, 0, 1] and the corresponding step function Φ (on the right).

The table in Fig. 64 shows data about lattice functions, which represent 64 triplets of DNA, their Walsh-Hadamard spectra (relative Hadamard matrix $[1, 1; 1, -1]^{(2)}$) and dyadic autocorrelations of corresponding step functions.

| Triplet | Its lattice function f | Walsh-Hadamard spectrum | Dyadic autocorrelation |
|---|---|---|---|
| CCC | [7, 0, 0, 0] | [7, 7, 7, 7] | [12.25, 0, 0, 0] |
| CCA | [6, 1, 0, 0] | [7, 5, 7, 5] | [9.25, 3, 0, 0] |
| CAC | [5, 2, 0, 0] | [7, 3, 7, 3] | [7.25, 5, 0, 0] |
| CAA | [4, 3, 0, 0] | [7, 1, 7, 1] | [6.25, 6, 0, 0] |
| ACC | [3, 4, 0, 0] | [7, -1, 7, -1] | [6.25, 6, 0, 0] |
| ACA | [2, 5, 0, 0] | [7, -3, 7, -3] | [7.25, 5, 0, 0] |
| AAC | [1, 6, 0, 0] | [7, -5, 7, -5] | [9.25, 3, 0, 0] |
| AAA | [0, 7, 0, 0] | [7, -7, 7, -7] | [12.25, 0, 0, 0] |
| CCT | [6, 0, 1, 0] | [7, 7, 5, 5] | [9.25, 0, 3, 0] |
| CCG | [6, 0, 0, 1] | [7, 5, 5, 7] | [9.25, 0, 0, 3] |
| CAT | [4, 2, 1, 0] | [7, 3, 5, 1] | [5.25, 4, 2, 1] |
| CAG | [4, 2, 0, 1] | [7, 1, 5, 3] | [5.25, 4, 1, 2] |
| ACT | [2, 4, 1, 0] | [7, -1, 5, -3] | [5.25, 4, 1, 2] |
| ACG | [2, 4, 0, 1] | [7, -3, 5, -1] | [5.25, 4, 2, 1] |
| AAT | [0, 6, 1, 0] | [7, -5, 5, -7] | [9.25, 0, 0, 3] |
| AAG | [0, 6, 0, 1] | [7, -7, 5, -5] | [9.25, 0, 3, 0] |
| CTC | [5, 0, 2, 0] | [7, 7, 3, 3] | [7.25, 0, 5, 0] |
| CTA | [4, 1, 2, 0] | [7, 5, 3, 1] | [5.25, 2, 4, 1] |
| CGC | [5, 0, 0, 2] | [7, 3, 3, 7] | [7.25, 0, 0, 5] |
| CGA | [4, 1, 0, 2] | [7, 1, 3, 5] | [5.25, 2, 1, 4] |
| ATC | [1, 4, 2, 0] | [7, -1, 3, -5] | [5.25, 2, 1, 4] |
| ATA | [0, 5, 2, 0] | [7, -3, 3, -7] | [7.25, 0, 0, 5] |
| AGC | [1, 4, 0, 2] | [7, -5, 3, -1] | [5.25, 2, 4, 1] |
| AGA | [0, 5, 0, 2] | [7, -7, 3, -3] | [7.25, 0, 5, 0] |
| CTT | [4, 0, 3, 0] | [7, 7, 1, 1] | [6.25, 0, 6, 0] |
| CTG | [4, 0, 2, 1] | [7, 5, 1, 3] | [5.25, 1, 4, 2] |
| CGT | [4, 0, 1, 2] | [7, 3, 1, 5] | [5.25, 1, 2, 4] |
| CGG | [4, 0, 0, 3] | [7, 1, 1, 7] | [6.25, 0, 0, 6] |
| ATT | [0, 4, 3, 0] | [7, -1, 1, -7] | [6.25, 0, 0, 6] |
| ATG | [0, 4, 2, 1] | [7, -3, 1, -5] | [5.25, 1, 2, 4] |
| AGT | [0, 4, 1, 2] | [7, -5, 1, -3] | [5.25, 1, 4, 2] |
| AGG | [0, 4, 0, 3] | [7, -7, 1, -1] | [6.25, 0, 6, 0] |
| TCC | [3, 0, 4, 0] | [7, 7, -1, -1] | [6.25, 0, 6, 0] |
| TCA | [1, 1, 4, 0] | [7, 5, -1, -3] | [5.25, 1, 4, 2] |
| TAC | [1, 2, 4, 0] | [7, 3, -1, -5] | [5.25, 1, 2, 4] |
| TAA | [0, 3, 4, 0] | [7, 1, -1, -7] | [6.25, 0, 0, 6] |
| GCC | [3, 0, 0, 4] | [7, -1, -1, 7] | [6.25, 0, 0, 6] |
| GCA | [2, 1, 0, 4] | [7, -3, -1, 5] | [5.25, 1, 2, 4] |
| GAC | [1, 2, 0, 4] | [7, -5, -1, 3] | [5.25, 1, 4, 2] |
| GAA | [0, 3, 0, 4] | [7, -7, -1, 1] | [6.25, 0, 6, 0] |
| TCT | [2, 0, 5, 0] | [7, 7, -3, -3] | [7.25, 0, 5, 0] |
| TCG | [2, 0, 4, 1] | [7, 5, -3, -1] | [5.25, 2, 4, 1] |
| TAT | [0, 2, 5, 0] | [7, 3, -3, -7] | [7.25, 0, 0, 5] |
| TAG | [0, 2, 4, 1] | [7, 1, -3, -5] | [5.25, 2, 1, 4] |
| GCT | [2, 0, 1, 4] | [7, -1, -3, 5] | [5.25, 2, 1, 4] |
| GCG | [2, 0, 0, 5] | [7, -3, -3, 7] | [7.25, 0, 0, 5] |
| GAT | [0, 2, 1, 4] | [7, -5, -3, 1] | [5.25, 2, 4, 1] |
| GAG | [0, 2, 0, 5] | [7, -7, -3, 3] | [7.25, 0, 5, 0] |
| TTC | [1, 0, 6, 0] | [7, 7, -5, -5] | [9.25, 0, 3, 0] |
| TTA | [0, 1, 6, 0] | [7, 5, -5, -7] | [9.25, 0, 0, 3] |
| TGC | [1, 0, 4, 2] | [7, 3, -5, -1] | [5.25, 4, 2, 1] |
| TGA | [0, 1, 4, 2] | [7, 1, -5, -3] | [5.25, 4, 1, 2] |
| GTC | [1, 0, 2, 4] | [7, -1, -5, 3] | [5.25, 4, 1, 2] |
| GTA | [0, 1, 2, 4] | [7, -3, -5, 1] | [5.25, 4, 2, 1] |
| GGC | [1, 0, 0, 6] | [7, -5, -5, 7] | [9.25, 0, 0, 3] |
| GGA | [0, 1, 0, 6] | [7, -7, -5, 5] | [9.25, 0, 3, 0] |
| TTT | [0, 0, 7, 0] | [7, 7, -7, -7] | [12.25, 0, 0, 0] |
| TTG | [0, 0, 6, 1] | [7, 5, -7, -5] | [9.25, 3, 0, 0] |
| TGT | [0, 0, 5, 2] | [7, 3, -7, -3] | [7.25, 5, 0, 0] |
| TGG | [0, 0, 4, 3] | [7, 1, -7, -1] | [6.25, 6, 0, 0] |
| GTT | [0, 0, 3, 4] | [7, -1, -7, 1] | [6.25, 6, 0, 0] |
| GTG | [0, 0, 2, 5] | [7, -3, -7, 3] | [7.25, 5, 0, 0] |
| GGT | [0, 0, 1, 6] | [7, -5, -7, 5] | [9.25, 3, 0, 0] |
| GGG | [0, 0, 0, 7] | [7, -7, -7, 7] | [12.25, 0, 0, 0] |

Figure 64. Lattice functions of 64 triplets, their Walsh-Hadamard spectra and dyadic autocorrelations.

The language of systems of Boolean functions can be also used to study the phenomenon of the symmetrical division of the set of 16 doublets into two equal subsets with 8 «strong» doublets and 8 «weak» doublets (they are marked by black and white colors in Fig. 4 and 64). This division is related with a degeneration of the genetic code and with encoding of amino acids by triplets (see above Fig. 5). Let us represent separate families of 16 doublets using the genetic matrices in Fig. 65, where 4 monoplets and 16 doublets have their dyadic-shift numerations, which were introduced due to the binary-oppositional attributes of nitrogenous bases of DNA.

**[C A; T G] =**

| | **0** | **1** |
|---|---|---|
| **0** | **C<br>0** | **A<br>1** |
| **1** | **T<br>1** | **G<br>0** |

**; [C A; T G]$^{(2)}$ =**

| | **00(0)** | **01(1)** | **10(2)** | **11(3)** |
|---|---|---|---|---|
| **00<br>(0)** | **CC<br>00 (0)** | **CA<br>01(1)** | **AC<br>10 (2)** | **AA<br>11 (3)** |
| **01<br>(1)** | **CT<br>01 (1)** | **CG<br>00 (0)** | **AT<br>11 (3)** | **AG<br>10 (2)** |
| **10<br>(2)** | **TC<br>10 (2)** | **TA<br>11(3)** | **GC<br>00 (0)** | **GA<br>01 (1)** |
| **11<br>(3)** | **TT<br>11 (3)** | **TG<br>10 (2)** | **GT<br>01 (1)** | **GG<br>00 (0)** |

Figure 65. Genetic matrices for nitrogeous bases and doublets of DNA (similar to Fig. 4).

Each of members of the complentary pair C and G has its binary numeration 0 inside the dyadic group of one-digit binary numbers $X_0$: 0, 1 (Fig. 65, on he left). Each of members of the second complementary pair A and T has its binary numeration 1 inside the same dyadic group. Correspondingly any of two members of the complementary pair C and G can be represented by the same Boolean function $f_{CG}$ (a 2-dimensional Boolean vector) and any of two members of the second complementary pair A and T can be represented by the same Boolean function $f_{AT}$ (Fig. 66).

| $X_0$ | $f_{CG}$ | $f_{AT}$ |
|---|---|---|
| 0 | 1 | 0 |
| 1 | 0 | 1 |

Figure 66. Boolean functions $f_{CG}$ and $f_{AT}$, which represent correspondingly members of the complementary pairs C-G and A-T.

The set of 16 doublets is divided into 4 subsets, each of which contains 4 doublets with identical two-digit binary numbers (Fig. 65, on the right):

- the subset with binary numeration 00 includes 4 strong doublets CC, CG, GC,GG; each of them can be represented by the lattice function $f_{00}$ of the system of two identical Boolean functions $f_{CG}$ and $f_{CG}$ (Fig. 67 and 68);
- the subset with binary numeration 01 includes 2 weak doublets CA, CG and 2 strong doublets CT, GT; each of them can be represented by the lattice function $f_{01}$ of the system of two different Boolean functions $f_{CG}$ and $f_{AT}$ (Fig. 68);
- the subset with binary numeration 10 includes 2 srong doublets AC, TC and 2 weak doublets AG, TG; each of them can be represented by the lattice function $f_{10}$ of two different Boolean functions $f_{AT}$ and $f_{CG}$ (Fig. 68);
- the subset with binary numeration 11 includes 4 weak doublets AA, AT, TA, TT; each of them can be represented by the lattice function $f_{11}$ of the system of two identical Boolean functions $f_{AT}$ and $f_{AT}$ (Fig. 68).

Fig. 67 shows an example of a construction of the lattice function $f_{00}$ of the system of two identical Boolean functions $f_{CG}$ and $f_{CG}$.

| $X_0$ | $f_{CG}$ | $f_{CG}$ | $f_{00}$ |
|---|---|---|---|
| 0 | 1 | 1 | 3 |
| 1 | 0 | 0 | 0 |

Fig. 67. The example of the representation of the subset of doublets CC, CG, GC, GG by means of the lattice function $f_{00}$ = [3, 0]

| Subsets of doublets | Lattice function | Walsh-Hadamard spectrum | Dyadic autocorrelation |
|---|---|---|---|
| CC, CG, GC, GG | $f_{00}$ = [3, 0] | [3, 3] | [4.5, 0] |
| CA, CG, CT, GT | $f_{01}$ = [2, 1] | [3, 1] | [2.5, 2] |
| AC, TC, AG, TG | $f_{10}$ = [1, 2] | [3, -1] | [2.5, 2] |
| AA, AT, TA, TT | $f_{11}$ = [0, 3] | [3, -3] | [4.5, 0] |

Figure 68. Lattice functions of 4 subsets of doublets, their Walsh-Hadamard spectra (relative to the Hadamard matrix [1, 1; 1, -1]) and dyadic autocorrelations.

In our spectral-logic approach, genetic lattice functions are connected with genetic information, which is duplicated under mitotic divisions of cells. Does an appropriate mathematical operation exist in the spectral logic for corresponding duplication of genetic lattice functions? Yes, such operation exists and it is based on a simple repetition of Walsh-Hadamard transformations since the square of symmetric Hadamard matrices satisfies to the following expression: $([1, 1; 1, -1]^{(n)})^2 = 2^n*E$, where E is identity matrix, (n) – the Kronecker power, n = 1, 2, 3,… . For example, $[x_0, x_1]*[1, 1; 1, -1]^2 = 2*[x_0, x_1]$ that is, this operation duplicates the initial vector $[x_0, x_1]$.

The geno-logic approach gives new abilities to create mathematical models of inherited cyclic processes. In genetics, it is well known the principle of selectivity of reading information from DNA, which allows a switching of the synthesis of proteins of one kind into other kinds on different stages of development of organisms by means of different variants of reading of information from the same DNA. By analogy the similar principle of selectivity of reading information from genetic lattice functions can be used in a simulation of cyclic processes, which is related with genetic lattice functions. One should remind here that the spectral logic analyses and creates digital devices with Boolean signals (0 or 1) in sets of input channels $X_i$ and output channels $Y_k$. The system of Boolean functions of output channels $Y_k$ define an appropriate lattice function and its step function, a Walsh-Hadamard spectrum of which is studied in the spectral logic. For example, such lattice function, which can characterise a current stage of an inherited structure, can be represented by the vector [4, 2, 0, 1] from Fig. 63. Its Walsh-Hadamard spectrum in relation to the Hadamard matrix $[1, 1; 1, -1]^{(2)}$ is represented by the vector $[4, 2, 0, 1]* [1, 1; 1, -1]^{(2)} = [7, 1, 5, 3]$. But components of this lattice function can be read in changed order of priorities, generated, for instance, by means of simple cyclic permutations: [2, 0, 1, 4], [0, 1, 4, 2], [1, 4, 2, 0], [4, 2, 0, 1], each of which has its own Walsh-Hadamard spectrum. Such cyclic sequences of genetic lattice functions can be used in a simulation of inherited cyclic processes. Separate cyclic sequences of genetic lattice functions can be combined by different ways into a whole network to model many mutually coordinated cyclic processes. In particular, here one can wait for a new knowledge on biological meaning of those linear codes, which are related with Boolean functions.

In the spectral logic, genetic lattice functions (Fig, 61, 63) are represented by $2^n$-dimensional vectors with integer non-negative entries since these entries represent decimal expressions of binary numbers. But one can ask about a type of geometric spaces, to which vectors of these lattice functions belong. For example, $2^n$-dimensional vectors can belong to spaces, which are related with Kronecker extensions of the algebra of complex numbers $([x_0, x_1; - x_1, x_0]^{(n)})$, or which are related with Kronecker extensions of the algebra of hyperbolic numbers $([x_0, x_1; x_1, x_0]^{(n)})$. Above we have shown that phenomenological structures of genetic alphabets of 16 doublets and 64 triplets, represented in black-and-white genomatrices, are connected with matrix representations of the Kronecker's exensions of algebras of hyperbolic and complex numbers, including coquaternions of Cockle and quaternions of Hamilton (Fig. 4, 22, 23, 26, 27, 28, 30). In light of this, one can think that vectors of genetic lattice functions can be interpreted as corresponding hypercomplex numbers with integer non-negative components. For example, the lattice function [5, 3] can be interpreted as the hyperbolic number 5+3*i, where $i^2=+1$ (the matrix representation of this hyperbolic number is [5, 3; 3, 5]). Multiplication of two hyperbolic numbers, each of which has only integer non-negative components, gives a new hyperbolic number with integer non-negative components. The set of hyperbolic numbers with integer non-negative components forms a semigroup in relation to multiplication. Corresponding semigroups - in relation to multiplication - take place in some other cases of the interpretation of $2^n$-dimensional vectors of lattice functions as hypercomplex numbers of different types, including the cases of the interpretation of 4-dimensional vectors with integer non-negative components as coquaternions of Cockle or as quaternions of Hamilton. In this way we propose to use associative algebras of hypercomplex numbers and also the important algebraic notion of semigroups for study genetic structures. In our opinion, for genetic sequences the so called «free semigroup», which is related with string concatenation, has a special importance (see its definition and properties in https://en.wikipedia.org/wiki/Free_monoid).

Of course, the described representations should be considered as initial materials for further researches of genetic systems and biological organisms from the standpoint of Boolean algebra of logic and spectral logic for deeper understanding of living matter and for developing of “algebra-logical biology”. One of many secrets of living matter is an extremely effective work of enzymes inside living bodies. As known biological enzymes can accelerate a chemical reaction in $10^{10}$-$10^{14}$ times [Varfolomeev, 2005]. Enzymes are millions of times more effective than catalysts in laboratory kinds of catalysis, which are known today. What makes the enzyme for 1 second, a conventional catalyst can make only for 200,000 years. We believe that such efficiency of enzymes in biological bodies is related with cooperative algebra-logical mechanisms of the geno-logic coding but not only with known aspects of chemical reactions. Here one can additionally mention that there are only 4 kinds of active centers of enzymes for a great number of enzymes. The set of these 4 kinds can be compared with 4 members of the dyadic group of 2-bit binary numbers 00, 01, 10, 11 by analogy with the case of 4 nitrogenous bases of DNA and RNA, which form complementary pairs, and by analogy with the case of 4 kinds of histones, which also form the complementary pairs H3-H4 and H2A-H2B.

From the described point of view, life is a complex system of logical functions and logical operations, which possesses properties of self-organization, self-development, noise-immunity, communications, etc. If the genetic system is closely related with Boolean algebra of logic, then it is natural to think that our sensory systems (visual, auditory, tactile and other) get and processing information from the world by means of Boolean variables and of systems of Boolean functions.

George Boole created his mathematics of logic to desribe the laws of thought: his book was titled «An Investigation of the Laws of Thoughts» (1854 year). Our reasoned statement about the existence of the geno-logic code shows that our genetically encoded body is created on the basis of the same laws of logic, on which our thoughts are constructed (the unity of the laws of thought and body). It gives a new material for a discussion about old problem: what is primary - thoughts or matter?

If our thought is built on the same laws, on which our body is built, then all intellectual activity of man, including a creation of theoretical sciences, is subordinated to logical laws of genetics. This intellectual activity reminds instinctive activities of some animals that are creating instinctively their building structures (such as bees with their honeycombs and spiders with their webs). In this light, human intellectual work reminds also the innate ability of animals to motor movements with a congenital coordination of a lot of muscle units, etc. This new direction of biological researches and comparative analizis includes, in particular, problems of musicology and so called "genetic music" (Gottfried Leibniz declared that music is arithmetic of soul, which computes without being aware of it). Some of the author's materials will be presented in his lecture on "Symmetry Festival-2016" (Vienna, Austria, July 18-23, http://festival.symmetry.hu ).

Here one can return to the mentioned fact that our body is a huge network of triggers of different types, different biological level, different functionality and material incarnation, including trigger subnets of genetic molecules, tubulin proteins, muscle fibers and neurons. The network of trigger devices of genetic molecules works to provide genetic information and genetic realization of all other trigger subnets of the body; the trigger subnet of tubulin proteins, which is very important for functional systems of the body tissue; the trigger subnet of muscle fibers, which provides the work of muscles; the trigger subnet of neurones, which is essential for information processing in the nervous system. But all these trigger subnets of an entire organism, which possess their relative autonomy, functional specificity and different material embodiment, are able to work in concert on the principles of Boolean algebra of logic by analogy with computers.

The trigger network of genetic molecules participates in the genetic realization of all other trigger networks of the body, and - in this sense - it is the most fundamental, which defines algorithms and relationships of all the trigger networks of a whole organism. From this perspective, biological evolution can be represented as a process of self-organization and self-development of systems of biological trigger networks.

With respect to effects of physiotherapy, including methods of ancient oriental medicine (acupuncture, etc.), one can think that medical intervention in the functioning of the organism acts in many aspects through its impact on the network of logical functions of the body, causing its restructuring. This is similar to the situation with a computer, when one click on the desired key of its keyboard is able to change all its work due to activation of another way in its logical networks. We believe that gerontological phenomena of aging and death are also closely interfaced with the system of trigger networks of genetic logical functions. Perhaps the Yin-Yang system of ancient Chinese book "I Ching" (see Appendix A), underlying the ancient Eastern medicine and diagnosis of the organism, in combination with the knowledge about the geno-logic code of DNA will be useful for further development of medicine.

The desire to reproduce in the technical device the operating principles of natural neural networks has given rise to an important technology of "neural computers". But trigger neural networks are just one of the types of genetically inherited trigger networks of the body (many organisms live perfectly without the nervous system). It seems that a similar desire to reproduce - in technical devices - operating principles of genetically inheritable sets of body's trigger networks is capable to generate a wider direction of "computers on genetically programmable systems of plenty of trigger networks" or briefly "geno-computers". Geno-computers differs from neurocomputers primarily in the fact that it is a union of many trigger networks endowed with different functions, different levels of autonomy, as well as the different characteristics of functional, spatial and temporal linkages with many other trigger networks of a unified device. By contrast to neurocomputers, which are created as imitators of properties of the nervous system, the creation of geno-computers is designed to imitate properties of biological bodies and their evolution, understood as a genetic process of self-organization and self-development of biological sets of trigger networks, including the emergence of new and new trigger networks (analogs of networks of neurons, muscle fiber networks, etc.).

There are many definitions of life proposed by various authors including E.Schroedinger, N.Bohr and I.Prigogine. One of them emphasizes the important role of the genetic system: "*life is a partnership between genes and mathematics*" [Stewart, 1999]. From the standpoint of our theory of genetic logic holography, the following new definition of life seems to be interesting: life is a complex system of genetic Boolean functions and their spectral logic with using Wals-Hadamard transformations and genetic logical holograms.

Materials about the geno-logic code and about living matter as a territory of systems of Boolean functions gives rise to many new questions. One of them is the question concerning a role of water in living bodies. As known, jellyfish consist of 99% water, but despite of this they live perfectly and their morphology implements heritable phenomena of phyllotaxis: tentacles, canals and zooids of some jellyfish exactly correspond to phyllotaxis laws [Jean, 1994, Chapter 12.3.3]. If such living bodies are genetically organized on the base of logic functions, then one can assume that water is also related with logic functions in it, which can be provided by systems of its molecular resonances and solitons (by analogy with computer systems, where logic functions exist on the base of electric signals). The author puts forward the hypothesis that structural water can be also carrier of logic functions and logic operations, which participate in some of water phenomena, for instance, homeopathic phenomena. It seems to be a new aspect of study of water, which is analyzed traditionally only from a physico-chemical standpoint without the idea of a participation of laws of algebra of logic in water properties.

## Appendix E. The geno-logic code and Fibonacci numbers

Inherited phyllotaxis laws of morphogenesis, widely known long ago, are related with Fibonacci numbers from the recurrent series of real numbers $F_{n+2} = F_n + F_{n+1}$ (Fig. 69). The ratio $F_{n+1}/F_n$ tends to the golden section $(1+5^{0.5})/2$ = 1,618… with increasing n.

| $F_n$ | 0 | 1 | 1 | 2 | 3 | 5 | 8 | 13 | 21 | 34 | 55 | 89 | 144 | … |
|---|---|---|---|---|---|---|---|---|---|---|---|---|---|---|
| n | 0 | 1 | 2 | 3 | 4 | 5 | 6 | 7 | 8 | 9 | 10 | 11 | 12 | … |

Figure 69. The additive series of Fibonacci numbers,

The algorithmic phyllotaxis structures occur at very different levels and branches of biological evolution: from alpha helixes of polypeptide chains till bodies of plants and animals (hundreds of publications are devoted to them, including the book with their references [Jean, 1994]). Some authors devoted their researches to Fibonacci numbers in genetic sequences [Beleza Yamagishi, Shimabukuro, 2008; Negadi, 2015; Perez, 1991]. The Fibonacci-stage scales are the bases of genetic music developed in the Moscow Conservatory [Darvas et al, 2012; Koblyakov, Petoukhov, Stepanyan, 2016; Petoukhov, 2008; Petoukhov, Petukhova, 2017].

In mathematics the matrix Q = [1, 1;  1, 0] is called «the Fibonacci matrix» since its integer powers "n" give matrices with Fibonacci entries only: $Q^n = [F_{n+1}, F_n; F_n, F_{n-1}]$. The Fibonacci matrix Q sometimes is called as the "growth matrix" since it was used many times in biomathematics for simulation of phyllotaxis phenomena [Jean, 1994]. From the standpoint of our geno-logic approach, it is natural to think that Fibonacci numbers and the Fibonacci matrix Q are related with the geno-logic code and correspondingly they have interesting mathematical interrelations with formalisms of spectral logic of Boolean functions described in [Karpovsky, Stankovic, Astola, 2008; Zalmanzon, 1989]. We reveal that it is true and such interrelations really exist. Let us show some examples. Below we use the following terms: vectors, all entries of which are Fibonacci numbers, are called "Fibonacci vectors"; vectors, all entries of which are Boolean variables 0 and 1, are called "Boolean vectors".

Rows of the Fibonacci matrix Q are Boolean vectors [1 1] and [1 0]. Multiplication of the [2*2]-matrix $Q^n$ with any of Boolean 2-dimensional non-zero vectors [0, 1], [1, 0] or [1, 1] gives a Fibonacci vector (Fig. 70).

| | | | | |
|---|---|---|---|---|
| $[0, 1]*Q = [1, 0]$ | $[0, 1]*Q^2 = [1, 1]$ | $[0, 1]*Q^3 = [2, 1]$ | $[0, 1]*Q^4 = [3, 2]$ | … |
| $[1, 0]*Q = [1, 1]$ | $[1, 0]*Q^2 = [2, 1]$ | $[1, 0]*Q^3 = [3, 2]$ | $[1, 0]*Q^4 = [5, 3]$ | … |
| $[1, 1]*Q = [2, 1]$ | $[1, 1]*Q^2 = [3, 2]$ | $[1, 1]*Q^3 = [5, 3]$ | $[1, 1]*Q^4 = [8, 5]$ | … |

Figure 70. Multiplications of Boolean vectors with the Fibonacci matrix $Q^n$

One can see from Fig. 71 that an additive series of Fibonacci vectors $V_{k+1}=V_k+V_{k-1}$ arises (Fig. 9) by analogy with the series of Fibonacci numbers (Fig. 69). In each of vectors $V_k$ sum of squares of its components is equal to a Fibonacci number: $F_n^2 + F_{n+1}^2 = F_{2n+1}$.

And what one can say about Walsh-Hadamard spectra of members of this series of Fibonacci vectors $V_k$ in relation to Hadamard matrix H=[1 1; 1 -1], that is, in other words, what type of vectors arises in the result of multiplication $V_k$*[1, 1; 1, -1]? The answer is that Walsh-Hadamard spectra of Fibonacci vectors $V_k$ are new Fibonacci vectors $W_p$, which form a new additive series $W_{p+1}=W_p+W_{p-1}$ (Fig. 71). In each of vectors $W_p$ sum of squares of its components is equal to a Fibonacci number with factor 2.

| $W_p$ | $V_0*H=[1, 1]$ | $V_1*H=[2, 0]$ | $V_2*H=[3, 1]$ | $V_3*H=[5, 1]$ | $V_4*H=[8, 2]$ | … |
|---|---|---|---|---|---|---|
| p | 0 | 1 | 2 | 3 | 4 | … |

Figure 71. The additive series f Fibonacci vectors $W_p$, which are Walsh-Hadamard spectra of Fibonacci vectors $V_k$ from Fig. 70.

And what one can say about dyadic derivatives of J.Gibbs [Stankovic, Astola, 2008], which are used in the spectral logic and which are expressed by the following formula:

$$f^{[1]}(x) = -\frac{1}{2}\sum_{r=0}^{n-1}(f(x \oplus 2^r) - f(x))2^r \qquad (40)$$

where $f^{[1]}(x)$ is the dyadic derivative of the first order, $f(x)$ is an analyzed lattice function ($2^n$-dimensional vector). Dyadic derivatives are closely related with Walsh functions [Stankovic, Astola, 2008].

Dyadic derivatives (40) of Fibonacci vectors $V_k$ and $W_p$ give new Fibonacci vectors $D(V_k)$ and $D(W_p)$, which form new additive series of Fibonacci vectors (in the case of $D(V_k)$, Fibonacci vectors have the weight factor 1/2) as Fig. 72 shows.

| | | | | | | | | |
|---|---|---|---|---|---|---|---|---|
| $V_k$ | [1, 0] | [1, 1] | [2, 1] | [3, 2] | [5, 3] | [8, 5] | [13, 8] | ... |
| $D(V_k)$ | [1,-1]/2 | [0, 0]/2 | [1, -1]/2 | [1, -1]/2 | [2, -2]/2 | [3, -3]/2 | [5, -5]/2 | ... |
| $W_p$ | [1, 1] | [2, 0] | [3, 1] | [5, 1] | [8, 2] | [13, 3] | [21, 5] | ... |
| $D(W_p)$ | [0, 0] | [1, -1] | [1, -1] | [2, -2] | [3, -3] | [5, -5] | [8, -8] | .. |

Figure 72. Dyadic derivatives $D(V_k)$ and $D(W_p)$ of vectors $V_k$ and $W_p$ from Fig. 70 and 71.

The wide theme of interrelations between $2^n$-dimensional Fibonacci vectors and their dyadic derivatives needs to be published separately. One should mention that the Fibonacci (2*2)-matrix Q can be generalized into Fibonacci matrices of higher orders, for example, by means of Kronecker multiplications of Q with the matrix [1, 1; 1, 1] [Petoukhov, 2003].

And how Fibonacci vectors are related with the spectral logic of systems of Boolean functions, which is described in [Karpovsky, Stankovic, Astola, 2008; Zalmanzon, 1989]? The spectral logic considers digital devices with input channels $X_0$, $X_1$,…$X_n$ and output channels $Y_0$, $Y_1$, …, $Y_m$. Signals in each of these channels can be only 0 or 1 (Boolean variables). Fig. 73 (left) shows a simple example of such digital device and two examples of a typical tabular representation of systems of Boolean functions $Y_0$, $Y_1$ and $Y_2$. For each of 8 possible combinations of values of input Boolean variables $X_0$, $X_1$, $X_2$, each of output Boolean functions $Y_0$, $Y_1$ and $Y_2$ takes one of the possible values 0 or 1. Such correspondence is written in rows of the table. This system of Boolean functions is characterized by one 8-dimensional vector S, which is called a step-function and each component of which is equal to decimal notation of 3-bit binary number $Y_0Y_1Y_2$ in the same row. Walsh-Hadamard spectra of such step functions of systems of Boolean functions (switching functions) are studied in the spectral logic as key characteristics for analysis and synthesis of digital devices.

The first table (Fig. 73, middle) shows a case of digital devices with the system of only two non-zero Boolean functions $Y_1$ and $Y_2$, step function of which is the Fibonacci vector $S_1$=[3, 2, 0, 0, 0, 0, 0, 0]. Its Walsh-Hadamard spectrum $S_1*[1, 1; 1, -1]^{(3)}$ is equal to the Fibonacci vector [5, 1, 5, 1, 5, 1, 5, 1]. The second table (Fig. 73, right) shows a case of digital devices with the system of three non-zero Boolean functions $Y_0$, $Y_1$ and $Y_2$, the step function of which $S_2$ = [5, 3, 0, 0, 0, 0, 0, 0] is the Fibonacci vector with increased values of its non-zero components. Its Walsh-Hadamard spectrum $S_2*[1, 1; 1, -1]^{(3)}$ is equal to another Fibonacci vector [8, 2, 8, 2, 8, 2, 8, 2].

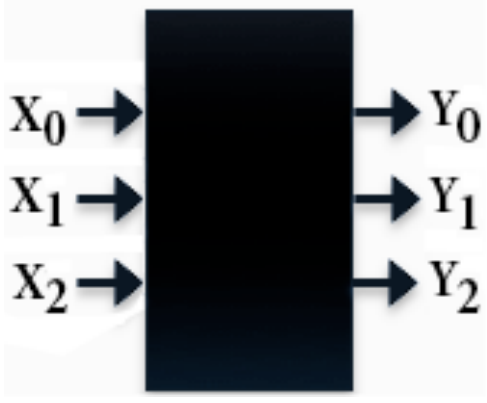

| $X_0$ | $X_1$ | $X_2$ | $Y_0$ | $Y_1$ | $Y_2$ | $S_1$ |
|---|---|---|---|---|---|---|
| 0 | 0 | 0 | 0 | 1 | 1 | **3** |
| 0 | 0 | 1 | 0 | 1 | 0 | **2** |
| 0 | 1 | 0 | 0 | 0 | 0 | **0** |
| 0 | 1 | 1 | 0 | 0 | 0 | **0** |
| 1 | 0 | 0 | 0 | 0 | 0 | **0** |
| 1 | 0 | 1 | 0 | 0 | 0 | **0** |
| 1 | 1 | 0 | 0 | 0 | 0 | **0** |
| 1 | 1 | 1 | 0 | 0 | 0 | **0** |

| $X_0$ | $X_1$ | $X_2$ | $Y_0$ | $Y_1$ | $Y_2$ | $S_2$ |
|---|---|---|---|---|---|---|
| 0 | 0 | 0 | 1 | 0 | 1 | **5** |
| 0 | 0 | 1 | 0 | 1 | 1 | **3** |
| 0 | 1 | 0 | 0 | 0 | 0 | **0** |
| 0 | 1 | 1 | 0 | 0 | 0 | **0** |
| 1 | 0 | 0 | 0 | 0 | 0 | **0** |
| 1 | 0 | 1 | 0 | 0 | 0 | **0** |
| 1 | 1 | 0 | 0 | 0 | 0 | **0** |
| 1 | 1 | 1 | 0 | 0 | 0 | **0** |

Figure 73. Examples of representations of digital devices in the spectral logic.

The Fibonacci step functions $S_2$ is obtained from the Fibonacci step function $S_1$ by means of the transformation $S_2 = S_1*Q_8$, where $Q_8 = ||q_{ij}||$ (here i,j = 1, 2,.., 8) is the Fibonacci (8*8)-matrix with the following entries: $q_{11}=1$, $q_{12}=1$, $q_{21}=1$, all other entries are equal to zero. It is obvious that for a case of a Fibonacci vector with higher values, for example, of the vector with Fibonacci numbers 55 and 34, we need to appeal to the system with more than three Boolean functions $Y_0$, $Y_1$ and $Y_2$ since the decimal value of the maximal 3-bit binary number 111 is equal to 7. Decimal number 55 is equal to 6-bit binary number 110111. Correspondingly we need a system with 6 Boolean functions $Y_0$, $Y_1$, $Y_2$, $Y_3$, $Y_4$ and $Y_5$ to represent number 55 as a component of a step function in tables of the spectral logic.

From the standpoint of the spectral logic, one can model Fibonacci phyllotaxis phenomena on the base of Fibonacci step functions (or their spectra) of genetic systems of Boolean functions. Iterative actions of Fibonacci matrices on Fibonacci vectors lead to increasing non-zero values in new and new Fibonacci vectors. To represent these increasing values in step functions of the spectral logic, systems with more and more quantities of Boolean functions should be organized. Such algorithmic growth of systems of Boolean functions can be interpreted as a part of ontogenesis of phyllotaxis structures in living bodies. We believe that the manifestation of Fibonacci numbers on very different levels and branches of biological evolution is related with the geno-logic coding. Fibonacci matrices, which are sometimes called the matrices of growth, act as matrices of breeding of systems of Boolean functions. Fibonacci systems of Boolean functions define some archetypes of physiologic patterns, including phyllotaxis.

In this approach, morphogenetic and other onto- and phylogenetic phenomena are interpreted as consequences of an algorithmic growth of genetic systems of Boolean functions. In engineering technologies, the spectral logic of systems of Boolean functions has wonderful achievements of analysis and synthesis of digital devices, which have elements of artificial intellect for a detection and a correction of errors; an adaptation to the external environment; a training in a course of work; an interrelation with other digital devices, etc. All these possibilities can now be transferred into the field of study and mathematical modeling of genetically inherited biological phenomena including bio-rhythmic processes and the ability of newborn animals to coordinated movements on the base of inherited motion algorithms without training (the genetic biomechanics). This transfer became possible due to the identification of the structural kinship of molecular-genetic systems with Walsh functions and mathematical formalisms of the spectral logic of systems of Boolean functions. It is this kinship is the foundation of our doctrine about geno-logic coding in living matter, which allows developing spectral-logic genetics by analogy with the theory of digital devices.

**Acknowledgments**. Described researches were made by the author in the frame of a long-term cooperation between Russian and Hungarian Academies of Sciences and in the frames of programs of "International Symmetry Association" (Hungary, http://symmetry.hu/). The author is grateful to Frolov K.V., Ganiev R.F., Darvas G., He M., Igamberdiev A., Kappraff J., Adamson G., Cristea P., Kassandrov V.V., Bakhtiarov K.I., Kulakov Y.I., Pavlov D.G., Petukhova E.S., Smolianinov V.V., Vladimirov Y.S. for their support. Special thanks to my collaborators Stepanyan I.V. and Svirin V.I for their softwares for calculations in the field of matrix genetics.